\documentclass[oneside,11pt,greek,english]{CUEDthesisPSnPDF}

\usepackage{amssymb,amsmath,bm}
\usepackage{setspace}
\usepackage{amsmath,amsfonts,amssymb,graphicx,dsfont}
\usepackage{cite}
\usepackage{amssymb,amsmath}
\usepackage{latexsym}
\usepackage{mathrsfs}
\usepackage{verbatim}
\usepackage{bm}
\usepackage{cite}

\def\bea{\begin{eqnarray}}
\def\eea{\end{eqnarray}}
\def\be{\begin{equation}}
\def\ee{\end{equation}}
\def\ba{\begin{array}}
\def\ea{\end{array}}

\def\nn{\nonumber}

\usepackage{watermark}
\usepackage{epstopdf}
\usepackage{abbrevs}
\usepackage{amsmath}
\usepackage{bm}
\usepackage[small, bf]{caption}
\usepackage{enumerate}
\usepackage{multirow}
\usepackage{relsize}
\usepackage{graphicx}
\usepackage{epsfig}
\graphicspath{{ThesisFigs/EPS/}{ThesisFigs/}}
\usepackage{epigraph}
\usepackage{array}
\usepackage{enumitem}
\usepackage{booktabs}
\usepackage{lettrine}
\usepackage{subfigure}
\usepackage{indentfirst}
\usepackage{setspace}
\DeclareMathAlphabet{\mathpzc}{OT1}{pzc}{m}{it}
\DeclareGraphicsExtensions{.eps, .ps, .png, .jpg, .pdf}
\newabbrev\degrees{$^{\circ}$}

\begin{document}


\maketitle

\setcounter{secnumdepth}{4}
\setcounter{tocdepth}{2}
\setlength{\epigraphwidth}{9.2cm}

\begin{declaration}
	\noindent I, Spyros Sypsas, declare that this thesis, titled ``Theoretical and Observational Constraints on Brane
Inflation and Study of Scalar Perturbations through the Effective Field Theory Formalism'', is the result of work undertaken between October 2009 and September 2013 under the supervision of Prof. Mairi Sakellariadou. I confirm that:

\begin{itemize}
\item This work was done wholly while in candidature for a research degree at this University.
\item Where I have consulted the published work of others, this is always clearly attributed.
\item This thesis contains the author's work published in
\begin{itemize}
\item[\cite{Gwyn:2010rj}] \emph{Cosmic strings from pseudo-anomalous Fayet-Iliopoulos
                        $U(1)_{FI}$ in $D3/D7$ brane inflation}, with Rhiannon Gwyn and Mairi Sakellariadou, {\bf JHEP 1010 075}, e-print \href{http://xxx.lanl.gov/abs/1008.0087}{{\tt arXiv:1008.0087}}, presented in Chapter~\ref{ch:paper-1},

\item[\cite{Gwyn:2011tf}] \emph{Theoretical constraints on brane inflation and cosmic
                        superstring radiation}, with Rhiannon Gwyn and Mairi Sakellariadou, {\bf JHEP 1109 075}, e-print \href{http://xxx.lanl.gov/abs/1105.1784}{{\tt
  arXiv:1105.1784}}, presented in Chapter~\ref{ch:paper-2},

\item[\cite{Gwyn:2012mw}] \emph{Effective field theory of weakly coupled inflationary
                        models}, with Rhiannon Gwyn, Gonzalo Palma and Mairi Sakellariadou, {\bf JCAP 1304 004}, e-print \href{http://xxx.lanl.gov/abs/1210.3020}{{\tt arXiv:1210.3020}}, presented in Chapter~\ref{ch:paper-3} and Appendix~\ref{app:several-massive-fields}.                                               
\end{itemize}
\item Parts of the author's unpublished work are discussed in Subsection~\ref{sec:EFT-unitary} and Appendix~\ref{app:slow-roll}, while ongoing work \cite{paper-4} is discussed in Subsection~\ref{sec:bispec}.
\end{itemize}

\vspace{1 in}

\begin{flushleft}
{\sc King's College London} \phantom{aaaaaaaaaaaaaaaaaaaaaaaaaaaaaaaa} Spyros K. Sypsas \\ \hspace{.9cm}{\sc October 2013} 
\end{flushleft}
\end{declaration}
\begin{acknowledgementslong}

Firstly, I would like to thank my supervisor, Mairi Sakellariadou, for her trust in me,
her guidance and support throughout these years. 
My collaborators Rhiannon Gwyn and Gonzalo Palma, both from whom I also
learned a lot, are gratefully acknowledged. 

I have benefited from discussions with almost all the staff members and students of 
the physics and math departments of King's College London but most of all I would like
to thank them for being my friends. These are Malcolm Fairbairn, Jean Alexandre, Eugene Lim, Bobby Acharya, John Ellis, Nick Mavromatos, Lev Kantorovich, Klaus Suhling, Alessandro De Vita and Sakura Schafer-Nameki -- the ``materialists'' -- Giovanni Peralta, Massimo Riello, Giovanni Doni, Luca Pavan, Joseph Bamidele, Federico Bianchini, Marco Caccin, Zhenwei Li, Gianmarco Zanda, Josep Relat Goberna, Dominic Botten, Lydia Sandiford, Andrea Floris, James Kermode, Alessio Comisso and Cono Di Paola -- the ``theorists" -- Nick Houston, Walter Tarantino, Maria Sueiro, Tevong You, Phillip Grothaus, Robert Hogan, John Heal, Chakrit Pongkitivanichkul, Julio Leite, James Brister, Tom Richardson, Thomas Elghozi and Achilleas Passias -- as well as Paul Le Long and Julia Kilpatrick. I would like to especially thank Nick Houston for his proofreading service and our QFT/gossiping sessions, and Giovanni Peralta for our countless ``Lamb" pints and his ``Goodenough" cooking abilities.

Amihay Hanany and Yang-Hui He have been teachers, friends and collaborators
and I thank them for that. I also wish to thank my Imperial College friends, Rak-Kyeong Seong, Giuseppe Torri and Christiana Pantelidou for making me feel like home at IC.

My Athens University friends, Dimitri Frantzeskakis, Fotis Diakonos, Vassos Achilleos, Lia Katsimiga and Alexandra Tzirkoti have always been a family and a constant source of inspiration. May our paths always braid.

Lastly, I would like to thank my parents, Kostas and Archontoula, and my brother Savvas, for being who they are. This thesis is dedicated to them.

\vspace{2cm}
\flushright{\it SKS}

\end{acknowledgementslong}
\begin{abstractslong}
In this thesis, consisting of two main parts, we study observational signatures of cosmic (super)strings in the context of D-brane inflation and properties of scalar perturbations on generic homogeneous inflating backgrounds. 

In the first part we study the production, nature and decay processes of cosmic superstrings in two widely used effective models of D-brane inflation, namely the $D3/D7$ and $D3/\bar{D}3$ models. Specifically, we show that the strings produced in $D3/D7$ are of local axionic type and we place constraints on the tension while arguing that the supersymmetry breaking mechanism of the model needs to be altered according to supergravity constraints on constant Fayet-Iliopoulos terms. Moreover, we study radiative processes of cosmic superstrings on warped backgrounds. We argue that placing the string formation in a natural context such as $D3/\bar{D}3$ inflation, restricts the forms of possible radiation from these objects.

Motivated by these string models, which inevitably result in the presence of heavy moduli fields during inflation, in the second part, using the Effective Field Theory (EFT) of inflation, we construct operators that capture the effects of massive scalars on the low energy dynamics of inflaton perturbations. We compute the energy scales that define the validity window of the EFT such as the scale where ultra violet (UV) degrees of freedom become operational and the scale where the EFT becomes strongly coupled. We show that the low energy operators related to heavy fields induce a dispersion relation for the light modes admitting two regimes: a linear and a non linear/dispersive one. Assuming that these modes cross the Hubble scale within the dispersive regime, we compute observables related to two- and three-point correlators and show how they are directly connected with the scale of UV physics.

\end{abstractslong}
\cleardoublepage
\phantomsection
\addcontentsline{toc}{chapter}{Contents}
\tableofcontents
\cleardoublepage
\phantomsection
\addcontentsline{toc}{chapter}{List of Figures}
\listoffigures
%
%
%
   \part{Introduction} \label{part:intro}
One of the cornerstones of human curiosity is the imposing question: {\it ``Where did we come from?''}. An answer may be sought via numerous paths and through many disciplines. To the deepest extent and abstraction, this question may be rephrased as: how was the universe created and evolved to what we see today, whereupon physics eventually becomes the main route towards an answer. Following the advance of General Relativity (GR) and Quantum Mechanics in the early 20-th century, people realised that the question of how the universe evolved may be tractable, both theoretically and experimentally. 

The standard cosmological scenario is currently believed to offer the answer. It is a model based on the theory of GR which describes, through Einstein's field equations, how spacetime evolves relatively to its matter content. Matter can be massless or massive, each type following its own evolution and giving its own contribution to the spacetime dynamics. Although technically complete, at least at a classical level, this model fails to answer a set of physical questions which partially consists of the following: why certain correlated areas of the sky when inversely evolved  appear to be causally disconnected, why is the universe so flat, and where are the stable topological relics produced in phase transitions that are supposed to have taken place during the early stages of the universe evolution. These are referred to as the horizon, flatness and unwanted relics problems respectively.

These issues can be cured in a simple way: a period of very fast expansion of spacetime during which the particle horizon increases slowly. That is to say, take a sphere with radius $R$ that stays almost constant while spacetime rapidly expands, so that curvature and relics are smoothed out. The horizon problem is also under control since patches of the observable universe can now be evolved backwards much faster to fit in a causally connected volume in the far past. 

Such an expansion can be modelled by a scalar field coupled to gravity. The idea of such a model was put forward in the early 80's \cite{Guth:1980zm,Linde:1981mu,Linde:1982zj,Linde:1982uu,Albrecht:1982wi} -- see also \cite{Brout:1977ix,Starobinsky:1980te,Kazanas:1980tx,Sato:1980yn} -- and since then it has become an integral part of early universe cosmology. 
Not long after this proposal, it was noted that inflation provides much more than a solution to the aforementioned problems. Mukhanov and Chibisov showed in \cite{Mukhanov:1981xt} -- see also \cite{Hawking:1982cz,Guth:1982ec,Starobinsky:1982ee,Bardeen:1983qw,Mukhanov:1985rz} -- that it also offers an explanation of the inhomogeneities of the universe by providing the quantum seeds of density perturbations which evolved towards the large scale structure that we observe today. These perturbations freeze when they cross the Hubble volume and they are imprinted in the Cosmic Microwave Background (CMB) as temperature fluctuations. Analysing the properties of these fluctuations has been one of the main aims of astrophysical surveys in the recent years, with {\sc Cobe}, {\sc Wmap} and the recent {\sc Planck}, providing revolutionary insights into the physics of the CMB and the early universe.

In parallel with the advances in inflation, there has been an explosion of ideas and theories of quantum gravity. Although progress is constantly made, albeit with an ``oscillatory" profile, none of these ideas have reached the long sought state of a theory of quantum gravity. Nevertheless, if such a theory exists, in the low energy limit it should account for the known particle physics and cosmology we observe. 
Hence, naturally, much effort has been made to study phenomenological aspects of these theories. 
Since inflation describes the evolution of spacetime at high energies, via a field theory approach, one of the most important and well studied, although still open, relevant questions is its embedding within such a unifying framework. 

Let us now set the general context in which our study will take place by briefly reviewing the standard cosmological model. We refer the reader to the classic textbooks \cite{steven2008cosmology,dodelson2003modern,turner1994early} for a concise introduction to cosmology.

\section*{Standard cosmological model} \label{sec:std-cosmo}
The standard cosmological model is based on GR and describes the evolution of a homogeneous and isotropic spacetime, given the fluid that dominates its energy density at each cosmological era, via the Einstein field equations
\be \label{efe}
G_{\mu\nu} \equiv R_{\mu\nu} - \frac{1}{2}R g_{\mu\nu} = \frac{1}{M_{\rm Pl}^2} T_{\mu\nu},
\ee
where $G_{\mu\nu}$ is the Einstein tensor, $R_{\mu\nu}$ and $R$ are the Ricci tensor and Ricci scalar respectively, and $T_{\mu\nu}$ the matter energy momentum tensor. 
One of the simplest homogeneous solution of \eqref{efe} is the Friedman-Lema\^itre-Robertson-Walken (FLRW) metric\footnote{We only write the flat version of the FLRW metric, since this is the one being strongly favoured by observational data -- see $e.g.$ \cite{Ade:2013zuv}.}, which may be parametrised as
\be \label{frw-metric}
ds^2 = -dt^2 + a(t)^2 \left( dr^2 + r^2 d\theta^2 + r^2 \sin^2\theta d\phi^2 \right),
\ee
where $a(t)$ is the \emph{scale factor}. Upon inserting this ansatz into the Einstein equations and specifying the fluid, $i.e.$ fixing the energy momentum tensor on the right-hand side (RHS) of \eqref{efe}, one can determine the scale factor $a$ as a function of time. 

The energy momentum tensor of a homogeneous perfect fluid reads
\be \label{tmn-fluid}
T_{\mu\nu} = {\rm diag}(\rho, p,p,p),
\ee
where $\rho$ is the energy density and $p$ the pressure of the fluid. The two basic equations following form \eqref{efe} are the Friedmann equation (the zero-zero component of the Einstein tensor)
\be \label{fried-eq}
H^2 = \frac{1}{3 M_{\rm Pl}^2} \rho, 
\ee
where the Hubble constant is defined as $H \equiv \dfrac{\dot a}{a}$, and the conservation equation (the Bianchi identity for the Riemann tensor)
\be \label{cons-eq}
\dot \rho + 3 H (\rho + p) = 0.
\ee
Thus, choosing the fluid amounts to specifying the equation of state $p = w \rho$.
The system \eqref{fried-eq},\eqref{cons-eq} is now solved by 
\be \label{mde-rde-sols}
\rho \propto a^{-3(1+w)} \qquad \text{and} \qquad a \propto t^{2/3(1+w)}.
\ee
For example, during an era dominated by ultra relativistic matter (RDE) with $w=1/3$ we have that $\rho \propto a^{-4}$, $a \propto t^{1/2}$, while for a pressureless matter dominated era (MDE) with $w=0$ we obtain $\rho \propto a^{-3}$, $a \propto t^{2/3}$. For a cosmological constant dominated era with $w=-1$, the solution reads 
\be \label{ro+a-infl}
\rho = {\rm const} \qquad \text{and} \qquad a \propto e^{\sqrt{\rho/3M_{\rm Pl}^2}t} = e^{Ht}.
\ee

Let us now quantify the horizon and flatness problems that we briefly mentioned above. Let us first specify the meaning of a \emph{horizon}. From the FRW metric \eqref{frw-metric}, we may calculate the maximum distance a photon can travel during a time interval $\Delta t = t - t_i$ and obtain 
$$
R_{\rm hor}(t) = a(t) \int_0^{ t} \frac{dt}{a(t)} = a(t) \int_0^{R_{\rm max}} dr,
$$
where we have set $t_i=0$. Depending on the cosmological era during which the photon travelled, the value of the horizon slightly changes but since the matter dominated era is the longest one, we may use \eqref{mde-rde-sols} -- with $w=0$ -- to obtain $R_{\rm hor} \sim H^{-1}$. In other words, the horizon length sets the size of the \emph{observable universe}. Therefore, if we observe a photon at time $t=t_{\rm present}$, we may assume that it must have travelled at most a distance $H_0^{-1}$, that is, the present value of the Hubble constant. Nevertheless, CMB observations suggest otherwise! 

{\sc Cobe} satellite was the first to confirm \cite{Smoot:1992td} that patches of the sky are correlated, $i.e.$ they have similar temperatures to a very high accuracy of five decimal points. From a microphysical perspective, this means that such regions should be causally connected in the far past, when the microphysical process responsible for these temperature anisotropies took place. We thus need to ensure that any present physical length scale $\lambda$ was less than the horizon scale $R_{\rm hor} \sim H^{-1}$ at the time of Last Scattering (LS), when the universe became transparent to radiation, and information, $i.e.$ light, started propagating freely, reaching our satellites and telescopes until the present time.

A physical length scale $\lambda$ evolves proportional to the scale factor, so at $t=t_{\rm LS}$ the largest physical scale that may exist today, the present horizon, had a value of $\lambda_H(t_{\rm LS}) = \dfrac{a_{\rm LS}}{a_0} R_{\rm hor}(t_0)$. During an MDE, the Hubble length evolved with a different law as $H^2 \sim \rho \sim a^{-3}$, leading to $H_{\rm LS}^{-1} =  \left( \dfrac{a_{\rm LS}}{a_0} \right)^{3/2} R_{\rm hor}(t_0)$. Comparing the physical to the observable universe volume at the time of last scattering, we find the unexpected result 
\be 
\dfrac{\lambda^3_H(t_{\rm LS})}{H_{\rm LS}^{-3}} \sim 10^6.
\ee
This asserts that at that time, there were about a million causally disconnected regions inside the volume that evolved to the present observable universe. In other words, the correlation of different parts of the sky that we observe lacks a microphysical explanation. This is known as the horizon problem.

In addition, observations suggest that our universe is very close to being flat. From the Friedmann equation we have that $\Omega - 1 = \mathpzc{k}/a^2H^2$, where $\Omega = \rho/\rho_{\rm cr}$, with the critical density defined as $\rho_{\rm cr} \equiv 3H^2M_{\rm Pl}^2$. $\mathpzc{k}$ sets the curvature of spacetime, with $\mathpzc{k} = 0$ denoting the flat case, so that flatness may be stated as $\Omega \sim 1$. Since $\Omega - 1$ evolves as $a^{2}$ during an RDE, we may compare its present value to the value of the same quantity at the Planck time, to find 
\be \label{fl-prob}
\dfrac{|\Omega-1|_{t_{\rm Pl}}}{|\Omega-1|_{t_0}} = \dfrac{a^2_{\rm Pl}}{a^2_0} \sim 10^{-60}.
\ee
This result implies that during the very early universe the matter density was extremely close to the critical density but not exactly the same. This is known as the flatness problem.

Both of these questions are addressed by inflation, a period of accelerated expansion. For the horizon problem we require that a physical length evolves faster than the Hubble scale, $i.e.$ $\frac{d}{dt}\left( \frac{a}{H^{-1}} \right) = \ddot a > 0 $. From the Friedmann equation in combination with the conservation equation we obtain
\be
\frac{\ddot a}{a} = - \frac{\rho}{6 M_{\rm Pl}^2} (1 + 3w),
\ee
so the acceleration condition $\ddot a>0$ leads to $w<-1/3$. 

For example, the case \eqref{ro+a-infl} with $w=-1$ and $a=e^{Ht}$, satisfies this constraint. Evading the horizon problem amounts to imposing
\be
\lambda_{H_0}(t_i) = H_0^{-1} \frac{a_f}{a_0} e^{-N} < H_i^{-1},
\ee
which is satisfied by $N \sim 70$, with $N = H \delta t$ the number of \emph{e-folds}, measuring how much the scale factor grew during a time interval $\delta t = t_f - t_i$. This choice also considerably ameliorates the flatness problem, which is an issue of fine tuning. The ratio \eqref{fl-prob} reads
\be \label{fl-prob-infl}
\dfrac{|\Omega-1|_{t_{f}}}{|\Omega-1|_{t_0}} = \dfrac{a^2_f}{a^2_0} = e^{-2N},
\ee
so by adjusting $N \sim 70$ we obtain the required value of $10^{-60}$. This implies that whatever the value of the matter density was before inflation, at the end of the process it will be almost identical to the critical one. 

However, the fact that we want this process to end at some point prevents us from using the cosmological constant as a source of inflation. Alternatively, the case \eqref{ro+a-infl} can be modelled by a homogeneous scalar field coupled to gravity
\be \label{infl-action} 
S = \int \sqrt{-g}dx^3 dt \left( \frac{M_{\rm Pl}^2}{2} R - \frac{1}{2}g^{\mu\nu}\partial_\mu \phi \partial_\nu \phi - V(\phi) \right),
\ee 
where $\phi=\phi(t)$, whose energy density is dominated by the potential term. The potential, which may depend on other fields, is chosen such that it provides a dynamical exit from inflation, by driving the inflaton field to zero after a certain period of time. The equation of motion following from the above Lagrangian about an FLRW background is the Klein-Gordon equation
\be \label{kg}
\ddot \phi + 3 H \dot \phi + V'(\phi) = 0,
\ee
where $V'\equiv \frac{dV}{d\phi}.$
The requirement of the kinetic energy being negligible compared to the potential (slow roll) and the flatness of the potential lead to the conditions
\be \label{e_v}
\epsilon_V = \frac{M_{\rm Pl}^2}{2} \left( \frac{V'}{V} \right)^2 \ll 1 \qquad \& \qquad \eta_V = M_{\rm Pl}^2  \frac{V''}{V} \ll 1.
\ee
Upon using \eqref{kg}, these parameters can be related to the \emph{slow roll parameters}
\be
\epsilon = \frac{|\dot H|}{H^2} \qquad \& \qquad \eta = \frac{\dot \epsilon}{\epsilon H},
\ee
that control the behaviour of the Hubble scale, as
\be \label{e_h}
\epsilon \simeq \epsilon_V \qquad \& \qquad \eta \simeq \eta_V - \epsilon_V.
\ee
From the definition of the Hubble constant we have that $\dfrac{\ddot a}{a} = H^2 + \dot H = H^2(1-\epsilon)$, which combined with the acceleration condition implies $\epsilon < 1$, in accordance with \eqref{e_v} and \eqref{e_h}. Therefore, $\epsilon =1$ signals the end of the inflationary period.

As already mentioned, one of the striking features of inflation, apart from successfully solving the shortcomings of the standard Big Bang model, is that it also provides the seed for the large scale structure as well as the CMB temperature anisotropies we see today. The quantum nature of inflation allows us to consider small deviations of the inflaton from its homogeneous vacuum expectation value (vev), that drives the expansion of spacetime, $i.e.$
\be \label{inflaton-intro}
\phi(t,x) = \phi_0(t) + \delta\phi(t,x).
\ee
Since the inflaton field dominates the matter content of the universe during inflation, these deviations represent small inhomogeneities in the energy density, which are transmitted to inhomogeneities of spacetime itself, finally manifesting themselves as temperature anisotropies in the CMB. The dynamics of these scalar quantum fluctuations will be the content of Part~\ref{part:eft} but let us here briefly discuss the predictions of slow roll inflation regarding their distribution in the CMB. 

A generic prediction of inflation is that these perturbations are \emph{Gaussian}, $i.e.$ they have vanishing odd correlators, hence obeying Gaussian statistics. An intuitive way to understand this is the following: the slow roll conditions \eqref{e_v} imply that the potential is almost flat, so that one may linearise the Klein-Gordon equation \eqref{kg} that governs the dynamics of the inflaton field and its fluctuations\footnote{Since we are dealing with a coupled system, that is, a scalar field on a gravitational background, one should bear in mind that in addition to the fluctuations of the inflaton field, there are also scalar perturbations of the metric itself. As we will see in Ch.~\ref{ch:prelims2}, if the inflaton fluctuations are linear and slow roll conditions are met, the metric perturbations are also linear, so the ``Gaussianity argument'' presented here still holds.} \eqref{inflaton-intro}. Going one step backwards, one realises that a linear equation of motion follows from a quadratic action. Now, correlation functions are encoded in the \emph{partition function}, which very loosely speaking is the exponential of the action, in such a way that the $n$-th order term in the expansion yields the $n$-point correlator. Therefore, if the action is quadratic, only even terms will appear in the expansion. To sketch it: $e^{x^2} = \sum x^{2n}/n!$.

As shown in the seminal paper of Maldacena \cite{Maldacena:2002vr}, any odd correlator of fluctuations in slow roll canonical\footnote{By canonical we mean an inflationary model with canonical kinetic terms $(\partial\phi)^2$.} inflation is proportional to the slow roll parameters, so that in the limit where these are small, the theory is indeed Gaussian\footnote{See however \cite{Chen:2006xjb,Chen:2008wn} for a way to generate non Gaussian signals in canonical slow roll modes by considering \emph{features} in the potential.}. In Ch.~\ref{ch:prelims2} and Ch.~\ref{ch:paper-3}, we will explore classes of (non canonical) slow roll inflationary models in which the fluctuations may significantly depart form Gaussian statistics. Such a deviation, if ever confirmed by observations, will hint upon deviations from canonical slow roll models \cite{Komatsu:2009kd}, which is an exciting prospect.

\begin{figure}[tbh] 
\begin{center}
\includegraphics[scale=0.8]{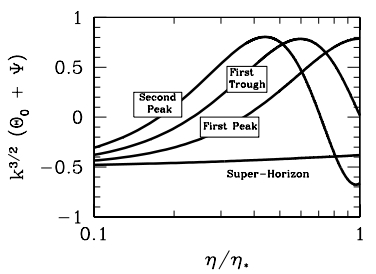}   
\end{center}
\caption[{\sf The CMB temperature fluctuation versus conformal time.}]{\sf The temperature fluctuation versus conformal time, with $\eta_*$ the time of recombination. A mode with small wavenumber, $i.e.$ large wavelength, lies outside the observable universe at the time of recombination, hence it is constant (super-horizon). The ``first peak" mode is one whose amplitude is at a maximum at $\eta_*$, leading to the first peak in the CMB temperature power spectrum. Another mode with slightly larger wavenumber has entered the horizon earlier and is thus captured at a later stage in the oscillation, when its amplitude is around zero (``first trough"). Finally, an even shorter mode is captured at its minimum amplitude (``second peak"). Figure taken from \cite{Dodelson:2003ip}.}
\label{fig:dode-1}
\end{figure}

\begin{figure}[tbh] 
\begin{center}
\includegraphics[scale=2.5]{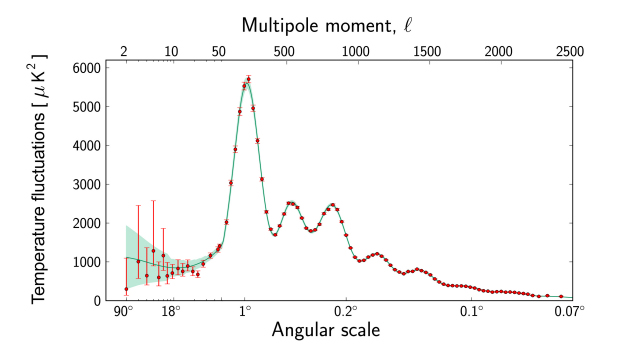}   
\end{center}
\caption[{\sf The CMB temperature power spectrum measured by {\sc Planck} mission.}]{\sf The temperature power spectrum measured by {\sc Planck} mission. Figure taken from \url{http://www.sciops.esa.int/index.php?project=planck&page=Planck_Legacy_Archive}.}
\label{fig:temperature-power}
\end{figure}

Another general feature of scalar fluctuations during inflation is that the they become time independent once their wavelength reaches the Hubble scale, a property of great importance, since it ``decouples'' the unknown physics that governs their dynamics outside the observable universe. After inflation ends, the Hubble scale starts growing during a radiation and then a matter dominated era, so that the frozen perturbations re-enter the observable universe at some point, when they start oscillating, driven by two competing forces: gravitational attraction and photon pressure. In overdense regions attraction wins and gravitational collapse leads to structure formation, while in underdense regions pressure forbids such a process. At the time of recombination, when photons start free-streaming throughout the universe, each wavenumber is captured at a specific phase of its oscillation (see Fig.~\ref{fig:dode-1}), and this produces the acoustic peaks and troughs in the temperature power spectrum, shown in Fig.~\ref{fig:temperature-power}.

\begin{figure}[tbh] 
\begin{center}
\includegraphics[scale=.8]{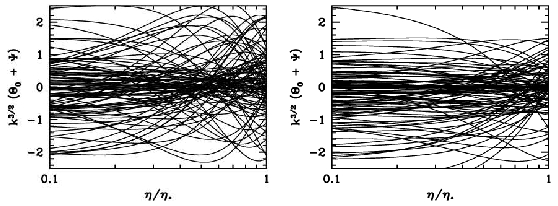}   
\end{center}
\caption[{\sf Noise-like CMB temperature fluctuations}.]{\sf Without the coherence of the oscillations implied by the inflationary mechanism, the expected temperature fluctuations should look like white noise, $i.e.$ without clear peak-trough structure. Axis as in Fig.~\ref{fig:dode-1}. Figure taken from \cite{Dodelson:2003ip}.}
\label{fig:dode-3}
\end{figure}

However, it is important to recall that there are infinite modes with similar wavenumbers, and since all of them are excited during inflation, the resulting power spectrum should lack clear peaks and troughs, resembling the power spectrum of white noise, as in Fig.~\ref{fig:dode-3}. Nevertheless, Fig.~\ref{fig:temperature-power}, which is an actual observation \cite{Planck:2013kta}, suggests that all modes with a given wavenumber are \emph{in phase} at the time of recombination. The fact that during inflation fluctuations freeze at super Hubble scales, offers a beautiful explanation of this result: since the modes start oscillating when they re-enter the observable universe, similar wavelenghts are in phase \cite{Dodelson:2003ip}. In other words, decomposing a mode $\zeta_{k_0}$ at horizon re-entrance time $\tau_0 = 0$, in sines and cosines, $\dot \zeta_{k_0} = 0$ implies that inflation excites only cosines! Therefore, the peak structure of the temperature power spectrum strongly supports the inflation idea.
%

Between the lines of the above discussion, lies an important implication, which provides a powerful tool for the study of inflationary perturbations. Amongst others, there are two distinct fundamental energy scales: one associated with the \emph{background} model which captures the microphysics of inflation and another associated with the dynamics of the \emph{perturbations} about this background, which freeze when they cross the Hubble scale, being imprinted in the CMB. The former is supposed to be around the Grand Unification (GUT) scale, $\Lambda_{\rm GUT} \sim 10^{16}$ GeV, whilst the latter, which is set by the Hubble scale, is significantly lower as can be deduced from \eqref{fried-eq} and the slow roll conditions. At first sight, given our poor understanding of ultra high energy physics and the lack of experimental probes, a description of the inflationary process at all scales would seem intractable. 
%
Fortunately, in order to describe a natural process such as inflation, one only needs to focus on a relatively small window of length scales which is set by the system under consideration itself.
 
Effective Field Theory (EFT) constitutes a powerful scheme that allows for an appropriate description of an important process in a physical system at a certain characteristic scale. Appropriate, because it focuses on the correct degrees of freedom that govern the dynamics of the system and important, for it isolates the most relevant processes at that energy while hiding any complicated, ``irrelevant'' details. The rough idea is that if there are quantities that are too small or too large compared to the characteristic scale of a system, we can set them equal to zero and infinity respectively and still get a correct estimate of the actual physics. Depending on one's desired accuracy, one can then include corrections to the rough estimate as powers of small ratios of the ignored scales and the characteristic length. 

An important concept that arises from this argument is the \emph{validity window} of an EFT which leads to the notion of \emph{ultra violet} (UV) \emph{completion}. The term validity window refers to the fact that an EFT is designed to address problems exhibiting a specific characteristic scale and processes of energies above this scale cannot be described by it. In other words, at energies higher than the characteristic scale, the corrections mentioned previously become of the same order as, or even more important than, the rough estimation. This fact manifests itself in many ways, which may be summarised via the following statement: {\it when an EFT begins to exhibit any kind of unphysical behaviour as the energy approaches a specific value, it needs to be UV completed.} The completion is another theory which ``takes over" control of the system when we want to consider processes at higher energies than this specific value, and this is what the prefix ``UV" implies. It is designed in such a way that it flows to the original EFT once the energy is lowered but the inverse is non applicable. 
Since the completion may be (and most of the times it actually is) an effective theory itself, a more appropriate term is an \emph{intermediate} UV completion, with the ``bare" term reserved for the ultimate theory that can describe physics at any scale. In the 70's, Wilson and Kogut \cite{Wilson:1973jj} revolutionised the way we view field theory by putting forward the idea that any theory should be regarded as an effective one, with all the unphysical behaviour, such as $e.g.$ divergencies of observable quantities, being an indication of the theory hitting its UV scale, $i.e.$ the upper bound of its validity window. 

Let us exemplify the above notions with a simple well known EFT, the Fermi theory of weak interactions. In the 30's, Fermi attempted to describe ${\mathrm \beta}$ decay by considering a contact four fermion interaction between a neutron, a proton, an electron and its antineutrino. This theory is quite accurate up to around 100 GeV, whilst at higher energies it deviates substantially from experimental results. As we now know, the correct theory of $\beta$ decay is the theory of weak interactions mediated by the W and Z bosons. These mediators, being massive, insert a scale into the problem so that for energies well below the W mass an effective description of neutron decay may apply. This is exactly the Fermi theory, which essentially treats the W boson as infinitely massive and removes it from the dynamics. The weak theory is an intermediate completion of the Fermi theory, which may itself be embedded into the electroweak theory, which can be further extrapolated upwards to a grand unified theory and so on and so forth, until a genuine UV complete theory is reached.

Since we will use effective field theory techniques to study both inflation in a UV complete framework and  perturbations about generic inflating backgrounds, let us close these introductory remarks and somehow ``formalise" these claims by reviewing the general principles of effective field theory. The reader is referred to \cite{Georgi:1994qn,Pich:1998xt,Polchinski:1992ed,Manohar:1996cq,Burgess:2007pt} for extensive reviews on EFT. 



\section*{Principles of Effective Field Theory} \label{sec:eff-field-theory}
An effective field theory is characterised by a Lagrangian which is a polynomial over operators $\mathcal{O}$ that respect the symmetry of the theory,
\be
L = \int dx^4 \sum_i c_i \mathcal{O}_i,
\ee
where $c_i$ are dimensionful couplings and spacetime dimension four is assumed. An operator is characterised by its \emph{scaling dimension} $d_i$ which governs its behaviour as the energy decreases\footnote{We assume a relativistic theory, where space and time both scale inversely proportional to energy. In a non relativistic situation, as the one we will encounter in Ch.~\ref{ch:paper-3}, the scaling dimension may differ from the mass dimension of an operator \cite{Polchinski:1992ed}.}, 
$$
E \to \lambda E \Longrightarrow \mathcal{O}_i \to \lambda^{d_i} \mathcal{O}_i \Longrightarrow \int dx^4 \mathcal{O}_i \to \lambda^{d_i - 4} \int dx^4 \mathcal{O}_i.
$$
Operators with $d >4$ are \emph{irrelevant} since they become more and more suppressed as the energy scales downwards, as opposed to \emph{relevant} operators, with $d<4$, whose presence becomes important in the same limit. Operators with $d=4$ are \emph{marginal}. The construction of this effective Lagrangian follows from a set of generic rules:
\begin{itemize}
\item[1.] The dynamics of a system at low energies (equivalently at large length scales) do not depend on high energy physics ($i.e.$ at small distances).
\item[2.] Upon identifying the set of characteristic scales $\Lambda_\star$ of a physical process, one replaces by zero and infinity the small $\Lambda_{\rm IR}$ and large $\Lambda_{\rm UV}$ remaining scales of the full problem, respectively. Upon relaxing this condition, finite corrections can then be incorporated as perturbations of the form $\left( \dfrac{\Lambda_\star}{\Lambda_{\rm UV}} \right)^n$.
\item[3.] The EFT describes the low energy physics to a given accuracy $\varepsilon$ in terms of a finite set of operators:
\be \label{eft-accuracy}
\left( \dfrac{\Lambda_\star}{\Lambda_{\rm UV}} \right)^{d_i-4} \gtrsim \varepsilon \quad \longleftrightarrow \quad d_i \lesssim 4 + \dfrac{\ln (\varepsilon)}{\ln \left( \dfrac{\Lambda_\star}{\Lambda_{\rm UV}} \right)}.
\ee
\item[4.] The EFT has the same IR (but different UV) behaviour as the full theory.
\item[5.] The only manifestation of UV physics lies in the low energy dimensionful couplings $c_i$ and the symmetries of the EFT.
\end{itemize}

These rules form the minimal set of axioms that prescribe the construction of an EFT describing physical processes of a system around a scale $\Lambda_\star$. However, such a construction is far from predictive in the sense that every result will depend on a set of unknown parameters. The third rule ameliorates this unpredictability by constraining this set to be finite; once the given accuracy $\varepsilon$ is decided, the number of unknown parameters $c_i$ is automatically fixed from \eqref{eft-accuracy}, since there exists only a finite number of operators of a given scaling dimension $d$. Thus, this set of \emph{principal} rules needs to be supplemented by an \emph{empirical} or a \emph{fundamental} input. 

The first term, empirical input, refers to a \emph{bottom-up} approach, where the set of $n$ dimensionful couplings is determined by $n$ experiments designed to probe these quantities. Then an $(n + 1)$-th measurement is a consistency check for the theory after which the EFT can be trusted, leading to actual predictions. If one wishes to further complete this EFT, one may search for theories that correctly reproduce these measurements at low energies.

What is meant by the second term, a fundamental input, is a different situation where one has full knowledge of the complete theory, yet the theory cannot be solved at low energies, $e.g.$ due to strong coupling effects like confinement in QCD. In that \emph{top-down} case, the EFT becomes a tool which allows for simplification of the technical aspects of the problem. Assuming full knowledge of the complete theory, the EFT contains no unknown couplings, and its predictability is guaranteed due to this input of UV information.

The action \eqref{infl-action} represents an effective theory in many aspects. First of all, GR itself is an effective theory and may be supplemented by higher dimensional operators, as we will see in Part~\ref{part:eft}. Furthermore, the matter Lagrangian contains free couplings and may itself be supplemented by $e.g.$ higher dimensional kinetic terms. 
The challenge for a UV complete theory is to naturally and dynamically provide a degree of freedom, or more generally a set of degrees of freedom that realise inflation for a certain finite period of the early universe evolution. In Part~\ref{part:strings}, we will study a class of such UV motivated models.

\section*{Scope \& structure of the thesis}
This thesis studies two classes of potential observational signatures of inflation: cosmic superstrings (Part~\ref{part:strings}) and non Gaussianities (Part~\ref{part:eft}). The first is related to background dynamics, while the second stems from perturbations along the inflationary background. Both these subjects are studied in an effective field theory framework. For the cosmic superstrings, we place ourselves in a known UV framework, which is type IIB string theory, and work within two effective four dimensional models, namely the $D3/D7$ and $D3/\bar{D}3$. This is a top-down EFT, since the four dimensional models exhibit free parameters descending from the various ways that dimensional reduction may be implemented. This freedom constitutes our poor understanding of string theory and in order to obtain a predictive effective theory, an empirical input is required. Alternatively, non Gaussian signatures of a large class of models are studied using a generic bottom-up EFT with unknown coefficients, which are directly related to observable quantities.

Theoretical understanding of UV models and their signals is crucial since it may provide insight into the complete theory under consideration, given that such signals are indeed detected by future experiments. For example, detection of cosmic superstrings and analysis of their properties could hint upon their UV origin, while detection of non Gaussianities may shed light to both the inflationary dynamics and the process of structure formation in the universe. In order for such information to be used accurately, it is important to know the exact relation between observable quantities and parameters of the theory.

Part~\ref{part:strings} is devoted to the study of observational signatures and theoretical constraints on brane inflationary models. 
More specifically, we find that observational bounds on the cosmic superstring tension that may form at the end of D-brane inflation, constrain unknown parameters of the models. In addition, a careful analysis of the supersymmetry breaking mechanism in combination with the compactification method, places theoretical constraints on both the inflationary process of each model and the potential signatures related to the decay of cosmic superstrings. 
%

In more detail, Ch.~\ref{sec:susy-primer} contains a short review of basic notions of supersymmetry, supergravity and string theory, which are relevant for this thesis. In Ch.~\ref{ch:paper-1}, based on the author's work \cite{Gwyn:2010rj}, we study the cosmic superstrings that are formed at the end of $D3/D7$ inflation. We argue that they are of local axionic type and we use observational bounds on the string tension to constrain the volume of the compact six dimensional manifold. We then argue that relatively recent developments in supersymmetry breaking mechanisms in string theory, imply that models of this kind are inconsistent with moduli stabilisation.

In Ch.~\ref{ch:paper-2}, based on the author's work \cite{Gwyn:2011tf}, we study the decay channels of superstrings that form at the end of brane-antibrane inflation in warped backgrounds in the context of the $D3/\bar{D}3$ model. After reviewing the construction of a bound state of fundamental strings and one dimensional branes, we argue that consistency of the compactification on a warped background forbids some of the radiative processes previously considered in the literature, while an estimation of the power of allowed radiation is not straightforward.

In Part~\ref{part:eft}, we change our approach to an EFT designed to capture the physics of scalar perturbations generated during inflation. We argue that in view of string inflationary models, a natural and generic assumption is that the inflaton may interact with heavy scalar fields of the UV theory. We then identify a class of operators in the low energy EFT of scalar perturbations, that capture the presence of these interactions in the UV. Finally, the effects of these operators on the low energy observables such as the two-point and three-point correlators are studied. We find that although these operators do not have dramatic effects on the shapes of three-point correlators of scalar perturbations, the observational bounds on these quantities are directly translated into bounds on the scale where UV degrees of freedom become dynamical.


In more detail, in Ch.~\ref{ch:prelims2} we review the relevant notions of cosmological perturbation theory and the construction of an EFT for the perturbations \cite{Creminelli:2006xe,Cheung:2007st} making several connections with gauge theory results about Goldstone bosons and spontaneous symmetry breaking. Subsection~\ref{sec:EFT-unitary} contains parts of the author's unpublished work. 

In Ch.~\ref{ch:paper-3}, based on the author's work \cite{Gwyn:2012mw}, we identify and study a certain class of operators modelling the presence of massive fields at high energies that may affect the inflationary dynamics. After computing the relevant scales of the problem, including the window of validity of the effective theory and the scale where the dispersion relation of the theory changes from linear to non linear, we calculate the effects of these operators on the power spectrum and the bispectrum of the scalar fluctuations. We find that observables related to the two-point and three-point correlators of the theory are directly related to the energy scale of the UV theory. 
Sec.~\ref{sec:bispec}, where the bispectra of the effective theory are computed, contains ongoing work \cite{paper-4}.

Both Parts, \ref{part:strings} and \ref{part:eft}, contain a partial introduction and a summary.
We finally present our collective conclusions and future directions in Part~\ref{part:conc}. App.~\ref{app:slow-roll} contains parts of the author's unpublished work, where known higher order slow roll corrections of the two-point action for the curvature perturbation are reproduced within the EFT formalism. App.~\ref{app:several-massive-fields} is relevant to Ch.~\ref{ch:paper-3} and contains additional calculations that have been suppressed throughout the main text. 

   \part{Top-down EFT of the background: inflation in string theory} \label{part:strings}

String theory is a unifying theoretical framework for the description of all elementary forces. Initially proposed in the 70's as a theory of strong interactions, it was soon abandoned due to the success of QCD\footnote{See \cite{Schwarz:2007yc,Mukhi:2011zz} for a nice historical account of the development of string theory.}. Its subsequent development took another turn however, when it was realised that it also contained gravitational degrees of freedom, and until today it is considered one of the most promising candidates for a theory of quantum gravity. As mentioned in the Introduction, a desirable feature of any UV complete theory is its ability to realise viable inflationary models, so there has been much effort to embed the inflationary process in a string framework.

In this part we will study topological defects, arising from the phase transition signalling the end of inflation, in the context of brane inflation. The main focus will be the formation, properties and decay mechanisms thereof, which may provide a unique observational window into the physics of very high energies. Since an integral part of string theory is supersymmetry (SUSY), we begin with a short review of supersymmetric gauge theories including SUSY breaking mechanisms. We then outline the basic concepts of string theory that will be used in what follows, including D-branes and moduli fields, closing with a discussion on cosmic strings and inflation in such a context. In Ch.~\ref{ch:paper-1} we focus on the formation of cosmic strings in the $D3/D7$ model and the study of inflation in combination with recent developments in SUSY breaking, whilst in Ch.~\ref{ch:paper-2} we study the radiative processes of such structures in the context of one of the best understood compactified examples of string theory. We close with our conclusions on these directions.

\chapter{Supersymmetry and string theory primer} \label{sec:susy-primer}

%
%
The effectiveness of the standard model in describing particle physics at energies up to (currently) a few TeV is unambiguous. However, there are many theoretical inconsistencies like for example the problem of large hierarchies. One such hierarchy is associated with the mass of the recently discovered Higgs boson. Calculating loop corrections, one finds that it has a power law dependence on the cutoff scale so one would expect the Higgs mass to be much higher than its measured value of 125 GeV. Hence, there must be some mechanism that cancels these quadratic divergencies. Supersymmetry is a symmetry that relates bosons to fermions and among other things provides a solution to the Higgs hierarchy problem. An intuitive way to see that is the following: since bosons and fermions contribute to loop corrections with different signs, one might hope that if a theory has the same number of fields of each kind and appropriate couplings, divergencies like the one mentioned might cancel, leaving a finite measurable quantity. Another intriguing feature of supersymmetry is that it allows for a unification of the three fundamental forces at the GUT scale, lying around $10^{16}$ GeV. 

Theoretically, supersymmetry is the only mixed internal-spacetime symmetry which can lead to  consistent particle physics theories. In \cite{Coleman:1967ad}, Coleman and Mandula showed that upon certain assumptions on the S-matrix, the only way that Poincare symmetry and an internal symmetry can coexist is the trivial, that is a direct product of the two, with no mixing of the generators. A few years later, the Poincare group was extended to include anticommuting generators \cite{Golfand:1971iw} and it was soon demonstrated \cite{Haag:1974qh} that there is a unique way to mix spacetime and internal symmetries in a non trivial way that evades the Coleman-Mandula no-go theorem: supersymmetry. 

Such an algebra may be written as
\bea \label{Q,Qdag}
\{Q_\alpha, Q^\dag_{\dot\alpha} \} &=& 2 \sigma^\mu_{\alpha \dot\alpha} P_\mu, \qquad \{Q_\alpha, Q_{\beta} \} =\{Q^\dag_{\dot \alpha}, Q^\dag_{\dot \beta} \} = 0, \\ && [ Q_\alpha,P_\mu ] = [Q^\dag_{\dot \alpha} ,P_\mu] = 0 \label{Q,P},
\eea
where $\alpha,\dot\alpha$ are spinor indices running from 1 to 2 and $\sigma^\mu_{\alpha \dot{\alpha}}=(1,\sigma^i), \; \bar{\sigma}^{\mu \alpha \dot{\alpha}}=(1,-\sigma^i)$, with $\sigma^i$ the Pauli matrices.
The commutator of the supercharges $Q, Q^\dag$ with the Poincare generators can be also computed as
\be \label{Q,Poincare}
[Q_\alpha, M^{\mu\nu}] = (\sigma^{\mu\nu})_\alpha^\beta Q_\beta, \qquad [Q_\alpha, P^\mu] = 0,
\ee
with $(\sigma^{\mu\nu})_\alpha^\beta = \frac{i}{4}(\sigma^\mu \bar{\sigma}^\nu - \sigma^\nu \bar{\sigma}^\mu )_\alpha^\beta, \;\; (\bar{\sigma}^{\mu\nu})_{\dot{\alpha}}^{\dot{\beta}} = \frac{i}{4}(\bar{\sigma}^\mu \sigma^\nu - \bar{\sigma}^\nu \sigma^\mu )_{\dot{\alpha}}^{\dot{\beta}}$.

From this algebra one can derive some crucial observations. Firstly, from \eqref{Q,Qdag} we may write the Hamiltonian $P^0$, as $H = \frac{1}{4} \sum \{Q_\alpha,Q^\dag_\alpha\}$. Now if the vacuum of a theory is supersymmetric it should be annihilated by the supersymmetry generators $Q_\alpha |0\rangle = 0$, implying that the vacuum energy of a supersymmetric theory must vanish, $\langle0|H|0\rangle = 0$. Indeed, as we will see in Sec.~\ref{sec:susy-break}, a non zero vacuum energy is a way to break supersymmetry. Secondly, from \eqref{Q,P} we see that the Hamiltonian commutes with the SUSY generators and in addition, from \eqref{Q,Poincare} it follows that the SUSY generators act on the states as lowering/raising operators of spin quantum numbers. These two statements suggest that supersymmetric particle states may be organised into \emph{supermultiplets}, which are collections of fermionic and bosonic fields that may be obtained by acting with the raising operators on a vacuum state of a given helicity or spin. In the next Section, we will consider the simplest SUSY theory and then gradually generalise to more complicated models, representing these supermultiplets as single objects that generalise ordinary fields. This formulation will be of use in Sec.~\ref{sec:p1-eff-4d-th}. For extensive reviews on supersymmetry the reader is referred to \cite{Drees:1996ca,Martin:1997ns,Lykken:1996xt,Sohnius:1985qm}, as well as the textbooks \cite{wess1992supersymmetry,terning2006modern}.

\section{Supersymmetric Lagrangians} \label{sec:susy-lag}
Let us start with a simple Lagrangian exhibiting supersymmetry:
\be \label{susyL-1}
\mathcal{L} = \partial_\mu \phi \partial^\mu \phi^* + i \psi^\dag_{\dot \alpha} \bar{\sigma}^{\mu \alpha \dot{\alpha}} \partial_\mu \psi_\alpha.
\ee
In order to obtain supersymmetry the scalar should map to the fermion and the fermion to the scalar. We thus need a transformation parameter that carries a spinor index, $i.e.$ an anticommuting Grassmann variable. The fermion  should have a transformation law that contains a derivative in order to match with the bosonic kinetic term, which has one derivative more. Since the derivative carries a spacetime index, we need to contract it with an object that carries analogous structure. 

Let us therefore consider the following transformation laws 
\be \label{boson-var}
\delta\phi = \epsilon^\alpha \psi_\alpha, \qquad \delta\phi^* = \epsilon^{\dag}_{\dot \alpha} \psi^{\dag \dot\alpha} ,
\ee
for the scalar boson and
\be \label{fermion-var}
\delta\psi_\alpha = -i(\sigma^\nu \epsilon^{\dag})_\alpha \partial_\nu \phi, \qquad \delta\psi_{\dot\alpha}^{\dag} = i(\epsilon \sigma^\nu)_{\dot\alpha} \partial_\nu \phi^* ,
\ee
for the fermion. After some algebra, using the Pauli identities
\be 
[\sigma^\mu \bar{\sigma}^\nu + \sigma^\nu \bar{\sigma}^\mu]_\alpha^\beta = 2 \eta^{\mu\nu}\delta^\beta_\alpha \qquad \text{and} \qquad [\bar{\sigma}^\mu \sigma^\nu + \bar{\sigma}^\nu \sigma^\mu]_{\dot\alpha}^{\dot\beta} = 2 \eta^{\mu\nu}\delta^{\dot\beta}_{\dot\alpha} ,
\ee
one can show that the variation of the Lagrangian can be written as a total derivative 
\be 
\delta\mathcal{L} = \partial_\mu ( \epsilon \sigma^\mu \bar{\sigma}^\nu \psi \partial_\nu \phi^* - \epsilon \psi \partial^\mu \phi^* + \epsilon^\dag \psi^\dag \partial^\mu \phi ), 
\ee
rendering the action supersymmetric. 

In order to check that we have correctly implemented the SUSY transformations we need to verify that the commutator of two SUSY transformations yields another transformation or in other words that the SUSY algebra is satisfied. By explicitly computing the commutator of two variations $[\delta_{\epsilon_1},\delta_{\epsilon_2}]$ on the scalar and the fermion, given by \eqref{boson-var} and \eqref{fermion-var} respectively, one can see that the algebra does not close off-shell. An intuitive way to understand this is to count bosonic and fermionic degrees of freedom on- and off-shell. In the former case, we have two bosonic and two fermionic degrees of freedom since the Dirac equation projects out one two-component fermion, while in the latter case, there is clearly a mismatch since the equations of motion do not provide any constraint. One of the consequences of the SUSY algebra \eqref{Q,Qdag}--\eqref{Q,Poincare}, is that fermionic and bosonic degrees are in one to one correspondence so this mismatch implies that the algebra does not close off-shell. We thus need to add an auxiliary complex scalar field that vanishes on-shell and provides the two missing off-shell bosonic fields. Therefore, the correct supersymmetric non interacting Lagrangian should read
\be \label{WZ-free}
\mathcal{L}_{\rm free} = \partial_\mu \phi \partial^\mu \phi^* + i \psi^\dag_{\dot \alpha} \bar{\sigma}^{\mu \alpha \dot{\alpha}} \partial_\mu \psi_\alpha + {\cal F}{\cal F}^*,
\ee
which is \eqref{susyL-1} supplemented by an extra non dynamical term involving the auxiliary field ${\cal F}$. The variations of the scalar fields remain as in \eqref{boson-var}, while the fermion transformations and the auxiliary field ones are given by 
\be \label{fermion-var-wz}
\begin{split}
& \delta\psi_\alpha = -i(\sigma^\nu \epsilon^{\dag})_\alpha \partial_\nu \phi + \epsilon_\alpha {\cal F} ,  \qquad \delta\psi_{\dot\alpha}^{\dag} = i(\epsilon \sigma^\nu)_{\dot\alpha} \partial_\nu \phi^* + \epsilon^{\dag}_{\dot\alpha} {\cal F}^* , \\ &
\delta {\cal F} = -i \epsilon^\dag_{\dot \alpha} \bar{\sigma}^{\mu \alpha \dot{\alpha}} \partial_\mu \psi_\alpha , \qquad \delta {\cal F}^* = i \partial_\mu \psi^\dag_{\dot \alpha} \bar{\sigma}^{\mu \alpha \dot{\alpha}} \epsilon_{\alpha}.
\end{split}
\ee
The collection of the fields $\{\phi,\phi^*,\psi,\psi^\dag,{\cal F},{\cal F} \}$ forms a \emph{chiral multiplet}. The Lagrangian \eqref{WZ-free} is known as the free Wess-Zumino model and it is the minimal supersymmetric model describing a free chiral multiplet. We will now consider the interacting Wess-Zumino model which will reveal general characteristics of SUSY theories. 

From \eqref{WZ-free}, the canonical dimensions of the various fields may be computed as $[\phi]=1,[\psi]=\frac{3}{2},[{\cal F}]=2$ (assuming spacetime dimension four). Therefore, the most general interaction term containing dimension four operators reads
\be \label{WZ-int}
\mathcal{L}_{\rm int} = -\frac{1}{2} W^{jk} \psi_j \psi_k + W^j{\cal F}_j + {\rm h.c.},
\ee
where summation over contracted indices is implied and h.c. stands for hermitian conjugate, $W^{jk}$ is a linear function of the complex scalar symmetric in its indices and $W^j$ is a quadratic one.

Using the transformations \eqref{boson-var}, \eqref{fermion-var-wz} to write the SUSY variation of the interaction Lagrangian, and requiring that it vanishes, one can derive constraints on the functions $W^{ij}$ and $W^j$. Specifically, from the terms of $\delta\mathcal{L}_{\rm int}$ containing four spinors, $W^{jk}$ is restricted to be a holomorphic function of the scalar fields, $i.e.$ it must depend only on $\phi$ and not $\phi^\dag$. A useful parametrisation of $W^{jk}$ is
\be \label{wjk}
W^{jk} = \frac{\partial^2}{\partial \phi_j \partial \phi_k} W,
\ee
where the function $W$, known as the \emph{superpotential}, reads
\be \label{superpotential}
W = \frac{1}{2} M^{jk} \phi_j \phi_k + \frac{1}{6} Y^{jkl} \phi_j \phi_k \phi_l,
\ee
with $M^{jk},Y^{jkl}$ being mass and Yukawa matrices respectively. As we will shortly see, the superpotential is related to the scalar potential of the model. 

From the one derivative terms of the SUSY variation of \eqref{WZ-int}, the function $W^j$ is constrained to obey
\be \label{Wj}
W^j = \frac{\partial W}{\partial \phi_j},
\ee
a relation that identically imposes the vanishing of the remaining terms in $\delta \mathcal{L}_{\rm int}$ which are linear in ${\cal F}$. Now the full Lagrangian can be written as
\be \label{WZ-full}
\mathcal{L}_{\rm WZ} = \partial_\mu \phi \partial^\mu \phi^* + i \psi^\dag_{\dot \alpha} \bar{\sigma}^{\mu \alpha \dot{\alpha}} \partial_\mu \psi_\alpha -\frac{1}{2} \left( W^{jk} \psi_j \psi_k + W^{*jk} \psi^\dag_j \psi^\dag_k \right) + V(\phi,\phi^*),
\ee
where the scalar potential reads
\be \label{scalar-potential-F}
V(\phi,\phi^*) = {\cal F}_j {\cal F}^{*j} = W^jW^*_j .
\ee
The last equality is obtained by integrating out the auxiliary field ${\cal F}$ and imposing the constraint equation ${\cal F}_j = - W^*_j$. Thus, \eqref{scalar-potential-F} relates the scalar potential $V(\phi,\phi^*)$ and the superpotential $W$ via \eqref{Wj}.

Let us now take one more step and generalise the previous construction to include a gauge symmetry. In order to write the correct theory that respects the SUSY algebra, it is instructive to count again the on- and off-shell degrees of freedom for a system of a gauge field and a four component spinor, the \emph{gaugino}. On-shell we have two bosonic degrees representing the two helicities of the gauge boson and two fermionic degrees as before. Off-shell there is again a mismatch since now the gauge field represents three degrees of freedom, while the fermion corresponds to four. The solution is again a non dynamical auxiliary field that describes one bosonic degree, usually denoted by $D$. A pure super Yang Mills (SYM) theory may now be written as
\be \label{SYM}
\mathcal{L}_{\rm SYM} = -\frac{1}{4}F_{\mu\nu}^a F^{\mu\nu a} + i \lambda^{\dag a} \bar{\sigma}^{\mu} D_\mu \lambda + \frac{1}{2} D^aD^a,
\ee
where $F_{\mu\nu}^a = \partial_\mu A_\nu^a - \partial_\nu A_\mu^a - g f^{abc} A_\mu^b A_\nu^c$ is the field strength, with $g$ the gauge coupling and $f^{abc}$ the structure constants of the gauge group and $D_\mu \lambda^a = \partial_\mu \lambda^a - g f^{abc} A_\mu^b \lambda^c$ is the covariant derivative. Bearing in mind that the SUSY variations should map the gauge field to the gaugino and vice versa, while having the correct index structure, and taking into account that the kinetic terms differ by one derivative we obtain
\be
\begin{split} \label{SYM-var}
& \qquad \qquad \qquad \qquad \qquad \delta A_\mu^a = -\frac{1}{\sqrt{2}} \left( \epsilon^\dag \bar{\sigma}_{\mu} \lambda^a + \lambda^{\dag a}\bar{\sigma}_{\mu} \epsilon \right), \\ & \delta \lambda_\alpha^a = -\frac{i}{2\sqrt{2}} (\sigma^\mu \bar{\sigma}^{\nu} \epsilon)_\alpha F_{\mu\nu}^a + \frac{1}{\sqrt{2}} \epsilon_\alpha D^a , \qquad  \delta \lambda_{\dot{\alpha}}^{\dag a} = - \frac{i}{2\sqrt{2}} (\epsilon^\dag \bar{\sigma}^{\nu} \sigma^\mu)_{\dot \alpha} F_{\mu\nu}^a + \frac{1}{\sqrt{2}} \epsilon^\dag_{\dot\alpha} D^a , \\ & 
\qquad \qquad \qquad \qquad \qquad
\delta D^a = -\frac{i}{\sqrt{2}} \left( \epsilon^\dag \bar{\sigma}^{\mu} D_\mu \lambda^a - D_\mu \lambda^{\dag a}  \bar{\sigma}^{\mu} \epsilon \right).
\end{split}
\ee

The collection $\{A_\mu, \lambda, \lambda^\dag, D \}$ forms a \emph{vector multiplet}. The pure SYM theory may be supplemented with interacting matter fields to yield a SUSY gauge theory. Observing that the mass dimension of the auxiliary field is $[D]=2$ and restricting to canonical dimension four singlet operators, the possible choices for interactions are
$$
(\phi^* T^a \psi)\lambda^a, \qquad \lambda^{\dag a}(\psi^\dag T^a \phi), \qquad (\phi^* T^a \phi) D^a,  
$$
with $T^a$ the generators of the gauge group, so that the complete Lagrangian of a SUSY gauge theory reads
\be
\mathcal{L} = \mathcal{L}_{\rm WZ} + \mathcal{L}_{\rm SYM} -\sqrt{2} g \left[ (\phi^* T^a \psi)\lambda^a + \lambda^{\dag a}(\psi^\dag T^a \phi) \right] + g (\phi^* T^a \phi) D^a.
\ee
We may again integrate out the auxiliary field via its constraint equation 
\be \label{D-field}
D^a = -g \phi^* T^a \phi,
\ee
so that the scalar potential \eqref{scalar-potential-F} receives a contribution from the gauge sector reading
\be \label{scalar-potential-full}
V(\phi,\phi^*) = {\cal F}_j {\cal F}^{*j} + \frac{1}{2} D^a D^a= W^jW^*_j + \frac{1}{2} g^2 (\phi^* T^a \phi)^2.
\ee

The first term, originating form the matter sector, is known as the $F$-term potential while the second one, from the gauge sector, is usually called the $D$-term potential. As mentioned previously, a SUSY vacuum should have a vanishing energy. If the scalar potential does not vanish for some field configuration, supersymmetry is spontaneously broken and depending on whether the $F$-term or the $D$-term potential is non zero, the SUSY breaking mechanism is dubbed $F$- or $D$-term breaking. 
Since at low energies there exists no observation of supermultiplets, SUSY breaking is an important aspect of any theory with phenomenological applications. In Ch.~\ref{ch:paper-1}, we will study a string theory inspired inflationary model whose predictions crucially depend on the SUSY breaking mechanism which is of the $D$-term type. We thus now proceed to review both mechanisms for completeness.

\subsection{Supersymmetry breaking} \label{sec:susy-break}
A primary example of $F$-term SUSY breaking is the O'Raifeartaigh model \cite{O'Raifeartaigh:1975pr}. It contains a sector with three chiral multiplets 
with a superpotential given by
\be \label{o'r}
W_{O'R} = -k^2 \phi_1 + m \phi_2 \phi_3 +\frac{y}{2} \phi_1 \phi_3^2,
\ee
where $k$ and $m$ are mass dimension one parameters and $y$ a dimensionless Yukawa coupling.
Since there are no $D$-terms, the scalar potential \eqref{scalar-potential-F} reads
\be \label{scalar-pot-o'r}
V = |{\cal F}_1|^2 + |{\cal F}_2|^2 + |{\cal F}_3|^2 = |k^2 - \frac{y}{2} \phi^{2}_3|^2 + |m \phi_3|^2 + |m \phi_2 + y\phi_1 \phi_3|^2 .
\ee
There is no solution where ${\cal F}_1$ and ${\cal F}_2$ are simultaneously zero, so this potential has a non zero minimum at $\phi_2 = \phi_3 = 0$ equal to $V = k^4$. This is actually a one parameter family of minima, since $\phi_1$ is left free, $i.e.$ it is a flat direction. This flat direction is lifted by one loop quantum corrections, which give a mass to $\phi_1$ stabilising the potential around $\phi_1 = 0$.

Alternatively, the Fayet-Iliopoulos model \cite{Fayet:1974jb} uses a non zero $D$-term so it takes place within a gauge theory. SUSY breaking is achieved by the addition of a term in the Lagrangian which is linear in the auxiliary field
\be
\mathcal{L}_{\rm FI} = \xi^2 D,
\ee
where $\xi$ has dimension of mass. Such a contribution is known as a Fayet-Iliopoulos (FI) term and may be added to the Lagrangian \eqref{SYM} only when the gauge group contains an abelian factor, since otherwise such a term is not supersymmetric. Hence, the Fayet-Iliopoulos mechanism exists for $U(1)$ gauge groups. The $D$-term potential reads
\be \label{scalar-pot-fi}
V = \frac{1}{2} D^2 - \xi^2 D + g D \sum_i q_i |\phi_i|^2,
\ee
where $q_i$ are the charges of the scalars under the $U(1)$, leading to a constraint equation for the auxiliary field given by
\be \label{D-field-FI}
D = \xi^2 - g \sum_i q_i |\phi_i|^2.
\ee
If the $\phi_i$ are stabilised to a zero vev, the minimum of the potential is set by the FI term as $V = \frac{1}{2} \xi^4$.

We now describe the superspace formalism that allows one to derive general rules for writing arbitrary supersymmetric Lagrangians in an elegant and concise way. We will exploit this formalism to write the Lagrangian of the brane inflation model that will be studied in Ch.~\ref{ch:paper-1}. 
 
\subsection{Superspace formalism} \label{sec:superspace}
Superspace formalism \cite{Salam:1974yz} is a notational device that simplifies the manipulations of the various fields of a SUSY theory by assembling all the fields of a supermultiplet in a single object called a \emph{superfield}. Superfields are defined on a \emph{superspace}, which is an extention of spacetime by anticommuting Grassmann coordinates $y^\mu = x^\mu - i \theta \sigma^\mu \bar{\theta}$, where $\{\theta_\alpha,\bar{\theta}_{\dot \alpha}\} = 0$. The ordinary scalar, vector and spinor components of the superfield are then recovered by Taylor expanding and integrating over these Grassmann subspace, following two simple integration rules for a Grassmann variable $\eta$, namely
\be
\int d\eta = 0 \qquad \text{and} \qquad \int \eta d\eta = 1.
\ee

A chiral superfield is defined as
\bea
\Phi(y) & \equiv & \phi(y) + \sqrt{2} \theta \psi(y) + \theta^2 {\cal F}(y) \nn \\ &=& \phi(x) - i \theta \sigma^\mu \bar \theta \partial_\mu \phi(x) + \frac{1}{4} \theta^2 \bar \theta^2 \Box \phi(x) + \sqrt{2} \theta \psi(x) \\ \nn &-& \frac{i}{\sqrt{2}} \theta^2 \partial_\mu \psi(x) \sigma^\mu \bar \theta + \theta^2 {\cal F}(x),
\eea
where we have Taylor expanded $y^\mu$ and used the anticommuting properties of the Grassmann coordinates to terminate the expansion at second order in $\theta$ and $\bar \theta$.
Now, the free Wess-Zumino Lagrangian \eqref{WZ-free} can be simply written as
\be
\int dx^4 \mathcal{L}_{\rm free} = \int dx^4  d\theta^4 \Phi^\dag \Phi,
\ee
where we have used the notation 
\be
d\theta^4 = d\theta^2 d\bar{\theta}^2, \quad d\theta^2 \equiv -\frac{1}{4} d\theta^\alpha d\theta^\beta \epsilon_{\alpha\beta}, \quad d\bar{\theta}^2 \equiv -\frac{1}{4} d\bar{\theta}_{\dot\alpha} d\bar{\theta}_{\dot\beta} \epsilon^{\dot{\alpha}\dot{\beta}}.
\ee

The interacting part of the Wess-Zumino model \eqref{WZ-int}, can be written in superspace by considering the superpotential \eqref{superpotential} as a function of chiral superfields. We have that
\be
\int d\theta^2 W(\Phi) = \int d\theta^2 \theta^2 W_2 = W_a{\cal F}^a - \frac{1}{2}W^{ab}\psi_a\psi_b - \partial_\mu \left( \frac{1}{4}W^a\bar{\theta}^2\partial^\mu\phi_a - \frac{i}{\sqrt{2}}W^a\psi_a\sigma^\mu\bar\theta \right),
\ee
where $W_2$ denotes the part of $W$ which is second order in the Grassmann variables. Therefore
\be
\int dx^4 d\theta^2 W(\Phi) = \int dx^4 {\cal L}_{\rm int}.
\ee

As a general rule, it can be shown that the product of two chiral superfields is again a chiral superfield and that the SUSY variation of the $\theta^2$ component of a chiral superfield is a total derivative, thus its integration over spacetime is a SUSY invariant. Furthermore, the $\theta^2\bar{\theta}^2$ component of any
superfield is again a total derivative so that any interaction can be written as
\be
\int dx^4 d\theta^4 {\cal K}(\Phi,\Phi^\dag).
\ee
The function ${\cal K}(\Phi,\Phi^\dag)$ is known as the \emph{K\"ahler function}.

We can also include a vector superfield so that gauge theories can be formulated in superspace. The vector superfield contains three scalars and a spinor in addition to the fields of a vector multiplet. Thus, one may choose a gauge where these extra degrees vanish, the \emph{Wess-Zumino} (WZ) \emph{gauge}, considerably simplifying the notation. The vector superfield in the WZ gauge is defined as
\be \label{WZ-vector}
V^a  =   \theta \sigma^\mu \bar \theta A^a_\mu + \theta^2 \bar \theta \lambda^{\dag a} + \bar \theta^2 \theta \lambda^a + \frac{1}{2} \theta^2 \bar \theta^2 D^a.
\ee
The gauge transformation that can restore the extra components is $V^a \to V^a + i \left( \Lambda^{a} - \Lambda^{\dag a} \right)$, under which the chiral superfield transforms as 
\be
\Phi \to e^{- 2 i g T^a \Lambda^a} \Phi.
\ee
We see that the ordinary gauge parameter has been promoted to a chiral superfield $\Lambda$, which contains exactly the three scalar and one spinor degrees of freedom that we dismissed by fixing the WZ gauge. The gauge invariant kinetic terms can be written as
\be
\int d\theta^4  \Phi^\dag e^{2 g T^a V^a} \Phi.
\ee
Finally, the field strength can be organised in a chiral superfield as
\be
W_\alpha^a = -i\lambda_\alpha^a + \theta_\alpha D^a - (\sigma^{\mu\nu} \theta)_\alpha F^a_{\mu\nu} - \theta^2 \sigma^\mu D_\mu\lambda^{\dag a},
\ee
where $\sigma^{\mu\nu}$ is defined below \eqref{Q,Poincare}.
The SYM Lagrangian \eqref{SYM} can now be written as
\be \label{SYM-superspace}
\mathcal{L}_{\rm SYM} = \frac{1}{4} \int d\theta^2 W^{a \alpha} W_\alpha^a.
\ee

These are the tools that we will mostly use in the rest of this Part, so we now close the discussion on supersymmetry and proceed to review basic notions of string theory. Instead of starting from the perturbative string and perform worldsheet calculations to arrive at D-branes and the various types of string theories, we will start from the low energy limit of string theory, which is supergravity and intuitively understand the features that we will need. A proper introduction to string theory is unachievable in a few pages and evades the scope of this thesis. We will only focus on notions that will be of use in Ch.~\ref{ch:paper-1} and Ch.~\ref{ch:paper-2}. The interested reader is referred to the textbooks \cite{green1988superstring,green1987superstring,polchinski1998string,polchinski2005string,johnson2003d,zwiebach2004first,becker2007string} for further exploration of this wide subject.
\section{Supergravity and String theory} \label{string-primer}
Supergravity (SUGRA) is a theory with local supersymmetry describing the dynamics of massless gravitons and their superpartners. It is the low energy limit of M-theory (a conjectural eleven dimensional UV complete theory) and under dimensional reduction it reduces to the various types of supergravities which are the low energy limits of the corresponding string theories. 

In this Section, we will start from the unique SUGRA theory in eleven dimensions and see how the reduction to ten dimensions yields Type IIA SUGRA. We will then review the D-brane spectrum of Type IIA string theory as well as T-duality, which will lead us to the Type IIB theory. We will finish this Section, with a discussion on other extended objects of string theory such as orientifold planes, as well as moduli fields, which are scalar fields that arise in lower dimensional effective models of string theory.

So far we have constructed the minimal version of supersymmetry with only one set of spinor supercharges $Q_\alpha,Q_{\dot \alpha}^\dag$ that act on a vacuum state of a given helicity as ladder operators and produce a whole multiplet. Nothing prevents us though from extending \cite{Haag:1974qh,Salam:1974za} this structure to include other sets of SUSY generators $Q_{a \alpha },Q_{a \dot \alpha}^{\dag}$, where $a$ runs from 1 to ${\cal N}$. Such extended supersymmetric theories are widely used, since they offer deep insight into the dynamics of gauge theories in extreme regimes $e.g.$ strongly coupled theories and confinement \cite{Seiberg:1994rs,Maldacena:1997re}. The resulting supermultiplets are extended as well, since now we have at our disposal more raising operators to act on a given helicity vacuum. From the extended algebra in four dimensions, one may conclude that starting from a vacuum of helicity $\lambda$, the maximal helicity state in the multiplet has $\lambda_{\rm max} = \lambda + \frac{{\cal N}}{2}$. Therefore, if one wants to construct a gravitational theory, where $|\lambda|\leq 2$ and $|\lambda + \frac{{\cal N}}{2}| \leq 2$, the maximal number of supercharges is ${\cal N} \leq 8$. Now in four dimensions, a spinor has dimension four, so the maximal SUGRA theory has $4 \times 8 = 32$ supercharges. This is a general bound on the number of supercharges for a theory that involves degrees of freedom up to spin two in any spacetime dimension $d$. That is, $d_S \times {\cal N} = 32$, where $d_S$ is the dimension of the spinor representation, taking the values $d_S = 2^{ \frac{d-2}{2} }$ for $d$ even, and $d_S = 2^{ \frac{d-1}{2} }$ for $d$ odd. Therefore, there is a maximal spacetime dimension for SUGRA and that is $d=11$, implying ${\cal N} = 1$. 

A supergravity theory must contain at least a second rank symmetric traceless tensor $g_{\mu\nu}$ representing the graviton and a spin $3/2$ superpartner $\psi_\mu^\alpha $, the gravitino, which is a vector-spinor. Let us now count the bosonic and fermionic components to see if there is any mismatch. Since we will consider massless particles, it is more convenient to organise our states in representations of the little group in eleven dimensions, which is $SO(9)$. The graviton has $\frac{9 \times 10}{2} - 1 = 44$ degrees, where the $-1$ comes from the vanishing trace condition. The gravitino transforms as a product of a vector and a spinor of $SO(9)$ so it has $9 \times 16 = 144$ components. There is another vanishing trace condition on the gravitino $\Gamma^\mu \psi_\mu^\alpha = 0$, $\Gamma^\mu$ being the Dirac matrices in spacetime dimension eleven, which removes one 16 component spinor leaving us with 128 fermionic degrees of freedom. We thus have a difference of 84 degrees. This can be matched by including a third rank antisymmetric tensor which in eleven dimensions has exactly 84 components. In conclusion, eleven dimensional SUGRA contains a graviton, a 3-form antisymmetric field and a gravitino. 

Let us now track these fields when we compactify one spatial dimension. The decomposition is
\be
\begin{split}
& g_{\mu\nu} (44) \to g_{\mu\nu} (35) \oplus C_\mu(8) \oplus \phi (1), \\
& C_{\mu\nu\rho} (84) \to C_{\mu\nu\rho}(56) \oplus B_{\mu\nu} (28), \\
& \psi^\alpha_\mu (128) \to \psi^{+\alpha}_\mu (56) \oplus \psi^{-\alpha}_\mu (56) \oplus \lambda^{+\alpha} (8) \oplus \lambda^{-\alpha} (8),
\end{split}
\ee
where the numbers in parenthesis denote the number of independent components of each field and the $\pm$ labels in the fermionic sector are used to denote spinors of opposite helicity. These are the fields of ten dimensional Type IIA SUGRA and Type IIA superstring theory, which is its UV completion. Now in complete analogy with electromagnetism, where a 1-form field couples to a zero dimensional matter particle, one would expect that the $p$-form fields present in SUGRA couple to extended $p-1$ dimensional objects. That is, the 1-form $C_\mu$ should couple to a particle, the 2-form $B_{\mu\nu}$ should couple to a string and the 3-form $C_{\mu\nu\rho}$ to a two dimensional membrane. 

Moreover, we may consider to what objects the dual form fields would couple to. In electromagnetism, the dual field strength $* F^{\mu\nu}=\frac{1}{2}\epsilon^{\mu\nu\kappa\lambda}F_{\kappa\lambda}$ is obtained from the field strength by exchanging the electric and magnetic fields. Hence, if $F_{\mu\nu}$ couples to an electrically charged particle, the dual tensor couples to a magnetically charged one, $i.e.$ a monopole. Generalising this terminology, the branes that couple to the fields that we mentioned are usually called electric branes, while those that couple to the dual fields are referred to as magnetic ones. An $n$-form field in $d$ dimensions has an $(n+1)$-form field strength, which is Hodge dual to a $(d-n-1)$-form field strength, implying a dual $(d-n-2)$-form gauge field. Hence, we may add to the previous brane spectrum a six dimensional extended brane, which is the magnetic dual of the particle corresponding to $C_\mu$, a four dimensional brane dual to the two dimensional brane that couples to $C_{\mu\nu\rho}$ and a five dimensional brane that is dual to the string. 
Indeed, all of these objects known as $p$-branes, $p$ denoting the spatial dimension, were shown \cite{Hughes:1986fa,Bergshoeff:1987cm} to be solutions of the SUGRA equations (see \cite{Duff:1996zn,Stelle:1996tz,Stelle:1998xg} for reviews on supermembranes in SUGRA).

In 1995, Polchinski demostrated \cite{Polchinski:1995mt} that these SUGRA $p$-brane solutions correspond to string theory extended objects called D$p$-branes\footnote{For reviews focused on the physics of D$p$-branes see \cite{Polchinski:1996fm,Bachas:1998rg,Johnson:2000ch}.}. D$p$-branes \cite{Dai:1989ua,Horava:1989ga} may be realised in perturbative string theory as loci of open string endpoints. In other words, they span the directions transverse to those, where the string endpoints obey Dirichlet boundary conditions\footnote{The prefix D in the term ``D-brane'' stands for Dirichlet.} of the form $\partial_\tau X^m(\sigma=0) = \partial_\tau X^m(\sigma=\pi) =0 $, where $m$ labels the directions with Dirichlet conditions, $\tau,\sigma$ are the worldsheet time and space coordinates respectively, and $X^m$ the string positions in the ambient ten dimensional space. The discovery of D-branes opened a way to embed gauge theories with various gauge groups in string theory, since the string endpoints attached to a brane carry vector degrees of freedom, and led to what is known as the \emph{second superstring revolution}, offspring of which are the idea of holography and AdS/CFT \cite{Susskind:1994vu,Witten:1998qj,Gubser:1998bc,Maldacena:1997re}. Moreover, D3-branes are also of cosmological importance, since they could play the role of our four dimensional universe in brane world cosmology scenarios \cite{Brax:2004xh,Langlois:2002bb}.

Type IIA superstring theory contains D$p$-branes with an even number of spatial dimensions $p$, which correspond to the membranes we found in the previous paragraph. In addition, there are the fundamental degree of freedom of string theory, the F-string and its magnetic dual, the so called NS5-brane, while the scalar $\phi$ corresponds to the dilaton. These three objects are common to all string theories and they comprise the Neveu Schwartz--Neveu Schwartz (NSNS) sector of the theory as opposed to the rest of the bosonic fields, constituting the Ramond--Ramond (RR) sector, which varies among different string theories. The fermionic fields arise as combinations of fields from the NS and R sectors and depending on the choice of their chiralities one may construct all five string theories, Type IIA and IIB, Type I, $E_8 \times E_8$ and $SO(32)$. 

In order to discuss Type IIB string theory which will be the UV complete framework for our discussion in Ch.~\ref{ch:paper-1} and Ch.~\ref{ch:paper-2}, we now review T-duality since we will also use it in Ch.~\ref{ch:paper-1} to reveal some of the properties of the inflationary model under consideration there.

T-duality is a symmetry of string theory which has no analogue in quantum field theory. At the level of the closed string we may intuitively understand it as follows: let us compactify one out of the 9 spatial dimensions, say $x^9$, on a circle of radius $R$ so that the momenta along that direction are quantised as $p^9 = n/R$, $n \in \mathbb{Z}$. In field theory, we may take the limit $R \to 0$, where we have a tower of infinitely massive particles labelled by $n$, that decouple from the dynamics of the pure eight dimensional modes. This is a Kaluza-Klein reduction, which results in a field theory in eight spatial dimensions.

In string theory, where the fundamental degree of freedom is one dimensional, a closed string may wind around the circle resulting in an extra contribution to the momentum along the compact direction $p^9 = n/R + w R/\alpha'$, where $\alpha'$ is a length dimension two quantity characterising the inverse string tension and $w \in \mathbb{Z}$. Now by taking the same $R \to 0$ limit, we see that momentum modes become infinitely heavy as before, but winding modes become massless so that a continuous tower of modes appears, manifesting itself as an effective non compact dimension. At the other limit, $R \to \infty$, winding modes become massive, while momentum modes become massless and again we obtain the uncompactified spectrum. This indicates a symmetry of the theory under the simultaneous transformations $n \leftrightarrow w$ and $R \leftrightarrow \alpha'/R$, which is known as T-duality.

At the level of the open string, T-duality along some spatial directions can be understood as exchanging Neumann and Dirichlet boundary conditions with respect to these coordinates. Recall from the previous paragraph that Dirichlet conditions signal the presence of D-branes. Therefore, a T-duality along a direction tangent to a D$p$-brane will reduce its dimensionality to $p-1$, since it will change a Neumann boundary condition to a Dirichlet one, while a T-duality along a normal direction will increase the spatial dimension to $p+1$. This definition, allows us to connect Type IIA theory with another string theory which is known as Type IIB and contains even p-form fields and odd dimensional branes, simply by considering the case where we perform a T-duality along a direction tangent to the even dimensional branes of Type IIA.
Since branes couple to RR form fields, this change in the dimensionality of a brane should also induce a transformation on the corresponding fields, which is indeed the case \cite{Dai:1989ua}. 

The field content of Type IIB string theory is given by
\be
\begin{split}
&  g_{\mu\nu} (35) \oplus B_{\mu\nu} (28) \oplus \phi (1), \\
&  C_0(1) \oplus C_{\mu\nu} (28) \oplus C_{\mu\nu\rho\sigma}(35) , \\
&  \psi^{\alpha}_\mu (112) \oplus \lambda^{\alpha} (16),
\end{split}
\ee
where the numbers in parenthesis denote the number of components of each form field and now the fermionic sector is chiral. The 4-form field should normally account for $70$ degrees of freedom but $C_{\mu\nu\rho\sigma}$ is constrained to be self dual, which removes half its components. The spectrum thus contains a D(-1)-brane\footnote{This notation refers to an object which is pointlike in spacetime, resembling an instanton configuration.} coupling to the 0-form and a D1-brane coupling to the 2-form, their magnetic duals, a D7- and a D5-brane respectively as well as a D3-brane coupling to the 4-form which is self dual, reflecting the self duality of a 5-form field strength in ten dimensions. Since the NSNS is the same as in Type IIA, the F-string and the NS5-brane are also part of the spectrum.

The content of both Type II theories may be understood from an open superstring perspective as follows: at the massless level, the open superstring contains a ten dimensional massless vector, which comprises a vector representation of the little group $SO(8)$, denoted by $\mathbf{8}_v$. Since the theory is supersymmetric, we also have an eight dimensional spinor but $SO(8)$ has two spinor representations of opposite chirality, denoted as $\mathbf{8}_s$ and $\mathbf{8}_c$. The closed string can now be obtained by tensoring the representations carried by the two endpoints. Since there are two spinors, we may choose to tensor either two spinors of the same chirality, or two of the opposite. The first choice leads to Type IIB theory while the second to Type IIA. Computing the bosonic tensor products we obtain indeed
$$\mathbf{8}_s \otimes \mathbf{8}_s = \mathbf{1} \oplus \mathbf{28} \oplus \mathbf{35}_+,$$ which are the RR fields of Type IIB and $$\mathbf{8}_c \otimes \mathbf{8}_s = \mathbf{8} \oplus \mathbf{56},$$ which are the RR fields of Type IIA. The tensor product of the  vector representation is common to both types and yields the NSNS fields $$\mathbf{8}_v \otimes \mathbf{8}_v = \mathbf{1} \oplus \mathbf{28} \oplus \mathbf{35}.$$

A useful way of obtaining different string theories is to truncate this spectrum to states which are invariant under global discrete symmetries. For example, consider string theory propagating on the quotient space $S^1/\mathbb{Z}_2$, where $\mathbb{Z}_2$ is a reflection symmetry with respect to the compact coordinate, $x^9 \to -x^9$. This space is known as an \emph{orbifold}, since the compact space $S^1$ is divided into orbits of the $\mathbb{Z}_2$ group and points that belong to the same orbit are folded into a single point in the orbifold space $S^1/\mathbb{Z}_2$. This particular construction is just a line segment with two fixed points at $0$ and $\pi$. The combination of such a discrete symmetry and a worldsheet parity transformation, which reverses left and right movers, is a generalisation of the orbifold known as an \emph{orientifold}, since the parity operation reverses the orientation of the worldsheet. 

One can show that the solutions to the string equations of motion imply that there is no momentum nor winding flow along the orientifold fixed points, so the string does not move in that direction. In analogy with the D-branes, this indicates the existence of \emph{orientifold planes}, which are subspaces of spacetime where the string endpoints can move. However, unlike D-branes, ${\cal O}$ planes are not dynamical. They are defined entirely by the action of the discrete group on the compact manifold and not by string boundary conditions. Since in these constructions the spectrum gets reduced, string theories on orbifold and orientifold spaces are easier to manipulate. Moreover, orientifold planes have a negative tension and are essential ingredients for engineering gauge theories on D-branes with $SO(N)$ and $Sp(N)$ gauge groups. In Ch.~\ref{ch:paper-2} we will see how the inclusion of orientifold planes in the brane inflationary model under consideration, leads to severe constraints on the form of radiation that a cosmic superstring may produce.

Let us now proceed to discuss the notion of moduli fields which are an unavoidable feature of phenomenological applications of string theory.

\subsection{Moduli fields}
As evident from the previous discussion, one peculiarity of superstring theory is that it requires a ten dimensional spacetime. Since, at large scales at least, our spacetime is unambiguously 
four dimensional, any phenomenological study in this framework should be placed in the appropriate dimensionality.

An idea of how one can effectively remove the extra six dimensions is to imagine that they are compactified in some internal spatial geometry, whose length scale evades our current observational resolution. Such a proposal was known long before the advent of string theory from the work of Kaluza and Klein, who attempted to unify electromagnetism and gravity by studying a five dimensional field theory compactified on a small circle. The relatively large number of extra dimensions of string theory however increases the possible choices of internal spaces that one has. In the previous Sections, we saw how supersymmetry can improve certain aspects of gauge theories and since it is an integral part of string theory, it is desirable that the compactification preserves some amount of supersymmetry. In the 90's, it was realised \cite{Candelas:1985en} that when string theories are placed on a product of four dimensional Minkowski space with certain compact spaces known as Calabi-Yau (CY) manifolds, the effective theory is indeed supersymmetric.

This internal manifolds manifest themselves in the effective four dimensional world through light scalar fields known as \emph{moduli}. These fields parametrise either geometrical features of these spaces, $e.g.$ the volume, the shape, or topological structures like their homology cycles, in which case they are usually referred to as closed string moduli. Another source of light scalars arises from D-branes. These configurations, when embedded in ten dimensional spacetime, admit tangential and transversal directions, with the latter appearing as scalar fields on the worldvolume theory of the D-brane. Intuitively these scalars can be understood as Goldstone bosons of the spontaneous symmetry breaking $SO(9,1) \mapsto SO(p,1) \times SO(9-p)$ triggered by the presence of a D$p$-brane on $M^{9,1}$, where $SO(p,1)$ is the Lorentz group on the worldvolume and $SO(9-p)$ the rotation group of the transverse space. These scalars are usually referred to as open string moduli. 

The fact that these fields parametrise the freedom of the internal space indicates that they should be massless. In supersymmetric theories, moduli fields have flat potentials to all orders in perturbation theory. It is expected that non perturbative contributions may eventually lift these flat directions and fix the moduli at vacuum expectation values but since not much is understood beyond perturbation theory, these fields are typically considered to be massless. Even though this is a desired property for an inflaton, it leads to several problems from a phenomenological perspective for many reasons. On the one hand, if they exist they should be copiously produced even in low energy processes, whilst they could also mediate fifth forces, which are experimentally unobserved. On the other hand, even if they are massive, there are constraints on their masses due to the cosmological moduli problem, which states that light moduli would overclose the universe \cite{Coughlan:1983ci,Ellis:1986zt,deCarlos:1993jw,Banks:1993en}. In the recent years, there has been much activity concerning the study of mechanisms that stabilise these moduli fields to a phenomenologically acceptable value \cite{Grana:2005jc,Douglas:2006es,Balasubramanian:2005zx,Denef:2007pq,Denef:2008wq}, leaving only the inflaton candidate as a light degree of freedom.

\section{Cosmic (super)strings} \label{sec:cosmic-superstrings}
Cosmic strings are one dimensional solitons that can be understood as continuous loci of zeroes of a scalar field serving as an order parameter\footnote{In a phase transition, the order parameter is defined as a quantity that has a non zero value in the ordered phase, while it vanishes in the disordered one (hence the name ``order parameter''). There is no a priori prescription for such a variable, anything that satisfies this condition can serve as an order parameter.} for phase transitions. Their possible role in early universe physics was first pointed out by Kibble \cite{Kibble:1976sj}, who showed that they are formed whenever there is a phase transition leaving a vacuum manifold with a non trivial first fundamental group. 

The fundamental group is a group of maps from the one dimensional sphere $S^1$ to a manifold ${\cal M}$. 
The reason why it is related to string defects is the following: consider the manifold ${\cal M}$ as being the vacuum manifold of the theory. Then, as we draw a closed path on the physical space, the scalar field spans a closed path in ${\cal M}$. If this path cannot be continuously deformed into a point, then there must be a ``hole" in ${\cal M}$, which corresponds to the string core in physical space. Thus, if the vacuum manifold has a non trivial fundamental group, $i.e.$ it contains holes, string defects are possible solutions of the model. The winding number of the string is then the integer that counts how many times the path in the vacuum manifold winds around the hole as we travel once around the circle in the physical space. 

For example, if the vacuum manifold is a circle, the first fundamental group is the group of integers $\mathbb{Z}$, representing the possible winding numbers of a string. Therefore, such one dimensional defects are expected to form in many inflationary models ending with a broken $U(1)$ symmetry. Furthermore, in \cite{Jeannerot:2003qv}, it was shown that if there was an intermediate GUT stage in the early universe, then most of the symmetry breaking paths via which the standard model gauge group may be reached, would lead to the production of one dimensional topological defects. Therefore, cosmic strings are a generic feature of phase transitions in the history of the universe.

\begin{figure}[tbh] 
\begin{center}
\includegraphics[scale=0.5]{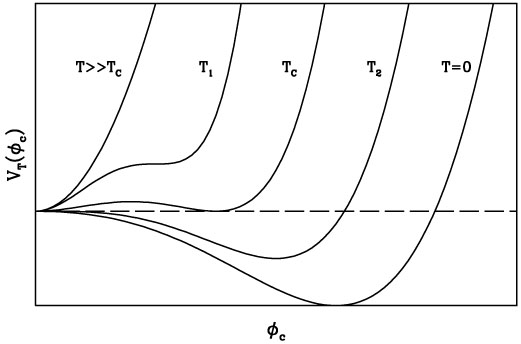}   
\end{center}
\caption[{\sf The vacua of the Abelian Higgs model.}]{\sf The vacua of the Abelian Higgs model \eqref{abelian-higgs-pt}. At high temperatures, the global minimum lies at a zero vev for the scalar $\phi$, whereas for temperatures lower that the critical one, the global minimum lies at a non zero vev. Hence, as the temperature drops, the model exhibits a phase transition from a disordered phase with a vanishing order parameter to an ordered phase with a non zero order parameter. Figure taken from \url{http://ned.ipac.caltech.edu/level5/March02/Gangui/Gangui1_1.html}.}
\label{fig:pt}
\end{figure}

As an example, let us review the formation of cosmic strings in such a cosmological phase transition. The reader is referred to \cite{vilenkin2000cosmic} for a complete account of cosmic strings and their properties. We will consider the Abelian Higgs model given by
\be \label{abelian-higgs-pt}
{\cal L} = \frac{1}{2} (D^\mu \phi)^\dag D_\mu \phi - \frac{1}{8} g^2 ( \phi^\dag \phi - \eta^2 )^2 + \frac{1}{8} F_{\mu\nu} F^{\mu\nu},
\ee
where $F_{\mu\nu}$ is the field strength for the $U(1)$ gauge field and $\phi$ a complex scalar.
At finite temperature $T$, this potential receives corrections of the form \cite{Weinberg:1974hy}
\be \label{pt-pot}
V(\phi) = \frac{1}{8} g^2 ( \phi^\dag \phi - \eta^2 )^2 + \frac{1}{24}(2 g^2 + 3 e^2) T^2 \phi^\dag \phi .
\ee
The vacuum structure of the model is depicted in Fig.~\ref{fig:pt}.
From \eqref{pt-pot}, we see that one extremum of the potential is at $\phi = 0$. The critical temperature at which the phase transition occurs, is the one where the $V(0) = \frac{g^2}{2} \eta^4$ vacuum becomes unstable and a new global minimum appears. A straightforward computation yields
\be
\frac{\partial^2 V}{\partial \phi^2}\Big|_{\phi = 0} = 0 \Rightarrow T_c = \eta \frac{1}{\sqrt{\frac{1}{3} + \frac{ e^2}{2 g^2}}}.
\ee
For temperatures below the critical one, the new minimum $V(\phi_c) = 0$ is found to be at 
$$
\phi_c = \eta \sqrt{1-\frac{T}{T_c}}.
$$
%

Therefore, as the universe cools down, it undergoes a phase transition at the critical temperature $T_c$, where the symmetry of the theory is spontaneously broken. The value of the field $\phi$ at the minimum of the potential serves as an order parameter since at the high temperature phase $\phi = 0$, while at the low temperature ordered phase $\phi \neq 0$. At zero temperature, where the symmetry is broken, the theory has solutions which are vortices extended in one dimension, $i.e.$ strings. The string ansatz reads \cite{Nielsen:1973cs}
\be
\phi = \eta f(r) e^{i n \varphi}, \qquad A_i = \alpha(r) n \partial_i \varphi ,
\ee
with asymptotics
\be
f(0) = 0, \quad \alpha(0) = 0, \quad f(\infty) = 1, \quad \alpha(\infty) = 1.
\ee
The functions $f,\alpha$ are only known numerically, since analytic results require solving a system of non linear coupled PDE's, the equations of motion of the two fields $\phi, A_\mu$.

From the asymptotic behaviour, we see that away from the string the order parameter is non zero and we are in the low $T$ phase of the theory, while near the string core the order parameter becomes zero and the theory is in the high $T$ phase. In other words, strings are one dimensional defects inside which the old phase is trapped\footnote{An amusing analogy is the ice cube. As we lower the temperature below the critical one, which is at $273 K$, the solid phase of water appears. Looking at any ordinary ice cube we will observe string defects inside which the old liquid phase is trapped.}.

 Depending on whether the broken symmetry is a gauge or a global one, the strings behave differently as far as their decay properties are concerned. The energy density of global strings falls as $1/r^2$ away from the core so they have long range interactions leading to their decay, whereas local strings have their energy density confined in the core so long range forces do not apply. Dynamically stable cosmic strings were thought to have a potential cosmological role, since they could provide the seeds of large scale structure as an alternative to inflation \cite{Battye:1997hu,Avelino:1997hy}. However, such a scenario would lack the coherence of the acoustic oscillations of density fluctuations, leading to a temperature power spectrum such as the one depicted in Fig.~\ref{fig:dode-3}. {\sc Cobe}, {\sc Boomer}an{\sc g} and {\sc Wmap} provided us with early versions of Fig.~\ref{fig:temperature-power}, thus disfavouring this approach, although strings were shown to be compatible with observational data \cite{Bouchet:2000hd}, as long as they have participated in the generation of primordial density perturbations in combination with inflation. 
 
Interest in cosmological applications of cosmic strings was revived \cite{Sakellariadou:2005wy,Kibble:2004hq}, when it was realised that strings and branes of string theory could have a cosmological role similar to that of field theory cosmic strings \cite{Copeland:2003bj,Polchinski:2004hb}. Even though these objects\footnote{See \cite{Polchinski:2004ia,Sakellariadou:2009ev,Sakellariadou:2008ie} for reviews on cosmic superstrings and their properties.} have a tension near the Planck scale, in \cite{ArkaniHamed:1998rs,ArkaniHamed:1998nn,Randall:1999vf,Randall:1999ee} it was shown how highly warped extra dimensions could lower this tension to phenomenologically acceptable values. More connections between cosmic strings and superstrings appeared when it was realised that at the end of brane inflation, D1-branes are generically formed providing a quantum analogue of cosmic strings \cite{Sarangi:2002yt,Jones:2002cv,Jones:2003da, Dvali:2003zj, Majumdar:2002hy}. This opened a possible ``observational window" into string theory, since the tension of these strings, given in terms of free parameters of the theory, is related with the amplitude of temperature fluctuations in the CMB, a quantity which was constrained by {\sc Cobe} \cite{Smoot:1992td}. Since we will use this bound in Ch.~\ref{ch:paper-1}, let us briefly review its origin (see $e.g.$ Ch.~10 of \cite{vilenkin2000cosmic}).

\subsection{Observational consequences of cosmic strings}
\begin{figure}[tbh] 
\begin{center}
\includegraphics[scale=0.7]{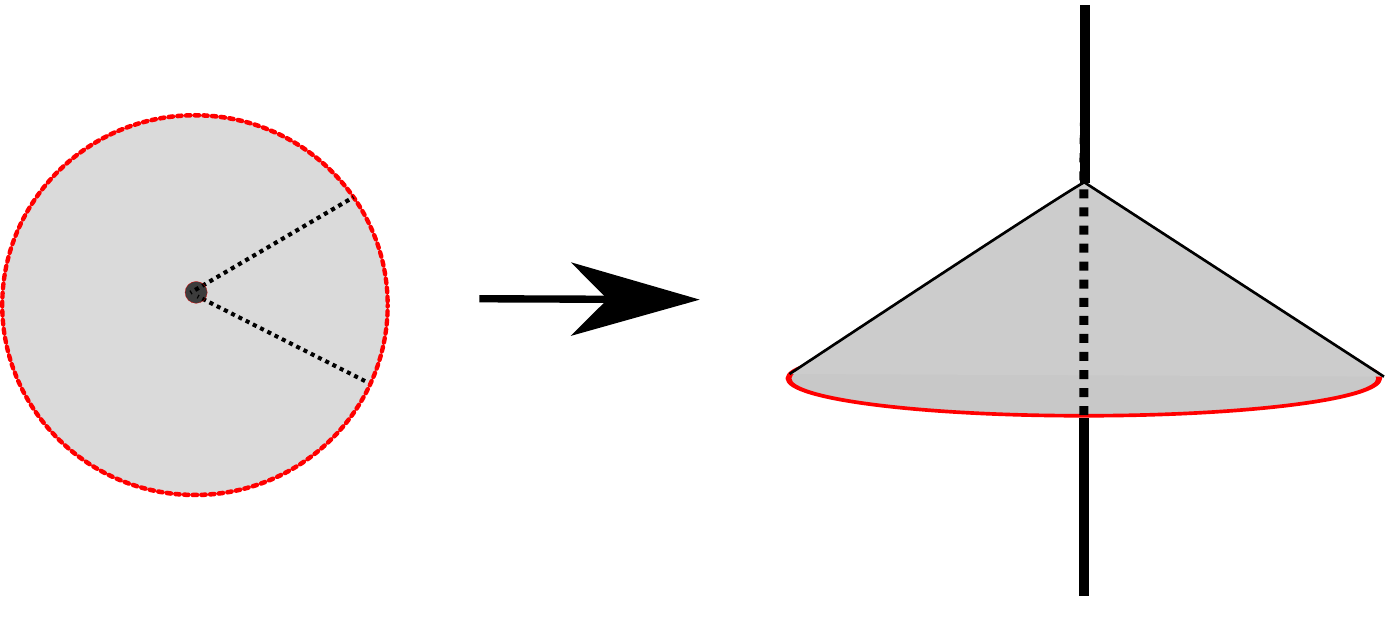}   
\end{center}
\caption[{\sf Spacetime around a straight cosmic string}.]{\sf A conical spacetime around a straight cosmic string depicted as a black point (left) and a thick black line (right). The cone results from the disk upon removing the wedge between the dashed lines and gluing the remaining edges. The angle of the wedge equals the deficit angle of the cone $\Delta = 8\pi G\mu$ (see \eqref{string-metric}).}
\label{fig:string-st}
\end{figure}

Around an infinite straight string, spacetime is locally flat but globally admits a cone geometry (see Fig.~\ref{fig:string-st}) with a metric given by
\be \label{string-metric}
ds^2 = -dt^2 + dz^2 + dr^2 + r^2(1-8 G \mu)d\theta^2,
\ee
where $G=\frac{1}{M_{\rm Pl}^2}$ is the gravitational constant and 
\be \label{string-mass}
\mu = 2\pi n \eta^2,
\ee
the linear mass density of the string, with $n$ the winding number and $\eta$ the vev of the scalar associated to the string.
Now let us think of two objects at rest. As a straight string, perpendicular to the line-of-sight, passes between them they start moving relative to each other. This can be understood intuitively from the procedure described in Fig.~\ref{fig:string-st}. By removing the wedge and gluing the two edges, we essentially bring spacetime points around the wedge surface closer to each other. Considering one object as a source and the other as an observer, the latter will detect a fluctuation in the energy stemming from the Doppler effect due to the relative motion. From the Doppler formula, the change in the frequency will be $\delta f = \delta v f$, where $\delta v$ is the change in the relative velocity. For the case at hand, we have \cite{vilenkin2000cosmic} $\delta v = 8\pi G\mu \gamma v$, where $\gamma$ is the Lorentz factor $\gamma = (1-v^2)^{-1/2}$, and $v \sim 0.55$ as suggested by simulations of cosmic string networks \cite{Allen:1990tv,Bennett:1989yp}. This process leads to a temperature fluctuation in the CMB, namely
\be \label{string-bound-0}
\frac{\delta T}{T} = 8\pi G\mu \gamma v \sim 8.5 G \mu.
\ee
Using the {\sc Cobe} bound on CMB temperature fluctuations \cite{Smoot:1992td}, we may obtain a constraint on the value of the string linear mass density $\mu$, which is related via \eqref{string-mass}, to the energy scale associated with the physical process responsible for the string formation. In Sec.~\ref{sec:cs-sols-paper-1}, we will use this bound in the context of brane inflation to deduce a constraint on a free parameter of the underlying physics, which in this case will be string theory. In fact, as already mentioned in the previous paragraph, this is a general way of constraining free parameters of inflationary models exhibiting cosmic string formation at the end of inflation.

Another potentially observable effect of a cosmic string is gravitational lensing. To understand the mechanism, we again refer the reader to Fig.~\ref{fig:string-st}. This time, let us imagine a static cosmic string lying between an observer and a source. Light rays coming from the source, upon reaching the boundary of the wedge, will bend due to the gluing procedure, eventually meeting at the spacetime location of the observer. Thus, the observer will receive a double image of the object. Contrary to gravitational lensing by other objects, the double image from a string lens is distortion free.

Even though there exist other observational signatures of cosmic strings and superstrings, like $e.g.$ gravitational waves, cosmic rays etc., we now close this chapter with a brief review of brane inflation, setting the context in which cosmic superstrings will be studied in Ch.\ref{ch:paper-1} and Ch.\ref{ch:paper-2}. The interested reader may find more details on cosmic (super)string observables in \cite{Hindmarsh:1994re,Copeland:2009ga} and references therein.

\section{Brane inflation} \label{sec:brane-inflation}
String inflationary models roughly fall into two large categories depending on whether the inflaton is a closed or open string modulus \cite{Cicoli:2011zz}.
We will focus on two widely studied models of open string moduli inflation, where our universe is considered as a bound state of D-branes spanning the transverse space of a compact six dimensional manifold with all moduli fields stabilised. The first proposal using membranes to describe inflation was reported in \cite{Dvali:1998pa}, where it was realised that the distance between two such branes can play the role of the inflaton field in four dimensions. This was subsequently generalised to include brane-antibrane models \cite{Burgess:2001fx,Dvali:2001fw,Alexander:2001ks}, as a special case of branes at angles \cite{Jones:2002cv,GarciaBellido:2001ky,Blumenhagen:2002ua}, as well as models with orbifold compactifications and orientifold planes \cite{Burgess:2001vr}. 

The main idea underlying brane inflation is roughly the following: when two membranes are parallel, the contributions from the exchange of RR, NSNS and dilaton fields exactly cancel so that no force between the two branes exists \cite{polchinski1998string,polchinski2005string}. This is a supersymmetric configuration with minimum energy. If somehow supersymmetry is broken ($e.g.$ by tilting the branes to an angle), this cancellation no longer occurs and an attractive force between the branes appears. Thus, their relative distance decreases until it reaches a critical point, where tachyonic modes appear and an inflationary waterfall stage occurs, leading to the collision of the two branes followed by the reheating process \cite{Felder:2000hj,Cline:2002it,Shiu:2002xp}. 

A typical potential between two D3-branes in $N$ extra dimensions \cite{Dvali:1998pa} is of the form 
\be
V(r) = T_3 \left( 1 - \frac{\beta}{r^{N-2}} \right),
\ee
where $T_3$ is the mass volume density of the D3-brane and $\beta$ is a mass dimension $2-N$ parameter. In a string framework where spacetime is ten dimensional so that $N=6$, the potential is a function of the inverse fourth power of the inflaton field, which is related to the distance between the branes as $\phi = \sqrt{T_3} r$. Thus, when the two branes are far form each other, this potential is very flat allowing for a slow roll of the inflaton towards zero, where the two branes collide, the potential becomes too steep and inflation stops.

Moreover, since the situation resembles that of hybrid inflation \cite{Linde:1993cn}, cosmic superstrings are expected to form at the collision point \cite{Sarangi:2002yt,Jones:2002cv,Jones:2003da, Dvali:2003zj, Majumdar:2002hy}, which can be used to constrain the parameters of such brane models.
Brane inflation has been a rich subject for the past fifteen years and the reader is referred to \cite{HenryTye:2006uv,Linde:2005dd,Burgess:2007pz,Baumann:2009ni,McAllister:2007bg,Cline:2006hu,Kallosh:2007ig,Cicoli:2011zz} for extensive reviews as well as the book \cite{Baumann:2014nda} for a more recent and wider analysis. 

Having set the framework and the technical background that we will need, we now proceed to discuss the $D3/D7$ inflationary model introduced in \cite{Dasgupta:2002ew,Dasgupta:2004dw,Haack:2008yb}, and further studied in \cite{Koyama:2003yc,Firouzjahi:2003zy,Brandenberger:2008if}.

\chapter{Cosmic superstrings in $D3/D7$ model of inflation} \label{ch:paper-1}

The $D3/D7$ model \cite{Dasgupta:2002ew,Dasgupta:2004dw,Haack:2008yb}, lies within a type IIB string theory context, with the main ingredients consisting of a D3- and a D7-brane. Ten dimensional spacetime is compactified on a $K3\times T^2/\mathbb{Z}_2$, where $K3$ is a four dimensional manifold, $T^2$ is a torus and $\mathbb{Z}_2$ is the orientifold operation $\mathbb{Z}_2=\Omega\cdot(-1)^{F_L}\cdot\mathcal{I}_{45}$, with $\Omega$ the orientation reversal on the worldsheet,  $\mathcal{I}_{45}$ the orbifold projection along the $T^2$ directions $x^4,x^5$, which are transverse to the D-branes, and $(-1)^{F_L}$ acts on the left moving worldsheet fermions. The $K3$ component of the internal manifold is wrapped by the D7-brane and it is transverse to the D3-brane, yielding an effective four dimensional gauge theory. Compactifications of string theory on this space have been extensively studied \cite{Tripathy:2002qw,Aspinwall:2005ad,Andrianopoli:2003jf,D'Auria:2004qv,Angelantonj:2003zx}, rendering this example one of the best understood cases with all the moduli fields stabilised.

The inflaton field in this construction is provided by the real part of the complexified coordinates of the D3-brane on the internal torus $T^2$. Initially, the D3- and the D7-branes are placed at a large distance forming a supersymmetric state such that no force between them is induced. Supersymmetry is broken by a flux $\cal F$ on the worldvolume of the D7-brane, which as we shall see corresponds to an effective four dimensional FI term. Due to the supersymmetry breaking, a potential in the effective theory is induced, which gives rise to an attractive force between the branes. The position of the D3-brane is a flat direction of this potential and at one loop level, it has the form of the Coleman-Weinberg correction \cite{Coleman:1973jx}, which slightly lifts the flat valley allowing for a slow roll phase of the inflaton down to its critical value. Inflation ends when the distance between the branes reaches a critical point, where the waterfall fields, coming from the strings stretched between the two branes, acquire tachyonic masses. In the following table we present the coordinates over which the branes, the internal manifold and the FI flux extend
$$
\begin{array}[c]{|c||c|c|c|c|c|c|c|c|c|c|}
\hline &x^0 & x^1 & x^2 & x^3 & x^4 & x^5 & x^6 & x^7 & x^8 & x^9 \\
\hline\hline D3&-&-&-&-& \times & \times & \times & \times & \times & \times \\
\hline D7&-&-&-&-& \times & \times &-&-&-&-\\
\hline \mathcal{F}& \times & \times & \times & \times & \times & \times &-&-&-&-\\
\hline K3& \times & \times & \times & \times & \times & \times &-&-&-&-\\
\hline T^2& \times & \times & \times & \times &-&-& \times & \times & \times & \times \\
\hline
\end{array}
$$
where a cross (dash) means that an object is pointlike (extended) in the corresponding dimension. 

An important aspect, which is generic in all models of inflation in string theory, is the stabilisation of the moduli fields, as well as other ingredients that are required either for consistency of the model or for making it realistic from a particle physics point of view\footnote{For example, a fully realistic model should account, in addition to inflation, for the standard model. This would enrich the set up with several stacks of D-branes sitting in different points of the internal space. See \cite{Kiritsis:2003mc,Lust:2004ks,Blumenhagen:2005mu} for D-brane particle physics models.}. As already mentioned, for a consistent compactification to occur, there must be a mechanism which induces a potential involving all the moduli of the theory so that they can be trapped in their respective minima. This can be obtained by considering non trivial RR backgrounds \cite{Dasgupta:1999ss,Giddings:2001yu,Kachru:2003aw}, as well as non-perturbative effects such as gaugino condensation or instantons \cite{Shifman:1987ia,Davies:1999uw,Novikov:1985ic,Shifman:1985ie} that take place in another\footnote{It is important for the gaugino condensation to take place in a different brane system than the FI one, since if this were not the case the volume modulus would be destabilised at the end of inflation where the D3 dissolves in the D7 \cite{Ganor:1996pe,Baumann:2006th,Baumann:2007ah}.} stack of D7-branes. 
Furthermore, there must be several stacks of D3-branes attached to the orientifold fixed planes in order to cancel their RR charges, as well as $\bar{\rm D}3$-branes used to uplift the anti-de Sitter vacuum to a de Sitter one \cite{Kachru:2003aw}, rendering the cosmological constant positive.

We now proceed to identify the several fields that arise from the reduction on the $K3$ manifold, trace the source of supersymmetry breaking and write down the effective Lagrangian which will be used to construct the cosmic string solutions of the model. 



\section{The effective four dimensional theory} \label{sec:p1-eff-4d-th}
The compactification of the model on $K3\times T^2/\mathbb{Z}_2$ preserves $\mathcal{N}=2$ supersymmetry, which is further broken to $\mathcal{N}=1$ by bulk 3-form fluxes that stabilise the closed string moduli \cite{Angelantonj:2003zx,Haack:2008yb}. A worldvolume flux on the D7 further breaks $\mathcal{N}=1$ supersymmetry spontaneously, resulting in the slow roll phase of inflation. Therefore, the action can be organised in terms of $\mathcal{N}=1$ superfields. $\mathcal{N}=2$ supersymmetry, ensures that once the matter content of the theory is specified, the K\"ahler potential and the superpotential can be uniquely defined. 

The moduli fields that arise from the closed string sector are
%
%
the $K3$ volume modulus, the $T^2$ complex structure modulus and the axion-dilaton modulus \cite{Andrianopoli:2003jf,D'Auria:2004qv,Angelantonj:2003zx}. These are respectively denoted as
\be \label{kahler-s}
s = {\rm Vol}(K3) - i C_{(4)} ,\quad
t = \frac{g_{12}}{g_{11}} + i \frac{\sqrt{\det g}}{g_{11}} ,\quad
u = C_{(0)} - i e^\phi,
\ee
where the $2\times 2$ matrix $g$ denotes the metric on the torus $T^2$, $C_{(4)}$ is the scalar that arises from the 4-form, when it has four legs along the $K3$ surface, and ${\rm Vol}(K3)$ represents the volume of $K3$.
The dimensional reduction on $K3$ may be described by a cubic K\"{a}hler potential of the form \cite{Haack:2008yb}
\begin{equation} \label{k2}
K=-\ln\Big\{-8{\rm Re}(s){\rm Im}(t){\rm Im}(u)-\frac{1}{2}{\rm Im}(u)\big({\rm Im}(y_3)\big)^2\Big\}.
\end{equation}  
%
%
%

The open string sector contributes another set of supermultiplets. The positions of the D3-brane on $T^2$ are parametrised by a complex scalar $y_3 = x_4 + i x_5$, and together with the four real scalars $\zeta_1=x_6+ix_7$ and $\zeta_2=x_8+ix_9$, corresponding to the four coordinates of the D3 on the $K3$ surface, are organised into additional hypermultiplets. These are respectively, the inflaton field (the real part of $y_3$), and the waterfall fields that produce the tachyonic instability signalling the end of inflation. Finally, there is the vector multiplet on the D7-brane associated with the FI term. 
The moduli $(t,u)$ and other open string moduli parametrising the positions of brane stacks can be stabilised by bulk 3-form fluxes, which give a contribution $W_0$ to the superpotential. 
Following \cite{D'Auria:2004qv,Angelantonj:2003zx,Haack:2008yb}, the $\mathcal{N}=2\rightarrow\mathcal{N}=1$ breaking is induced by stabilising the complex structure and the axion-dilaton multiplets at vacuum expectation values $t=u=-i$, hence, the K\"ahler potential \eqref{k2} reads
\begin{equation} \label{kalher-before-gs}
K =-\log\Big(4(s + \bar{s}) + \frac{1}{8} (y_3-\bar{y}_3)^2 \Big).
\end{equation}

Therefore, upon integrating out the relevant multiplets, we are left with the following $\mathcal{N}=1$ fields 
\begin{equation} \label{final-fields}
(A_m,\lambda_a,D);(y_3,\Psi_3,F_3);(\zeta_1,\Psi_1,F_1);(\zeta_2,\Psi_2,F_2);(s,\chi_a,F_s).
\end{equation}
The K\"ahler modulus $s$ can be stabilised by gaugino condensation taking place on a stack of D7-branes. The non perturbative superpotential underlying this mechanism is
\begin{equation}
W = W_0 + W_{np} = W_0 + A(t,u,y_3,\zeta_i) e^{-cs} \label{sstabilizing},
\end{equation}
where $c$ is a positive constant, $W_0$ is the constant, flux-induced superpotential, and $A$ is some suitable function, which in principle may depend on any matter field in the theory \cite{Haack:2008yb}. An important consequence of this construction is that the inflation mechanism is no longer of pure $D$-term type, since now the inflaton, which is a position modulus, is involved in the volume stabilising superpotential \eqref{sstabilizing}, which gives rise to $F$-term contributions in the scalar potential. This might lead to serious obstructions to inflation, $i.e.$ the $\eta$ problem where the inflaton gets a large mass of order of the Hubble parameter. The shift symmetry of the K\"ahler potential \eqref{k2}, as far as the inflaton field is concerned\footnote{Recall that the inflaton is given by the real part of $y_3$ which cancels out in \eqref{k2}. What is meant by a shift symmetry is the fact that $K \left( {\rm Re} (y_3) + c \right) = K \left( {\rm Re} (y_3) \right)$. }, is crucial as it protects the flatness of the inflaton direction in the potential. However, a contribution to the inflaton mass from the non perturbative superpotential is unavoidable but at least tunable. As shown in \cite{Haack:2008yb}, this symmetry survives the partial SUSY breaking and protects the inflaton from developing a large mass, even when quantum corrections are taken into account.

\subsection{The Fayet-Iliopoulos term}
The FI term of the model is sourced by a constant flux $\cal F$ on the worldvolume of the D7-brane, as can be seen by writing the action of the $D3/D7$ system and integrating over the compact space to obtain the four dimensional effective theory \cite{Dasgupta:2002ew,Dasgupta:2004dw}. Let us argue that this construction implies that the $U(1)_{FI}$ symmetry associated with the FI term is anomalous, an observation which will be crucial for the cosmic strings of the model.

  
 
 In compactified models of string theory, FI terms are generated from the Green-Schwarz (GS) mechanism \cite{Green:1984sg}, which in four dimensions contributes a term like $C \wedge F $ in the worldvolume action \cite{Burgess:2003ic}. The field $C$ is either the RR 2-form $C_2$, or a 2-form coming from the dimensional reduction of $C_4$, and $F$ the 2-form field strength. In type IIB string theory, such a term arises from the dimensional reduction of the Chern-Simons piece of the worldvolume action. Let us see how it is generated in the $D3/D7$ model with a constant flux along the $K3$ directions.

The Chern-Simons part of the D7 woldvolume action reads
\be \label{G-S-term}
\int_{\mathcal{M}_8} C \wedge [e^{F}],
\ee
where the notation means that the exponential is expanded and for each contribution of order $n$ in the 2-from, $C$ is a RR $(8-2n)$-form, so that the whole term is an 8-form integrated over the worldvolume of the D7-brane. We thus obtain a contribution $C_{K\Lambda MN}F_{P\Sigma}F_{TY}\epsilon^{K\Lambda MNP\Sigma TY}$ in ten dimensions. Placing the indices of the forms on the internal manifold as $C_{\kappa\lambda m n}F_{\rho\sigma} F_{ty}\epsilon^{\kappa\lambda mn \rho \sigma ty}$, where $m,n,t,y$ run along the $K3$ directions and $\kappa,\lambda,\rho,\sigma$ along the non compact spacetime, and integrating over the $K3$, we obtain the desired term in four dimensions, $i.e.$ $\int dx^4 C_{\kappa\lambda}F_{\rho\sigma} \epsilon^{\kappa\lambda \rho\sigma}$. 

Upon integrating by parts and taking the Hodge dual of the 3-form field strength, this term reduces to an axionic coupling of the form $\int dx^4 \partial_\mu \phi A^{\mu}$, where $\phi$ is the dual scalar of $C_{\mu\nu}$. Now recall that the RR 4-form must be self-dual in ten dimensions, a constraint which implies that the scalar $\phi$ should be the same degree of freedom as the scalar that arises in four dimensions when $C_4$ has all its legs along the internal space. That is the axionic partner $C_{(4)}$ of the volume modulus ${\rm Vol}(K3)$, which is the imaginary component, $s_I$, of the complex scalar $s$ in \eqref{kahler-s}; so $\phi = s_I$. 

Under a gauge transformation, $A_\mu \to A_\mu +  \partial_\mu \lambda$, the axion shifts by $s_I \to s_I +  \delta_{\rm GS} \lambda$, with $\delta_{\rm GS}$ a constant, so the gauge invariant combination is $\partial_\mu s_I - \delta_{\rm GS}A_\mu$. Supersymmetry then implies that the K\"ahler potential \eqref{kalher-before-gs} should read 
\begin{equation} \label{kahler-pot-gs}
K = - \log\Big( 4 (S + \bar{S}) + (Y_3 - \bar{Y}_3)^2 - 4 \delta_{\rm GS} V \Big),
\end{equation}
in order to maintain supergauge invariance. This extension produces an FI term in the four dimensional superspace Lagrangian, which reads 
\be \label{field-dep-fi}
\int d^4\theta \frac{\partial K}{\partial V}\Big|_{V=0} V = \frac{\delta_{\rm GS}}{S + \bar{S}} V.
\ee

This axion couples to the field strength $F$ as $s_I \epsilon^{\mu\nu\rho\sigma}F_{\mu\nu}F_{\rho\sigma}$, thus contributing to the gauge anomaly a term $\delta_{\rm GS} \lambda \epsilon^{\mu\nu\rho\sigma}F_{\mu\nu}F_{\rho\sigma}$, which needs to be somehow compensated for, since we want an anomaly free theory. As shown in \cite{AlvarezGaume:1983ig}, when the trace of the $U(1)$ charge operator on the matter sector does not vanish, the theory has mixed gravitational and gauge anomalies which are proportional to this trace. Therefore, if the coefficient $\delta_{\rm GS}$ is appropriately fixed, they may serve as the desired counterterm. In \cite{Atick:1987gy}, it was shown that the suitable form of the FI term is
\begin{equation} \label{fi-gs}
\xi = \delta_{\rm GS} g^2 M_{\rm Pl}^2, \;\;  \delta_{\rm GS} = \frac{{\rm tr} Q}{192\pi^2}.
\end{equation}

In order to see that for the model under consideration the trace of the generator of the $U(1)_{FI}$, under which the waterfall scalars are charged, is indeed non vanishing, it is instructive to perform a T-duality.
The constant flux $\cal F$ that spans the $K3$ directions can be written as
\begin{equation}
A_7=x^6\mathcal{F}_{67},\; A_6=x^7\mathcal{F}_{76},\;A_8=x^9\mathcal{F}_{98},\;\text{and}\; A_9=x^8\mathcal{F}_{89},
\end{equation}
and T-dualising along directions $x^6$ and $x^8$, we obtain 
\begin{equation} \label{t-dual-coordinates}
x'^6=2\pi \alpha'A^6=2\pi\alpha'x^7\mathcal{F}_{76}\quad\text{and}\quad x'^8=2\pi \alpha'A^8=2\pi\alpha'x^9\mathcal{F}_{98} .
\end{equation}
As we saw in Sec.~\ref{string-primer}, T-dualities along tangential directions of a D-brane reduce its dimensionality while transversal T-dualities increase it, thus duality along $x^6,x^8$ will yield a $D5/D5$ system. 
From the form of the dual coordinates \eqref{t-dual-coordinates}, we see that the resulting brane system will be tilted by an angle $\theta_1=\tan^{-1}(2\pi\alpha'\mathcal{F}_{76})$ in the plane $x^6,x^7$ and
by an angle $-\theta_2=\tan^{-1}(2\pi\alpha'\mathcal{F}_{89})$ in the plane $x^8,x^9$. Complexifing the positions, $\zeta_1=x^6+ix^7$ and $\zeta_2=x^8+ix^9$, we can write the rotations on the planes $x^6,x^7$ and $x^8,x^9$ in the form of $U(1)$ transformations, $\zeta_1'=e^{i\theta_1}\zeta_1$ and $\zeta_2'=e^{-i \theta_2}\zeta_2$. From this form, one can identify the angles $\theta_1,\theta_2$ as the charges of these fields under the $U(1)_{FI}$. Equation \eqref{fi-gs} implies that the FI term should be $\xi \propto {\rm tr}Q = \theta_1 - \theta_2$, and this is indeed the FI term computed in \cite{Dasgupta:2002ew,Dasgupta:2004dw} for the $D3/D7$ model. This is consistent with the analysis of \cite{Berkooz:1996km}, where it was shown that for $\theta_1 \neq \theta_2$, this
configuration of branes intersecting at angles does not preserve supersymmetry. Since T-duality does not affect the supersymmetry of the set up, this picture is actually equivalent to the $D3/D7$ system.

From a four dimensional point of view, the $K3$ coordinates $x^6,x^7,x^8,x^9,$ appear as the waterfall scalar fields \cite{Dasgupta:2002ew}, which produce a tachyonic instability of the inflationary vacuum, driving the inflaton field $\phi$ towards a new vacuum at $\phi = 0$. Therefore, the above discussion implies that the waterfall fields are charged under the anomalous $U(1)_{FI}$ which breaks at the end of inflation, where cosmic strings are expected to form. In summary, the ten dimensional term \eqref{G-S-term} yields a four dimensional axion, which produces a field dependent FI term \eqref{field-dep-fi} and contributes to the anomaly; this contribution is counterbalanced by the non vanishing trace of the $U(1)_{FI}$ charges of the matter sector comprising the waterfall fields that couple to the cosmic superstrings. Such symmetries are usually referred to as \emph{pseudo-anomalous}, to indicate the fact that the theory is eventually anomaly free.

%

We are now in a position to write down the four dimensional Lagrangian that we will use to construct the cosmic superstrings of this model. The fields that we have, as listed in \eqref{final-fields}, are the two chiral multiplets charged under the anomalous symmetry $U(1)_{FI}$, which we denote as $Z_i(\zeta_i, \psi_{i \alpha}, F_i)$; a chiral superfield $S(s, 2 s_R \chi_\alpha, F_s)$ containing the K\"ahler modulus $s$ given in \eqref{kahler-s}, with the real part of $s$, $s_R = {\rm Vol}(K3) = g^{-2}$ giving the effective four dimensional coupling; a gauge multiplet $V(A_\mu, s_R^{- \frac{1}{2}} \lambda_\alpha, D)$ which is the vector superfield associated with the FI term. Since we will focus on cosmic string solutions at the end of inflation, the inflaton chiral multiplet will be suppressed in what follows. According to the rules of Sec.~\ref{sec:superspace}, the supersymmetric Lagrangian for such a collection of fields may be written as 
\begin{eqnarray}
\label{BDDaction}
{\cal L} & = & \left ( Z_i^\dag e^{ 2 q_i V} Z_i + {\cal K}(S,\bar{S}) \right ) \Big|_{\theta^2 \bar \theta^2} + \left (\frac{1}{4} S  W^\alpha W_\alpha + W(Z_i, S) \right) \Big|_{\theta^2} + \text{h.c.},
\end{eqnarray}
where ${\cal K}(S,\bar{S})$ is the K\"ahler function given by an expansion in terms of derivatives of the K\"ahler potential \eqref{kahler-pot-gs}. The charges of the waterfall fields $\zeta_i$ are denoted as $q_{1,2} = \theta_1,-\theta_2$. Evaluating the bosonic terms in \eqref{BDDaction} one by one we obtain
\be
\label{kineticterms+strength}
\begin{split}
& Z_i^\dag e^{ 2 q_i V} Z_i \Big|_{\theta^2 \bar \theta^2} = |D_{\mu} \zeta_i|^2  - |F_i|^2 - q_i D |\zeta_i|^2 ,\\ 
& \frac{1}{2}S W^\alpha W_\alpha \Big|_{\theta^2} = \frac{s_R}{4} F_{\mu \nu} F^{\mu \nu} - \frac{s_I}{4} F_{\mu \nu} \tilde F^{\mu \nu} + \frac{s_R}{2} D^2 .
\end{split}
\ee
The contribution from the K\"ahler function may be obtained as 
\be
\label{kuzenko}
{\cal K} = - K_{i \bar j} \left ( \partial ^ \mu \bar \phi^{\bar j} \partial_\mu \phi^i - \bar F^{\bar j} F^i \right ),
\ee
where
\be
K_{i \bar j} \equiv \frac{\partial^{2}K(\phi,\bar \phi)}{\partial \phi^{i} \partial \bar \phi^{\bar j}},
\ee
with $\phi$ and $\bar \phi$ being the bottom components of the chiral fields $\Phi$ and $\bar \Phi$, and with $K(\phi, \bar \phi)$ denoting the K\"ahler potential. Once this formula is used for the potential \eqref{kalher-before-gs}, the correction \eqref{kahler-pot-gs} may be implemented by the following substitutions:
%
\be
\partial_\mu s  \rightarrow  \partial_\mu s - \frac{i}{2} \delta_{\rm GS} A_\mu, \quad
  \Box s  \rightarrow  \Box s - \frac{1}{2} \delta_{\rm GS} D, \quad {\cal K}  \rightarrow  {\cal K} + \frac{\partial K}{\partial V}\Big|_{V=0} V.
\ee
Therefore, the bosonic contribution from the K\"ahler function reads
\be
\label{kahlerpotential}
{\cal K} = \frac{1}{16 s_R^2} ( \partial_\mu s_R)^2 + \frac{1}{16 s_R^2} ( \partial_\mu s_I - \frac{i}{2} \delta_{\rm GS} A_\mu)^2 - \frac{1}{16 s_R^2} |F_s|^2 .
\ee
Combining all the contributions, the bosonic part of the Lagrangian \eqref{BDDaction} reads 
\begin{eqnarray}
\label{bos-action}
{\cal L}_{\rm bos} &\!\! = \!\! & \nn |D_{\mu} \zeta_i|^2  + \frac{1}{16 s_R^2} ( \partial_\mu s_R)^2 + \frac{1}{16 s_R^2} ( \partial_\mu s_I - \frac{i}{2} \delta_{\rm GS} A_\mu)^2 + \frac{s_R}{4} F_{\mu \nu} F^{\mu \nu} - \frac{s_I}{4} F_{\mu \nu} \tilde F^{\mu \nu} \\ && + V_D + V_F ,
\end{eqnarray}
where the $D$-term and $F$-term potentials read 
\be \label{d-term}
V_D = -\frac{1}{2s_R} \left( \theta_1|\zeta_1|^2-\theta_2|\zeta_2|^2 + \frac{\delta_{\rm GS}}{s_R} \right)^2 \quad \text{and} \quad V_F = - |F_i|^2 - \frac{1}{16 s_R^2} |F_s|^2,
\ee
with $F^{i} = \frac{\partial W}{\partial \phi_{i}}$, $W$ the superpotential given in \eqref{sstabilizing} and $\phi_i$ with $i = 1,2,s$, denoting the scalar component of the corresponding chiral superfield that $F_i$ belongs to.
The transformation laws for the fermions that leave the total action invariant are
\be \label{fermions}
\begin{split}
& \delta \psi_{1,2a} = \sqrt{2} F_{1,2} \kappa_a + i\sqrt{2} \sigma^\mu \bar{\kappa}_a D_\mu \zeta_{1,2}, \\
& \frac{1}{\sqrt{s_r}} \delta\lambda_a = i D \kappa_a + \frac{1}{2} \sigma^\mu \bar{\sigma}^\nu \kappa_a  F_{\mu\nu}, \\
& 2s_R \delta\chi_a = \sqrt{2} F_s \kappa_a + i\sqrt{2} \sigma^\mu \bar{\kappa}_a (\partial_\mu s - \frac{i}{2} \delta_{\rm GS} A_\mu). 
\end{split}
\ee
These transformations will be used in what follows to deduce the supersymmetry properties of the cosmic superstrings which we now describe.

\section{Constructing the cosmic superstring solutions} \label{sec:cs-sols-paper-1}

Cosmic superstrings are expected to form at the end of brane inflation, upon breaking of a $U(1)$ symmetry \cite{Sarangi:2002yt} and as we have previously shown, the $U(1)_{FI}$ that breaks at the end of $D3/D7$ inflation is a pseudo-anomalous one. Cosmic superstrings from pseudo-anomalous $U(1)$'s, although resulting from the breakdown of a local symmetry, are expected to have global properties \cite{Harvey:1988in,Casas:1988pa}, in the sense that their energy is not confined to the string core and long range interactions are induced. This is because these strings couple to axion fields and the gauge field cannot cancel the contributions of both axions and Higgs fields to the string energy. As a result, they decay soon after their formation. 

At first sight, this would reconcile the cosmic superstrings of the model with observational data, circumventing the need to further complicate it by $e.g.$ adding more branes in order to make the strings semilocal \cite{Vachaspati:1991dz,Dasgupta:2004dw}. As pointed out though in Sec.~\ref{sec:p1-eff-4d-th}, the superpotential \eqref{sstabilizing} used to stabilise the volume modulus, depends on other matter fields of the theory through the function $A$. Therefore, on a string background, where the waterfall scalars vary in space, the vacuum expectation value of $s$ will have a spatial dependence as well. Such a configuration alters the nature of the strings yielding their energy confined to the core, so that the long range interactions initially expected do not occur \cite{Davis:2005jf}. Defects of this form are referred to as \emph{local axionic strings}. 

The Lagrangian \eqref{bos-action} of the low energy theory of the $D3/D7$ model is the same as the one found in \cite{Davis:2005jf}, where it was shown that it contains local axionic strings. Following this analysis, 
the $D$-term potential \eqref{d-term} is minimised by
\begin{equation}
s_R = \infty \quad \text{or} \quad |\zeta_1| = 0 , \quad |\zeta_2| = \eta , \quad s_R = \frac{\delta_{\rm GS}}{\theta_2\eta^2},
\end{equation}
with $\delta_{\rm GS}$ given by \eqref{fi-gs}. Our cosmic string ansatz thus reads 
%
%
%
\be \label{stringansatz}
 \zeta_1  =  0 , \quad
 \zeta_2  =  \eta f(r) e^{i n \varphi}, \quad
 s  =  \frac{\delta_{\rm GS}}{\theta_2 \eta^2 \gamma(r)^2} + 2 i n \delta_{\rm GS} \varphi, \quad
 A_\varphi  =  n \frac{u(r)}{r},
\ee
with the asymptotic behaviour $$f(0) = u(0) = 0, \; \gamma(0)\sim 1/2,  \quad\text{and}\quad f(\infty) = u(\infty) = \gamma(\infty) = 1. $$

%

Let us now study the properties of such defects. Firstly, in order to see if the strings preserve supersymmetry, one may perform a supersymmetric variation of the fermionic sector according to the transformations \eqref{fermions}. Then one may check for zero modes travelling along the string. Presence of such zero energy solutions may put severe constraints in the model, since if these modes are chiral, $i.e.$ they are either right or left movers, they may stabilise a closed string loop via angular momentum conservation and form a vorton \cite{Davis:1988ij}. Stable vortons are catastrophic, since they dominate and overclose the universe soon after their formation. In order for such modes to correspond to physical states they must be well behaved \cite{Jackiw:1981ee} in the two regimes $r \to 0$ and $r \to \infty$, with $r$ the distance from the string core. 

Inserting the ansatz \eqref{stringansatz} into the transformations \eqref{fermions}, we obtain 
\be \label{fermi-trans}
\begin{split}
& \delta \psi_{1\alpha} = \sqrt{2} F_1 \kappa_\alpha , \\
& \delta \psi_{2\alpha} = \sqrt{2} F_2 \kappa_\alpha + i \sqrt{2} \eta \kappa_a^* e^{i(n\mp 1)\varphi} \left( f'\pm n\frac{f}{r} (1 - \theta_2 u) \right), \\
& \delta \lambda_\alpha = i \sqrt{\frac{\delta_{\rm GS}}{\theta_2 \eta^2 \gamma^2}} \kappa_\alpha \left( \frac{\theta_2^2 \eta^4 \gamma^2 }{\delta_{\rm GS}} \left( f^2 - \gamma^2 \right) \mp \frac{n}{r} u'  \right), \\
& \delta \chi_\alpha = \frac{1}{\sqrt{2}}  \frac{\theta_2 \eta^2}{\delta_{\rm GS}} \left( F_s \kappa_\alpha - 2 i \delta_{\rm GS} \kappa_\alpha^* e^{\mp i\varphi} \left( \frac{\gamma'}{\theta_2\eta^2\gamma} \mp  n \gamma^2 \frac{1-u}{r} \right) \right),
\end{split}
\ee
where the upper (lower) signs correspond to $\alpha = 1(2)$.
Since the variation of the vacuum configuration is non vanishing, supersymmetry is completely broken on the vortex background. Now, in order for the fermions $(\psi,\lambda,\chi)$ to correspond to physical states we may look at their behaviour near the string core at $r \to 0$. The model \eqref{bos-action} with the ansatz \eqref{stringansatz} was solved numerically in \cite{Davis:2005jf} for a standard racetrack superpotential \cite{Krasnikov:1987jj} of the form $$W_{np} = \zeta_1 \left[ h_1\left(\frac{\zeta_2}{\eta}\right)^{n_1} e^{-\frac{3 s}{2N_1}} -h_2 \left(\frac{\zeta_2}{\eta}\right)^{n_2} e^{-\frac{3 s}{2N_2}} \right],$$ arising from gaugino condensation. At large $r$, the fields fall exponentially, while inside the string core the ansatz reads
\be \label{small-r-sol}
f \sim C r^{|n|} \;,\;\; u \sim \frac{r^2}{1-2 |n| C^2 \eta^2 \log r} \;,\;\; \gamma \sim \frac{C}{\sqrt{1-2 |n| C^2 \eta^2 \log r}}, 
\ee
where the constant $C$ satisfies $C \sim 1 - |n| \eta^2 \log \left( \frac{M_D}{M_F}\right)$, with $M_D,M_F$ the masses of the scalars $\zeta_2$ and $s$ arising from the $D$-term and $F$-term potential respectively, which set the inner and outer core of the string $r_D \sim M_D^{-1},\;r_F \sim M_F^{-1}$. Plugging these solutions into \eqref{fermi-trans}, we see that all the modes behave properly at zero apart from the $\chi$ fermion. Its equation reads
\be 
\lim_{r \to 0} \chi_\alpha = \frac{1 \mp {\rm sgn} (n)}{-\eta r \log r},
\ee
with sgn denoting the sign function. We thus see that only one mode is normalisable, either $\chi_{1}$ for $n>0$ or $\chi_{2}$ for $n<0$, resulting in a chiral degree of freedom. 

In \cite{Davis:2005jf}, the constraints for vorton formation where shown to be evaded for $\frac{M_F}{ M_D} < 10^{-2}$ which is trivially satisfied due to the exponential suppression of the $F$-term. Therefore there is no vorton formation in the model, leaving us with the standard constraints from the cosmic superstring tension. As is common in $D$-term inflation, cosmic superstrings have a tension $\mu \sim \xi$, where $\xi$ is the FI term that sets the inflationary scale, so the {\sc Planck} bound reads $ G \xi < 10^{-7}$ \cite{Ade:2013xla}, with $G=\frac{1}{8 \pi M_{\rm Pl}^2}$. As we argued, the FI term is given by the GS parameter \eqref{fi-gs} so that the previous constraint reads $g^2 \delta_{\rm GS} < 10^{-7}$. A typical value for $\delta_{\rm GS}$ is $\sim 1/10$ \cite{Dine:1987xk,Dine:1987gj} so that the tension bound is only satisfied if $g < 10^{-3}$. Such a value leads to a spectral index of order one \cite{Haack:2008yb} which is observationally disfavoured \cite{Ade:2013uln}. 

The model can be made compatible with observations at the price of complicating its structure by either adding extra branes in order to make the strings semilocal \cite{Vachaspati:1991dz,Hindmarsh:1991jq} as in \cite{Urrestilla:2004eh,Dasgupta:2004dw,Chen:2005ae}, since the upper bound on the tension of semilocal strings is higher than for local abelian strings \cite{Urrestilla:2007sf}; or by suppressing the string production by taking higher order corrections to the K\"ahler potential into account \cite{Seto:2005qg,Seto:2007zzb,BasteroGil:2006cm}. As we will now discuss, there is however another problem which seems to obstruct the inflationary process. 

\section{Obstruction to consistent moduli stabilisation}
In analogy with the standard field theory procedure, where a gauge field must couple to a current of matter particles for the action to be gauge invariant, when one makes supersymmetry local to obtain a supergravity theory, one has to construct a supercurrent multiplet that couples to the gravity multiplet. This is the Ferrara-Zumino (FZ) current multiplet \cite{Ferrara:1974pz} which contains the energy momentum tensor, the Noether current associated with sypersymmetry and the R current as component currents. In \cite{Komargodski:2009pc,Komargodski:2010rb}, it was argued that the FZ multiplet fails to be gauge invariant in presence of a constant FI term. The authors then proposed a new superfield compatible with gauge invariance, the $S$ multiplet, which contained extra degrees of freedom compared with the FZ one, arriving at the following statement: in a supergravity theory, the FI term must be field dependent and the moduli space must be non compact\footnote{See also \cite{Dienes:2009td} for an independent study on FI terms in supergravity.}. These extra degrees of freedom in the $S$ multiplet, which are the K\"ahler moduli of the compactification manifold, are crucial since they are the degrees of freedom underlying the resolution of both constraints. 

Let us now see how this discussion is adapted in the case of the model under consideration.
Roughly speaking the situation is as follows:
the moduli space of the theory, to a first approximation, is the internal space probed by the mobile D3-brane, $i.e.$ $K3 \times T^2$. Although this is a compact space, the moduli fields fibre over the $K3$ base rendering the space non-compact, 
 in accordance with the aforementioned constraint. Moreover, the FI term \eqref{fi-gs} is field dependent since it is proportional to the low energy coupling parameter which in turn is given by the real part of the K\"ahler modulus $s$ in \eqref{kahler-s}. Once the modulus is stabilised, both of these characteristics are lost. The $S$ multiplet reduces to the FZ multiplet and the theory becomes inconsistent.

The only case in which the K\"ahler fields can be frozen to their vacuum expectation values, is if one first fixes the open string moduli. The reason is that the moduli space of the worldvolume theory on the D3 then degenerates to a point (since its positions are frozen) and compactness loses its meaning. In the inflationary case, where the open string moduli contain the light inflaton, this means that the volume modulus may only be stabilised after inflation ends, when the open string moduli are heavy. This implies a contradiction: on the one hand, the K\"ahler moduli must be heavy and fixed in their vev's, as they give rise to an FI term which should be constant in order to break SUSY and start inflation, and on the other hand, the K\"ahler moduli should be lighter than the inflaton during slow roll for consistency of the model. 
Therefore, the $D3/D7$ model, at least in its present form, is inconsistent. This argument extents to any $D$-term inflationary model in a string theory context where moduli stabilisation proceeds along the same lines. However, this does not mean that $D$-term models are problematic in general but that moduli stabilisation should be performed in a way that respects the coupling to SUGRA. 
An example of a model evading the above argument is the so called fluxbrane inflation \cite{Hebecker:2011hk,Hebecker:2012aw}, where the moduli are stabilised above the SUSY breaking scale. 

Having discussed the properties of cosmic superstrings in a $D$-term model and having seen that these models have subtleties as far as moduli stabilisation is concerned, we now shift our focus to the study of observational signatures of cosmic superstrings in warped backgrounds in the context of brane-antibrane inflation, which gives rise to an $F$-term inflaton potential.

\chapter{Radiative processes of cosmic superstrings on warped backgrounds} \label{ch:paper-2}

Cosmic superstrings, $i.e.$ fundamental strings and D1-branes, or F- and D-strings for short, are sourced by NSNS and RR 2-forms, which give rise to scalar fields (axions), when placed on a compact internal space. These axions naturally couple to these objects and hence, strings can radiate scalar particles. In this Chapter we will study their dominant decay channel, 
focusing on $F$-term models on warped backgrounds. A primary example of this class is the $\mathds{K}$L$\mathds{M}$T model \cite{Kachru:2003sx}, where a slow roll phase corresponds to the movement of a D3-brane towards a $\bar{\rm D}3$-brane. The necessity of warping is due to two reasons. First, as already mentioned in the Introduction, it lowers the tension of superstrings from the Planck scale down to phenomenologically acceptable values. In addition, the warping resolves the so called $\eta$-problem which is the observation that in the presence of $F$-terms, the inflaton field naturally acquires a mass of the order of the Hubble scale, obstructing the slow roll phase\footnote{As first pointed out in \cite{Kachru:2003sx}, the stabilisation of the K\"ahler modulus generically induces an additional mass term to the inflaton potential. This is the same problem that we mentioned in Sec.~\ref{sec:p1-eff-4d-th}, that the minimum of the potential will generically depend on open moduli fields such as the inflaton. Contrary to the $\eta$-problem, this obstruction may be cured by fine-tuning, $e.g.$ it may be cancelled by appropriate counterterms which arise as corrections to the minimal K\"ahler potential.}. As we will see in Sec.~\ref{sec:orientifold-constraints}, the warped geometry also places severe constraints on the possible forms of radiation by strings. 

In what follows, we start by reviewing the results of \cite{Firouzjahi:2007dp} for superstrings on warped backgrounds and then we consider superstrings in the $\mathds{K}$L$\mathds{M}$T inflationary model and study the possible forms of axionic radiation. We argue that the warped compactification drastically alters these results.

\section{Gravitational vs axionic radiation on warped backgrounds} \label{sec:hasan}
%
Let us first consider the decay of a cosmic D-string loop to RR scalar particles. The Einstein frame action for a D1-brane on a warped background of the form
\begin{eqnarray} \label{warped-metric}
ds^2 & = & h^2 \eta_{\mu \nu} dx^\mu dx^\nu + g^{(6)}_{mn} dy^m dy^n~, 
\end{eqnarray}
where $\mu,\nu \in \{0 - 3\}$ and $m,n \in \{4 - 9 \}$,
reads

%
\begin{eqnarray} \label{Sd}
S_{\rm D} & = & \frac{1}{2 \kappa_{10}^2} \int d^{10} x \sqrt{- g^{(10)}} \left ( R^{(10)}
- \frac{ g_s}{12} F_3^2\right ) - \frac{\mu_{1}}{g_s} \int dt dx \sqrt{-
  \gamma} + \mu_1 \int dt dx C_2~,
\end{eqnarray}
where $F_3$ is the 3-form field strength of the RR 2-form $C_2$ that couples to the string, $\gamma_{ab}$ is the induced metric on the string worldsheet and $\kappa_{10}^2 = \frac{(2 \pi \sqrt{\alpha'})^8}{4 \pi}$ is the ten dimensional gravitational constant. 
Performing a dimensional reduction and writing $g^{(4)}_{\mu\nu} = h^2 g_{\mu\nu} $ we have that $$\int d^{10} x \sqrt{- g^{(10)}} \to \int d^{4} x \int d^{6} y h^4 \sqrt{- g} \sqrt{g^{(6)}}, $$ $$ R^{(10)} = h^{-2} R + \ldots \quad \text{and} \quad F_3^2  \to  h^{-6} F_3^2 , $$ where all the quantities on the RHS are four dimensional. Written in terms of $g_{\mu\nu}$, the induced metric $\gamma_{ab}$ also picks up a factor of $h^2$. Focusing on the zero modes, we obtain a four dimensional effective theory
\begin{eqnarray}
S_{\rm D} & = & \frac{M_{\rm Pl}^2}{2} \int d^{4} x \sqrt{- g} \left ( R
- \frac{ \beta g_s}{12} F_3^2\right ) - \mu_{eff} \int dt dx \sqrt{-
  \gamma} + \mu_1 \int dt dx C_2~,
\end{eqnarray}
where
\be
\mu_{\rm eff} = h^2 \mu_1 g_{\rm s}^{-1}, \quad M_{\rm Pl}^2 = \frac{1}{\kappa_{10}^2} \int d^6 y \sqrt{g^{(6)}} h^2 (y)~, \quad \beta = \frac{\int d^6y\sqrt{g^{(6)}} h^{-2}(y)}{\int d^6 y \sqrt{g^{(6)}} h^2 (y)} ,
\ee
are the effective string tension, with $\mu_1 = (2 \pi \alpha')^{-1}$, and the four dimensional Planck scale respectively, while $\beta$ takes into account the different scaling of the gravitational and kinetic parts of the action with respect to the warp factor. 

Upon rescaling $C_2 \to \frac{2}{\sqrt{\beta g_s} M_{\rm Pl}} C_2$, this Lagrangian is of the same form as the one used in \cite{Vachaspati:1984gt,Vilenkin:1986ku} (see also \cite{Sakellariadou:1990ne,Sakellariadou:1991sd}) for field
theory cosmic strings, the difference being the contribution of the warp factor. So using these results, the power of radiation of RR particles reads \cite{Firouzjahi:2007dp}
\be 
\label{power-RR-g}
P_{\rm RR} = \frac{\Gamma_{\rm RR} \mu_1^2}{ \pi^2 g_{\rm s} \beta
  M_{\rm Pl}^2},
\ee
where $\Gamma_{\rm RR} \sim {\cal O}(50)$.

For an F-string sourced by the NSNS 2-form $B_2$, the relevant part of the Type IIB Einstein frame action reads 
\be
S_{\rm F} = \frac{1}{2 \kappa_{10}^2} \int d^{10}x \sqrt {- g^{(10)}} \left
[ R^{(10)} - \frac{1}{12 g_s} H_{(3)}^2 \right ] - \frac{\mu_{1}}{g_s} \int dt dx \sqrt{-
  \gamma} + \mu_1 \int d^2 \sigma B_{2}^{\rm NS},
\ee
where $H_3$ is now the 3-form field strength of the NSNS $B_2$ field. Reducing to four dimensions, we obtain
\be
S_{\rm F} = \frac{M_{\rm Pl}^2}{2} \int d^{4} x \sqrt{- g} \left ( R
 - \frac{\beta}{12 g_s} H_{(3)}^2 \right ) - \mu_{eff} \int dt dx \sqrt{-
  \gamma} + \mu_1 \int d^2 \sigma B_{2}^{\rm NS},
\ee
and the power of radiation of NSNS particles can be computed as previously, only now the rescaling of the $B$ field reads $B_2 \to \frac{2\sqrt{g_s} }{\sqrt{\beta} M_{\rm Pl} } B_2$, leading to
\be
\label{power-NS}
P_{\rm NSNS} = \frac{\Gamma_{\rm NS} \mu_1^2 g_{\rm s}}{ \pi^2 \beta
  M_{\rm Pl}^2},
\ee
where $\Gamma_{\rm NS}$ is a numerical factor of the same order as $\Gamma_{\rm RR}.$ We see that compared to the RR radiative power \eqref{power-RR-g}, this form of radiation is suppressed by $g_s$.
%
%

Finally, the power of gravitational radiation per solid angle from a cosmic string loop may be computed from the energy momentum tensor associated with the action as \cite{weinberg1972gravitation}
\be \label{gr-power}
\frac{d P}{d \Omega} = \frac{G \omega^2}{\pi} [T_{\mu \nu}^
  * (\omega, \vec{k}) T^{\mu \nu} (\omega, \vec{k}) -
  \frac{1}{2} |T^\nu _\nu ( \omega, \vec{k})|^2],
\ee  
where
\be  
  T_{\mu \nu}
(\omega, \vec{k}) =   \int^{\infty}_0 dt e^{i
  \omega t} \int d^3 x e^{- i \vec{k}  \vec{x}} T_{\mu \nu}
(\vec{x},t).
\ee
This computation was performed in \cite{Vachaspati:1984gt} and for our case the result will be modified by the presence of the warp factor as
\be  \label{Pg}
P_{\rm g} = \Gamma_{\rm g} G \left( \frac{h^2 \mu_1}{g_s} \right)^2.
\ee

We may now compare axionic to gravitational radiation to obtain
\be 
\label{RRratio}
\frac{P_{\rm RR}}{P_{\rm g}} = \Big(\frac{8\Gamma_{\rm RR}}{\pi
  \Gamma_{\rm g}} \Big ) \frac{g_{s}}{\beta h^4} \quad \text{and} \quad \frac{P_{\rm NSNS}}{P_{\rm g}} = \Big(\frac{8\Gamma_{\rm NS}}{\pi
  \Gamma_{\rm g}} \Big ) \frac{g_{s}^3}{\beta h^4}.
\ee
At first sight, this analysis asserts that although in a flat background, where $\beta = h =1$, axionic and gravitational radiation might be of the same strength, in presence of a warp factor $h \neq 1$, RR particle emission dominates the decay process. Nevertheless, there is a caveat in these arguments and that is the constraints that the warp factor places on the form fields that can exist in four dimensional spacetime.

Since it is more appropriate to place our discussion in the context of brane inflation, where cosmic superstrings are expected to appear naturally, we will first review the $D3/\bar{D}3$ model \cite{Kachru:2003sx}, which takes place on a compact version of the Klebanov-Strassler (KS) throat, as well as the construction of a $(p,q)$ string, a bound state of $p$ F-strings and $q$ D-strings, on such a geometry.

\section{Brane-antibrane inflation on a throat geometry}
The only warped background with all moduli stabilised that has been constructed so far is the Giddings-Kachru-Polchinski (GKP) compactification \cite{Giddings:2001yu} which may be roughly described as a (KS) throat, a \emph{deformed conifold} \cite{Klebanov:2000hb}, attached to a compact Calabi-Yau manifold. The conifold \cite{Candelas:1989js}, topologically resembles a cone geometry with an $S^2 \times S^3$ base, which may locally be thought of as the subspace of $\mathbb{C}^4$ defined by 
$$
w_1^2 + w_2^2 + w_3^2 + w_4^2 = 0.
$$
The singularity at the apex of the cone, where the $S^2$ and $S^3$ radii shrink to zero, may be smoothed by either blowing up the $S^2$ or the $S^3$ spheres. The deformed conifold corresponds to the stabilisation of the $S^3$ radius to a minimum finite value which may be described as the submanifold satisfying 
$$
w_1^2 + w_2^2 + w_3^2 + w_4^2 = z,
$$
with $z$ representing the volume modulus of the $S^3$ which is stabilised by a superpotential involving 3-form fluxes. This smooth non compact space is then glued to some Calabi-Yau manifold yielding a six dimensional compact space with a highly warped region (throat) which can be used as a background for phenomenological applications.

This setup is similar to the well known Randall-Sundrum (RS) models \cite{Randall:1999vf,Randall:1999ee} with the IR and UV cutoff branes replaced by the smooth geometry of the resolved conifold at the apex and the CY compact space respectively. It can be thus described by an $AdS_5$ space with a radial coordinate confined between $r_{\rm IR}$ and $r_{\rm UV}$ representing the IR and UV cutoff scales. In the same way the warping produces a hierarchy in the RS scenario by lowering the value of the higher dimensional Planck scale, the warping in the GKP compactification reduces the string tension by powers of $\frac{r_{\rm IR}}{R}$, $R$ being the AdS scale. The background 3-form fluxes that stabilise the IR radius of the $S^3$, satisfy
\be 
\frac{1}{ (2 \pi)^2 \alpha'} \int_{S^3} F = M \quad \text{and} \quad \frac{1}{ (2 \pi)^2 \alpha'} \int_{S^3} H = - K , \qquad M,K \in \mathbb{Z}, 
\ee
leading to a warp factor $\frac{r_{\rm IR}}{R} = e^{-\frac{2 \pi K}{3 g_s M}}$ \cite{Giddings:2001yu}, which can be very small if $K \ll g_s M$.
\begin{figure}[tbh] 
\begin{center}
\includegraphics[scale=0.4]{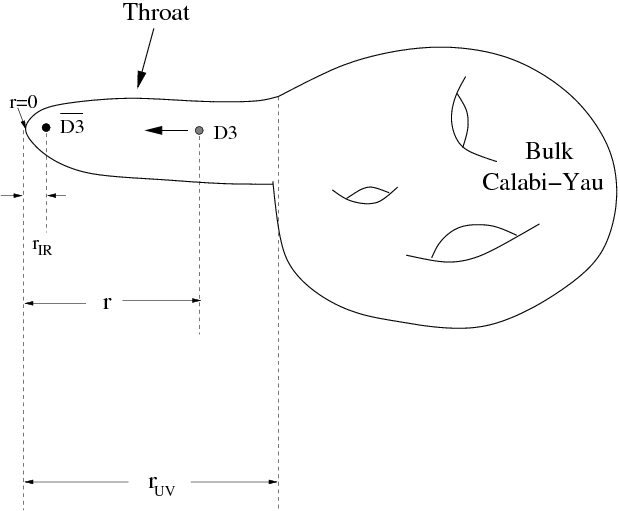}   
\end{center}
\caption[{\sf The $D3/\bar{D}3$ inflationary setup}.]{\sf The $D3/\bar{D}3$ inflationary setup. The KS throat is glued to a compact CY, with the $\bar{\rm D}3$-brane sitting at a finite distance from the tip and the D3-brane moving towards it.}
\label{fig:d3d3}
\end{figure}

The $D3/\bar{D}3$ model, depicted in Fig.~\ref{fig:d3d3}, consists of a $\bar{\rm D}3$-brane that sits at the tip of the KS throat which is the minimum of its potential. A mobile D3-brane is placed on the compact CY space and feels an attractive potential due to the antibrane of the form 
\be
V \sim 2 T_3 \frac{r_{\rm IR}^4}{R^4} \left( 1- \frac{r_{\rm IR}^4}{ r^4 }\right),
\ee
with $T_3 = \frac{1}{ (2 \pi)^3 g_s \alpha'^2}$ the D3-brane tension and $r$ the radial distance of the two branes, to be identified with the inflaton, $\phi = \sqrt{T_3} r$. This potential is quite flat for large values of the field and its movement along the throat corresponds to the slow roll phase of inflation which ends when the two membranes collide. 

Cosmic superstrings are expected to form at this point \cite{Copeland:2003bj} and they couple to gravitons and axion fields descending from the dimensional reduction of the Type IIB model on the internal manifold. For this case, the relevant construction is a $(p,q)$ string on the deformed conifold which is a bound state of $p$ F-strings and $q$ D-strings with the cases of F- or D-strings, obtained by setting $q=0$ or $p=0$ respectively. Let us now review this construction and see how the gravitational radiation power \eqref{Pg} gets modified. We will start by considering the $(p,q)$ string realised as a D3-brane wrapping a 2-cycle of the deformed conifold with suitable electric and magnetic fluxes \cite{Firouzjahi:2006vp} and then we will demonstrate that consistency of this set up, places severe constraints on the type of axionic radiation allowed in the context of $D3/\bar{D}3$ inflation.

\section{A $(p,q)$ string on a KS background}
A $(p,q)$ string in a KS throat can be constructed by wrapping a
D3-brane on a 2-cycle which is stabilized by
suitable fluxes \cite{Firouzjahi:2006vp}. The action is given by
\be \label{Spq}
S_{\rm D3} = - T_3 \int d^4 x \sqrt{- |m_{ab}|} + \mu_3 \int  \left( C_2 \wedge {\cal F} + \frac{1}{2}
C_0 {\cal F} \wedge {\cal F} \right)~,
\ee
where $m_{ab} = g_{ab} + {\cal F}_{ab}$, with ${\cal F}_{ab} = B_{ab} +  2 \pi \alpha' F_{ab}$, $\mu_3 = g_s T_3$ is the D3-brane charge and the integral and $a,b$ indices
run over the four-dimensional worldvolume $0,1,2,3$, with $2$ and
$3$ denoting the coordinates on the 2-cycle the D3-brane wraps. $C_2$ is the RR 2-form of the Type IIB superstring.
The metric is the same as in \eqref{warped-metric} while the
necessary fluxes are given by
\be \label{fluxes}
F_{23} = \frac{q}{2}, \quad 
\tilde F^{01} = - \frac{p}{4 \pi}~, \quad
B_{23} \neq 0, \quad B_{01} = 0 , \quad C_{01} = 0,
\ee
where $\tilde F^{\mu\nu}$ denotes the conjugate momentum of the electric
field and the integers $p,q$ number the NSNS and RR units of charge respectively. 

Let us now compute the power of gravitational radiation using \eqref{gr-power}.
Noting that only the DBI part of \eqref{Spq} is involved in the computation since the Chern-Simons term is topological, $i.e.$ it does not involve the metric and using the definition $$ T^{\mu\nu} = \frac{2}{\sqrt{-g}} \frac{\delta S}{\delta g_{\mu\nu}}, $$ and the Jacobi formula for the differentiation of a determinant $$\delta (\det A ) = \det A \; {\rm tr} \left( A^{-1} \delta A \right), $$ we obtain
\begin{eqnarray} \label{tmn}
       T^{\mu \nu}_{\rm DBI} & = &  T_3 \sqrt{ \frac{m}{g} } m^{\mu\nu} ,
\end{eqnarray}
where $m^{\mu\nu}$ is the inverse of the matrix
\be
m_{\mu\nu} = \left( 
\begin{array}{cccc}
- h^2 & - 2 \pi \alpha' F_{10} & 0 & 0 \\ 2 \pi \alpha' F_{10} & h^2 & 0 & 0 \\
0 & 0 & g_{22} & B_{23} + 2 \pi \alpha' F_{23} \\ 0 & 0 & - B_{23} - 2 \pi \alpha' F_{23} & g_{33}
\end{array} \right) ,
\ee
and $m,g$ denote the determinants of the respective matrices.

Now in order to check if this construction correctly captures the physics of a $(p,q)$ bound state one may compute the temporal component of the stress energy tensor which should match the $(p,q)$ string tension. From \eqref{tmn} we have that 
\begin{eqnarray}
  T^{00} & = &   T_3 h^4 \frac{(g_{22} g_{33} +
    {\cal F}_{23}^2)^{1/2}}{(h^4 - \lambda^2 F_{10}^2)^{1/2}} ,
\end{eqnarray}
which is in agreement with the Hamiltonian found in \cite{Firouzjahi:2006vp}. Upon minimising this energy on the KS background, one obtains \cite{Firouzjahi:2006vp}
\be
T_{(p,q)} = \frac{h^2}{2 \pi \alpha'} \sqrt{\frac{q^2}{g_s^2} +
  \left (\frac{bM}{\pi} \right )^2 \sin^2 \left (\frac{\pi (p - q
    C_0)}{M} \right ) } = \sqrt{T_{\rm D}^2 + T_{\rm F}^2}~,
\label{pq-gr-rad}
\ee
with $T_{\rm D}, T_{\rm F}$ denoting the tensions of the D-string and
F-string, respectively. 
This expression reduces to the flat space one,
$T_{(p,q)} = T_{F1} \sqrt{(q^2/g_s^2) +
  p^2}$~\cite{Schwarz:1995dk}, in the limit $M \to \infty$, $b = h = 1$ and $C_0 =
0$ \cite{Firouzjahi:2006vp}. From \eqref{Pg}, the gravitational power $P_{\rm g}$ for a $(p,q)$ string is given by 
\be  \label{Pg-pq}
P_{\rm g} = \Gamma_{\rm g} G T_{(p,q)}^2 = \Gamma_{\rm g} G \left( T_{\rm D}^2 + T_{\rm F}^2 \right).
\ee

We thus see that the DBI part of the action has the
same effect as the usual Nambu-Goto action, except that the tension
of the string is modified. 
%
Moreover, as evident from \eqref{Pg-pq}, the radiative power of a $(p,q)$ string is the same as that of an
F-string network and a D-string network considered separately.

\section{Constraints from the compactification} \label{sec:orientifold-constraints}
%
%
We now consider the compactification of
the KS solution given in \cite{Giddings:2001yu,Dasgupta:1999ss}, since this is the case for the $D3/\bar{D}3$ model. 

Let us start with the Type IIB action in the string frame 
\be \label{IIB-action}
\begin{split}
& S_{\rm IIB} = \frac{1}{2 \kappa_{10}^2} \int d^{10} x \sqrt{- g^{(10)}}  \left\{ e^{-2\phi} \left[ R^{(10)} +4 \left( \nabla\phi \right)^2 \right] - \frac{|F_1|^2}{2} - \frac{|G_3|^2}{2 \cdot 3!}  - \frac{|\tilde F_5|^2}{4 \cdot 5!} \right\} \\ 
& + \frac{1}{8 i \kappa_{10}^2} \int  e^{\phi}C_4 \wedge G_3 \wedge \bar{G}_3 + S_{\rm loc},
\end{split}
\ee
where $\kappa_{10}$ is the ten dimensional Newton's constant as in \eqref{Sd}, $G_3 = F_3 - \tau H_3$ with $\tau = C_0 + i e^{-\phi}$, $F_{n+1}$ denotes the field strength of the RR n-form $C_n$, while $H_3 = dB_2$ is the NSNS field strength. The 5-form $\tilde F_5$ is defined as $\tilde F_5 = F_5 -\frac{1}{2}C_2 \wedge H_3 + \frac{1}{2} B_2 \wedge F_3.$ The local part of the action $S_{\rm loc}$ contains contributions from any localised source, such as D-branes or orientifold planes, that might be present in the theory. 

Considering a warped geometry of the form \eqref{warped-metric}, in \cite{deWit:1986xg,Maldacena:2000mw,Giddings:2001yu}, it was shown that Einstein's equation can be written as
\be \label{warp-constraint}
\nabla^2_{(6)} h^4(y) = h^2 \frac{G_{mnk}G^{mnk}}{12 {\rm Im}\tau} + \frac{1}{4h^6} \left( \partial_m \alpha \partial^m \alpha + \partial_m h^4 \partial^m h^4 \right) + \frac{\kappa_{10}^2}{2}h^2 \left( T^m_m - T^\mu_\mu \right)_{\rm loc},
\ee
where $T_{\rm loc}$ is the energy momentum tensor derived from the variation of the local action $S_{\rm loc}$ with respect to the metric and $\alpha$ a function on the internal coordinates that contributes in the self dual 5-form $F_5$. Assuming for the moment that the RHS of \eqref{warp-constraint} is non negative and multiplying with $h^4$ and integrating over the internal compact manifold ${\cal M}$ we obtain 
\be 
\int_{{\cal M}} d^6y \sqrt{g^{(6)}} h^4 \nabla^2_{(6)} h^4 \geq 0,
\ee
so that after partial integration we have that
\be 
\int_{{\cal M}} d^6y \sqrt{g^{(6)}} \left(\nabla_{(6)} h^4 \right)^2 \leq 0.
\ee
The last inequality is only possible when the warp factor is constant which implies that the RHS of \eqref{warp-constraint} should also vanish. Therefore, the only way that non zero fluxes can exist, is in the presence of localised sources with negative tension so that $\left( T^m_m - T^\mu_\mu \right)_{\rm loc} < 0$, evading the above argument. Such localised branes are known as orientifold planes which represent subspaces fixed under the orientifold action \cite{Grimm:2004uq}
\begin{equation} \label{O-action}
{\cal O} =  (-1)^{F_{\rm L}} \Omega_p \sigma,
\end{equation}
where $\sigma$ is an isometric holomorphic involution\footnote{A map $\sigma$ is an involution if $\sigma^2= {\rm id}$. If it also preserves the metric it is called isometric and if it maps holomorphic functions to holomorphic functions it is holomorphic.}, $\Omega_P$ is the world-sheet
parity and $F_L$ is the space-time fermion number in the left-moving
sector. The involution $\sigma$ 
acts non trivially on the holomorphic 3-form\footnote{A $n$ complex dimensional CY manifold has a globally defined holomorphic form usually denoted as $\Omega$, which for our case is a 3-form. The action of the involution on this form is equivalent to its action on the CY internal space.} $\Omega$, as either $\sigma \Omega = - \Omega$ or $\sigma \Omega = \Omega$. 

The first choice leads to ${\cal O}3/{\cal O}7$ orientifold planes while the second implies the existence of ${\cal O}5/{\cal O}9$ ones. This can be easily understood from the dimensionality of the fixed points of the $\sigma$ action. Firstly, since the four dimensional Minkowski space is left invariant under the $\sigma$ action, the fixed subspace has spatial dimension at least three. Writing the holomorphic 3-form as $\Omega \propto dy^1 \wedge dy^2 \wedge dy^3$ and assuming that $\sigma$ is a reflection with respect to some complex coordinates $y$ we see that if we reflect one or three coordinates we get $\sigma \Omega = - \Omega$, while reflecting none or two coordinates leads to $\sigma \Omega = \Omega$. Now suppose that $\sigma$ is a reflection with respect to one complex dimension. The fixed subspace of the three complex dimensional internal manifold is thus a complex plane, $i.e.$ a four dimensional hypersurface. Adding the three non compact spatial dimensions to the four internal ones we obtain an ${\cal O}7$ plane. In analogy, reflecting all three complex coordinates leads to an ${\cal O}3$ plane while the cases of zero or two coordinates lead to the ${\cal O}5/{\cal O}9$ system. Since the GKP compactification, used in the $D3/\bar{D}3$ model, contains ${\cal O}3/{\cal O}7$ orientifold planes, in what follows we adopt the relevant $\sigma$ action. 

Let us now study the orientifold action on the various form fields in order to decide the spectrum. Under the worldsheet parity $\Omega_p$ the NSNS symmetric forms, $i.e.$ the metric and the dilaton $g_{\mu\nu}, \phi$ are even, while the antisymmetric form $B_2$ is odd. In the RR sector, we have the opposite situation since we need to take into account the exchange of worldsheet fermions. Since the RR 2-form is in the symmetric representation and the 0- and 4-forms in the antisymmetric one, $(\phi,g_{\mu\nu},C_2)$ are even while $(C_0,B_2,C_4)$ are odd under $\Omega_p$. The operator $(-)^{F_L}$ leaves the NSNS fields invariant and reverses the signs in the RR sector.
These operations are summarised in the following table: 
$$
\begin{array}[c]{c|cccccc}
 &C_0 & C_2 & C_4 & B_2 & g_{\mu\nu} & \phi \\
\hline\hline \Omega_p  & - & + & - & - & + & + \\
 (-)^{F_L}  & - & - & - & + & + & + \\
 \hline \Omega_p (-)^{F_L}  & + & - & + & - & + & +
\end{array}
$$
Therefore, in order for the $B_2$ and $C_2$ fields to be even under the combined orientifold action \eqref{O-action}, they should obey 
\begin{eqnarray} \label{sb2d2}
\sigma B_2 \, \, =\, \, - B_2
&&\ \ \mbox{and}\ \ \ \ \sigma C_2 \, \, = \, \, -
C_2.
\end{eqnarray}
Since the involution $\sigma$ leaves the four dimensional Minkowski space invariant, \eqref{sb2d2} implies that these 2-forms should have legs only in the internal manifold. We may thus conclude that in four dimensions, the NSNS and RR 2-forms that couple to the F- and D-strings respectively are projected out of the spectrum. This means that massless\footnote{See \cite{Dufaux:2012np} for massive particle production from cosmic superstrings.} radiation from these sources in a brane inflationary model on a warped background cannot be considered and gravitational radiation is the dominant channel\footnote{However, see \cite{Avgoustidis:2012vc} for an argument that gravitational radiation from cosmic superstrings is suppressed compared to that of field theory strings due to the presence of the extra dimensions.}. 

However, for a $(p,q)$ string which is actually a wrapped D3-brane
with fluxes, as in \cite{Firouzjahi:2006vp}, the situation may be
different. In order to see which fields that couple to the string worldvolume
are allowed, let us decompose the RR forms that couple to the D3-brane as
\be \label{2-4-decomp}
\begin{split}
& \quad \qquad B_2(x^M) = b_0(x^\mu) \alpha_2(y^m), \qquad C_2 (x^M) =  c_0(x^\mu) \beta_2(y^m),\\ 
& C_4(x^M) = c_2 (x^\mu) \wedge \gamma_2(y^m) + d_1(x^\mu) \wedge \gamma_3(y^m) + \tilde{d}_1(x^\mu) \wedge \tilde{\gamma}_3(y^m) + d_0(x^\mu) \gamma_4(y^m) ,
\end{split}
\ee
where the indices denote the order of the form fields, and the dependence on non compact or compact coordinates, $x^\mu$ and $y^m$ respectively, has been made explicit. We have also imposed the orientifold constraint \eqref{sb2d2} and neglected the components of the 2-forms having legs along the non compact dimensions.
According to the orientifold action, the only modes that are allowed are the scalars $b_0, c_0$ arising from the NSNS and RR 2-forms, provided that the corresponding internal 2-forms $\alpha_2,\beta_2$ are odd under the involution, as well as the $c_2$ arising from the RR 4-form, provided that $\gamma_2$ is in the even eigenspace of the involution $\sigma$. However, the analysis of Sec~\ref{sec:hasan} does not trivially apply for these modes, since in the Type IIB superstring the kinetic term for the 4-form mixes with the RR and NSNS 2-forms and their corresponding field strengths as in \eqref{IIB-action}. Therefore, in order to decide the axionic radiation, one has to solve the coupled equations of motion on the KS background and compute the power spectrum.

We have not mentioned the possibility of dilatonic radiation, since in the GKP compactification it is fixed by the background fluxes. 
Moreover, as shown in \cite{Frey:2006wv}, its wave function is highly localised in the throat, so that it may be consistently integrated out of the effective four dimensional theory. 

We now close this Part with a short summary and proceed to the study of scalar perturbations on general inflationary backgrounds.
\chapter{Summary of Part \ref{part:strings}}

In this Part, we placed ourselves in the UV complete framework of string theory and studied the cosmic superstrings produced at the end of inflation in two effective models, namely $D3/D7$ and $D3/\bar{D}3$ brane inflation.
In Ch.~\ref{ch:paper-1}, we showed that the supersymmetry breaking mechanism, which is of $D$-term type, leads to an anomalous $U(1)$ symmetry that spontaneously breaks down at the point where the two branes approach each other and inflation ends. This anomaly is cancelled by a counterterm in the Lagrangian which is the four dimensional analogue of the GS mechanism. The theory supports cosmic string solutions that do not have long range interactions and the usual constraints on the string tension apply, while vorton formation is trivially satisfied by the model. 

A closer look at the stabilisation process reveals possible subtleties of the model. The fact that the K\"ahler modulus, which controls the volume of the internal space, is constant during inflation in combination with the FI term being a function of this modulus, renders the current status of the model incompatible with general arguments about the inconsistency of constant FI terms in supergravity theories.

In Ch.~\ref{ch:paper-2}, we studied cosmic superstrings on warped backgrounds. We argued that when these objects are placed in a natural context, such as brane-antibrane inflation on a highly warped geometry, the fields that couple to the string are projected out of the massless spectrum so that radiation from these sources is not possible. Radiation from higher form fields that couple to the string might be possible but a proper computation of the axionic radiation power spectrum was not attempted.

   \part{Bottom-up EFT of scalar perturbations generated during inflation} \label{part:eft}
   In this Part, we will focus on a bottom-up EFT approach to the study of scalar inflationary perturbations, whose properties are encoded in the CMB. The main idea behind the construction is the following: let us think of inflation as an unknown UV scalar quantum field theory on a time dependent gravitational background. General covariance, that is invariance of GR under arbitrary local spacetime reparametrisations, can be thought of as a gauge symmetry\footnote{Attempts of quantising gravity in exact analogy with gauge theory are well known to be notoriously hopeless. Here, we are dealing with a quantum field theory on a classical gravitational background and thinking in analogy with gauge theory will help our intuition.} of this theory. 

Focusing on the dynamics of perturbations of the scalar field along its background solution, one can easily see that time reparametrisations are not a symmetry any more, as opposed to spatial coordinate transformations which remain unbroken. In other words, the evolution of the background results in the spontaneous breakdown of time reparametrisations. 

One can now construct an effective field theory of the perturbations around this symmetry breaking pattern using the principles of EFT as outlined in the Introduction; that is, to identify the set of operators that are invariant under the reduced symmetry of the system, and then write down a Lagrangian as an infinite power series in these operators, where, on dimensional grounds, the higher powers are suppressed by powers of some UV scale. 

This effective Lagrangian now constitutes a parametrisation of any UV theory that respects this symmetry breaking pattern. A specific possible completion can be chosen by adjusting the unknown parameters of the effective theory. As pointed out in the Introduction, such a construction on its own has no predictability at all. However, using experiments one can in principle constrain these unknown couplings and get an insight into the UV complete theory. The more accurate the observation, the closer to completion one gets and hopes that, as time passes by, the way towards the correct UV theory of inflation (if of course inflation is correct in first place) is slowly paved. 

The effective field theory for inflation is a formalism that addresses questions about perturbations of matter and spacetime during an inflationary period of expansion. Since it combines concepts of gravity and gauge theory, we will devote the next Chapter to a brief discussion of cosmological perturbation theory and the notion of gauge transformations on the gravity side, as well as the spontaneous breakdown of chiral symmetry in a simple example, which will be insightful to certain aspects of the EFT construction for inflation \cite{Creminelli:2006xe,Cheung:2007st}, which we also review. In Ch.~\ref{ch:paper-3}, we argue that the presence of massive scalar fields during inflation is natural from a UV point of view, and we construct a class of operators that capture effects of such scalars on the low energies dynamics of the perturbations. We then compute the two-point and three-point correlation functions, identifying the signatures of such a scenario. 

Although inflation is not confined to the use of scalar fields\footnote{For a nice exposition of such alternatives see for example \cite{Maleknejad:2012fw} and references therein.}, in this thesis we will examine models with a single scalar field as the inflaton. 

\chapter{Cosmological perturbations and EFT of inflation} \label{ch:prelims2}

%

\section{Cosmological perturbation theory}
One of the fundamental questions of cosmology is about the origin of the large scale structure that we observe today. Our current understanding is that massive structures were formed from small density inhomogeneities which, through the Jeans instability, evolved to become self-gravitating objects. Inflation predicts an almost homogeneous universe, so the question becomes {\it ``How did these inhomogeneities form?"}. The modern consensus is that their origin is due to the quantum nature of inflation. That is, quantum fluctuations of the inflaton grew during inflation to become the classical ``seeds" of large scale structure \cite{Mukhanov:1981xt}. From the form of the Einstein field equations, matter fluctuations are inevitably transmitted to the gravitational sector, thus consistency requires inflaton and metric perturbations to be studied together. Cosmological perturbation theory addresses the evolution of these scalar, vector and tensor  quantum fluctuations during inflation. We now briefly review the formalism of cosmological perturbations including only the most relevant parts needed for the rest of the thesis. The reader is referred to \cite{Mukhanov:1990me,Kodama:1985bj,Malik:2008im,Riotto:2002yw,Baumann:2009ds} for classic reviews on the subject.
\subsection{Matter and metric fluctuations}
As already mentioned, since the dynamics of spacetime are sourced by matter, perturbations of the latter inevitably translate to perturbations of the former:
\be \label{schematic-pert}
\delta\phi \Longleftrightarrow \delta T^{\mu\nu} \Longleftrightarrow \delta G^{\mu\nu} \Longleftrightarrow \delta g^{\mu\nu},
\ee
where $\phi$ represents the matter content, $T^{\mu\nu}$ the energy-momentum tensor of matter, $G^{\mu\nu}$ the Einstein tensor, $g^{\mu\nu}$ spacetime and $\delta$ a fluctuation of a quantity with respect to its background value. As we will see in Sec.~\ref{sec:gauge-trans}, this relation between matter and metric fluctuations is not only a requirement of the dynamics of the theory but also a consequence of general covariance. 

Let us parametrise the linearised metric perturbations 
\be \label{perturbed-metric}
g_{\mu\nu} = g_{\mu\nu}^{(0)} + \delta g_{\mu\nu},
\ee
according to their transformation under spatial rotations. A symmetric tensor in $d=n+1$ spacetime dimensions has $N_{\rm g} = \dfrac{1}{2}(n+1)(n+2)$ degrees of freedom, which contain $N_{\rm s},N_{\rm v},N_{\rm t}$ scalar, vector and tensor degrees of freedom respectively. A vector can be decomposed\footnote{This decomposition comes with various names: in a group theory language it would be the reduction of a representation into irreducible parts; in a vector calculus framework one would call this a Helmholtz decomposition, or scalar-vector-tensor decomposition; in the language of differential geometry it is similar to the Hodge decomposition.} as
$$
V_i = \partial_i V + \bar{V}_i \quad ; \quad \partial^i\bar{V}_i = 0,
$$
whilst the tensor decomposition reads
$$
g_{ij} = -2\delta_{ij} \psi + 2\partial_{ij}E + 2(\partial_{j}E_i + \partial_{i} E_j) + h_{ij}\;; \quad \partial^i E_{i} = 0 \quad \& \quad \partial^i h_{ij} = 0,
$$
where $\partial_{ij} = \partial_{i}\partial_{j}$, so that the pure vector degrees of freedom are $N_{\rm v} = 2(n-1)$, whilst the pure traceless/transverse tensor degrees of freedom are $N_{\rm t} = \frac{1}{2}n(n+1)-n-1 = \frac{1}{2}[n(n-1)-2]$. For $n=3$ we thus have 
\be \label{dof-count}
N_{\rm g} = N_{\rm s} + N_{\rm v} + N_{\rm t} \Rightarrow 10 = 4 + 4 + 2.
\ee
In single scalar field inflation, vector perturbations decay and we neglect them in what follows. To the $N_{\rm s}$ scalars, we must also add the fluctuation of the inflaton field $\delta\varphi$, so we are finally left with five scalars and two tensor fields parametrised as\footnote{This is a parametrisation for first order perturbation theory. For the second order formulation see \cite{Acquaviva:2002ud,Bartolo:2003bz,Malik:2003mv}.}
\be \label{metric+inflaton-fluctuations}
g_{\mu\nu} =  \left( 
\begin{array}{cc}
- 1 - 2 \phi &  a^{-1} \partial_i V \\ a^{-1} \partial_i V & a^2 (1 - 2 \psi)\delta_{ij} + 2 a^2 \partial_{ij}E + a^2 h_{ij}
\end{array} \right)\;,\quad \varphi(x,t) = \phi_0(t) + \delta\varphi(x,t),
\ee
where $\phi_0(t)$ is the background solution of Einstein's equations, $\delta\varphi(x,t)$ represents a fluctuation along this field trajectory, and $g_{\mu\nu}^{(0)}$ of \eqref{perturbed-metric} has been chosen as the FLRW metric \eqref{frw-metric}, which we rewrite here for convenience
\be \nn 
ds^2 = -dt^2 + a(t)^2 \left( dr^2 + r^2 d\theta^2 + r^2 \sin^2\theta d\phi^2 \right).
\ee
%
%

As already explicit in \eqref{metric+inflaton-fluctuations}, perturbations are parametrised as differences of quantities between a \emph{background} spacetime $\mathcal{M}_0$, representing a homogeneous and isotropic solution of Einstein's equations and a \emph{physical}, perturbed spacetime $\mathcal{M}$. Since we are considering differences of functions evaluated at points that belong to different manifolds, we should also specify a \emph{map} that uniquely assigns a point on $\mathcal{M}$ to a point on $\mathcal{M}_0$, while preserving the differential structure of $\mathcal{M}_0$. We thus need to specify a \emph{diffeomorphism} $\mathcal{D}: \mathcal{M}_0 \mapsto \mathcal{M}$, that is, a homeomorphism that is differentiable at each point $q\;\in \mathcal{M}_0$. 
This map is of course not unique, leaving us the freedom to specify it as we wish. Such a freedom is usually referred to as a \emph{gauge choice}, borrowing terminology from field theory. We devote the next paragraph to a brief discussion of gauge transformations of this kind.

\subsubsection{Gauge transformations} \label{sec:gauge-trans}
In the notation of the previous paragraph, we choose a coordinate system $x^\mu_0(p_0),\;p_0\;\in \mathcal{M}_0$ on the background manifold $\mathcal{M}_0$ and a diffeomorphism $\mathcal{D}: \mathcal{M}_0 \mapsto \mathcal{M}$, which induces a coordinate system $x = \mathcal{D}(x_0)$ on the physical manifold $\cal M$. On the physical manifold, let us denote a set of functions including scalar, vector or tensor quantities, $e.g.$ the metric and the inflaton considered in \eqref{metric+inflaton-fluctuations}, as $\mathcal{Q}(x^\mu)$. A gravitational theory on $\cal M$ is thus defined by assigning specific values to these functions. A background model is defined by setting the functions $\cal Q$ equal to some background functions $\mathcal{Q}_0(x_0)$ that are defined on  $\mathcal{M}_0$. These are constant, prescribed functions that describe the evolution of the unperturbed background spacetime, and as such their functional form remains the same irrespective of the coordinate system $x$ induced on $\cal M$ by $\mathcal{D}$. 
Next, we define the perturbation $\delta \mathcal{Q}(p)$ on $\mathcal{M}$ as
%
\be \label{deltaQ}
\delta \mathcal{Q}(p) = \mathcal{Q}(p) - \mathcal{Q}_0 \left( \mathcal{D}^{-1}(p) \right).
\ee

A second diffeomorphism $\tilde{\mathcal{D}}: \mathcal{M}_0 \mapsto \mathcal{M}$ would define a second coordinate system $\tilde x$ on $\cal M$ and the new perturbation $\delta \tilde{\mathcal{Q}}(p)$ would read as in \eqref{deltaQ} with $\mathcal{D}^{-1}$ replaced with $\tilde{\mathcal{D}}^{-1}$. Gauge transformations correspond to diffeomorphisms between $\mathcal{M}_0$ and $\cal M$, which form the group of diffeomorphisms of the physical manifold $\rm{Diff}(\mathcal{M})$, with multiplication operation corresponding to composition of maps. 

Another mathematically equivalent approach is the following: let $x^\mu,\;\tilde x^\mu$ be two coordinate systems on the physical manifold $\cal M$ and define a set of functions $\mathcal{Q}_0$ on $\cal M$ that fix the background dynamics.
The perturbation of a function $\cal Q$ is then defined as 
\be \label{dQ-in-x}
\delta \mathcal{Q}\left( x^\mu(p) \right) = \mathcal{Q}\left( x^\mu(p) \right) - \mathcal{Q}_0\left( x^\mu(p) \right),
\ee
whilst in the second coordinate system $\tilde x^\mu$, the same quantity reads
\be \label{dQ-in-xtilde}
\delta \tilde {\mathcal{Q}}\left( \tilde{x}^\mu(p) \right) = \tilde{\mathcal{Q}}\left( \tilde{x}^\mu(p) \right) - \mathcal{Q}_0\left( \tilde{x}^\mu(p) \right),
\ee
since $\tilde{\mathcal{Q}}_0\left( \tilde{x}^\mu(p) \right) = \mathcal{Q}_0\left( \tilde{x}^\mu(p) \right)$.
The transformation 
\be 
\delta \mathcal{Q}\left( x^\mu(p) \right) \mapsto \delta \tilde {\mathcal{Q}}\left( \tilde{x}^\mu(p) \right) ,
\ee
is a gauge transformation. 
Let us see for example the transformation of a homogeneous scalar field under a time reparametrisation. Upon parametrising the transformation between the two coordinate systems as 
\be \label{trans-gender}
t \to t + \xi,\quad x^i = x^i + \partial_i \epsilon + \bar{\epsilon}_i; \quad \partial^i\bar{\epsilon}_i=0,
\ee
from \eqref{dQ-in-x}, \eqref{dQ-in-xtilde} we can deduce the transformation law for the perturbation of a scalar function $\phi_0(t)$:
%
\be \label{trans-law-pert-general}
\delta \tilde {\phi}\left( x , \tilde{t} \right) = \delta \phi(x,t) - \xi \dot \phi_0(t).
\ee
From the invariance of the perturbed metric \eqref{metric+inflaton-fluctuations} under such a reparametrisation, we may further deduce the transformation laws for all the scalar quantities, which we list for the sake of completeness:
\be \label{potential-trans}
\phi \to \phi - \dot\xi,\quad V \to V + \frac{\xi}{a} -a \dot\epsilon, \quad E \to E - \epsilon, \quad \psi \to \psi + \xi H.
\ee
There are now two ways to proceed with a calculation: either one chooses a gauge and works bearing in mind that any final result representing an observable quantity should be gauge invariant, or one formulates the theory in terms of gauge invariant expressions throughout the entire computation. Keeping gauge invariance manifest might be an appealing feature but as usual it is computationally much more involved. In what follows, we will adopt the former path of gauge fixing.

From \eqref{trans-law-pert-general} we see that there is a special value of $\xi$ that results in $\delta \tilde {\phi}\left( x , \tilde{t} \right) = 0$, namely
\be \label{special-value}
\xi = \dfrac{\delta \phi(x,t)}{\dot{\phi}_0(t)}. 
\ee
For such a choice, a perturbation is present in one coordinate system and vanishes in another, a fact implying that the perturbation is not a physical one but a so called \emph{gauge mode}. The importance of considering inflaton and metric fluctuations simultaneously becomes now clear: neglecting the former, or the latter, might lead to gauge modes being treated as physical perturbations and vice versa. When both are taken into account, a physical perturbation never disappears from the dynamics; it is just hidden in a different degree of freedom in each gauge. 
We now exemplify the above discussion, outlining two gauge choices that will be used in what follows.  

\subsection{Comoving and spatially flat gauge}
There are many gauge choices that one can make, each with its own advantages. The degrees of freedom that will be considered, that is the set of functions $\mathcal{Q}$ in the previous notation, are the metric $g^{\mu\nu}$ and a scalar field $\phi(t,x)$ and we will focus on the case where the background metric is the FLRW metric \eqref{frw-metric},
since this is the appropriate choice for the study of inflation.
The five scalar perturbations of the metric and the inflaton of \eqref{metric+inflaton-fluctuations} can be reduced to two, say $\psi$ and $\delta\varphi$, by a coordinate transformation of the form $x^i \to x^i + \partial^i \epsilon$ and the use of the perturbed Einstein equations, so that
\be \label{generic-gauge}
g_{ij} = e^{2\mathcal{\psi}(t,x)}e^{2\rho(t)}\delta_{ij} \;\; ,\;\; \varphi(x,t) = \phi_0(t) + \delta\varphi(x,t),
\ee
respectively, where $\rho \equiv \ln a$ and $\psi$ is the curvature perturbation, since to first order in the fluctuations it represents the scalar curvature of the metric, namely
$$
R^{(3)} = 4 \dfrac{\nabla^2\psi}{a^2}.
$$
The \emph{comoving gauge} is defined as the one where the inflaton perturbation $\delta\varphi(x,t)$ is set to zero and the spatial metric becomes conformally flat:
\be \label{unitary-gauge}
\varphi(x,t) = \phi_0(t)\;,\quad g_{ij} = e^{2\mathcal{R}(t,x)}e^{2\rho(t)}\delta_{ij},
\ee
where $\mathcal{R}=\psi_{\rm com}$ is now the \emph{comoving curvature perturbation} to be defined in the next paragraph. From \eqref{special-value} we can specify the gauge transformation $t\mapsto t + \xi_{\rm com}$, that brings us from an arbitrary slicing to the comoving one:
\be \label{to-comoving-special-value}
\xi_{\rm com} = \frac{\delta\varphi}{\dot\phi_0} \quad \Longrightarrow \quad \delta\varphi_{\rm com} = 0.
\ee
The \emph{flat gauge} is defined as the one where the spatial metric is flat, $i.e.$ $\psi_{\rm flat}=0$ but the scalar perturbation $\delta\varphi(x,t)$ is now present, that is
\be \label{flat-gauge}
\varphi(x,t) = \phi_0(t) + \delta\varphi(x,t) \;,\quad g_{ij} = e^{2\rho(t)}\delta_{ij}.
\ee
From the transformation law \eqref{potential-trans} of the $\psi$ potential, 
we see that the gauge parameter $\xi_{\rm flat}$ that brings us from an arbitrary slicing to the flat one is
\be \label{to-flat-special-value}
\xi_{\rm flat} = -\frac{\psi}{H}   \quad \Longrightarrow \quad  \psi_{\rm flat} = 0.
\ee
Note that in both cases there is one scalar degree of freedom manifested either as a curvature perturbation in the comoving gauge or as a matter fluctuation in the flat one.
Let us now define some gauge invariant quantities that will shed some light on the notation that we will use later. 
\paragraph*{The comoving curvature perturbation}
Since both the curvature perturbation $\psi$ and the inflaton fluctuations $\delta \varphi$ vary under gauge transformations, one can define a gauge invariant quantity
\be \label{com-curv-pert}
\mathcal{R} = \psi + H \frac{\delta \varphi}{\dot\phi_0},
\ee
that represents the curvature perturbation in the comoving gauge \eqref{unitary-gauge}, hence the term comoving curvature perturbation.

\paragraph*{Matter fluctuations in the spatially flat gauge}
Another gauge invariant quantity can be constructed as
\be \label{flat-infl-pert}
Q = \delta\varphi + \frac{\dot\phi_0}{H} \psi,
\ee
representing the fluctuation of the inflaton field in the spatially flat gauge \eqref{flat-gauge}, where $\psi_{\rm flat} = 0$.  Another quantity of interest is
\be 
\pi \equiv \dfrac{\delta\varphi(t,x)}{\dot\phi_0(t)}, \label{pi-def}
\ee
whose relation to $Q$ and $\mathcal{R}$ is\footnote{In Maldacena's paper \cite{Maldacena:2002vr} the letter $\zeta$ is used to denote what here we call $\cal R$. In a more usual notation, $\zeta$ denotes the curvature perturbation on hypersurfaces of uniform energy density. On superhorizon scales the two quantities coincide. 
}
\be 
\pi_{\rm flat} = \dfrac{Q}{\dot\phi_0} = \dfrac{\mathcal{R}}{H}, \label{pi-Q-R}
\ee
since in this gauge we have that $\psi_{\rm flat} = 0$, while by definition $\pi_{\rm com} = 0$.
Using \eqref{trans-law-pert-general}, we obtain its transformation law under a general time diffeomorphism $t \mapsto \tilde t = t + \xi(t,x)$ as
\be \label{pi-trans} 
\tilde{\pi}(\tilde t,x) = \pi(t,x) - \xi( t,x).
\ee 

Relations like \eqref{pi-Q-R} can be used when one changes the gauge. For instance, this happens when one computes quantities like correlation functions of scalar perturbations which translate to observable temperature fluctuations in the CMB. The field $\cal R$ is known \cite{Maldacena:2002vr,Lyth:2004gb,Langlois:2005qp,Pimentel:2012tw,Senatore:2012ya} to become independent of time at superhorizon scales, a fact that allows one to compute self-correlators of the comoving curvature perturbation, without specifying in detail the processes that occur on scales $\lambda \gg H^{-1}$ where interesting, yet unknown physics becomes important. It is thus useful to express all the final results in the comoving gauge where $\cal R$ is the dynamical degree of freedom. 

Further exploration of this relation, beyond the linear approximation, will reveal an important feature of $\pi$ which we now derive.
Let us start from the flat gauge and perform a time reparametrisation \cite{Maldacena:2002vr}, 
\be \label{time-diff} 
t_{\rm flat} \mapsto  t_{\rm com} = t_{\rm flat} + \xi_{\rm com}(x, t_{\rm flat}),
\ee 
under which $\delta\varphi$ shifts according to \eqref{trans-law-pert-general} and with $\xi_{\rm com}$ defined in \eqref{to-comoving-special-value}.
Due to this temporal reparametrisation, the metric acquires non diagonal contributions, which have to be cancelled by a spatial counter-transformation. Indeed, using the transformation law for the metric tensor in the the Arnowitt-Deser-Misner (ADM) form \cite{Arnowitt:1962hi}
\begin{equation}
ds^2=-N^2dt^2+\gamma_{ij}(N^idt+dx^i)(N^jdt+dx^j), \label{adm}
\end{equation}
where $N$ denotes the \emph{lapse} function and $N^i$ the \emph{shift} vector, so that
\begin{equation}
\begin{split}
g_{00}&=-N^2+\gamma_{ij}N^iN^j,\quad g_{0i}=\gamma_{ij}N^j,\quad g_{ij}\equiv\gamma_{ij}=a^2(t)\delta_{ij} , \\
g^{00}&=-\dfrac{1}{N^2},\quad g^{0i}=\dfrac{N^i}{N^2}, \quad g^{ij}=\gamma^{ij}-\dfrac{N^iN^j}{N^2} \quad \text{with} \quad \sqrt{-g}=N\sqrt{-\gamma},
\end{split}
\end{equation}
and including terms up to second order in the perturbations\footnote{For example, we write $(\partial \xi_{\rm com})^2N^2 \sim (\partial \xi_{\rm com})^2 $ on the grounds that $N^2 = 1 + \mathcal{O}(\delta N)$, while by definition $(\partial \xi_{\rm com})^2$ is already of order $\mathcal{O}(\delta\varphi^2)$.} we obtain
\be \label{metric-time-trans}
 g_{ij}^{\rm com}(t_{\rm com}) \equiv \tilde g_{ij}^{\rm flat}(\tilde t_{\rm flat}) =  e^{2\rho(t_{\rm com} - \xi)} \left(\delta_{ij} - e^{-2\rho(t_{\rm com} - \xi)} \partial_i \xi \partial_j \xi - \partial_i \xi N_j - \partial_j \xi N_i \right).
\ee
Note that we have dropped the label ``com" in $\xi_{\rm com}$ but retained it in the metric and time variable to make clear the relation between the tensors in the two gauges.
We see that the non diagonal terms of the $\tilde{g}_{ij}$ metric are second order in the fluctuations. Under a spatial diffeomorphism  $\tilde x^i \mapsto \tilde x'^i = \tilde x^i +\epsilon^i$ the metric transforms as
\be \label{metric-spatial-trans}
\tilde g'_{ij} =  \tilde g_{ij} + \partial_i \epsilon_j + \partial_j \epsilon_i.
\ee
The parameter $\epsilon_i$ to second order in the perturbation is fixed by the requirement that the non diagonal terms cancel \cite{Maldacena:2002vr}. That is,
\be \label{spatial-trans-condition}
\partial_i \epsilon_j + \partial_j \epsilon_i - e^{-2\rho(t)} \partial_i \xi \partial_j \xi - \partial_i \xi N_j - \partial_j \xi N_i = 2\alpha \delta_{ij}.
\ee
This equation can be solved by decomposing $\epsilon_i = \partial_i \epsilon + \bar{\epsilon}_i\;,\;\partial^i\bar{\epsilon}_i=0$, as in \eqref{trans-gender}, and operating with the trace $\delta^{ij}$ along with $\partial^{ij} \equiv \partial^i\partial^j$ and $\partial^i$.
After a straightforward calculation one obtains
%
%
%
\be \label{alpha}
\begin{split}
\alpha = & \frac{1}{4} \left( \partial^{-2}\partial^{kj} - \delta^{kj} \right) e_{kj}, \\ \epsilon =  \frac{1}{4}\partial^{-2} \left( 3 \partial^{-2} \partial^{kj} -  \delta^{kj} \right) & e_{kj}, \quad
\bar{\epsilon}_i = - \partial^{-2} \left( \partial^{-2} \partial^{kj} \partial_i - \delta_i^k \partial^j  \right)e_{kj}, \\ \!\!\!\!\!\! e_{kj} = & e^{-2\rho} \partial_k \xi \partial_j \xi + \partial_{(k} \xi N_{j)} , 
\end{split}
\ee
where $\partial_{(i} \xi N_{j)} = \partial_{i} \xi N_{j} + \partial_{j} \xi N_{i}$. 
%
%
Combining \eqref{metric-time-trans},\eqref{metric-spatial-trans},\eqref{spatial-trans-condition} and matching the result with the comoving gauge \eqref{unitary-gauge}, we obtain a relation between $\xi_{\rm com}$ and the comoving curvature perturbation, namely
\be \label{R-to-pi}
\mathcal{R}(\xi) = \rho \big[t - \xi[\tilde t,x + \epsilon] \big] - \rho(t) + \alpha[\xi( t,x)].
\ee
This formula, relating the curvature perturbation on a comoving slice at time $t+\xi(t,x)$ to the one on a flat slice at time $t$, is essentially the $\delta N$ formalism \cite{Starobinsky:1982ee,Starobinsky:1986fxa,Salopek:1990jq,Sasaki:1995aw,Lyth:2004gb,Lyth:2005fi,Sugiyama:2012tj} typically used to compute classical non Gaussianities on superhorizon scales. In that case, $\rho$ is viewed as the number of e-folds $N={\rm ln}a$ and the difference in $N$ between the two slices characterised by time coordinates $t$ and $t+\xi$, reads 
\be \label{dN}
\delta N = N[\phi_0 + \delta\varphi] - N[\phi_0].
\ee
The background inflaton field $\phi_0(t)$ can be set equal to $t$, since it is the clock of the system, $i.e.$ its evolution specifies the time direction. Note that on superhorizon scales, the gradient contribution $\alpha[\xi]$, given by \eqref{alpha}, vanishes so that \eqref{R-to-pi} exactly matches \eqref{dN}.

In order to correctly calculate $\mathcal{R}$ to order $n$ in $\xi$ one needs to iterate this Taylor expansion to the same degree $n$ wherever $\xi$ appears. For example, to second order we have
\be \label{R-to-pi-2nd-order}
\mathcal{R}^{(2)}(\xi) = \rho \Big[t - \xi\big[t-\xi(\tilde t,x),x \big] \Big]-\rho(t) + \alpha[\xi] = -H\xi + H\xi \dot\xi +\frac{1}{2} \xi^2 \dot H + \alpha[\xi].
\ee

Now from \eqref{R-to-pi-2nd-order} one can deduce that $\pi$ realises time diffeomorphisms in a non linear manner. In order to see this, let us adopt the following notation, inspired by the iteration of the Taylor expansion that led to \eqref{R-to-pi-2nd-order}:
\be 
\pi_1 = \pi(t)\;,\;\; \pi_2 = \pi \left( t - \pi_1 \right),\;\; \pi_3 = \pi \left( t - \pi_2 \right) ,\; \ldots \;,\;\; \pi_n = \pi \left( t - \pi_{n-1} \right),
\ee
where in each iteration the transformation \eqref{time-diff} with $\xi^0 = \pi$ is applied. From the definition of the Taylor expansion we can write the general $n$-th order term as
\be 
\pi_n =  e^{- \pi_{n-1} D_t} \pi,
\ee
where $D_t \equiv \dfrac{d}{dt}$ is the generator of time shifts and $D_t^n \equiv \dfrac{d^n}{dt^n}$. Now the comoving curvature perturbation to order $n$ in $\pi$ can be written as
\be 
\mathcal{R}_n =  \left( e^{-\pi_{n} D_t} - 1 \right) \rho.
\ee

To summarise, $\pi$ has the following (equivalent) properties: it appears explicitly in the matter content as an inflaton perturbation in the flat gauge, while it is hidden in the metric as a curvature perturbation in the comoving gauge; it is a non linear realisation of the broken symmetry; it shifts, proportionally to the parameter under a broken transformation. In other words, its behaviour resembles that of a Goldstone boson. 

This is the key observation for the whole construction of the EFT. To state it clearly: {\it we will think of $\pi$ as the Goldstone mode that arises due to the spontaneous breakdown of time diffeomorphisms by the evolution of the background vacuum expectation value $\phi_0(t)$ of the inflaton field}. 

It is quite interesting to note that the full transformation $(\tilde t, \tilde x_i) = \left( t + \xi , x_i + \epsilon_i(\xi) \right)$, with $\epsilon_i$ given in \eqref{alpha}, corresponds to a conformal transformation with a scaling factor $\cal R$ as in \eqref{R-to-pi}. In \cite{Hinterbichler:2012nm}, it was shown that this Weyl rescaling, can be compensated by an appropriate transformation of the comoving curvature perturbation which now provides a non linear realisation of the conformal group in three dimensions, $SO(4,1)$. Hence, $\cal R$ can be also thought of as the Goldstone mode that arises due to the spontaneous breakdown of $SO(4,1)$ by the time dependence of the background\footnote{According to the standard counting, one might expect such a symmetry breaking to yield four Goldstone bosons corresponding to the three special conformal generators plus the dilation. As shown in \cite{Low:2001bw} though, the counting rule for Goldstone bosons for spacetime broken symmetries is different and for $SO(4,1) \mapsto E_3$ specifically there is indeed only one Goldstone mode, the dilaton.}. In the limit where spacetime is exactly de Sitter this symmetry corresponds to the isometry group of dS. In a general setting where slow roll corrections are taken into account, this $SO(4,1)$ represents the conformal group on three-dimensional hypersurfaces. This symmetry was used in \cite{Creminelli:2012ed} to deduce consistency conditions between correlation functions of different order, generalising those found in \cite{Maldacena:2002vr} -- see also \cite{Assassi:2012zq}. 

Since the situation for inflation is in close analogy with the physics of Goldstone bosons, we will devote the next section to review an example of spontaneous symmetry breaking that will reveal an important feature of the $\pi$ dynamics, highlighting the connections with the cosmological case along the way.

\section{Chiral symmetry breaking, pions and the equivalence theorem} \label{sec:pions}
As an illustrative example we will discuss the non Abelian Higgs mechanism due to $SU(2)\times SU(2)$ chiral symmetry breaking. The simplest action with the required symmetry is
\be \label{su2xsu2-action-1}
S(\mathbf{A_\mu},\bm{\phi}) = \int d^4x \;{\rm tr}\; \Big[ \frac{1}{4} \mathbf{F}_{\mu\nu}\mathbf{F}^{\mu\nu} + (\mathbf{D}_\mu \bm{\phi})^\dag\mathbf{D}^\mu \bm{\phi} + r \bm{\phi}^\dag \bm{\phi} +  \frac{\lambda}{4} (\bm{\phi}^\dag \bm{\phi})^2 \Big],
\ee
where $\phi$ is a scalar transforming in the $(1/2,1/2)$ representation of $SU(2) \times SU(2)$. Bold symbols denote three-dimensional vectors in group space, for example $\mathbf{A}_\mu \equiv (A_\mu^1,A_\mu^2,A_\mu^3)$. A dot product is then the standard inner product in $\mathbb{R}^3$, $\bm{A}_\mu\cdot \bm{\tau} = \sum_\alpha A_\mu^\alpha \tau^\alpha$ with the group indices $\alpha$ running form 1 to 3 and $\tau$ denoting the Pauli matrices obeying the usual commutation relations $[\tau^i,\tau^j] =2 i \epsilon_{ijk}\tau^k$. The trace in the action runs over the group indices $\alpha$. In this notation $\mathbf{D}_\mu = \mathbb{I}\partial_\mu + \dfrac{ie}{2} \mathbf{A}_\mu$ is the covariant derivative and  $\mathbf{F}_{\mu\nu} = \partial_\mu \mathbf{A}_\nu - \partial_\nu \mathbf{A}_\mu - e \mathbf{A}_\mu \times \mathbf{A}_\nu$ is the field strength, where $ (\mathbf{A}_\mu \times \mathbf{A}_\nu)_k = \sum \epsilon_{ijk} A_\mu^i A_\nu^j$ is the $k$ component of the outer product in $\mathbb{R}^3$. 

We will break the $SU(2) \times SU(2)$ chiral symmetry down to the diagonal $SU(2)$ subgroup. A convenient parametrisation for the scalar is
\be \label{phi} 
\bm{\phi} = \frac{1}{\sqrt{2}}(\bm{\mathbb{I}} \sigma + i \bm{\pi}\cdot \bm{\tau}),
\ee
with $\sigma,\bm{\pi}$ real fields. Note that a consequence of the commutation relations of the Pauli matrices is that
\be 
(\bm{X} \cdot \bm{\tau})(\bm{Y} \cdot \bm{\tau}) = (\bm{X} \cdot \bm{Y})\cdot \bm{\tau} + i (\bm{X} \times \bm{Y}) \cdot \bm{\tau}.
\ee
Therefore under an infinitesimal gauge transformation $\bm{g}(x)=1-i\bm{\omega}\cdot\bm{\tau}/2$ the fields change as
\be \label{gauge-trans-su2} 
\delta \bm{A}_\mu = \frac{1}{e} \partial_\mu \bm{\omega} - \bm{A}_\mu \times \bm{\omega}, \quad
\delta \sigma = \frac{1}{2} \bm{\omega} \cdot \bm{\pi},  \quad
\delta \bm{\pi} = -\frac{1}{2} \sigma \bm{\omega} +\frac{1}{2}  \bm{\omega}\times \bm{\pi}. 
\ee
%
%
In these variables the scalar part of the action \eqref{su2xsu2-action-1} reads
\be \label{su2xsu2-action-2}
S_{\rm sc} = \frac{1}{2}\int d^4x \Big[ \left( \partial_\mu \sigma - \frac{e\bm{\pi}\cdot\bm{A}_\mu}{2} \right)^2 +  \left( \partial_\mu \bm{\pi} - \frac{e\sigma\bm{A}_\mu}{2} - \frac{e\bm{A}_\mu \times \bm{\pi}}{2} \right)^2 + V(\sigma^2 + \bm{\pi}^2) \Big],
\ee
where $V(X) = rX + \frac{\lambda}{12} X^2$.
Note that in these variables the potential implies $O(4)$ symmetry since $(\sigma,\bm{\pi})$ is a vector in $\mathbb{R}^4$ and $SU(2) \times SU(2) \cong O(4)$. Upon setting $\langle \sigma \rangle=u$ the $O(4)$ breaks to $O(3)$ which is the subgroup of three-dimensional rotations of the subspace normal to the $\sigma$ direction. In terms of the covering group, $O(3)$ is the diagonal $SU(2)$ with elements $(g,g),\;g \in SU(2).$ If the gauge field was not present, $\bm{\pi}$ would be a massless Goldstone boson triplet parametrising the residual $SU(2)$. In our case though, due to the gauging of the chiral symmetry by $\bm{A}_\mu$, the would be Goldstone mode $\bm{\pi}$ can be removed form the dynamics by performing a suitable gauge transformation. Indeed from \eqref{gauge-trans-su2}, one can see that setting $\bm{\omega} = 2\bm{\pi}/\sigma$ results in the action 
\be \label{su2xsu2-action-3}
S(\bm{A}_\mu,\sigma) = \frac{1}{2}\int d^4x \Big[  \frac{1}{4 } \mathbf{F}_{\mu\nu}\mathbf{F}^{\mu\nu} + \left( \partial_\mu \sigma \right)^2 + \frac{e^2}{8}\sigma^2\bm{A}_\mu^2  + V(\sigma^2) \Big],
\ee
in which $\bm{\pi}$ is gauged away. This is in exact correspondence with the gauge parameter \eqref{to-comoving-special-value}, that was used to remove the scalar fluctuation $\pi$ from the matter sector of the inflationary dynamics. As a result of the non zero value of $\sigma$, the gauge field acquires a mass $M_A=\dfrac{e u}{2}$, and since we are in three spatial dimensions it now has three independent polarisations. The longitudinal one is the form in which the hidden degree of freedom $\pi$ manifests itself. This gauge, where the Goldstone mode is hidden in the gauge field, is the \emph{unitary gauge} which has only physical degrees of freedom propagating. Recall that this is the analogue of the comoving gauge \eqref{unitary-gauge} in the cosmological set up, where $\pi$ represents the inflaton fluctuations. 
Before passing to this discussion let us comment on another useful concept arising from the spontaneous breaking of gauge symmetries, that is the equivalence theorem\footnote{We will restrict the  proof of the equivalence theorem to the Feynman -- 't Hooft gauge, following \cite{Lee:1977eg}. For the proof in a general $R_\xi$ gauge see \cite{Chanowitz:1985hj,Gounaris:1986cr}.} \cite{Cornwall:1974km,Vayonakis:1976vz,Lee:1977eg}.

The action \eqref{su2xsu2-action-3} is no longer gauge invariant since we fixed $\tilde{ \bm{\pi}}=\bm{\pi} + \delta \bm{\pi}=0$ (unitary gauge). Even though the unitary gauge has the advantage of involving strictly physical degrees of freedom there are other useful gauge choices in which interesting dynamics of the hidden Goldstone mode can emerge. Gauge invariance can be restored by performing a gauge transformation, restoring the $\bm{\pi}$ field. This is essentially the St\"uckelberg procedure \cite{Stueckelberg:1938zz,Stueckelberg:1900zz} -- see \cite{Ruegg:2003ps} for a modern review -- which corresponds to adding a scalar degree of freedom playing the role of a Goldstone mode, which non linearly realises the gauge symmetry. In the language of the cosmological perturbations we essentially used the St\"uckelberg trick when we promoted the parameter $\xi^0$ of the gauge transformation \eqref{time-diff} to the field $\pi$ with the transformation law \eqref{pi-trans}.

Let us write the action in a general $R_\xi$ gauge by imposing a constraint as a Lagrange multiplier in the path integral using the gauge function $F(\bm{A}_\mu,\bm{\pi})=\partial^\mu \bm{A}_\mu+\frac{1}{2} z \xi \bm{\pi}$, with $z$ an arbitrary parameter to be fixed appropriately \cite{'tHooft:1971rn}. We have that\footnote{In order to be consistent we should have also added the Faddeev-Popov ghost contribution $S_{\rm gh}$ \cite{Faddeev:1967fc}. Since we only want to illustrate the main ideas behind Goldstone bosons and the equivalence theorem, in what follows we will neglect the ghost part of the action.}
$$
S = S_{\rm F} + S_{\rm sc} + S_{\rm gf}.
$$
The gauge fixing contribution to the action reads
\be \label{Sgf}
S_{\rm gf} = \frac{1}{2e^2\xi}\int d^4x \left( e\partial^\mu \bm{A}_\mu - \frac{1}{2} z \xi \bm{\pi} \right)^2,
\ee
which sets the mass of the scalar mode to $M_\pi = \dfrac{z\sqrt{\xi}}{2e}$.
By choosing the Feynman -- 't Hooft gauge $\xi =1$ we impose the constraint 
\be \label{gauge-constraint}
\partial^\mu \bm{A}_\mu=\frac{1}{2}M_\pi\bm{\pi}.
\ee
Now $z$ is fixed such that the interaction term between the gauge field and the would be Goldstone boson $\bm{\pi}$ in \eqref{su2xsu2-action-2} is cancelled, that is $z=e^2 u$. Note that this choice also renders the mass of the scalar equal to the mass of the gauge field, $M_\pi = M_A = \dfrac{e u}{2}$.

The longitudinal mode $\bm{A}_\mu^L$ of the gauge field defined in Fourier space is given by 
\be \label{longi-mode}
\bm{A}_L(k) = \epsilon_L^\mu \bm{A}_\mu(k),
\ee
where $\epsilon_L^\mu = \frac{1}{M_\pi}(|\bm{k}|,0,0,E_k)$ and $k^\mu=(E_k,\bm{k})$ is the four-momentum carried by the massive gauge field. An important observation is that the longitudinal vector becomes more and more parallel to the four-momentum as the energy increases. This can be shown by computing the difference of the two as a power series in $M_\pi/E_k$. By expanding 
$$E_k = k - \dfrac{M_\pi}{2}  \big[ M_\pi/k + \mathcal{O}\left(M_\pi^2/k^2 \right) \big]$$ 
for $k\gg M_\pi$, we obtain 
\be 
\epsilon_L^\mu - k^\mu/M_\pi = \mathcal{O}\left( M_\pi/k \right).
\ee
Combining this result with the constraint \eqref{gauge-constraint} in momentum space and the definition \eqref{longi-mode} we find that
\be \label{equi-theorem}
\bm{A}_L(k) = \bm{\pi} + \mathcal{O}\left(M_\pi/k \right).
\ee
This is the equivalence theorem, which states that at high enough energies $s \gg M_\pi^2$ the scattering matrix for longitudinal modes equals the scattering matrix of processes involving the Goldstone mode. The mass of the gauge field reveals its origin due to spontaneous symmetry breaking of the chiral group. The scalar particle becomes more and more massless compared to the center of mass energy of the scattering process, approaching a true Goldstone mode. We will see that this theorem lies at the heart of the simplicity offered by the effective field theory of inflation, to which the next section is devoted, since it allows for a slow roll expansion of the effective action.


\section{EFT of inflationary perturbations} \label{sec:eft}
Having set the field theory aspects of the construction, we now focus on the effective action for inflationary perturbations. Following \cite{Cheung:2007st}, we begin by classifying the operators that are consistent with the reduced symmetry, which we then use to construct an action in the unitary gauge. Recall that the unitary -- or comoving -- gauge is the one where the adiabatic scalar perturbation is set to zero, hence rendering the transformation properties of operators with tangential or transverse indices with respect to the three dimensional spatial slices, easily identifiable. Next, we comment on the limits of the unknown couplings, where this general effective action parametrises known inflationary models. Finally, the action is presented in the spatially flat gauge. This choice is more convenient from a computational point of view and as such it will be the one used in Ch~\ref{ch:paper-3}, where we will generalise our study to a certain class of operators that capture effects of \emph{natural} intermediate effective field theories exhibiting a mass hierarchy. 

\subsection{The effective action in the unitary gauge}
Using the principles of effective field theory we will construct the action in the unitary gauge, where the scalar Goldstone mode is absent. The first step is to identify the operators that are consistent with the reduced symmetry. One can then write the effective action as a polynomial of infinite order over these operators with arbitrary coefficients. 

\subsubsection{Invariant operators}
Obviously, all operators that respect the full spacetime diffeomorphism group will be present. These are powers of the Riemann tensor $R_{\mu\nu\rho\sigma}$ together with its covariant derivatives and their contractions, including for example the Ricci tensor $R_{\mu\nu}=g^{\rho\sigma}R_{\mu\rho\nu\sigma}$ and the Ricci scalar $R=g^{\mu\nu}R_{\mu\nu}$. Moreover, since temporal diffeomorphisms are violated, generic functions of time are allowed in the effective Lagrangian. Thus, the coefficient of any operator will be in general a function of time $f(t)$. The unitary gauge is defined as the coordinate system, where the function $\tilde t$, that breaks temporal diffeomorphisms ($e.g.$ a rolling scalar), coincides with time such that the extra degree of freedom contained in $\tilde t$ is absent. Hence, the gradient of $\tilde t$ in the unitary gauge becomes $\partial_\mu \tilde t = \delta^0_\mu$, and consequently any tensor is allowed to have free upper 0 indices. For example, $g^{00},\;\partial_\mu g^{00}$ as well as $R^{00},\;\partial_\mu R^{00}$ (properly contracted) are acceptable choices. 

Another object out of which Lagrangian operators can be constructed, is the normal unit vector on hypersurfaces of constant time $\Sigma_{t}$, defined as $\dfrac{\partial_\mu \tilde t}{\sqrt{\partial_\sigma \tilde t \partial^\sigma \tilde t}}$. In a $(-,+,+,+)$ signature, the gradient $\partial_\mu \tilde t$ is time-like and the normalisation constant should be chosen as
\be \label{normal-vec} 
n_\mu = \frac{\partial_\mu \tilde t}{\sqrt{-g^{\mu\nu}\partial_\mu \tilde t\partial_\nu \tilde t}}.
\ee
This vector can be used to build a projection operator on the hypersurfaces $\Sigma_t$, namely
\be 
h_{\mu}^{\phantom{a}\nu} = g_{\mu}^{\phantom{a}\nu} + n_{\mu}n^{\nu}.
\ee
That is, the projection onto the hypersurface $\Sigma_t$ of a vector field in the tangent space of the spacetime manifold $\cal M$ at a point $p \in \cal M$ is 
\be \label{projection}
h_{\mu}^{\phantom{a}\nu}u_\nu = u_\mu + (n \cdot u)n_\mu \in T_p\Sigma_t,
\ee
where $T_p \Sigma_t$ is the tangent space of the submanifold $\Sigma_t$. From \eqref{projection}, it follows that the tensor $h_{\mu\nu} = g_{\mu\sigma}h_{\nu}^{\phantom{a}\sigma}$ is the induced metric on $\Sigma_t$, since it defines the inner product on $ T_p\Sigma_t $. Having the induced three dimensional metric $^{(3)}h_{\alpha\beta}$ one can use polynomials of the contractions of the three-dimensional Riemann tensor $^{(3)}R_{\alpha\beta\gamma\delta}$ and its covariant derivatives, as operators in the effective action.

Another tensor that one can construct from the normal vector is the extrinsic curvature of $\Sigma_t$, whose entries are defined as the directional derivatives of $n$ along a unit tangent vector $\hat u \in T_p\Sigma_t$
\be 
K_{\mu\nu} = h_{\mu}^{\phantom{a}\sigma}\nabla_\sigma n_\nu.
\ee
This is the only way that covariant derivatives of the normal vector enter in the Lagrangian since from the definition we have that $n^\nu \nabla_\sigma n_\nu = 0$, whilst the other contraction can be written like
\be 
n^\sigma \nabla_\sigma n_\nu = -\frac{1}{2g^{00}} h_{\nu}^{\phantom{a}\sigma} \nabla_\sigma g^{00},
\ee
thus contributing $g^{00}$ and $\partial_\mu g^{00}$ terms that have already been accounted for. 

Finally, let us observe that using the three-dimensional Riemann tensor and the extrinsic curvature tensor at the same time is redundant since the two are related by the Gauss-Codazzi identity
\be 
^{(3)}R_{\alpha\beta\gamma\delta} = h_{\alpha}^{\phantom{a}\mu} h_{\beta}^{\phantom{a}\nu} h_{\gamma}^{\phantom{a}\rho} h_{\delta}^{\phantom{a}\sigma} R_{\mu\nu\rho\sigma} - K_{\alpha\gamma} K_{\beta\delta} + K_{\alpha\delta} K_{\beta\gamma}.
\ee
In addition, the projection operator can be used to express any three-dimensional quantity in terms of ambient spacetime objects. For example, the covariant derivative of a tensor as a three-dimensional quantity can be written as the projection of the four dimensional analogue,
\be 
\nabla_\alpha G_{\beta\gamma} = \nabla_\alpha (h_{\beta}^{\phantom{a}\nu}h_{\gamma}^{\phantom{a}\mu}G_{\mu\nu}) =  h_{\beta}^{\phantom{a}\nu}h_{\gamma}^{\phantom{a}\mu}h_{\alpha}^{\phantom{a}\sigma} \nabla_\sigma G_{\mu\nu},
\ee
where we used the fact that the three-dimensional covariant derivative of the induced metric vanishes. Hence, we can avoid explicit use of objects intrinsic to $\Sigma_t$. 

Summarising, the effective action can be symbolically written as \cite{Cheung:2007st}
\be \label{general-eff-action}
S = \int dx^3 dt \sqrt{-g} \mathcal{G}^{\infty}(g_{\mu\nu},R_{\mu\nu\rho\sigma},K_{\mu\nu},\nabla_\mu,t),
\ee
where $\mathcal{G}^{\infty}$ is a polynomial of infinite order in the tensors which is allowed to have terms with free upper zero indices.

This is the most general Lagrangian that one can write down for a field theory on a generic time dependent background. Since our aim is the construction of a field theory on a quasi de Sitter background, which locally can be described by the homogeneous and isotropic FLRW metric, we will restrict ourselves to the effective field theory on an FLRW spacetime. In view of the general discussion of EFT in the Introduction, the requirement of a specific background is the first input we give to this effective theory and as a result we will see that we immediately gain predictability by fixing two of the arbitrary time dependent couplings. 

\subsubsection{Minimal UV input: requirement of FLRW background} \label{sec:FRW}
Considering the unperturbed action \eqref{infl-action}, which we rewrite here for convenience,
\be \label{background-action} 
S = \int \sqrt{-g}dx^3 dt \left( \frac{M_{\rm Pl}^2}{2} R - \frac{1}{2}g^{\mu\nu}\partial_\mu \varphi \partial_\nu \varphi - V(\varphi) \right),
\ee 
on a flat FLRW metric, with the usual notation $g=\det g_{\mu\nu}$, for a homogeneous background scalar field $ \phi_0(t) $, 
we obtain the standard equations of FLRW cosmology,
\be \label{frw-phi-equations}
\dot\phi_0^2 = -2\dot H M_{\rm Pl}^2, \quad V(\phi_0) = (3H^2 + \dot H)M_{\rm Pl}^2,
\ee
with $H=\dfrac{\dot a}{a}$ the Hubble constant. Thus, the unperturbed matter action reads
\be \label{frw-0-action}
S^{(0)}_m = \int dx^3dt \sqrt{-g}\left[-(3H^2 + \dot H)M_{\rm Pl}^2  +  \dot H M_{\rm Pl}^2 g^{00} \right].
\ee
As advertised, this action is in the form of a polynomial over the aforementioned tensors with the first two arbitrary coefficients of the $g^{00}$ terms fixed. Note though that \eqref{frw-0-action} is just the zero-th order term in the perturbations and in reality it is followed by the infinite set of terms denoted by $\mathcal{G}^{\infty}$ in \eqref{general-eff-action}. Had we required for the canonical scalar field to be the full UV complete theory, the predictability of our effective action would have been the highest possible since that would mean that $S^{(0)}_m$ is the full action and $\mathcal{G}^{\infty} = 0$. In the next paragraph, we will discuss how one can recover known models of inflation by assigning specific values to the arbitrary effective couplings of $\mathcal{G}^{\infty}$. 

Since we have extracted the zero-th and first order terms in the metric perturbations
\be 
\delta g^{00} = g^{00} + 1,
\ee
we can rewrite the general effective action \eqref{general-eff-action} in a way that the FLRW background is manifest and all the corrections are at least quadratic in the perturbations. Defining the perturbation of an arbitrary tensor $T$ as
\be \label{random-pert-tensor}
\delta T = T - T^{(0)},
\ee
with $T^{(0)}$ denoting the background FLRW value of $T$, the effective action reads
\be \label{frw-eff-action}
\begin{split}
S & = \int dx^3dt \sqrt{-g} \Big[ \frac{M_{\rm Pl}^2}{2}R - (3H^2 + \dot H)M_{\rm Pl}^2  +  \dot H M_{\rm Pl}^2 g^{00}  \\ &+ \mathcal{G}^{\infty}(g^{00}+1,\delta R_{\mu\nu\rho\sigma},\delta K_{\mu\nu},\nabla_\mu,t) \Big],
\end{split}
\ee
where $\mathcal{G}^{\infty}$ now starts at quadratic order.
At this point let us comment on the inclusion of Riemann tensor powers in connection with the approach of Weinberg \cite{Weinberg:2008hq} to the effective field theory of inflation.
\subsubsection{Weinberg's approach} \label{sec:weinberg}
In \cite{Weinberg:2008hq}, the author starts with an effective unperturbed Lagrangian whose leading term is the Einstein Hilbert action coupled to matter \eqref{background-action} and the first corrections consist of all generally covariant terms with four spacetime derivatives. As shown in \cite{Elizalde:1994sn,Elizalde:1994nz} the possible choices of covariant four derivative terms are included in the following combinations:
\bea \label{weinberg-lagrangian}
\mathcal{L}^{(1)} &=& f_1 \left( g^{\mu\nu}\varphi_{;\mu}\varphi_{;\nu} \right)^2 + f_2 g^{\mu\nu}\varphi_{;\mu}\varphi_{;\nu} \Box \varphi + f_3 \left( \Box \varphi \right)^2 + f_4 R^{\mu\nu}\varphi_{;\mu}\varphi_{;\nu} + f_5 R g^{\mu\nu}\varphi_{;\mu}\varphi_{;\nu} \nn \\ &+& f_6 R \Box \varphi + f_7 R^2 + f_8 R^{\mu\nu} R_{\mu\nu} + f_9 C^{\mu\nu\rho\sigma}C_{\mu\nu\rho\sigma} + f_{10} \epsilon_{\mu\nu\kappa\lambda}C^{\mu\nu}_{\phantom{aa}\rho\sigma}C^{\kappa\lambda\rho\sigma},
\eea
where a semicolon denotes the covariant derivative, $C_{\mu\nu\rho\sigma}$ is the Weyl tensor and the coefficients are functions of the scalar $f_i \equiv f_i(\varphi)$. As shown by Ostrogradski long ago \cite{Ostrogradski}, a higher derivative theory like this leads to extra dynamical degrees of freedom and ghost instabilities. For example, time derivatives of the ADM shift vector and lapse function do not appear in the first order Lagrangian but do appear in the higher order terms, rendering these Lagrange multipliers dynamical. 

This situation is not special to the Lagrangian \eqref{weinberg-lagrangian} but a rather general feature of effective field theory. In fact, we will face the same problem in our discussion on ghosts in Sec.~\ref{sec:ghosts}, where we will have more to say about ``spurious" degrees of freedom but let us here briefly comment on how to cure such a pathology. The excitation of non dynamical degrees of freedom and ghosts simply tells us that the theory is not effective anymore and forces upon us an (intermediate) UV completion. In order to avoid such a case, we need to restrict our description to energies within the validity window of the EFT at hand. Consequently, the ratio of the characteristic energy of the system (hidden in derivative terms -- revealed in Fourier space) to the UV scale -- the upper value of the validity range -- provides us with a small parameter allowing for an expansion of the leading term Lagrangian. All the dynamical modes should therefore respect such an expansion \cite{PhysRevD.41.3720}. As Weinberg points out, this 
means that one may consider the dynamics of the theory at order $n$ in derivatives, only after ensuring that all the degrees of freedom are ``on-shell" with respect to the leading action, eliminating in this way all the higher derivative terms together with their spurious contributions.

The equations of motion for the leading action \eqref{background-action} are the Einstein field equations which after expressing the Ricci scalar through the trace of the matter energy momentum tensor $T_{\mu\nu}$ read
\be
R_{\mu\nu} = \frac{1}{M_{\rm Pl}^2} \left( T_{\mu\nu} -\frac{1}{2}Tg_{\mu\nu} \right) \; , \quad \Box\varphi = \frac{M_{\rm Pl}^2}{M^2} V'(\varphi),
\ee
where $M$ is a mass scale that has been used to make the scalar dimensionless, $i.e.$ $\varphi = \varphi_c/M$, with $\varphi_c$ the canonically normalised field. Using these equations, $i.e.$ imposing the ``on-shell" condition with respect to the leading term in the action, \eqref{weinberg-lagrangian} simplifies to 
\be \label{weinberg-lagrangian-2}
\mathcal{L}^{(1)} = f_1(\varphi) \left( g^{\mu\nu}\varphi_{;\mu}\varphi_{;\nu} \right)^2 + f_9(\varphi) C^{\mu\nu\rho\sigma}C_{\mu\nu\rho\sigma} + f_{10}(\varphi) \epsilon_{\mu\nu\kappa\lambda}C^{\mu\nu}_{\phantom{aa}\rho\sigma}C^{\kappa\lambda\rho\sigma},
\ee
where we have redefined the function $f_1$. In the unitary gauge \eqref{unitary-gauge}, where the perturbation of the scalar field is set to zero, the first term in \eqref{weinberg-lagrangian-2} reads $f_1(\phi_0) \left( g^{00}\dot\phi_0^2 \right)^2$. If one further sets $\phi_0(t)=\tilde t$, so that the background scalar field plays the role of time, $i.e.$ it is a clock, this term becomes the first (quadratic) contribution of $g^{00}$ to the arbitrary polynomial $\mathcal{G}^{\infty}$ in \eqref{frw-eff-action}. 
\subsubsection{Effective action for the comoving curvature perturbation} \label{sec:EFT-unitary}
We will now derive the effective action for the comoving curvature perturbation $\mathcal{R}$. In order not to overload the notation we will consider only extrinsic curvature perturbations and not contributions from the Weyl tensor, or the Riemann tensor in the notation of \eqref{frw-eff-action}, since the latter yield similar terms with the former which can be merged by redefining the arbitrary coefficients. An extensive analysis of such terms and their observational signatures in the flat gauge can be found in \cite{Bartolo:2010bj,Bartolo:2010di,Bartolo:2010im,Anderson:2014mga}.

Starting from the unitary gauge action \eqref{frw-eff-action} and truncating to second order in the perturbations we have
%
%
%
\be
\begin{split}
S_2 & = \int dx^3 dt \sqrt{-g} \bigg[ \frac{M_{\rm Pl}^2}{2} R + M_{\rm Pl}^2 \dot H g^{00} - M_{\rm Pl}^2 (3 H^2 + \dot H)+ \frac{1}{2!} M_2^4 (1+ g^{00})^2  \\  
& -  \frac{1}{2} \bar M_1^3 (1+ g^{00}) \delta K^{\mu}_{\mu}  -  \frac{1}{2} \bar M_2^2 ( \delta K^{\mu}_{\mu} )^2  -  \frac{1}{2} \bar M_3^2 \delta K^{\mu}_{\nu} \delta K^{\nu}_{\mu}    \bigg],    \end{split}
\label{starting-action}
\ee
where $M_n,\bar{M}_n$ are time dependent, mass-dimension one, arbitrary coefficients parametrising departures from a standard relativistic perfect fluid. After performing an ADM decomposition of the metric, as in \eqref{adm}, the pure gravitational part of the action reads
\be \label{e-h-action}
S_{\rm EH} =  \frac{M_{\rm Pl}^2}{2} \int dx^3 dt \sqrt{\gamma} N \left\{ R^{(3)}  + N^{-2} (K^{ij} K_{ij} - K^2) \right\} ,
\ee
where the extrinsic curvature is given by
\be \label{Kij}
K_{ij} = \frac{1}{2} \left( \dot \gamma_{ij} - \gamma_{j k}\nabla_{i} N^k - \gamma_{i k} \nabla_{j} N^k  \right)\;\;\text{and}\;\; K = \gamma^{ij}K_{ij},
\ee
with covariant derivatives $\nabla_iN^k=\partial_iN^k + \Gamma^{k}_{ij}N^j $ being constructed with respect to the spatial metric $\gamma_{ij}$ as in \eqref{unitary-gauge}. The affine connection in the comoving gauge reads
\be
\Gamma^{k}_{ij} = \frac{1}{2}\gamma^{kl} (\partial_{i} \gamma_{l j} + \partial_{j} \gamma_{i l} - \partial_{l} \gamma_{ij}) =   \delta_{j}^{k} \partial_{i} \mathcal{R} + \delta_{i}^{k} \partial_{j} \mathcal{R} - \delta_{i j}\partial^k \mathcal{R} ,
\ee
%
which may be inserted in the definition of $K_{ij}$, yielding
\be
K_{ij} = a^2  \frac{e^{ 2 \mathcal{R} } }{2} \left( 2 \delta_{ij} (H + \dot{\mathcal{R}} - N^k \partial_k \mathcal{R}) - \delta_{j k} \partial_{i} N^k -  \delta_{i k} \partial_{j} N^k  \right),
\ee
so that
\be \label{curv-traces}
\begin{split}
K_{ij} K^{ij} &= \frac{1}{3}\left(K^2-( \partial_{k} N^k)^2\right) +  \frac{1}{2} \left(  \partial_i N^j \partial_j N^i +  \partial^l N^k \partial_l N_k \right),  \\
K^2 &= 9(H + \dot{\mathcal{R}} - N^k \partial_k \mathcal{R})^2 - 6 (H + \dot{\mathcal{R}} - N^k \partial_k \mathcal{R}) \partial_{k} N^k + ( \partial_{k} N^k)^2.
\end{split}
\ee
%
%
The three-dimensional Ricci scalar is given by
\be \label{ric-3}
R^{(3)} = -2 e^{-2\mathcal{R}} \left( 2 \frac{\partial^2 \mathcal{R}}{a^2}  + \frac{(\partial \mathcal{R})^2}{a^2} \right),
\ee
where $(\partial \mathcal{R})^2 = \delta^{i j} \partial_i \mathcal{R} \partial_j \mathcal{R}$ and $\partial^2 = \delta^{i j} \partial_i  \partial_j $.
Now substituting \eqref{curv-traces} and \eqref{ric-3} into \eqref{e-h-action}, we obtain
\be
\begin{split}
S_{\rm EH} &=  \frac{M_{\rm Pl}^2}{2} \int dt dx^3 N a^3 e^{ 3 \mathcal{R}} \bigg\{ -4 a^{-2} e^{ -2 \mathcal{R}} \partial^2 \mathcal{R}  -  2 a^{-2} e^{ -2 \mathcal{R}} (\partial \mathcal{R})^2  \\
&- \frac{1}{ N^2 } \bigg( 6(H + \dot{\mathcal{R}} - N^k \partial_k \mathcal{R})^2 - 4 (H + \dot{\mathcal{R}} - N^k \partial_k \mathcal{R}) \partial_{k} N^k  + \left( \partial_{k} N^k \right)^2 \bigg)  \\ 
 &+   \frac{1}{2 N^2 } \left(  \partial_i N^j \partial_j N^i +  \delta_{i j } \partial^l N^i \partial_l N^j \right)   \bigg\} .
\end{split}
\ee
The matter sector is straightforward; we have
\be
S_{\rm m} = -  M_{\rm Pl}^2  \int dt dx^3 N a^3 e^{ 3 \mathcal{R}} \bigg\{  \dot H  \frac{1}{N^2} + 3 H^2 + \dot H \bigg\} .
\ee
The contribution of the $M_2^4$ term in \eqref{starting-action} is simply
\be
\frac{1}{2!} \int dt dx^3 N a^3 e^{ 3 \mathcal{R}} M_2^4 \left(1 - \frac{1}{N^2}  \right)^2.
\ee
We may also consider contributions from the extrinsic curvature. Having in mind \eqref{random-pert-tensor}, the extrinsic curvature perturbation reads
\be
\delta K^{i}{}_{j}  =   \frac{1}{2 N} \left( 2 \delta^{i}{}_{j} (H + \dot{\mathcal{R}} - N^k \partial_k \mathcal{R}) - \delta_{j k}\partial^{i} N^k -  \partial_{j} N^i  \right) - H \delta^{i}{}_{j},
\ee
whilst its trace is given by
\be
\delta K  =   \frac{1}{ N} \left( 3 (H + \dot{\mathcal{R}} - N^k \partial_k \mathcal{R}) -   \partial_{i} N^i  \right) - 3 H.  
\ee
%
One can now substitute everything back into \eqref{starting-action} to obtain an action involving the ADM lapse and shift functions. The next step is to integrate out these non dynamical degrees of freedom. This is performed by varying this action with  respect to $N$ and $\partial_iN^i$, which yields a system of algebraic coupled equations for $\partial_iN^i$ and $N$ respectively. Writing the lapse function as $N=1+\delta N$ and decomposing $N^k = \partial^k \psi_1 + \bar{N}^k$ with $\nabla\bar{N}=0$, the algebraic system can be decoupled and the ADM variables can be obtained as functions of the comoving curvature perturbation which is the dynamical degree of freedom in this gauge. We obtain
\be
\begin{split}
& \partial_iN^i = -\frac{2M_{\rm Pl}^2\dfrac{\partial^2 \mathcal{R}}{a^2}\left(H (2M_{\rm Pl}^2+3 \bar{M}_2^2+\bar{M}_3^2)-\bar{M}_1^3\right)}{C(M,\bar{M})}  \\ 
& \qquad\qquad\qquad\quad\qquad\qquad - \frac{\dot{\mathcal{R}} \left(2(\dot H M_{\rm Pl}^2 -2 M_2^4) (2M_{\rm Pl}^2+3\bar{M}_2^2 +\bar{M}_3^2)-3 \bar{M}_1^6\right)}{C(M,\bar{M})}, \\
& \delta N = \frac{2 M_{\rm Pl}^2\dfrac{\partial^2 \mathcal{R}}{a^2} (\bar{M}_2^2+\bar{M}_3^2) + 2  \dot{\mathcal{R}}(\bar{M}_3^2-M_{\rm Pl}^2) \left( \bar{M}_1^3-H(2M_{\rm Pl}^2+3 \bar{M}_2^2+\bar{M}_3^2) \right) }{C(M,\bar{M})}, \label{ADM-functions}
\end{split}
\ee
where we have defined
$$C(M,\bar{M}) \!=\! \bar{M}_1^6 - 2(\dot H M_{\rm Pl}^2 -2 M_2^4) (\bar{M}_2^2+\bar{M}_3^2) + 2 H^2 (M_{\rm Pl}^2 - \bar{M}_3^2) \left[(2M_{\rm Pl}^2 +3 \bar{M}_2^2+\bar{M}_3^2) - \frac{2\bar{M}_1^3}{H} \right].$$
The quadratic action for $\mathcal{R}$ thus reads
\be
S_2=\int dx^3 dt a^3\Bigg\{C_{\dot{\mathcal{R}}^2}\dot{\mathcal{R}}^2 + M_{\rm Pl}^2 \left( \frac{(\partial\mathcal{R})^2}{a^2} + C_{\dot{\mathcal{R}}\partial^2\mathcal{R}}\frac{\dot{\mathcal{R}}\partial^2\mathcal{R}}{a^2} \right) + M_{\rm Pl}^4 C_{(\partial^2\mathcal{R})^2}\frac{(\partial^2\mathcal{R})^2}{a^4}  \Bigg\},
\ee
with the following coefficients:
\be
\begin{split}
C_{\dot{\mathcal{R}}^2} &= \frac{(\bar{M}_3^2-M_{\rm Pl}^2) \left(2(\dot H M_{\rm Pl}^2 -2 M_2^4) (2M_{\rm Pl}^2 + 3 \bar{M}_2^2+\bar{M}_3^2)-3 \bar{M}_1^6\right)}{C(M,\bar{M})}, \\
C_{\dot{\mathcal{R}}\partial^2\mathcal{R}} &= \frac{4 (\bar{M}_3^2-M_{\rm Pl}^2) \left(H(2M_{\rm Pl}^2 + 3\bar{M}_2^2 + \bar{M}_3^2) -\bar{M}_1^3\right)}{C(M,\bar{M})}, \label{C_dot} \\
C_{(\partial^2\mathcal{R})^2} &= 2 \frac{ (\bar{M}_2^2 + \bar{M}_3^2)}{C(M,\bar{M})}.
\end{split}
\ee
The term $aC_{\dot{\mathcal{R}}\partial^2\mathcal{R}}\dot{\mathcal{R}}\partial^2\mathcal{R}$ can be integrated by parts twice
%
%
to give 
$$aC_{\dot{\mathcal{R}}\partial^2\mathcal{R}}\dot{\mathcal{R}}\partial^2\mathcal{R}= \frac{1}{2}\frac{d}{dt} \left( aC_{\dot{\mathcal{R}}\partial^2\mathcal{R}} \right) (\partial\mathcal{R})^2.$$
Therefore, the coefficient of the spatial kinetic term in the Lagrangian is
\be \label{Cpz}
C_{(\partial\mathcal{R})^2} = 1+\frac{1}{2a}\frac{d}{dt}\left( aC_{\dot{\mathcal{R}}\partial^2\mathcal{R}} \right) =  1+\frac{1}{2} \left( \dot C_{\dot{\mathcal{R}}\partial^2\mathcal{R}} + HC_{\dot{\mathcal{R}}\partial^2\mathcal{R}} \right),
\ee
and the effective second order action in the unitary gauge thus reads
\be \label{final-R-action}
S_2=\int dx^3 dt a^3 \Bigg\{C_{\dot{\mathcal{R}}^2}\dot{\mathcal{R}}^2 + M_{\rm Pl}^2 C_{(\partial\mathcal{R})^2} \frac{{(\partial\mathcal{R})^2}}{a^2}  + M_{\rm Pl}^4 C_{(\partial^2\mathcal{R})^2}\frac{(\partial^2\mathcal{R})^2}{a^4}  \Bigg\},
\ee
where $C_{(\partial\mathcal{R})^2}$ is given by \eqref{C_dot} and \eqref{Cpz}.

\subsection{Single field inflationary models as limits of the effective action} \label{sec:known-models}

In this paragraph, we summarise how various limits of the general Lagrangian \eqref{starting-action} reproduce known models of single field inflation. As evident from the discussion in Sec.~\ref{sec:FRW}, the minimal slow roll single field inflation with a canonical kinetic term \eqref{background-action}, corresponds to the limit 
\be 
M_n = 0,\;\;\forall \;\; n.
\ee
%
In order to see what kind of models the $M_n$ coefficients parametrise it is easier to consider them separately by setting the curvature coefficients $\bar{M}_n$ to zero. We then have 
%
\be
 C_{\dot{\mathcal{R}}^2} = \frac{\dot H M_{\rm Pl}^2 -2 M_2^4 }{H^2}, \quad
 C_{\dot{\mathcal{R}}\partial^2\mathcal{R}} = -\frac{2}{H}, \quad
 C_{(\partial^2\mathcal{R})^2} = 0,
\ee
so that the unitary gauge quadratic action becomes
\be \label{final-R-action-Mb=0}
S_2 = M_{\rm Pl}^2 \int dx^3 dt a^3 \epsilon \Bigg\{\frac{1}{c_{\rm s}^2}\dot{\mathcal{R}}^2 - \frac{{(\partial\mathcal{R})^2}}{a^2} \Bigg\},
\ee
where $\epsilon=\dfrac{|\dot H|}{H^2}$ is the slow roll parameter and with the speed of sound\footnote{We call the phase velocity, $c_{\rm s}=\frac{\omega(p)}{p}$, speed of sound. See \cite{Christopherson:2008ry} for a discussion of different notions of propagation speed used in the literature.} defined as 
\be \label{sof} 
\dfrac{1}{c_{\rm s}^2} = 1 + \dfrac{2 M_2^4 }{|\dot H| M_{\rm Pl}^2}.
\ee
This is the action for the perturbations of k-inflation \cite{ArmendarizPicon:1999rj}, which has a non minimal kinetic term for the inflaton, derived by Garriga and Mukhanov in \cite{Garriga:1999vw}. Note also from \eqref{ADM-functions} that the ADM variables in the $\bar{M}=0$ limit read 
\be 
\begin{split}
& \frac{\partial^2 \psi_1}{a^2} = -\dfrac{\partial^2\mathcal{R}}{a^2 H} + \frac{\epsilon}{c_{\rm s}^2}\dot{\mathcal{R}}, \\
& \delta N = \frac{ \dot{\mathcal{R}} }{H}, \label{ADM-functions-Pxf}
\end{split}
\ee
where we have set $N^i = \gamma^{ij}\partial_j \psi_1$. These are the ADM constraints found in \cite{Seery:2005wm}.
By further computing the third order action one finds full agreement with the results of \cite{Seery:2005wm} upon setting\footnote{The $M_1$ coefficient in \eqref{starting-action} is already fixed by the requirement of an FLRW background to $M_1^4 = |\dot H| M_{\rm Pl}^2 $.}
\be \label{Mn-for-Pxf}
M_n^4 = X^n \frac{\partial^n P}{\partial X^n}\Big|_{\phi=\phi_0(t)},
\ee 
where $X = g^{\mu\nu}\partial_\mu \phi \partial_\nu \phi$ and $P(X,\phi)$ is a generic function of the minimal kinetic term and the inflaton field.\footnote{The relation \eqref{Mn-for-Pxf} is evident already from the unitary gauge Lagrangian \eqref{starting-action} before the ADM constraints are implemented \cite{Cheung:2007st}.} Thus, we may conclude that the $M_n$ coefficients offer a parametrisation of non canonical $P(X,\phi)$ models.

Note that the appearance of $\epsilon$ in front of the quadratic action is in accordance with the interpretation of the comoving curvature perturbation as a Goldstone boson of the symmetry breaking $SO(4,1)\mapsto E_3$. It has a non trivial action only in the case where the de Sitter isometry is broken by the background, while in an exact de Sitter space it would be a pure gauge mode that could be eliminated by a conformal rescaling. 

This overall slow roll suppression implies that in the de Sitter limit, where the Hubble rate $H$ becomes a constant, the speed of sound \eqref{sof} vanishes. When extrinsic curvature terms are included though, as in \eqref{starting-action}, the limit of exactly de Sitter spacetime is well defined \cite{Cheung:2007st}, 
since then the Goldstone modes acquire a spatial kinetic term from the last contribution in \eqref{final-R-action}. These operators offer a parametrisation of non conventional models of inflation such as ghost inflation \cite{ArkaniHamed:2003uy,ArkaniHamed:2003uz}, admitting a non relativistic dispersion relation for the fluctuations. A similar dispersive behaviour can be achieved by the inclusion of higher derivative operators, without even considering extrinsic curvature contributions, but as we will see in the next section this case is distinct from the one discussed here. Having seen the effects of extrinsic curvature terms, henceforth, we will only consider terms including $\delta g^{00}$ and higher derivative operators.

\subsection{The effective action in the spatially flat gauge} \label{subsec:eft-flat}
We now move on to the spatially flat gauge where a slow roll expansion will be naturally implemented. One can follow the same steps as before: 
\begin{itemize}
\item start from the unitary gauge action and express all the quantities involved in terms of their flat gauge counterparts via the temporal gauge transformation \eqref{time-diff}, so that the degree of freedom that we are interested in, $\pi$, is explicit in the action. This is equivalent to the St\"uckelberg trick, as previously mentioned in the discussion above \eqref{Sgf}. 
\item Integrate out the ADM constraints and finally arrive to an action involving the dynamical degree of freedom \cite{Cheung:2007sv}. 
\end{itemize}
Since this computation is conceptually identical to the one described in the previous paragraph, here we only sketch the important steps. A detailed presentation appears in App.~\ref{app:slow-roll}. 

Starting from the action in the unitary gauge \eqref{starting-action} and expressing the time coordinate $t_{\rm com} = t_{\rm flat} + \pi$, as prescribed by the transformation that takes us from flat to comoving gauge \eqref{time-diff}, bearing in mind the definition \eqref{pi-def}, we arrive at the following action
\be  \label{S[pi,g]}
\begin{split}
S[\pi,g] &= M_{Pl}^2\int dx^3 dt \sqrt{-g}  \Big[ \frac{1}{2} R - 3H^2(t+\pi) - \dot H(t+\pi) \\ &+ \dot H(t+\pi) \left((1+\dot \pi)^2g^{00} +2(1+\dot\pi)\partial_i \pi g^{0i} + g^{ij}\partial_i \pi\partial_j \pi\right) \\ &+  \frac{M_2 ^4(t+\pi)}{2!M_{Pl}^2} \left((1+\dot \pi)^2g^{00} +2(1+\dot\pi)\partial_i \pi g^{0i} + g^{ij}\partial_i \pi\partial_j \pi\right)^2+\ldots \Big] ,
\end{split}
\ee
where we have made use of the relation between the metric tensors in the two gauges, namely
\be 
g^{00}_{\rm com} = (1+\dot \pi)^2g^{00}_{\rm flat} +2(1+\dot\pi)\partial_i \pi g^{0i}_{\rm flat} + g^{ij}_{\rm flat}\partial_i \pi\partial_j \pi.
\ee
Imposing the ADM constraints and Taylor expanding the time dependent quantities as 
\begin{equation} \label{taylor}
C(t+\pi)=C(t)+\dot C(t)\pi+\frac{1}{2}\ddot C(t)\pi^2+\ldots,
\end{equation}
we obtain an action for the Goldstone mode $\pi$, whose quadratic part reads
\be  \label{Seft2}
 S_2[\pi] = M_{\rm Pl}^2\int dx^3 dt a^3(t) \frac{|\dot H|}{c_{\rm s}^2} \Big[\dot\pi^2 - c_{\rm s}^2\frac{(\partial\pi)^2}{a^2} +3 \epsilon H^2\pi^2 \Big], 
\ee
with the speed of sound defined in \eqref{sof}. Note that \eqref{taylor} allows for the computation of the action to any desired order in slow roll. In App.~\ref{app:slow-roll}, we compute the quadratic part to higher order, making contact with known results in the literature ($e.g.$ \cite{Burrage:2011hd,Ribeiro:2012ar}).
An equivalent route to obtain \eqref{Seft2}, is to start with \eqref{final-R-action-Mb=0} and use the relation \eqref{R-to-pi-2nd-order} for $\xi^0 = \pi_{\rm flat}$, which at the linear level reads\footnote{In what follows we will often omit the label ``flat" of the Goldstone mode, assuming that we always work in the flat gauge where the degree of freedom $\pi$ is dynamical.} $\mathcal{R}=-H\pi_{\rm flat}$.

An immediate observation is that the canonically normalised Goldstone boson, $\pi_c = \frac{M_{\rm Pl}\sqrt{|\dot H|}}{c_{\rm s}}\pi$, appears to have a mass $\mathcal{M}_\pi \sim \sqrt{\epsilon}H$. In order to understand the origin of this mass it is instructive to look at \eqref{S[pi,g]} and observe that after canonically normalising $g^{00}_c = M_{\rm Pl}g^{00}$ through the Einstein-Hilbert contribution, the interaction term between the Goldstone boson $\pi$ and the ``gauge" boson $g^{00}$ is given by $\omega_{\rm mix} \sim \sqrt{\epsilon}H/c_{\rm s} \geq \mathcal{M}_\pi.$ This is in exact analogy with the gauge theory example of Sec.~\ref{sec:pions}. There, we saw that the action, written in the Feynman -- 't Hooft gauge via the gauge fixing \eqref{Sgf}, contains a mass term for the Goldstone boson equal to its coupling to the gauge boson. Using the equivalence theorem \eqref{equi-theorem}, we deduced that for energies higher than this mass scale the Goldstone mode decouples from the gauge dynamics and becomes the only relevant degree of freedom. 

Translated to the cosmological set up, this decoupling implies that for energies higher than $\omega_{\rm mix}$ the interaction with gravity and the mass of $\pi$ can be neglected. Since the characteristic scale of the system is $H$, the requirement of $H \gg \omega_{\rm mix}$ translates to imposing $\epsilon \ll 1$. In other words, in the inflationary context, the equivalence theorem implies the slow roll condition \cite{Cheung:2007st}. Having observed this nice analogy with gauge theory, in what follows we will work in this decoupling limit, $i.e.$ to first order in slow roll, where the effective action to cubic order in the perturbation reads
\be  \label{Seft-simple}
S[\pi] = \int dx^3 dt a^3 \Big\{ M_{\rm Pl}^2 \dot H (\partial_\mu\pi)^2 + 2M_2^4 \left(\dot \pi^2 + \dot\pi^3 - \dot\pi \frac{(\partial_i\pi)^2}{a^2} \right) -\frac{4}{3}M_3^4 \dot\pi^3 + \ldots \Big\}.
\ee
Before passing to the study of higher derivative operators in the effective action, let us summarise the progress made so far in the literature and highlight a few points of this formulation of inflationary perturbations.

First of all, using the principles of effective field theory, outlined in the Introduction, one can treat all single field inflationary models in a unified way. For instance, as discussed in Sec.~\ref{sec:known-models}, non canonical models like DBI or k-inflation \cite{Alishahiha:2004eh,Silverstein:2003hf,ArmendarizPicon:1999rj}, as well as more exotic models such as ghost inflation \cite{ArkaniHamed:2003uy,ArkaniHamed:2003uz}, can be collectively parametrised by the action \eqref{starting-action}. Although models endowed with Galilean symmetry \cite{Nicolis:2008in,Deffayet:2009wt,Deffayet:2009mn,deRham:2010eu,Burrage:2010cu,Kobayashi:2010cm} are not captured by \eqref{starting-action}, the idea of organising the IR properties of these models in a Lagrangian containing all operators respecting Galilean symmetry has been used in \cite{Creminelli:2010qf} to derive and study an effective action with $\pi$ as a dynamical field. The case of multi-field models has been discussed in the context of the EFT in \cite{Senatore:2010wk}, whilst in \cite{Khosravi:2012qg} the same subject has been analysed using Weinberg's approach, briefly outlined in Sec.~\ref{sec:weinberg}. Dissipation effects during inflation have been discussed in \cite{LopezNacir:2011kk}, where contact was made with models like trapped inflation \cite{Kofman:2004yc,Kofman:1997yn,Traschen:1990sw,Green:2009ds}, in which dissipation, introduced via couplings of the inflaton to other fields, makes inflation possible even in a steep potential. This formalism has also been extended to a supersymmetric context in \cite{Baumann:2011nk,Baumann:2011nm}. Finally, the idea of EFT has also been applied to the study of perturbations about general effective fluids \cite{Baumann:2010tm,Ballesteros:2012kv}, including dark energy and dark matter \cite{Gubitosi:2012hu,Hertzberg:2012qn} -- see \cite{Piazza:2013coa} for a review of the EFT of both early and late time acceleration -- as well as large scale structure \cite{Carrasco:2012cv}. 

Furthermore, the Goldstone boson action unravels correlations between different operators of the theory which are not manifest in the unitary gauge Lagrangian. For example, from \eqref{Seft-simple}, we see that the coefficient  $M_2^4$ of the $(g^{00}+1)^2$ term, appears in front of both the quadratic and cubic interaction. Its effect on the quadratic part is a reduction of the propagation speed, as shown in \eqref{sof}, while on the cubic part it is an enhancement of non Gaussianity. This demonstrates through symmetry arguments the well known fact that non Gaussianity is inversely proportional to the speed of sound. From \eqref{Seft-simple}, one may also deduce that the cubic interaction stems from two operators with different coefficients, an observation that led to the use of the orthogonal template \cite{Senatore:2009gt} in combination with the equilateral one.
 

   \chapter{Inclusion of higher derivative operators in the EFT of inflation} \label{ch:paper-3}

In the previous Chapter, we studied the effective action \eqref{frw-eff-action} ignoring arbitrary operators of the form $f(\Box)$. The purpose of this Chapter based on \cite{Gwyn:2012mw}, is to physically motivate specific operator insertions of this class and to present their possible effects on the low energy observables of inflation. Along the way we will further clarify aspects of this EFT, such as the validity of the effective action in the presence of these operators.

\section{Mass hierarchies and intermediate completions} \label{sec:intro-heavy-fields}
%
As discussed in Part~\ref{part:strings}, string theory is a candidate of a unifying quantum theory and provides a framework in which phenomenological questions may be addressed. Consistency of the theory requires ten spacetime dimensions, thus dimensional reductions on internal spaces are required for a realistic treatment of any physical problem. Moduli fields are a typical manifestation of our UV ignorance which have to be stabilised at large masses in order to be reconciled with low energy phenomenology.
%
Therefore, when studying inflation in a such a UV context, it is natural to include massive fields which in general should couple to the light inflaton.

These massive fields can span a wide range of heavy masses depending on $e.g.$ the mechanism that is responsible for their stabilisation. The measure of how heavy or light a degree of freedom is, depends on the characteristic scale of the specific problem. In the inflationary fluctuations context, all observable quantities are calculated at the energy scale set by the Hubble constant $H$, which geometrically describes the curvature of de Sitter space.

Given $n$ collections of fields with a hierarchical mass structure $ M_n \gg M_{n-1} \gg \ldots M_1 \gg H $ one can run a renormalisation group flow to an intermediate scale $M_{k} \gg \Lambda_m \gg M_{k-1} $ at which all fields of mass $M \gg \Lambda_m $ can be integrated out. The remaining fields are the dynamical degrees of freedom that describe the most important processes of the system at energies $\Lambda_m$. 

In what follows, we may some times refer to the effective theory at an intermediate scale $\Lambda_{m}$ as the completion of the theory at a lower scale $\Lambda_{m-1}$ or as an intermediate UV-completion, although the term UV-completion usually refers to the full finite theory including all the fields. It is important to stress that when such a term is (ab)used, what is meant is always an effective field theory in the sense described in the Introduction.

\subsection{Physically motivated operators}
As we shall see in Sec.~\ref{sec:integration-1-massive-field}, the process of integrating out heavy fields results in the replacement of the low energy couplings with operators of the form
\be \label{first-operator}
\frac{\beta^2}{M^2 - \Box},
\ee
where $\beta$ is a mass-dimension one coupling, $M$ is the heavy mass and $\Box = \frac{1}{\sqrt{g}}\partial_\mu \sqrt{g} g^{\mu\nu} \partial_\nu $ is the D' Alembertian operator. This is to say that the contribution of a term $(g^{00}+1)^n$ in the effective action \eqref{starting-action} should be replaced with
\be \nn
\mathcal L_{\rm EFT}^{(n)} \propto   \left[ (g^{00}+1) \frac{M^2 }{ M^2 - \Box} \right]^{n-1} (g^{00}+1). \label{basic-modification}
\ee
The notation $\dfrac{1}{\hat{O}}$ is used throughout the text to denote the inverse of an operator defined as $\dfrac{1}{\hat{O}} \hat{O} f(x) = f(x)$ and most of the time we will be working in Fourier space writing $\dfrac{1}{\Box} = \dfrac{1}{\omega^2 - p^2}$ and neglecting the friction term $3H \partial_t$ in the D' Alembertian.

The dimensionful coupling $\beta$ in \eqref{first-operator} is parametrised as $\beta_n^2 = M^2 \frac{M_n^4}{M_{\rm Pl}^2 |\dot H|}$ such that in the limit $M \rightarrow \infty$ any UV effect vanishes and one recovers the operators of \eqref{starting-action},
$$
\mathcal L_{\rm EFT}^{(n)} \propto \frac{M_n^4}{M_{\rm Pl}^2 |\dot H|}(g^{00}+1)^n.
$$
In sufficiently low energies, such an insertion can be expanded as
\be \label{first-expansion}
\frac{M^2}{M^2 - \Box} = 1 + \frac{\Box}{M^2} + \left( \frac{\Box}{M^2} \right)^2 \ldots,
\ee
which to leading order yields back the effective action we have been considering so far. The leading correction $\dfrac{\Box}{M^2}$ has been studied in \cite{Baumann:2011su}, where it was shown to be irrelevant for observations. Assuming a relativistic dispersion relation of the form $\omega = c_{\rm s} p,$ with $ c_{\rm s} \ll 1$, the expansion \eqref{first-expansion} requires 
\be \label{low-energy-cond-0}
p^2 \ll M^2 \quad \text{or equivalently} \quad \omega^2 \ll M^2c_{\rm s}^2,
\ee
which defines the low energy regime. 
Now recall that at the core of the EFT formalism lies the spontaneous breakdown of time reparametrisation invariance, a fact which allows for a non relativistic regime where a hierarchy between energy and momentum may exist. This means that the expansion \eqref{first-expansion} can be replaced by
\be
\frac{ M^2 }{ M^2 - \Box} = \frac{ M^2 }{M^2 - \nabla^2}  \left( 1 - \frac{\partial_t^2 }{M^2 - \nabla^2} + \cdots \right), \label{second-expansion}
\ee
from which the low energy condition can be deduced as
\be \label{low-energy-cond}
\omega^2 \ll p^2 + M^2.
\ee
In such a non relativistic regime, contrary to \eqref{first-expansion} and \eqref{low-energy-cond-0}, one can have $p^2 \gg M^2$ or $\omega^2 \gg M^2 c_{\rm s}^2$ without exciting massive fields. As we will show in the next section and further demonstrate in Sec.~\ref{sec:ghosts}, the high energy scale, at which additional degrees of freedom become operative, is set by 
\be \label{luv}
\Lambda_{\rm UV} = \dfrac{M}{c_{\rm s}}.
\ee
Worrisome is the fact that in both cases suspicious higher time derivatives appear which may lead to extra dynamical degrees of freedom and ghosts. We address this issue in Sec.~\ref{sec:ghosts}, where the UV scale will be derived from the requirement that no propagating ghost states appear in low energies. 
We will therefore consider the following modification of the action \eqref{starting-action}
\be
\mathcal L_{\rm EFT}^{(n)} \propto   \left[ (g^{00}+1) \frac{ M^2 }{ M^2 - \nabla^2} \right]^{n-1} (g^{00}+1). \label{eft-modified}
\ee

Let us close these introductory remarks by stressing once more that we will be discussing a class of higher derivative operators that capture effects of massive fields coupled to the inflationary process at some high energy scale. In the context of the EFT of \cite{Cheung:2007st} one may study any kind of operator one wishes with the only constraint being the symmetry of the problem. 
Although it might not always be the most fruitful choice to bound oneself by physical intuition, we will restrict our study to the case of the operators \eqref{eft-modified}, the presence of which has a clear physical meaning.

\section{Dispersion relation and characteristic scales} \label{sec:dispersion+scales}
We now consider the action \eqref{Seft-simple} to cubic order, which after implementing the modification \eqref{eft-modified} reads 
\be 
\begin{split}
S &= -M_{\rm Pl}^2 \int d^3xdt a^3 \dot H \bigg[   \dot \pi \bigg( 1 + \frac{2 M_2^4 }{M_{\rm Pl}^2 |  \dot H| }  \frac{M^2}{M^2 - \tilde \nabla^2} \bigg) \dot \pi -   (\tilde \nabla \pi)^2  \bigg] \\
& +  \int d^3xdt a^3 \bigg[   2 M_2^{4}  \bigg(   \dot \pi^2 -  (\tilde \nabla \pi)^2   \bigg) \frac{M^2}{M^2 - \tilde \nabla^2} \dot \pi   -  \frac{4}{3} M_3^{4}   \bigg(  \dot \pi  \frac{M^2}{M^2 - \tilde \nabla^2}  \bigg)^2 \dot \pi    \bigg], \label{EFT-new-physics-1}
\end{split}
\ee 
where $\tilde\nabla \equiv \dfrac{\nabla}{a}$. The Goldstone boson has acquired a non trivial kinetic term with a strong scale dependence. As a consequence, the dispersion relation characterising the free theory is
\be
\omega(p) =  \sqrt{ \frac{M^2 + p^2}{ M^2 c_{\rm s}^{-2} + p^2} } p , \label{full-modified-dispersion}
\ee
with the speed of sound, defined in the long wavelength limit, given by \eqref{sof} and $p \equiv k/a$ denoting the physical momentum. The low energy condition \eqref{low-energy-cond} now translates to
\be
p^2 \ll \frac{ M^2 }{c_{\rm s}^{2}} \quad \text{and}  \quad \omega^2 \ll \frac{ M^2 }{c_{\rm s}^{2}}, \label{puv}
\ee
where the second inequality follows from the first and the dispersion relation \eqref{full-modified-dispersion}. We thus see that the scale that sets the UV cutoff of \eqref{EFT-new-physics-1} is indeed \eqref{luv} for both energy and momentum. Above this scale, the expansion~(\ref{second-expansion}) breaks down and the system has to be explicitly UV-completed in such a way that it incorporates the states characterised by the mass scale $M c_{\rm s}^{-1}$ as new degrees of freedom. 
In the limit \eqref{puv}, which we henceforth assume, the dispersion relation may be expanded as 
\be
\omega^2 (p) =   c_{\rm s}^2 p^2  + \frac{   ( 1  - c_{\rm s}^2 )   }{  \Lambda_{\rm UV}^2  } p^4 + \mathcal O (p^6) , \label{low-energy-dispersion}
\ee
the term proportional to $p^6$ being subleading.

Up to this point the system seems to be characterised by the two scales $\Lambda_{\rm UV}$ and the Hubble rate $H$ and the dimensionful couplings $M_n$. There exists another important scale and that is $M$ itself. Even though $M$ and $\Lambda_{\rm UV}$ are related through \eqref{luv} they are completely independent from an IR point of view, in the sense that knowing one can not determine the other. The speed of sound is a relatively\footnote{Here ``relatively" refers to the fact that the speed of sound is bounded from below by observations.} 
free parameter, knowledge of which would require UV input into the problem hence naturally reducing the unknown, dimensionful or dimensionless parameters.

In order to see what this scale signifies in the IR we may inspect the low energy dispersion relation \eqref{low-energy-dispersion} to observe that when $p^2 \ll M^2$, the  quadratic term dominates whereas in the range $M^2 \ll p^2 \ll   \Lambda_{\rm UV}^2$ the quartic term takes over. The associated threshold energy scale is 
\be \label{lnew}
\Lambda_{\rm new} = M c_{\rm s}, 
\ee
found by evaluating $\omega^2$ at $p^2 = M^2$. The label ``new"\footnote{First coined in \cite{Baumann:2011su}.} becomes clearer if we Taylor expand the dispersion in the two regimes:
\bea
\omega (p) &=&   c_{\rm s} p \qquad\qquad\quad\; ; \quad p^2 \ll M^2, \label{linear-dispersion} \\
\omega (p) &=&   \frac{p^2}{\Lambda_{\rm UV}} +  \frac{\Lambda_{\rm new}}{2}\quad ; \quad M^2 \ll p^2 \ll \Lambda_{\rm UV}^2. \label{non-linear-dispersion}
\eea
Rewriting the non linear part as 
\be \label{schrodinger-dispersion}
\omega = \dfrac{p^2}{2M_{\rm eff}}  + M_{\rm eff}c_{\rm s}^2,
\ee
while bearing in mind \eqref{lnew}, we identify the energy spectrum of a non relativistic particle of mass $M_{\rm eff} = \dfrac{\Lambda_{\rm UV}}{2}$ with a rest energy $M_{\rm eff}c_{\rm s}^2 = \dfrac{\Lambda_{\rm new}}{2}$. Note that it is important to consider the rest energy instead of the mass $M_{\rm eff}$ since our mode propagates with a phase velocity less than unity. 

This effective mass stems from the fact that the process of integrating out heavy modes is analogous to the insertion of a ``medium" through which the IR mode propagates. This medium is responsible for the reduction of the speed of sound and as a result the mode behaves as if it were massive. Even though the Lagrangian of $\pi$ does not contain a mass term, at least in lowest order in slow roll, $M_{\rm eff}$ appears at the level of the dynamics as a reminiscent of the UV physics that we have hidden in the IR medium. 
The linear dispersion relation would thus be more appropriately described by the analogy with a phonon. It is important to stress that $\Lambda_{\rm new}$ is not a spurious scale, but a real independent physical one which is inserted in the problem via the change in the dispersive behaviour of the IR mode. Depending on its value relative to $H$, the observables change as we will demonstrate later. The label ``new" refers to the dynamical change in the description of the IR field: it sets the scale where particle-like excitations with a non linear dispersion \eqref{schrodinger-dispersion} start dominating over phonon-like ones with a definite speed of sound \eqref{linear-dispersion}.

Let us now adopt this idea of the light mode propagating through a medium provided by the heavy field, and see how the effective mass, appearing in \eqref{schrodinger-dispersion}, arises from such a perspective. Given a medium with a refractive index and a group refractive index defined as
\be \label{refr-ind-defs}
n = \frac{1}{v_{\rm ph}} = \frac{1}{\omega/p}, \qquad n_{\rm g} = \frac{1}{v_{\rm g}} = \frac{1}{\partial \omega/ \partial p},
\ee
respectively, where $v_{\rm ph},v_{\rm g}$ denote the phase and group velocities, one may associate an effective mass to the light mode, which reads
\be \nn 
M_{\rm eff} = n n_{\rm g} \omega .
\ee
Using the dispersion relation \eqref{full-modified-dispersion}, we obtain
\be \label{ref-indices}
n(p) = \sqrt{\frac{\Lambda_{\rm UV}^{2} + p^2}{M^{2} + p^2}}, \quad n_{\rm g}(p) = n(p) \frac{\Lambda_{\rm UV}^{2} + p^2}{\Lambda_{\rm UV}^{2} + p^2n(p)^2}, \quad M_{\rm eff}(p) = n_{\rm g}(p) p.
\ee
We thus see that the phase speed $v_{\rm ph}$ and the effective mass depend on momenta, which is to be expected since the dispersion \eqref{full-modified-dispersion}, rewritten as $\omega = v_{\rm ph}(p)p$, is characteristic of a dispersive medium. 

We will now calculate these quantities at the Hubble scale, which is the characteristic scale of our system, assuming $p<\Lambda_{\rm UV}$, since this is where our effective theory is valid. From the dispersion relation \eqref{full-modified-dispersion}, we may obtain the momentum $p_*$ corresponding to $\omega(p_*) = H$ as a function of the dimensionless ratio $x \equiv \dfrac{H}{\Lambda_{\rm new}}$, namely
\be \label{pstar-of-x}
p_*^2(x) = \frac{M^2}{2}\left( \sqrt{1+4x^2}-1 \right),
\ee
so that \eqref{refr-ind-defs},\eqref{ref-indices} yield
\be \label{phase-v-of-x}
v_{\rm ph}^*(x) = c_{\rm s} \sqrt{\frac{1 + \sqrt{1 + 4x^2}}{2}},
\ee
as well as
\be \label{ref-indices-of-x}
v_{\rm g}^*(x) = c_{\rm s} \sqrt{\frac{2 + 8 x^2}{1+\sqrt{1 + 4 x^2}}} \quad \text{and} \quad M^*_{\rm eff}(x) = \Lambda_{\rm UV} \frac{ x}{ \sqrt{1+4 x^2}}.
\ee
Note that using the formulae for the effective mass and the group velocity at the Hubble scale, we may rewrite \eqref{pstar-of-x} as
\be \label{pstar-of-x-Meff}
p_*(x) = v^{*}_{\rm g}(x) M^{*}_{\rm eff}(x),
\ee
which essentially follows form \eqref{ref-indices} evaluated at $\omega = H$.
Now in the limit $x<1$, that is in the linear regime \eqref{linear-dispersion}, these expressions reduce to  
\be \label{ref-indices-linear-regime}
p_* = \frac{H}{c_{\rm s}}, \quad v_{\rm ph}^* = v_{\rm g}^*  = c_{\rm s}, \quad M^*_{\rm eff} = \frac{H}{c_{\rm s}^2},
\ee
as expected, while in the dispersive regime \eqref{non-linear-dispersion}, where $x>1$, we obtain
\be \label{ref-indices-non-linear-regime}
p_* = \sqrt{H \Lambda_{\rm UV}}, \quad v_{\rm ph}^* = \sqrt{\frac{H}{\Lambda_{\rm UV}}}, \quad v^*_{\rm g} = 2 v_{\rm ph}^*, \quad M^*_{\rm eff} = \frac{\Lambda_{\rm new}/2}{c_{\rm s}^2},
\ee
which fully agrees with \eqref{schrodinger-dispersion}. Let us finally observe that from \eqref{ref-indices-linear-regime} and \eqref{ref-indices-non-linear-regime}, the ratio of the effective masses in the two regimes $x<1$ and $x>1$, reads
\be \label{mass-ratio-x}
\frac{M^*_{\rm eff; \; linear}}{M^*_{\rm eff; \; non\;linear}} \propto x,
\ee
where the labels ``linear" and ``non linear" have been chosen according to the dispersion relation in each case.

These relations suggest that in presence of a heavy field -- corresponding to an IR dispersive medium of characteristic length $M^{-1}$, once integrated out -- the light mode admits two physical descriptions: 
{\it when its wavelength at the Hubble scale lies below the characteristic length of the medium, it may be interpreted as a particle excitation with a dispersion relation of the form \eqref{schrodinger-dispersion}; alternatively, when the mode reaches the Hubble scale with a wavelength greater than the characteristic length of the medium, it may be realised as a sound wave (or phonon excitation) with a linear dispersion relation of the form \eqref{linear-dispersion}}. 

\begin{figure}[htb]
\begin{center}
\includegraphics[scale=.6]{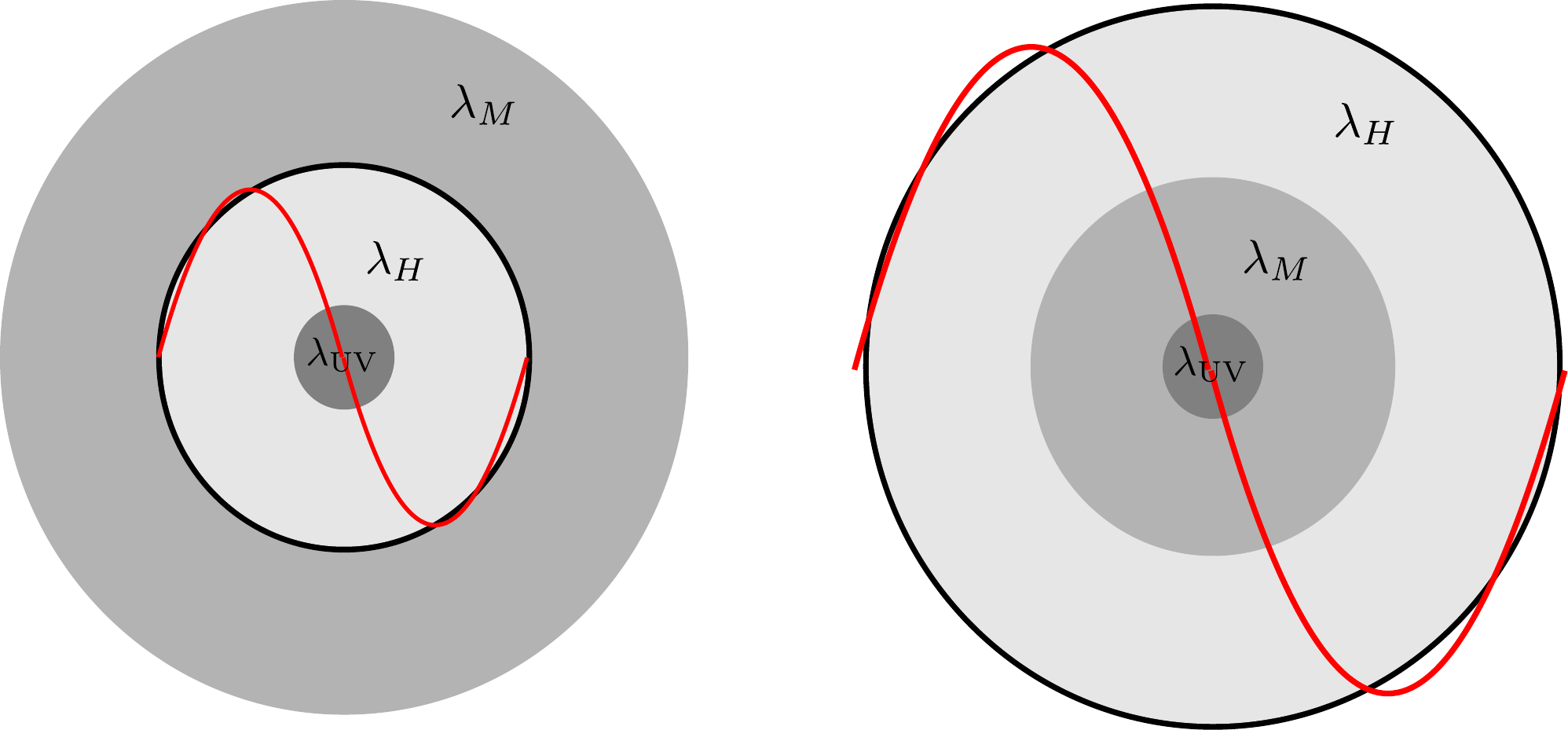} 
\caption[{\sf Goldstone mode through an effective medium}.]{\sf The Hubble volume lies either within the dispersive medium $\lambda_H < \lambda_M$ (left panel) or outside $\lambda_M < \lambda_H$ (right panel). In the former case, a mode (red wavy line) reaching the Hubble wavelength $\lambda_H=1/\sqrt{H\Lambda_{\rm UV}}$ (see \eqref{ref-indices-non-linear-regime}) feels the effective medium, whose characteristic length is $\lambda_M = 1/M$, and admits a particle interpretation, while in the latter case, where $\lambda_H=c_{\rm s}/H$ (see \eqref{ref-indices-linear-regime}), the mode does not feel the medium, hence exhibiting a phonon behaviour. Depending on which behaviour the mode has at the Hubble scale, where it freezes, the information contained in the observables stemming from the correlation functions changes. The UV length scale $\lambda_{\rm UV} = \Lambda_{\rm UV}^{-1}$ is also depicted.} 
\label{fig:media}
\end{center}       
\end{figure}

From \eqref{mass-ratio-x}, we see indeed that when $x<1$, the phonon mode is energetically preferable, while in the case $x>1$, a particle is the favourable excitation. Note that the two extreme limits $x \to 0$ and $x \to \infty$, correspond to $M \to \infty$ and $M \to 0$, respectively. Therefore, the phonon excitation is described by the effective Lagrangian \eqref{Seft-simple}, while the particle one corresponds to the higher derivative extended EFT \eqref{EFT-new-physics-1}. In what follows, we will see that \eqref{EFT-new-physics-1} is a straightforward generalisation of \eqref{Seft-simple}, with the speed of sound $c_{\rm s}$ replaced by the phase velocity $v_{\rm ph}$ -- see \eqref{ref-indices-linear-regime} and \eqref{ref-indices-non-linear-regime}. Thus, the predictions of the two Lagrangians are identical in form: they both depend on the phase velocity, only that in each case, this quantity is related to a different set of unknown parameters. We summarise our claim in Fig.~\ref{fig:media}. 
%


Finally, let us clarify that just as for the case described by the effective theory \eqref{Seft-simple}, its extended version \eqref{EFT-new-physics-1}, taken on its own, provides no explicit information about the value of the UV cutoff scale $\Lambda_{\rm UV}$ at which the effective field theory breaks down. In other words, the theory (\ref{EFT-new-physics-1}) may be taken literally as it reads all the way up to momenta $p \gg M c_s^{-1}$, for which the dispersion relation (\ref{full-modified-dispersion}) becomes $\omega^2 (p) \simeq  p^2$, consistent with a Lorentz invariant spectrum of massless particles. 
However, to keep our discussion on firm physical grounds, we assume a UV cutoff $\Lambda_{\rm UV}$, consistent with the existence of a UV regime where the Goldstone boson interacts with one or more heavy fields.

\subsection{The symmetry breaking scale}
Following \cite{Baumann:2011su}, we now discuss the scale at which the temporal reparametrisation invariance is broken and the Goldstone boson appears in the dynamics. As already mentioned in the discussion around \eqref{schematic-pert}, the presence of the scalar perturbation $\delta \phi$, which we have been referring to as $\pi$, generates a perturbation of the energy momentum tensor, which is proportional to $\dot\pi$ (see $e.g.$ \cite{Riotto:2002yw}). Since $\dot\pi$ is dimensionless, the proportionality constant $\delta T^{00}/\dot\pi$ should represent an energy density associated with the scalar perturbation, and it is this energy that we call symmetry breaking scale.

Restricting ourselves to the decoupling limit, the time component of the unitary gauge metric reads $g^{00} = -1 - 2\dot\pi + \mathcal{O}(\dot\pi^2)$ so that $\dfrac{\partial}{\partial g^{00}} = -\dfrac{1}{2}\dfrac{\partial}{\partial \dot\pi}$. Thus, the perturbed temporal component of the energy momentum tensor in the spatially flat gauge is given by\footnote{The same expression can be obtained by computing the Noether current associated to the broken time reparametrisation invariance \cite{ArkaniHamed:2003uy,Baumann:2011su}.}
\be 
T^{00} = -\frac{2}{\sqrt{-g}}\frac{\delta S^{(2)}}{\delta g^{00}} = 2 M_{\rm Pl}^2 |\dot H|  \left( \frac{\Lambda_{\rm UV}^2 + p^2}{ M^2 + p^2} \right) \dot\pi,
\ee
where $S^{(2)}$ is the quadratic part of the action \eqref{EFT-new-physics-1}. According to our discussion on the previous paragraph, on dimensional grounds, this should read
\be 
T^{00} = 2 M_{\rm Pl}^2 |\dot H| \left( \frac{\Lambda_{\rm UV}^2 + p^2}{ M^2 + p^2} \right)  \dot\pi = \Lambda_{\rm sb} p^3 \dot\pi.
\ee
Using the non linear dispersion relation \eqref{non-linear-dispersion} to compute $p$ at $\omega = \Lambda_{\rm sb}$ and considering the new physics regime limit $M \ll p \ll \Lambda_{\rm UV}$, 
we obtain~\cite{Baumann:2011su}
\be 
\Lambda_{\rm sb} = \left( \frac{2 M_{\rm Pl}^2 | \dot H |}{\Lambda_{\rm UV}^4 } \right)^{2/7} \Lambda_{\rm UV}. \label{symmetry-breaking-scale}
\ee

This result is consistent with the intuitive expectation that the symmetry breaking scale should be given in terms of background quantities. In order to see that, it is instructive to compute this quantity for the low derivative EFT and then compare to our effective medium interpretation. Following the same procedure, one finds \cite{Baumann:2011su} that the symmetry breaking scale of the EFT \eqref{Seft-simple} reads $\Lambda_{\rm sb}^4 = 2 M_{\rm Pl}^2 | \dot H | c_{\rm s}$. Now recall our discussion on the previous section, where we claimed that all the quantities computed with \eqref{Seft-simple} should generalise to those of \eqref{EFT-new-physics-1}, upon using the phase velocity \eqref{ref-indices} evaluated at the relevant scale, that is $\omega = \Lambda_{\rm sb}$ for our purpose here, in place of $c_{\rm s}$ defined in \eqref{sof}. Indeed, computing the symmetry breaking scale using the formula 
\be 
\Lambda_{\rm sb}^4 = 2 M_{\rm Pl}^2 | \dot H | v_{\rm ph}(x), \label{symmetry-breaking-scale-v-ph}
\ee
we recover, in the $x>1$ limit, the result \eqref{symmetry-breaking-scale}. Therefore, the symmetry breaking scale does depend solely on background quantities, only now the light mode propagates along a modified background, which is manifested through its effect on the phase velocity.

The expression \eqref{symmetry-breaking-scale}, allows us to see that the value of $\Lambda_{\rm sb}$ compared to $\Lambda_{\rm UV}$ depends on the ratio $ M_{\rm Pl}^2 | \dot H | /  \Lambda_{\rm UV}^4$. For instance, if the UV physics in charge of modifying the low energy dynamics of curvature perturbations is also responsible for producing inflation, it is perfectly feasible to have $ \Lambda_{\rm UV}^4 \sim  M_{\rm Pl}^2 | \dot H | $, implying $\Lambda_{\rm sb}  \sim \Lambda_{\rm UV}$. 

\subsection{The strong coupling scale} \label{sec:strong-coupling}
Another important dimensionful quantity of the problem is the scale $\Lambda_{\rm sc}$ at which the effective theory becomes strongly coupled. This inevitably sets an upper cutoff of the effective theory from a practical point of view: if a theory exhibits strong coupling behaviour, tree level calculations are of the same order with loop corrections, hence perturbative control is lost. 

Before proceeding with the computation of $\Lambda_{\rm sc}$ for the theory \eqref{EFT-new-physics-1}, let us discuss the strong coupling regime of the EFT \eqref{Seft-simple} in the absence of heavy propagators. In \cite{Cheung:2007st}, the scale at which the Goldstone boson self interactions become of order one was derived by requiring tree level unitarity of a scattering process $\pi\;\;\pi \; \rightarrow \; \pi\;\;\pi$ and was found to be\footnote{Note that in \cite{Baumann:2011su} the strong coupling scale differs from \eqref{standard-sc} by a factor of $1/4\pi \sim 0.1$. This will not affect our discussion though.}
\be
\Lambda_{\rm sc}^4 = 16 \pi^2 M_{\rm Pl}^2 |\dot H | \frac{c_{\rm s}^5}{ 1 - c_{\rm s}^2} . \label{standard-sc}
\ee
A possible caveat of this approach is the fact that the relevant quantity involved in observations for inflation, as pointed out in \cite{Maldacena:2002vr}, is not a scattering amplitude but a correlation function. Perturbative calculations should thus be performed using 
the $in - in$ formalism\footnote{The $in-in$ formalism was developed in the early 60's by Schwinger and others \cite{Schwinger:1960qe,Bakshi:1962dv,Bakshi:1963bn,Keldysh:1964ud}. Its relevance for cosmology was noticed in \cite{Jordan:1986ug,Calzetta:1986ey} while its application to correlators of cosmological perturbations was established in \cite{Maldacena:2002vr,Weinberg:2005vy} -- see \cite{Seery:2007we,Koyama:2010xj,Chen:2010xka} for recent reviews.} as opposed to the $in - out$ formalism used for scattering amplitudes. Nevertheless, as pointed out in \cite{Adshead:2009cb}, the $in-in$ correlation function may be formulated in terms of $in-out$ scattering amplitudes. Hence, the latter method gives a correct estimation of the scale $\Lambda_{\rm sc}$. 

The requirement of the theory remaining weakly coupled at horizon crossing, $H \ll \Lambda_{\rm sc}$, translates through \eqref{standard-sc} into a lower bound on the speed of sound $c_{\rm s} > 10^{-2}$ consistent with the observational bound \cite{Hinshaw:2012fq,Bennett:2012fp}. However, for small values of the speed of sound ($c_{\rm s}^2 \ll 1$), Goldstone boson modes described by (\ref{starting-action}) may appear strongly coupled at energies well below the cutoff energy scale $\Lambda_{\rm UV}$ at which UV degrees of freedom become excited. 
Furthermore, \eqref{standard-sc} is strictly valid only for the case in which $\omega^2 = c_{\rm s}^2 p^2$, characteristic of the standard EFT picture. In fact, in \cite{Baumann:2011su}  it was found that a modification of the dispersion relation will generally alleviate the strong coupling problem by making $\Lambda_{\rm sc}$ larger than the  value of (\ref{standard-sc}). In that study though, a general analysis incorporating the scale dependence of self-interactions consistent with the modification of the dispersion relation was not taken into account. In what follows we incorporate this aspect into the analysis of strong coupling and show that the conditions for the theory to remain weakly coupled are satisfied all the way up to an energy scale of the same order as, or larger than, the natural cutoff $\Lambda_{\rm UV}$ of the new physics regime.

To proceed, we will closely follow the analysis of \cite{Baumann:2011su}. First, by normalising the Goldstone boson as $\pi_n = ( 2 M_{\rm Pl}^2 \epsilon H^2  )^{1/2} \pi$ the quadratic part of the extended EFT action \eqref{EFT-new-physics-1} may be written as
\be
S = \frac{1}{2}  \int dx^3dt  \bigg[    \dot \pi_n \left(   \frac{   \Lambda_{\rm UV}^2 -  \nabla^2  }{M^2 -  \nabla^2}  \right) \dot \pi_n -   ( \nabla \pi_n)^2   \bigg] . \label{gaussian-eft-no-a}
\ee
Notice that we have fixed $a=1$ for the sake of simplicity, assuming that our discussion involves processes at energy scales much larger than $H$. Now, the conjugate momentum $P_{\pi} \equiv \partial \mathcal L / \partial \dot \pi_n $ of this free field theory is given by
\be
P_{\pi}  = \frac{\Lambda_{\rm UV}^2  - \nabla^2}{   M^2 -   \nabla^2} \dot \pi_n .
\ee
This implies that the commutation relation $[\pi_n , P_{\pi}] = i \delta$ reads
\be
\left[ \pi_n  ({\bf x}_1) \, ,  \dot \pi_n  ({\bf x}_2) \right] = i  \frac{M^2 -  \nabla^2_2}{   \Lambda_{\rm UV}^2 -   \nabla^2_2} \delta ({\bf x}_1 - {\bf x}_2 )  , \label{commutation-new}
\ee
where $\nabla^2_2$ stands for a Laplacian operator written in terms of the coordinate ${\bf x}_2$. 
%
%
In order to satisfy these commutation relations we may consider the quantisation of the free field $\pi_n (x)$ in terms of creation and annihilation operators $\hat a^{\dag}({\bf k})$ and $\hat a({\bf k})$ satisfying \\ $[\hat a({\bf k}_1) , \hat a^{\dag}({\bf k}_2) ] = \delta({\bf k}_1 - {\bf k}_2)$. We find
\be
\pi_n (x) = \frac{1}{(2 \pi)^{3/2}} \int d^3 p \left[ \pi_n (p) \hat a({\bf p}) e^{- i \omega t +  i {\bf p } \cdot {\bf x} }  + \pi_n (p)^{*} \hat a^{\dag}({\bf p}) e^{+ i \omega t -  i {\bf p } \cdot {\bf x} }   \right] ,
\ee
where $\pi_n (p)$ corresponds to the field amplitude in Fourier space, given by
\be
\pi_n (p) = \sqrt{ \frac{M^2 +p^2}{   \Lambda_{\rm UV}^2  +  p^2 } } \frac{1}{\sqrt{2\omega(p)}} = v_{\rm ph}(p) \frac{1}{\sqrt{2\omega(p)}}, \label{new-amplitude}
\ee
where $v_{\rm ph}(p) $ is the phase velocity defined in \eqref{ref-indices}. After canonically normalising $\pi_c = \pi_n / v_{\rm ph}$, we obtain the standard normalisation $\pi_c (p) = \frac{1}{\sqrt{2\omega(p)}}$. Note though that the functional form of $\pi_n (p)$ differs substantially from $1/\sqrt{2 c_{\rm s} p}$, which is the $p < M$ limit of \eqref{new-amplitude}. Apart from this modification due to the dispersive background, the quantisation of the quadratic action proceeds in the usual way. 

Let us now move on to consider the interacting part of the theory. Notice that the relevant quartic interaction due to $M_2^4$, coming from the non linear self-interactions in the EFT is given by\footnote{Another relevant interaction that could contribute to this analysis is the one proportional to $\mathcal L_{\rm int } \propto \dot \pi^2  \Sigma (\tilde\nabla^2)  \dot \pi^2 $. However, in the new physics regime one has $\omega^4 \ll p^4$, implying that this interaction will be substantially suppressed compared to (\ref{strong-coupling-interaction}).}
\be
\mathcal L_{\rm int} =  \frac{ 1}{16 M_{\rm Pl}^2 \epsilon H^2}  ( \nabla \pi_n)^2     \frac{\Lambda_{\rm UV}^2 }{M^2 -  \nabla^2}  ( \nabla \pi_n)^2 , \label{strong-coupling-interaction}
\ee
after taking into account the normalisation $\pi_n = ( 2 M_{\rm Pl}^2 \epsilon H^2  )^{1/2} \pi$.
Then we can analyse the effect of this interaction on the tree level scattering of two $\pi$ fields into two final $\pi$'s in the center of mass reference frame. The main point to be kept in mind when performing this computation is that the new amplitude (\ref{new-amplitude}) for the quantised Goldstone boson field implies that each external leg of the relevant diagram will come with an additional factor 
\be
\sqrt{ \frac{M^2 + p_i^2}{   \Lambda_{\rm UV}^2 +   p_i^2 } } =  \frac{ \omega_i }{ p_i },
\ee
where $p_i$ is the momentum carried by the particle represented by the $i$-th external leg of the diagram and we have made use of the dispersion relation \eqref{full-modified-dispersion}. 

After a straightforward computation,
we find that the scattering amplitude of this interaction, in the centre of mass (c.m.) frame, is given by
\be
\begin{split}
\mathcal A (p_1 , p_2 \to p_3 , p_4) & = \frac{ c_{\rm s}^{-2} p^4}{2 M_{\rm Pl}^2 | \dot H | } \frac{\omega^4}{p^4} \times \\
& \times \bigg[ 1 + \frac{M^2 \cos^2 \theta}{M^2 + 2 p^2 (1 - \cos \theta)} + \frac{M^2 \cos^2 \theta}{M^2 + 2 p^2 (1 + \cos \theta)} \bigg], \label{scatt-amplitude}
\end{split}
\ee
where $\theta$ is the angle of scattered particles with respect to the impact axis. By recalling that $\left( M^2 - \nabla^2 \right)^{-1}$ can be interpreted as the propagator of a heavy field, the first, second and third terms in the square bracket of \eqref{scatt-amplitude} may be thought of as contributions coming from the interchange of a heavy boson through the $s$, $t$ and $u$ channels, respectively. However, it is important to stress here that since we have broken Lorentz invariance, this computation is frame dependent\footnote{The Mandelstam variables are Lorentz scalars but recall that the requirement $\omega < \Lambda_{\rm UV}$ breaks boost symmetry.}. The factor $c_{\rm s}^{-2}$ appearing in front of \eqref{scatt-amplitude}, may be interpreted as the phase velocity \eqref{ref-indices} evaluated at the c.m. frame. In what follows we will thus denote it as $v_{\rm ph}$ since at the end we will switch to the lab frame. 
%

The previous result may be expressed as a partial wave expansion
\be
\mathcal A (p_1 , p_2 \to p_3 , p_4)=  16 \pi \left( \frac{\partial \omega}{\partial p} \frac{\omega^2}{p^2} \right) \sum_\ell (2 \ell + 1) P_{\ell} (\cos \theta) a_\ell,
\ee
where the $P_{\ell} (\cos \theta)$ are Legendre polynomials. Using the orthogonality properties of Legendre polynomials, the lowest order coefficient $a_0$ me be obtained as
\be
\begin{split}
a_0 &= \frac{ v_{\rm ph}^{-2}}{32 \pi M_{\rm Pl}^2 | \dot H | }  \omega^2 p^2 \frac{\partial p}{\partial \omega} \int_{-1}^{1} d\cos \theta \left( 1 + \frac{2 ( M^2 + 2 p^2) M^2 \cos^2 \theta}{( M^2 + 2 p^2)^2  - 4 p^4 \cos^2 \theta } \right) \\
&= \frac{ v_{\rm ph}^{-2}}{16 \pi M_{\rm Pl}^2 | \dot H | }  \omega^2 p^2 \frac{\partial p}{\partial \omega}  \bigg[ 1 + \frac{ M^2 ( M^2 + 2 p^2)}{2 p^4} \left(  \frac{ M^2 + 2 p^2}{4 p^2} \log \Big(1 + \frac{4 p^2 }{ M^2}  \Big) - 1 \right)  \bigg]. \label{a-0-result-1}
\end{split}
\ee
In order to preserve the unitarity of the tree level scattering process under consideration, the optical theorem leads to the constraint $a_\ell + a_{\ell}^{*} \leqslant 1$. Our main interest is to assess the unitarity of the EFT above $\Lambda_{\rm new}$, where 
%
%
the log term in the square bracket of \eqref{a-0-result-1} becomes negligible, leading to
\be
a_0 \simeq \frac{  \Lambda_{\rm UV}^{3/2 } \omega^{5/2}}{32 \pi M_{\rm Pl}^2 | \dot H | } v_{\rm ph}^{2}.   
\ee
Then, using the constraint $a_\ell + a_{\ell}^{*} \leqslant 1 \; \Longrightarrow \;{\rm Re}(a_\ell) < \frac{1}{2}$,  for the particular case $\ell = 0$, we find that the theory remains weakly coupled as long as
\be
\omega^{5/2} < 8 \pi v_{\rm ph}^{2} \left[ \frac{ \Lambda_{\rm sb} }{ \Lambda_{\rm UV}} \right]^{7/2} \Lambda_{\rm UV}^{5/2} .
\ee
Bearing in mind the quartic dispersion relation of the new physics regime, $\omega \simeq \frac{p^2}{\Lambda_{\rm UV}}$, we can now evaluate the phase velocity at energy $\omega = \Lambda_{\rm sc}$ to obtain 
\be \label{phase-v-at-sc}
v_{\rm ph} = \sqrt{\frac{\Lambda_{\rm sc}}{ \Lambda_{\rm UV}} },
\ee
from which we deduce that the strong coupling scale is set by
\be \label{strong-coupling-scale}
\Lambda_{\rm sc} = (8 \pi)^{2/3}  \left[ \frac{ \Lambda_{\rm sb} }{ \Lambda_{\rm UV}} \right]^{7/3} \Lambda_{\rm UV} ,
\ee
where $\Lambda_{\rm sb}$ is the symmetry breaking scale \eqref{symmetry-breaking-scale}. Equation~(\ref{strong-coupling-scale}) admits a variety of regimes depending on the values of the scales $\Lambda_{\rm new}$, $\Lambda_{\rm UV}$ and $\Lambda_{\rm sb}$.  For instance, if we take $\Lambda_{\rm sb} \sim \Lambda_{\rm UV}$ 
then $\Lambda_{\rm sc}$ is of order $\Lambda_{\rm UV}$. 
Thus we see that the non trivial modifications characterising the new physics regime $M^2 \ll p^2 \ll \Lambda_{\rm UV}^2$ imply that the interactions of the theory scale differently with energy, changing significantly the value at which the EFT becomes strongly coupled. Let us finally note that as was the case for the symmetry breaking scale \eqref{symmetry-breaking-scale}, the result \eqref{strong-coupling-scale} follows directly from \eqref{standard-sc}, upon replacing $c_{\rm s}$ with the phase velocity \eqref{phase-v-at-sc}.

So far we have seen that the extended EFT \eqref{EFT-new-physics-1} implies that $\Lambda_{\rm new} \sim M c_{\rm s}$ and $\Lambda_{\rm UV} \sim M c_{\rm s}^{-1}$. Hence, a suppressed speed of sound automatically induces a hierarchy $\Lambda_{\rm new} \ll \Lambda_{\rm UV} ,$ without however giving any information about the relative values of $\Lambda_{\rm sc}$ and $\Lambda_{\rm sb}$ with respect to $\Lambda_{\rm UV}$, since they strongly depend on the specific UV realisation of the inflationary model at hand. However, if the UV physics allowing for the existence of a new physics regime is also valid for the description of the background inflationary model, it is reasonable to expect all these scales to be of the same order, namely
\be
\Lambda_{\rm UV} \sim \Lambda_{\rm sb} \sim \Lambda_{\rm sc} .
\ee
This is a natural assumption which injects UV information into the EFT reducing again the free parameters. 

A similar assumption is that horizon crossing happens in the non linear regime \eqref{non-linear-dispersion}, implying that $x \equiv \dfrac{H}{\Lambda_{\rm new}} > 1.$ Such a choice is justifiable from a UV perspective. In \cite{Dvali:1995mj,Linde:1996cx} it was pointed out that any scalar degree of freedom present during inflation should have a mass at least one order of magnitude greater than the Hubble scale in order to soften the cosmological moduli problem. Assuming $M \sim 10 H$ and a speed of sound of the order $10^{-2}$, compatible with observational bounds \cite{Ade:2013ydc}, \eqref{lnew} yields $x \sim 10.$  
As we shall discuss later, depending on the value of this ratio the theory leads to different predictions. Something already evident is, for example, the crucial dependence of the strong coupling scale on the dispersion relation, whose form is ultimately controlled by $x$, or the values of the momentum and the phase velocity at the Hubble scale, which are again $x$ dependent -- see \eqref{ref-indices-linear-regime},\eqref{ref-indices-non-linear-regime}.

To summarise, the extended EFT \eqref{EFT-new-physics-1} is characterised by three independent scales 
\be \label{scales}
\left\{ \Lambda_{\rm UV},H,\Lambda_{\rm new} \right\},
\ee
all of which consistently appear in the effective action \eqref{EFT-new-physics-1} and the dispersion relation \eqref{full-modified-dispersion}. Our analysis is strictly valid only for inflationary models with a single scalar degree of freedom driving inflation; models with multiple inflaton fields will inevitably introduce a larger set of scales into the problem. We now proceed to show how the operator insertions considered in \eqref{eft-modified} arise in low energies when one integrates out heavy scalar fields.

\section{Integration of massive fields} \label{sec:integration-1-massive-field}

Actions containing an arbitrary number of differential operators are manifestly non local, potentially suffering from classical instabilities and the appearance of ghosts at the quantum level. Ostrogradski found that theories which depend non trivially on more than one time derivatives ({\it i.e.} in such a way that the higher derivatives cannot be removed by integration by parts or field redefinitions) are unstable, with their Hamiltonians unbounded from below~\cite{Ostrogradski, Woodard:2006nt}. Upon quantisation the instability persists, manifested by the appearance of negative norm states or ghosts, which in turn translates into loss of unitarity. Although \eqref{eft-modified} seems to contain ghost states in its spectrum, as we will discuss in the next section this is not the case. 

Another related issue is that such a theory has an infinite dimensional phase space so that an infinite number of initial conditions is required for a solution. However, when the infinite sequence of differential operators stems from the expansion of an inverse derivative\footnote{Non locality of this form has been dubbed \emph{derived non locality} in \cite{Eliezer:1989cr}.} this is not the case, since an equation of the form $\hat{O}^{-1} f = g$ has a finite dimensional phase space; when brought to the form $f = \hat{O} g$, its solution requires $n$ initial conditions, $n$ being the differential order of $\hat{O}$. Potential problems of this kind arise when the non local operators are analytic functions of $\nabla$. 
This ``harmless" non locality, is expected from an effective field theory point of view. Indeed, when the theory in question corresponds to an effective field theory derived from a local theory by integrating out one or more fundamental dynamical variables, it is not valid to consider the resulting non local terms as limits of higher-derivative analytic operators \cite{Eliezer:1989cr, Bennett:1997wj}, implying that there are no problems either with instabilities or ghosts, as long as the theory remains within its domain of validity.

Let us now explicitly relate the non local form of action \eqref{eft-modified} to the presence of additional degrees of freedom that become operative at high energies\footnote{The potentially large influence that heavy fields could have on the low energy dynamics of inflation was first emphasised by Tolley and Wyman in ref.~\cite{Tolley:2009fg}. For other recent approaches studying the effects of heavy fields on the low energy dynamics of inflation, see for instance refs.~\cite{Cremonini:2010sv, Jackson:2010cw, Cremonini:2010ua, Jackson:2011qg, Shiu:2011qw, Avgoustidis:2012yc, Pi:2012gf, Gao:2012uq, Burgess:2012dz,Gao:2013ota,Noumi:2012vr,Noumi:2013cfa,Gong:2013sma}.}. As shown in \cite{Achucarro:2010da}, despite the fact that heavy modes exponentially vanish outside the Hubble horizon, their effects, like $e.g.$ oscillatory features in the power spectrum~\cite{Achucarro:2012fd}, may be captured by the IR dynamics only when one integrates out these fields instead of just truncating them in the effective action. Following this route, we find that the theory at hand becomes ill defined only if one insists in assuming its validity at energies of order $\Lambda_{\rm UV}$, where a second degree of freedom inevitably becomes excited. The result is that at low energies the theory \eqref{eft-modified} is safe from any pathology related to non locality. 

As an illustrative example we will perform the integration over a single heavy field of mass $M$. A general treatment for an arbitrary number of heavy modes with multiple mass hierarchies is presented in App.~\ref{app:several-massive-fields}.
Bearing in mind that $\delta g^{00} = g^{00}  + 1$, our starting effective action will be
\be
S_{\mathcal F} =  \frac{1}{2}  \int \!\!  d^3xdt a^3  \bigg\{     \dot { \mathcal F}^2 - (\nabla \mathcal F)^2  -  M^2  \mathcal F^2    - 2 \alpha^3 \mathcal{F}  \delta g^{00} -  \beta^2 \mathcal F^2 \delta g^{00}   - \frac{2}{3!} \gamma \mathcal{F}^3 \bigg\}  , \label{heavy-field-unitary-gauge}
\ee
where the unitary gauge is assumed, $\alpha,\beta,\gamma$ represent couplings of mass dimension one, and we have included up to cubic contributions. Note that we have intentionally neglected a term $ \sim \mathcal {F}  (\delta g^{00})^2$, since it yields the same low energy couplings as the ones already contained in \eqref{heavy-field-unitary-gauge}. Furthermore, the cubic term $\mathcal F^3$, leading to a non linear equation of motion for the heavy field, is to be treated perturbatively. Such non linearities, present in any interacting theory, are usually treated in the interaction picture, which allows us to use the linear solution when computing the non linear terms within some perturbative scheme like the $in-in$ formalism. 

\begin{figure}[htb]
\begin{center}
\includegraphics[scale=.55]{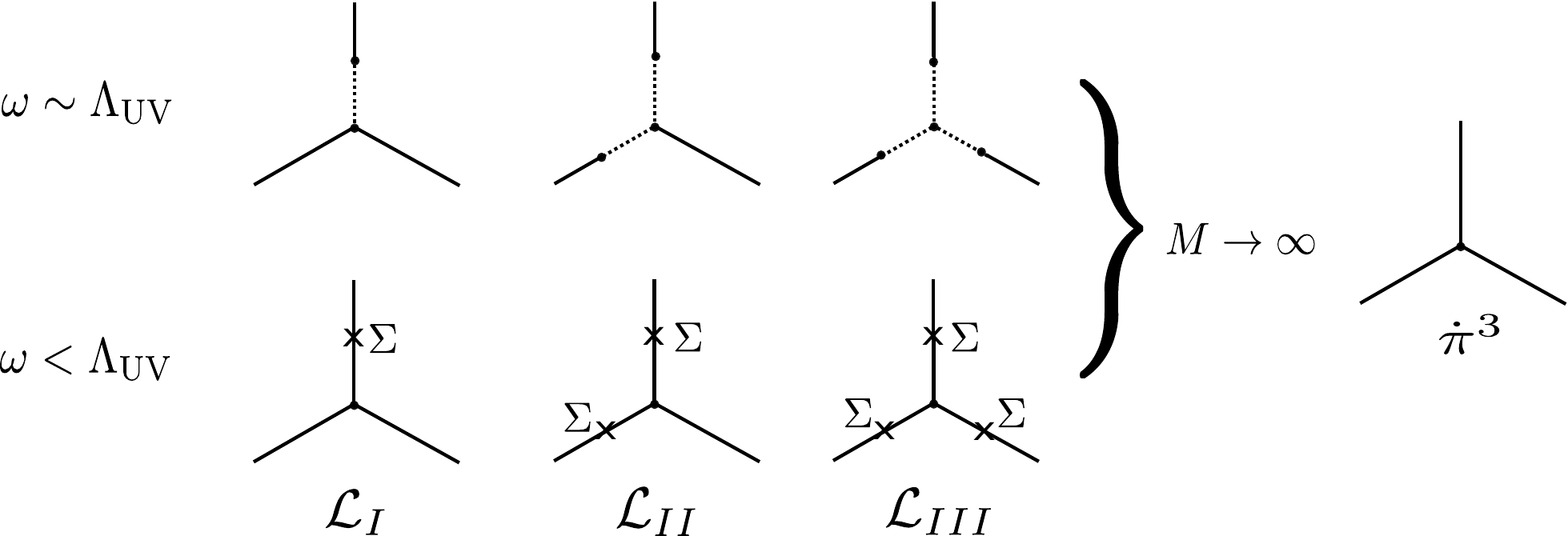} 
\caption[{\sf The interaction vertices of the effective action}.]{\sf The interaction vertices of the effective action \eqref{EFT-new-physics-3}. In the top series the vertices of the full theory are depicted, where solid lines correspond to $\pi$ and dashed lines to a heavy scalar. In the bottom series, the massive field is integrated out and its effects are parametrised, at low energies, by the insertion of the operator $\Sigma$. The limit of $M \to \infty$, where the theory flows to the low-derivative EFT \eqref{Seft-simple}, is also depicted.} 
\label{fig:feyn}
\end{center}       
\end{figure}

The linear equation of motion of the heavy field $\cal F$ reads
\be
\ddot { \mathcal F} +3H \dot {\mathcal F}  + \left(M^2 -  \nabla^2  + \beta^2 \delta g^{00} \right)  {\mathcal F} =  -  \alpha^3  \delta g^{00}. \label{equation-F}
\ee
We are interested in studying the low energy regime of the system, where the second-order time variation of the heavy field $\ddot {\mathcal F}$ is subleading with respect to the term $( M^2 - \nabla^2 ) \mathcal F$, consistent with an expansion of the form \eqref{second-expansion} and the low energy condition \eqref{low-energy-cond}. 
If this is granted, we may simply disregard the kinetic term and solve for $\mathcal F$ by rewriting the equation of motion \eqref{equation-F} as
\be
 \left[1  +  \delta g^{00} \frac{ \beta^2 }{M^2 -  \nabla^2} \right] (M^2 -  \nabla^2 ) {\mathcal F} =  -  \alpha^3  \delta g^{00}. \label{equation-F-2}
\ee
It is important to recognise that at low energies the massive scalar field $\mathcal F$ has no dynamics, in the sense that its value is completely determined by the source $ - \alpha^3 \delta g^{00}$ at the right hand side of (\ref{equation-F-2}). In other words, the heavy field $\mathcal F$ plays the role of a Lagrange multiplier, carrying with it the scale dependence implied by the $\nabla^2$ operator, allowing us to explicitly write it in terms of $\delta g^{00}$ as
\be
\begin{split}
{\mathcal F}   & =  \frac{ \alpha^3 }{\nabla^2 - M^2 -  \beta^2 \delta g^{00} } \delta g^{00} \\
  & =   -  \frac{ \alpha^3 }{M^2 -  \nabla^2} \left[1  +  \delta g^{00} \frac{ \beta^2 }{M^2 -  \nabla^2} \right]^{-1} \delta g^{00} \\
  & =    - \frac{\alpha^3}{M^2 -  \nabla^2} \sum_{n=0}^\infty (-1)^{n} \left[ \delta g^{00} \frac{ \beta^2 }{M^2 -  \nabla^2} \right]^n \delta g^{00} ,
\end{split}
\label{solution-F-g00}
\ee
where in the last step we made use of the formal expansion $(1 + z)^{-1} = \sum_n (-1)^n z^n$, valid for $|z| < 1$.
Neglecting the kinetic term at the level of the equations of motion is equivalent to having dropped them in the action, thus inserting (\ref{solution-F-g00}) into the action \eqref{heavy-field-unitary-gauge}, we recover the contribution to the EFT for the Goldstone boson \eqref{eft-modified}, now stemming from the heavy field:
%
%
%
\be
\begin{split}
S & =   \int  d^3xdt a^3 \Bigg \lbrace M_{\rm Pl}^2 \dot H \delta g^{00} + \frac{M_{\alpha\beta}^4}{2} \sum_{n=1}^{\infty}  (-1)^n  \bigg[  \delta g^{00}   \frac{  \beta^2 }{M^2 -  \nabla^2   }  \bigg]^{n}  \delta g^{00} \\ & + \gamma \left( \frac{\alpha^3}{M^2 -  \nabla^2} \sum_{n=0}^\infty (-1)^{n} \left[ \delta g^{00} \frac{ \beta^2 }{M^2 -  \nabla^2} \right]^n \delta g^{00} \right)^3 \Bigg  \rbrace , \end{split}
\label{resulting-action-F}
\ee
%
%
%
where we have also included a kinetic term for $\delta g^{00} =  g^{00} +1 $ according to \eqref{starting-action} and parametrised the ratio of the two couplings as $M_{\alpha\beta}^2 = \dfrac{\alpha^3}{\beta}$. This action matches \eqref{EFT-new-physics-1} upon setting
\be \label{heavy-field-eft-matchings}
\alpha^6 = M_2^4 M^2,  \quad \beta^2 = \frac{1}{3}\dfrac{M_3^4}{M_2^4} M^2, \quad M_{\alpha\beta}^2 = \sqrt{3} \dfrac{M_2^4}{M_3^2},
\ee
apart from the term proportional to $\gamma$. 
For example, having in mind that in the decoupling limit we have $\delta g^{00} \sim -2 \dot\pi - \dot \pi^2$ and assuming that there are no additional sources of deviations from standard single field inflation other than the heavy field $\mathcal F$, the speed of sound, in the long wavelength limit, reads
\be
\frac{1}{c_{\rm s}^2} = 1 + \frac{2 \alpha^6}{ M_{\rm Pl}^2 |\dot H| M^2}, \label{sof-single-heavy-field}
\ee
where a canonical normalisation $\pi_c = M_{\rm Pl}\sqrt{|\dot H|} \pi$ is implied. Using \eqref{heavy-field-eft-matchings}, this is in agreement with \eqref{sof}.
In order to incorporate the missing coupling, let us rewrite the action \eqref{EFT-new-physics-1}, this time including the $\gamma$ contribution:
\be
\begin{split}
S &= M_{\rm Pl}^2 |\dot H| \int d^3xdt  a^3 \bigg[    \dot \pi \left( 1 +  \Sigma (\tilde\nabla^2)  \right) \dot \pi -   (\tilde \nabla \pi)^2     +     \big[   \dot \pi^2 -  (\tilde \nabla \pi)^2   \big] \Sigma (\tilde\nabla^2) \dot \pi    \\
& -    \frac{2 c_3}{3(1 - c_{\rm s}^{2} )}  \dot \pi   \Sigma (\tilde\nabla^2) \left( \dot \pi \Sigma (\tilde\nabla^2)\dot \pi \right)  -    \frac{2\tilde{c}_3}{3(1 - c_{\rm s}^{2} )^2} \left ( \Sigma (\tilde\nabla^2) \dot \pi \right ) \left ( \Sigma (\tilde\nabla^2) \dot \pi \right )  \left ( \Sigma (\tilde\nabla^2)\dot \pi \right )   \bigg],  
\end{split}
\label{EFT-new-physics-3}
\ee
where we have defined 
\be \label{sigma,c3-def}
\Sigma (\tilde\nabla^2) \equiv  (1 - c_{\rm s}^{2} )  \frac{\Lambda_{\rm UV}^{2} }{M^2 - \tilde\nabla^2},\quad c_3 \equiv c_{\rm s}^2 \dfrac{M_3^4}{M_2^4},\quad \tilde{c}_3 \equiv c_{\rm s}^4 \dfrac{ M^2_2}{M^3}\gamma.
\ee
As advertised, the action \eqref{eft-modified} offers an IR parametrisation of UV degrees of freedom, where the scale dependent low energy self couplings $M_n^4(p^2) = M_n^4 \dfrac{M^2}{M^2 + p^2}$ of the Goldstone mode originate in the mediation of heavy fields. The low derivative EFT \eqref{Seft-simple} is reached in the limit where all the dimensionful couplings approach infinity, $i.e.$ $(M,\alpha,\beta) \to \infty$, while $M_{\alpha\beta}^2 = \text{const}$. In the language of Fig.~\ref{fig:media}, this limit corresponds to shrinking the characteristic length of the effective medium, $\lambda_M = 1/M$, to zero. Note that in this limit the $\tilde{c}_3$ contribution of \eqref{EFT-new-physics-3}, arising from a cubic self interaction of the heavy field, vanishes. In Fig.~\ref{fig:feyn}, we sketch the interaction vertices of the higher-derivative EFT before and after integrating out the heavy field, that is, above and below the scale $\Lambda_{\rm UV}$.

The action \eqref{resulting-action-F} is a general EFT that can parametrise UV effects missed by the lowest derivative theory \eqref{Seft-simple}. As a specific completion, let us mention the scenario where the inflaton rolls down a flat, yet windy valley\footnote{Although the expression ``windy valley", given a different accent, may well describe a physical landscape, what is meant here is a flat trajectory in field space deviating from a straight line.} of the potential  \cite{Achucarro:2010jv, Achucarro:2010da, Cespedes:2012hu, Achucarro:2012sm, Achucarro:2012yr, Cespedes:2013rda}. This is a common feature of string theory models where the vacuum expectation value of moduli fields depend on the light inflaton.
In such a case the dimensionful couplings of the effective theory \eqref{heavy-field-unitary-gauge} are given by $\alpha^3 = \dot \theta \dot \phi_0$, $\beta^2 = \dot\theta^2$ and $\gamma = 0$, where $\phi_0$ is the time dependent background inflaton value satisfying $\dot \phi_0^2 = - 2 \dot H M_{\rm Pl}^2 $ and $\dot \theta$ is the angular velocity characterising the turns of the multi-field trajectory in the scalar field target space. As shown in \cite{Cespedes:2012hu}, in order to integrate out $\mathcal F$ it is also important to assume the adiabaticity condition $ | \ddot \theta / \dot \theta | \ll M$, which ensures that the background dynamics of the turning trajectory are consistent with the low energy condition \eqref{low-energy-cond}. In the notation of the previous paragraph this would be a slow roll condition $ \dfrac{\dot\beta}{H \beta} \ll \dfrac{M}{H} $ on the coupling of the trivalent vertex $\mathcal{F}^2 \delta g^{00}$. 
The speed of sound now reads
$$
\frac{1}{c_{\rm s}^2} = 1 + \frac{4 \dot \theta^2}{M^2},
$$
in consistency with \eqref{sof-single-heavy-field}, implying that the $M_n^4$ coefficients of the EFT action \eqref{starting-action} may be written as~\cite{Achucarro:2012sm}
\be \label{Mns-in-curved-valley}
M_n^4 = (-1)^{n} |\dot H | M_{\rm Pl}^2 n!  \left( \frac{1 - c_{\rm s}^2}{4 c_{\rm s}^2}  \right)^{n-1} .
\ee
The realisation of a curved field trajectory imposes a relation among the arbitrary couplings of \eqref{heavy-field-unitary-gauge} which now depend on the angular velocity $\dot\theta$. This is an input of UV information and, as expected, it immediately results in the reduction of the unknown parameters via \eqref{Mns-in-curved-valley}. 

As a final remark, before passing to the phenomenological study of the theory, we now discuss the role of the higher time derivatives that we have been neglecting so far. It is shown that the expected ghost states signal the need of the effective theory for an explicit completion.  

\section{Ghosts as IR signals of intermediate completions} \label{sec:ghosts}
As already argued, the expansion \eqref{second-expansion} is only possible if Lorentz invariance is broken, which in the present context is a consequence of the broken time translation invariance induced by the background. In Fourier space, \eqref{second-expansion} may be expressed as
\be
 \frac{1 }{ M^2 + p^2 - \omega^2} = \frac{1}{M^2  + p^2} \left( 1  + \frac{ \omega^2 }{M^2  + p^2} + \cdots \right). \label{prop-expansion-momentum}
\ee
Let us now consider the quadratic action for $\pi$ obtained from \eqref{resulting-action-F}, but this time keeping the time derivative $\partial_t$ to all orders, that is
\be
S_\pi^{(2)} = M_{\rm Pl}^2 \int d^3xdt  a^3 |\dot H| \left[ \dot \pi \left(  1 +  \frac{  2 \tilde\alpha^2  }{ M^2 - \Box }  \right) \dot \pi -  (\tilde \nabla \pi)^2 \right], \label{S-2-pi}
\ee
where we have defined $ \tilde\alpha^2 \equiv \dfrac{ \alpha^6  }{M_{\rm Pl}^2 |\dot H| }$, which from \eqref{sof-single-heavy-field} can be written as $ \tilde\alpha^2 = \frac{1 - c_{\rm s}^2}{2c_{\rm s}^2}M^2$.
From this expression, we can read off the propagator $G(p^2)$ of the low energy Goldstone boson in momentum space, which is found to be
\be \label{propagator}
G(p^2) \propto \frac{1}{ \Gamma(p^2)} , \qquad  \Gamma(p^2) = p^2 - \omega^2  -  \frac{  2 \tilde\alpha^2  \omega^2 }{ M^2 + p^2 - \omega^2  }  .
\ee
This propagator has two poles, at values $\omega_{+}^2$ and $\omega_{-}^2$, given by
\be \label{frequencies}
 \omega^{2}_{\pm} = \frac{M^2 }{2 c_{\rm s}^2} +  p^2   \pm \frac{M^2 }{2  c_{\rm s}^{2}}  \sqrt{  1 +  \frac{4 p^2 ( 1 - c_{\rm s}^{2}) }{M^2 c_{\rm s}^{-2}}   }  .
\ee
A particle state characterised by a propagator with two or more poles is condemned to include ghosts in its spectrum \cite{Biswas:2005qr, Barnaby:2007ve}, in close connection with our previous discussion on non locality. To illustrate this statement, consider a propagator with two simple poles at values $\mathbf{p}^2 = -m_1^2$ and $\mathbf{p}^2 = -m_2^2$ with $\mathbf{p}^2 = p^2 - \omega^2$ and $m_2^2 > m_1^2 > 0$, so that its denominator can be written as $\Gamma(-\mathbf{p}^2) = (\mathbf{p}^2 + m_1^2) (\mathbf{p}^2 + m_2^2) f(-\mathbf{p}^2)$, with $f$ a finite piece. Since there are no more zeros of $\Gamma$ in the interval $-m_2^2 < \mathbf{p}^2 < -m_1^2$, $f$ is either positive or negative implying that the two residues differ in sign \cite{Biswas:2005qr}. 
The negative one represents a spurious state with negative norm, $i.e.$ a ghost.

Now if we restrict the theory to momenta $p \ll M c_{\rm s}^{-1}$, corresponding to the low energy regime \eqref{low-energy-cond}, we find that $\omega_{+}$ and $\omega_{-}$ of \eqref{frequencies} are well approximated by
\be
\begin{split}
& \omega_+^2 (p) = M^2 c_{\rm s}^{-2} + \mathcal O (p^2) ,  \\
& \omega_-^2 (p) =  c_{\rm s}^2 p^2 + \frac{(1 - c_{\rm s}^2)^2 }{M^2 c_{\rm s}^{-2}} p^4 + \mathcal O (p^6) , \label{low-energy-frequency-resumed-theory}
\end{split}
\ee
where $\mathcal O (p^2)$ and $\mathcal O (p^6)$ denote subleading higher-order terms. The low energy dispersion relation $\omega_-$ coincides with the one in \eqref{low-energy-dispersion}, which was found excluding time derivatives, apart from a factor of $(1 - c_{\rm s}^2)$ in front of the quartic piece of the expansion. Since this term is only relevant for $c_{\rm s}^2 \ll 1$, we see that the difference between these two expressions is marginal, justifying the approximation by which one drops higher-order time derivatives. Therefore, as long as we focus on low energy processes for which $p \ll \Lambda_{\rm UV}$ and $\omega \ll \Lambda_{\rm UV}$, where $\Lambda_{\rm UV}$ coincides with the one defined in \eqref{luv} $(\Lambda_{\rm UV}=M/c_{\rm s})$, intermediate particle states are characterised by well-defined propagators (away from the dangerous -- ghost -- pole $\omega_+$) and the effective field theory (\ref{S-2-pi}) remains ghost free. The appearance of ghosts when the energy approaches the cutoff scale $\Lambda_{\rm UV}$ is a manifestation of the necessity of the theory for explicit completion, incorporating the heavy field $\cal F$ as a dynamical degree of freedom. 

This is essentially the statement that when we integrate \emph{over} a massive field, we do nothing more than just rewriting the path integral as
\be 
\mathcal{Z} = \int [D\pi] [D\mathcal{F}] e^{S[\mathcal{F},\pi]} = \int [D\pi] e^{S_{\rm eff}[\pi]},
\ee
where
\be 
e^{S_{\rm eff}[\pi]} = \int [D\mathcal{F}] e^{S[\mathcal{F},\pi]}.
\ee
The crucial step, which allows us to claim that we really integrate \emph{out}, as opposed to integrate  over the massive field\footnote{See \cite{Castillo:2013sfa} for a discussion on the difference of fields and particle states in time dependent backgrounds.}, is to restrict the effective theory to momenta $p<\Lambda_\star$, where $\Lambda_\star$ is the scale at which the heavy modes are excited. For our case, as demonstrated in \eqref{low-energy-frequency-resumed-theory}, $\Lambda_\star = \Lambda_{\rm UV}$, and $p<\Lambda_\star$ is exactly what \eqref{puv} imposes. Therefore, $\Lambda_{\rm UV}=M c_{\rm s}^{-1}$ is indeed the scale where the heavy degree of freedom becomes dynamical, and if one keeps neglecting this mode as the energy approaches this value, one ends up with ghosts, encoding the missing degree of freedom in the IR.

Having addressed the possible subtleties of our construction we now proceed with the dynamics of the theory and its prediction regarding the two- and three-point correlators of the comoving curvature perturbation.

\section{Dynamics of curvature perturbations} \label{sec:dynamics}
Truncating the action \eqref{EFT-new-physics-3} to quadratic order, we obtain \eqref{gaussian-eft-no-a}, which we rewrite here restoring the scale factor $a$
\be 
S = \frac{1}{2}  \int d^3xdt  a^3 \bigg[    \dot \pi_n \left(   \frac{  \Lambda_{\rm UV}^2 -  \tilde \nabla^2  }{M^2 -  \tilde \nabla^2}  \right) \dot \pi_n -   ( \tilde \nabla \pi_n)^2   \bigg] , \label{gaussian-eft}
\ee
where $\pi_n = ( 2 M_{\rm Pl}^2 \epsilon H^2  )^{1/2} \pi$ is the normalised mode and $\tilde\nabla = \nabla/a$ with eigenvalue $p=k/a$, the physical momentum. The equation of motion reads
\be \label{pi-full-eom}
\ddot\pi_n + H \left( 1 - 2\frac{\dot \omega}{H \omega} \right) \dot \pi_n + \omega^2 \pi_n = 0,
\ee
where $\omega$ is given by \eqref{full-modified-dispersion}. This equation is of the general Riccati type and exact solutions are not known. Although a solution would reveal the dynamics of the light mode throughout the whole energy range, it is important to recall that our theory has an upper cutoff set by $\Lambda_{\rm UV}$. We will therefore be interested in the regime where $p \ll \Lambda_{\rm UV}$, allowing ourselves to expand the action and obtain a simplified equation of motion
\be
\ddot \pi_n + 3 H \dot \pi_n + 2H \frac{p^2}{M^2+p^2} \dot \pi_n + \left( \frac{p^4}{ \Lambda_{\rm UV}^2} + c_s^2 p^2 \right) \pi_n = 0,  \label{eq-motion-pi-lim-1}
\ee
which can be written exactly as \eqref{pi-full-eom} but with the low energy dispersion relation given by \eqref{low-energy-dispersion}. Passing to conformal time and redefining $\pi_n = \dfrac{u}{a^2} \sqrt{1 + \frac{M^2}{p^2}}$, the above equation can be brought in the form
\be
u'' + \left[ \frac{H^2k^4}{ \Lambda_{\rm UV}^2}\tau^2 + c_s^2 k^2 -\frac{2}{\tau^2} -\frac{1}{\tau^2} \frac{  4 + \frac{M^2}{k^2} \frac{1}{H^2\tau^2}  }{ \left( 1+\frac{M^2}{k^2} \frac{1}{H^2\tau^2} \right)^2 }  \right] u = 0,  \label{eq-motion-u-lim-1}
\ee
Unfortunately, an analytic solution is not known for this equation either. It is important to stress here that this would be an ideal case where the theory would be completely solved throughout its whole window of validity. We are thus forced to limit our discussion to the regimes where $p \ll M$ or $p \gg M$. 

The first case, zeroth order in $\left( p/M \right)^2$, corresponds to the limit $M \to \infty$ giving us back the standard lower derivative EFT \eqref{Seft-simple}. 
In what follows, we will calculate the three-point correlators in the dispersive regime as functions of momenta. We will see that the low energy observables associated with two-point and three-point correlators are directly related with the scale of UV physics.  

Equations similar to \eqref{eq-motion-pi-lim-1} but without the complicated time dependence of the friction term have been studied in \cite{Ashoorioon:2011eg,Chialva:2011hc} to zeroth order in $\left( M/p \right)^2$, in a general context of modified dispersion relations, as well as in 
\cite{Bartolo:2010bj,Bartolo:2010di,Bartolo:2010im} in the context of extrinsic curvature operators of the form \eqref{starting-action}. When $p \gg M$, the quartic piece dominates the dispersion relation which now reads $\omega^2 \sim \dfrac{p^4}{\Lambda_{\rm UV}^2}$, allowing us to consider the following equation of motion \cite{Baumann:2011su}
%
\be
\ddot \pi_n + 5 H \dot \pi_n + \frac{p^4}{ \Lambda_{\rm UV}^2} \pi_n = 0,  \label{eq-motion-pi}
\ee
for the light mode, again following from \eqref{pi-full-eom}. The solution for the comoving curvature fluctuation $\mathcal{R} = -H \pi$, after imposing Bunch-Davies vacuum and the commutation relations \eqref{commutation-new} in the infinite past, reads \cite{Baumann:2011su}
\be \label{pi-solution}
\mathcal{R}_k(\tau) = -\frac{H^2 }{( 2 M_{\rm Pl}^2 \epsilon  )^{1/2}} \sqrt{\frac{\pi}{8}} \frac{k}{\Lambda_{\rm UV}} (-\tau)^{5/2}  H_{5/4}^{(1)} (x), \quad\text{with} \quad x = \frac{H}{2 \Lambda_{\rm UV} } k^2 \tau^2   ,
\ee
where $\tau = - ( H a)^{-1}$ is the conformal time and $H^{(1)}$ denotes the Hankel function of the first kind. 

The power spectrum, $i.e.$ the two-point correlator, is defined as
\be 
\langle \hat{\mathcal{R}}_{\mathbf k_1} \hat{\mathcal{R}}_{\mathbf k_2} \rangle = (2 \pi)^3 P_{\mathcal{R}}(k) \delta \left( {\mathbf k_1} + {\mathbf k_2} \right),
\ee
where $P_{\mathcal{R}}(k) = |\mathcal{R}_k|^2$.
The field operator $\hat{\mathcal{R}}$ in Fourier space is defined as
\be \label{field-op}
\hat{\mathcal{R}}_{\mathbf k}(\tau) = \mathcal{R}_{k}(\tau)\hat a_{\mathbf k} + \mathcal{R}^*_{ k}(\tau) \hat a^\dag_{-\mathbf k},
\ee
where $\mathcal{R}_{\mathbf k}$ denotes the Fourier mode of the field with wavevector $\mathbf k$ and $\hat a^\dag,\; \hat a$ are the usual creation and annihilation operators obeying the canonical commutation relation 
\be
[\hat a_{\mathbf k},\hat a^\dag_{-\mathbf k'}] = (2\pi)^3 \delta({\mathbf k} + {\mathbf k}').
\ee
Considering the limit of the solution \eqref{pi-solution} as $\tau \to 0$ 
\be \label{R^0}
\mathcal{R}_k^{0} \sim - \dfrac{\sqrt{2}\Gamma(5/4)}{\sqrt{\pi}}\dfrac{H}{ M_{\rm Pl} \sqrt{\epsilon}}\left( \dfrac{\Lambda_{\rm UV}}{H} \right)^{1/4}\dfrac{1}{k^{3/2}},
\ee
we obtain the dimensionless power spectrum
\be
\mathcal{P}_{\mathcal{R}} = \frac{k^3}{2 \pi^2} |\mathcal{R}_k |^2  = \frac{\Gamma^2(5/4) }{\pi^3} \frac{H^2}{M_{\rm Pl}^2 \epsilon} \sqrt{\frac{\Lambda_{\rm UV}}{H}}, \label{power-spec-new-phys-0}
\ee
where $\Gamma(5/4) \simeq 0.91$ is the Gamma function evaluated at $5/4$. Using the power spectrum \eqref{power-spec-new-phys-0}, we can further compute the tensor-to-scalar ratio $r = \dfrac{\mathcal P_{\gamma}}{\mathcal P_{\mathcal{R}}}$ to obtain
\be
r = \frac{2 \pi \epsilon}{\Gamma^2(5/4)} \sqrt{\frac{H}{\Lambda_{\rm UV}} } . \label{r}
\ee

Before proceeding to analytic calculations of the bispectrum, in order to ensure that qualitative features of the theory, like $e.g.$ scale invariance, are insensitive to our simplifying new physics regime, $p_* \gg M$, let us comment on  $( M/p )^2$ corrections to the power spectrum \eqref{power-spec-new-phys-0}.

\subsection{Corrections to the power spectrum} \label{sec:power-corrections}
Expanding the Lagrangian \eqref{gaussian-eft} to first order in $( M/p )^2$, we obtain
\bea \label{1st-order-lagrangian}
\mathcal{L} &=&  \underbrace{a^3 \bigg[   \dot \pi_n \frac{\Lambda_{\rm UV}^2}{p^2}  \dot \pi_n -   (\tilde \nabla \pi_n)^2   \bigg]}_{\mathcal{L}_0} - \underbrace{a^3   \dot \pi_n \frac{\Lambda_{\rm UV}^2}{p^2} \frac{M^2}{p^2} \dot \pi_n }_{\mathcal{L}_{\rm int}},
\eea
with the $\mathcal{L}_0$ piece -- which is the same as in \eqref{eq-motion-pi} -- leading to the solution \eqref{pi-solution}.
%
%
%
%


Corrections to the power spectrum \eqref{power-spec-new-phys-0} coming from $\mathcal{L}_{\rm int}$ in \eqref{1st-order-lagrangian}, can be computed using the $in-in$ formalism \cite{Maldacena:2002vr,Weinberg:2005vy}, according to which the expectation value of an operator $\hat{O}$ is evaluated as
\be
\langle \hat{O} \rangle = \langle 0| \left[ \bar{\mathcal{T}} \exp \left\{ i\int_{-\infty}^0 d\tau'\hat H_{I}(\tau') \right\} \right] \hat{O} \left[ \mathcal{T} \exp \left\{ -i\int_{-\infty}^0 d\tau'\hat H_{I}(\tau') \right\} \right] |0\rangle ,
\ee
with $\mathcal{T},\bar{\mathcal{T}}$ standing for time ordering, anti-ordering respectively.
Using the Baker-Campbell-Hausdorff formula one can expand the previous expression as
\be
\begin{split}
\langle \hat{O} \rangle(\tau) & = \langle 0| \Bigg\{ \hat{O}(\tau) - i \int_{-\infty}^\tau d\tau_1[\hat H_{I}(\tau_1),\hat{O}(\tau)] \\ & -  \int_{-\infty}^
{\tau} d\tau_1\int_{-\infty}^
{\tau_1}d\tau_2\left[\hat H_{I}(\tau_2),[ \hat H_{I}(\tau_1),\hat{O}(\tau)] \right] + \ldots \Bigg\} |0 \rangle. \label{in-in-0}
\end{split}
\ee
We will focus on the tree level corrections consisting of the first line of \eqref{in-in-0}, where the operator under consideration is $\hat{O} = \hat{\mathcal{R}}_{\mathbf k_1} \hat{\mathcal{R}}_{\mathbf k_2} $, with $\hat{\mathcal{R}}_{\mathbf k}$ defined in \eqref{field-op}.
The tree level correction to the power spectrum \eqref{power-spec-new-phys-0} will thus be 
\be
(2\pi)^3 \delta(k_1 + k_2) \Delta P_\mathcal{R} = - i \lim_{\tau \rightarrow 0} \int_{-\infty}^\tau d\tau' \langle 0| [\hat H_{I}(\tau'),\hat{\mathcal{R}}_{\mathbf k_1} \hat{\mathcal{R}}_{\mathbf k_2}(\tau)] |0 \rangle . \label{power-spec-correction-0}
\ee
Since the time ordered product is the normal product plus all possible contractions, with a contraction defined as 
$$
[\hat{\mathcal{R}}_{k_1}(\tau'),\hat{\mathcal{R}}_{k_2}(\tau'')]=(2\pi)^3 \mathcal{R}_{k_1}(\tau') \mathcal{R}^*_{k_2}(\tau'')\delta(k_1+k_2),$$
the only terms that survive the vacuum projection are the fully contracted terms. 

The interaction Hamiltonian can be extracted by the interacting part of the Lagrangian \eqref{1st-order-lagrangian} and reads 
\be
H_{I} (\tau)= 2\frac{M_{\rm Pl}^2\epsilon}{H^2} \dfrac{\Lambda_{\rm UV}^2 }{H^2} \dfrac{M^2 }{H^2} \frac{1}{k^4} \frac{ \mathcal{R}'^*(\tau) \mathcal{R}'^*(\tau)}{\tau^6}. \label{Hint}
\ee
By expanding the commutator in \eqref{power-spec-correction-0}, using the Hamiltonian \eqref{Hint}, we finally obtain
\be
\Delta \mathcal{P}_\mathcal{R} = 2 \frac{k^3}{2 \pi^2} \text{Im} \int_{-\infty}^0 d\tau H_{I}(\tau) \mathcal{R}^{0}(k)  \mathcal{R}^{0}(k), \label{power-spec-correction}
\ee
where the limit of $\mathcal{R}(\tau)$ as $\tau\rightarrow 0$ is given by \eqref{R^0}.
Upon redefining the integration variable $\tau \mapsto z= k\tau$, the change in the power spectrum reads
\be
\Delta \mathcal{P}_\mathcal{R} = \dfrac{\Gamma^2(5/4)}{4\pi^2}\dfrac{H^2}{M_{\rm Pl}^2 \epsilon} \sqrt{\dfrac{\Lambda_{\rm UV}}{H}} \dfrac{M^2}{H^2}\text{Im} \int_{-\infty}^0 \frac{dz}{z^6} \frac{d \mathcal{R}_s^*(z)}{dz}\frac{d \mathcal{R}_s^*(z)}{dz},
\ee
where we defined
\be \label{zeta(z)}
\mathcal{R}_s^* (z) \equiv  z^{5/2}H_{5/4}^{(2)} \left( \frac{H}{2 \Lambda_{\rm UV} } z^2 \right).
\ee
%
The imaginary part of the integral is equal to $-\dfrac{H}{\Lambda_{\rm UV}}$, so to first order in $M^2/p_*^2 = M^2/H \Lambda_{\rm UV}$, the power spectrum is given by
\be \label{first-order-power-hankel}
\mathcal {P}_\mathcal{R} = \frac{\Gamma^2(5/4) }{\pi^3} \frac{H^2}{M_{\rm Pl}^2 \epsilon} \sqrt{\frac{\Lambda_{\rm UV}}{H}} \left[ 1 - \dfrac{\pi}{4} \dfrac{M^2}{H \Lambda_{\rm UV}} \right].
\ee

Thus, corrections of this kind yield unobservable shifts in the value of the power spectrum which is still scale invariant. By extracting the overall momentum dependence of the integral via the change of variables $z =k \tau$, scale invariance can be shown to persist to all orders in perturbation theory. Note again, that since we relaxed the condition of perturbing to zeroth order in $M/p$, the scale $M$ appears in the observables, albeit in a trivial way. We now proceed to the study of the three-point contributions.

\subsection{Estimation of the bispectrum amplitudes}
A practical way to parametrise non Gaussianities, that is non vanishing cubic correlators, is through the non linearity parameter $f_{\rm NL}$. Although traditionally $f_{\rm NL}$ is defined as the deviation of the primordial curvature perturbation from the Gaussian ansatz\footnote{The factor of $\frac{3}{5}$ is a historical convention, since non Gaussianities were first considered using the Newtonian potential $\Phi$ which, during a matter dominated era, equals the curvature perturbation modulo a factor of $\frac{3}{5}$. Note that this definition of $f_{\rm NL}$ leads to local type non Gaussianity, while the one we will use, $i.e.$ a parametrisation of the amplitude of the three-point function, is suitable for equilateral type non Gaussianity, characteristic of higher derivative theories.}, namely
\be 
\mathcal{R} = \mathcal{R}_{\rm G} + \frac{3}{5} f_{\rm NL} (\mathcal{R}_{\rm G}^2 - \langle \mathcal{R}_{\rm G}^2 \rangle ),
\ee
one may use it as a general parametrisation of the amplitude of the cubic contributions to the Lagrangian of the curvature fluctuations. writing the three-point correlator as
\be 
\langle \hat{ \mathcal{R} }_{\mathbf k_1} \hat{ \mathcal{R} }_{\mathbf k_2} \hat{ \mathcal{R} }_{\mathbf k_3} \rangle =(2\pi)^3 \delta(\mathbf{k}_1 + \mathbf{k}_2 + \mathbf{k}_3) f_{\rm NL} S(k_1,k_2,k_3) ,
\ee 
$f_{\rm NL}$ quantifies how strong the cubic (non Gaussian) contribution is -- compared to the quadratic (Gaussian) one -- and may thus be approximated by weighing the cubic Lagrangian with respect to the quadratic one. Namely,
\be \label{eq-ng-zeta}
\frac{\mathcal L^{(3)}}{\mathcal L^{(2)}}\Bigg|_{\omega = H} \sim f_{\rm NL} \mathcal{R} ,
\ee 
where all the length scales are evaluated at the Hubble scale, which is where the modes freeze, and as already mentioned sets the characteristic scale of the system as far as experimental measurements are concerned. The function $S(k_1,k_2,k_3)$ encodes the momentum dependence of the three-point correlator and will be the main subject of Sec.~\ref{sec:bispec}.

In order to compute the non linearity parameter using \eqref{eq-ng-zeta}, we need to calculate the amplitudes of the kinetic and the four cubic operators comprising the effective theory \eqref{EFT-new-physics-3} at $\omega = H$. These are, 
\be \label{cubics}
\begin{split}
& \qquad \qquad \mathcal L^{(2)} = a^3 M_{\rm Pl}^2  |\dot H|\dot \pi \left( 1 + \Sigma(\tilde\nabla^2) \right) \dot \pi, \\
\mathcal L_{I}^{(3)} &=   a^3 M_{\rm Pl}^2  |\dot H|   \dot \pi^2   \Sigma (\tilde \nabla^2) \dot \pi ,    \quad
\mathcal L_{II_1}^{(3)} =  - a^3 M_{\rm Pl}^2 |\dot H|  (\tilde \nabla \pi)^2   \Sigma (\tilde \nabla^2) \dot \pi    ,  \\
\mathcal L_{II_2}^{(3)} &= - a^3 M_{\rm Pl}^2  |\dot H|  \frac{2 c_3}{3(1 - c_{\rm s}^{2} )}  \dot \pi   \Sigma (\tilde\nabla^2) \left( \dot \pi \Sigma (\tilde\nabla^2)\dot \pi \right)    , \\
\mathcal L_{III}^{(3)} &= -  a^3 M_{\rm Pl}^2 |\dot H|  \frac{2\tilde{c}_3}{3(1 - c_{\rm s}^{2} )^2} \left ( \Sigma (\tilde\nabla^2) \dot \pi \right ) \left ( \Sigma (\tilde\nabla^2) \dot \pi \right )  \left ( \Sigma (\tilde\nabla^2)\dot \pi \right )   , 
\end{split}
\ee
%
%
where $\Sigma$ was defined in \eqref{sigma,c3-def}. 
We will assume that around $H$, the dispersion relation is quartic, $i.e.$ $\Lambda_{\rm new} \ll H$, so that the operator $\Sigma$ may be approximated by  
$$ 
\Sigma (\tilde\nabla^2) \to -  (1 - c_{\rm s}^{2} ) \frac{ \Lambda_{\rm UV}^2 }{\tilde\nabla^{2}} .
$$
From the dispersion relation in the new physics regime \eqref{non-linear-dispersion}, we can deduce the momentum that corresponds to the Hubble scale, $p_*^2 \simeq H \Lambda_{\rm UV}$, as in \eqref{ref-indices-non-linear-regime}. This allows us to consider the following replacements
\be \label{horizon_cross_replacements_nl}
- \nabla^2 / a^2 \to H \Lambda_{\rm UV} , \qquad   \partial_t \to H , \qquad  \Sigma(\tilde\nabla^2) \to (1 - c_{\rm s}^{2} ) \frac{\Lambda_{\rm UV}}{H},
\ee
when evaluating the ratio \eqref{eq-ng-zeta}.
Neglecting the  $a^3 M_{\rm Pl}^2  |\dot H| (1 - c_{\rm s}^{2} )$ common factor, since it will cancel out in the final ratio, the quadratic piece reads
\be \label{quadratic-cot}
\mathcal L^{(2)}\Big|_{\omega = H} \simeq   H \Lambda_{\rm UV} \pi^2.
\ee 
Using the relation $\mathcal{R} = -H \pi$, we may obtain the cubic contributions in \eqref{cubics} as 
\be \label{cubics-hor}
\begin{split}
\mathcal L_{I}^{(3)}\Big|_{\omega = H} =   H \Lambda_{\rm UV} \pi^2 \mathcal{R}   &,   \quad
\mathcal L_{II_1}^{(3)}\Big|_{\omega = H}  =   \Lambda_{\rm UV}^2 \pi^2  \mathcal{R}  ,  \\
\mathcal L_{II_2}^{(3)}\Big|_{\omega = H} =  \frac{2c_3}{3}   \Lambda_{\rm UV}^2 \pi^2 \mathcal{R}   &, \quad
\mathcal L_{III}^{(3)}\Big|_{\omega = H} =   \frac{2\tilde{c}_3}{3} \frac{\Lambda_{\rm UV}^3}{H} \pi^2 \mathcal{R}   , 
\end{split}
\ee
%
and substituting these expressions into \eqref{eq-ng-zeta}, we see that the generic prediction is
\be
f_{\rm NL} \propto \frac{\Lambda_{\rm UV}}{H}. \label{f-NL-prediction}
\ee

Comparing with the analogous formulae from the lower derivative EFT \eqref{Seft-simple},
\be
\mathcal{P_R}  \simeq \frac{1}{8 \pi^2} \frac{H^2}{M_{\rm Pl}^2 \epsilon c_{\rm s}}  , \qquad   r \simeq 16 \epsilon c_{\rm s} ,  \qquad f_{\rm NL} \sim \frac{1}{c_{\rm s}^2} ,  \label{predictions-0}
\ee
we see that $c_{\rm s}^{-2}$ is replaced by the ratio $\frac{\Lambda_{\rm UV}}{H}$.
Bearing in mind the discussion on the effective medium and \eqref{ref-indices-non-linear-regime}, this comes as no surprise: since we have assumed that the modes freeze within the dispersive medium, the speed of sound $c_{\rm s}$, which is the phase velocity in the long wavelength limit, is replaced by the phase velocity of the new physics regime. Thus, the correct way to interpret \eqref{f-NL-prediction} is
\be
f_{\rm NL} \propto \frac{1}{v_{\rm ph}^{*2}} . \label{f-NL-prediction-phace-vel}
\ee
Furthermore, this result is consistent with our expectations: our approximation to study the theory to zero-th order in $M/p$ has effectively fixed the scale $M$ or $\Lambda_{\rm new}$ relative to $H$; hence $\Lambda_{\rm UV}$ and $H$ are the only two remaining free scales characterising the problem, and as such they should appear at the level of the observables. It is important to underline here that there is no extra degeneracy in the low energy observables due to the presence of heavy fields, since in both sets \eqref{predictions-0} and \eqref{power-spec-new-phys-0}, \eqref{r}, \eqref{f-NL-prediction}, three measurements are required to fix three unknown parameters of the theory. In the former case, these are $\{H,\epsilon,c_{\rm s}\}$, while in the latter $\{ H,\epsilon,\Lambda_{\rm UV} \}$.
The observational bound on the non linearity parameter $f_{\rm NL}$ now translates into a bound\footnote{See \cite{Assassi:2013gxa} for a related discussion on which quantities are actually probed by measurements of $f_{\rm NL}$.} on the value of $\Lambda_{\rm UV}$ relative to $H$. 

In order to roughly estimate this ratio it is instructive to compute the scaling dimension of the three-point operators in \eqref{cubics}. The scaling dimension of $\pi$ can be deduced from the requirement of the kinetic term to be a marginal operator. Performing a rescaling of the energy $\omega \to \lambda \omega$, the momentum scales as $p \to \sqrt{\lambda} p$ as a result of the non linear dispersion $\omega \sim p^2$, so that $\pi$ scales as\footnote{We see that even though at the level of the dispersion relation the theory is similar to ghost inflation \cite{ArkaniHamed:2003uz}, it has different IR behaviour. This is evident from the scaling dimension of the Goldstone mode which for ghost inflation is $[\pi] = 1/4$.} $\pi \to \lambda^{3/4} \pi$. The dimensions of the three-point operators are thus 
\be \label{scaling-dims}
[\mathcal{L}_{I}^{(3)}] = 7/4, \quad [\mathcal{L}_{II_1}^{(3)}] = [\mathcal{L}_{II_2}^{(3)}]= 3/4, \quad [\mathcal{L}_{III}^{(3)}] = -1/4.
\ee
As we will see in Sec.~\ref{sec:bispec}, the operators $[\mathcal{L}_{I}^{(3)}]$ and $[\mathcal{L}_{III}^{(3)}]$ are indeed suppressed and enhanced\footnote{The negative scaling dimension of the $[\mathcal{L}_{III}^{(3)}]$ operator might look alarming but we have to recall our assumption that the modes freeze within the new physics regime. We are thus not allowed to consider the limit $\omega \to 0$, in which case the non linearity parameter is infinite, in our set up. In order to do that, we would have to first decouple the heavy fields by taking the $M \to \infty$ limit, in which case the dangerous $[\mathcal{L}_{III}^{(3)}]$ coupling would vanish -- see the discussion below \eqref{EFT-new-physics-3}.} by $H/\Lambda_{\rm UV}$ and $\Lambda_{\rm UV}/H$ respectively, compared to the other two.
Now we may assume that at the UV cutoff of our theory, $\Lambda_{\rm UV}$, the non Gaussianity is of order one. This is consistent with our discussion of the strong coupling scale in Sec.~\ref{sec:strong-coupling}: strong coupling means that the two- and three-point functions are of the same order and in the case where $\Lambda_{\rm UV} \sim \Lambda_{\rm sc}$, this translates to $\dfrac{ \mathcal L_{II_1}^{(3)} }{ \mathcal L^{(2)} }\Bigg|_{ \omega = \Lambda_{\rm UV}} \sim 1$. Then from \eqref{eq-ng-zeta}, \eqref{f-NL-prediction},
we deduce that
\be \label{scaling-argument}
\dfrac{ \mathcal L_{II_1}^{(3)} }{ \mathcal L^{(2)} }\Bigg|_{ \omega = H} = \left( \dfrac{ H }{ \Lambda_{\rm UV} }\right)^{3/4} \dfrac{ \mathcal L_{II_1}^{(3)} }{ \mathcal L^{(2)} }\Bigg|_{ \omega = \Lambda_{\rm UV}} \quad \Longrightarrow \quad \dfrac{\Lambda_{\rm UV}}{H} \mathcal{R} \sim \left( \dfrac{ H }{ \Lambda_{\rm UV} }\right)^{3/4},
\ee
since the scaling dimension of the operator under consideration is [${ \mathcal L_{II_1}^{(3)}}]=3/4$. Normalising the power spectrum as $\mathcal{P_R}^{1/2} \sim 10^{-5}$, we expect that\footnote{The same number can be obtained using any of the three-point operators of \eqref{cubics} in combination with the correct scaling dimension.} 
\be \label{scaling-value}
10^2 < \dfrac{\Lambda_{\rm UV}}{H} < 10^3.
\ee
%

So far, we have seen how the magnitude of $f_{\rm NL}$ can be large in the new physics regime where $M^2 \ll p^2 \ll \Lambda_{\rm UV}^2$ and how the information hidden in the observables of our theory may differ from the low derivative EFT. We now proceed to the calculation of the bispectrum in order to identify possible observational signatures of heavy fields. 
\subsection{The bispectrum shape} \label{sec:bispec}
The \emph{shape} functions \cite{Babich:2004gb}, that is, the bispectra associated with the cubic operators \eqref{cubics} as functions of momenta, may be obtained via the $in-in$ formalism briefly outlined in Sec.~\ref{sec:power-corrections}.
In order to highlight the method, we will present the computation regarding the operator ${ \mathcal L_{II_1}^{(3)} }$ in some detail and quote the results for the other three. 

In the limit $p^2 \gg M^2$, where momentum dominates over the mass $M$, the Hamiltonian in momentum space is given by
\be \label{3pt-hamiltonian-nonlin}
\hat H_{II_1} \!=  - \int d^3x \hat{\mathcal{L}}_{II_1} = \frac{1}{(2\pi)^6}\frac{M_{\rm Pl}^2 \epsilon}{H^2} \dfrac{\Lambda_{\rm UV}^2}{H^2} \int \frac{d^3q_1 d^3q_2 d^3q_3}{\tau^3} \frac{q_1^2 - q_2^2 - q_3^2}{2 q_1^2} \hat{\mathcal{R}}'_{q_1} { \hat{\mathcal{R}} }_{q_2} { \hat{\mathcal{R}} }_{q_3} \delta \left({\mathbf q}\right),
\ee
where ${\mathbf q} = \sum {\mathbf q}_i$, and from \eqref{in-in-0}, the tree level correction to the three-point correlator reads
\be
(2 \pi)^3 \delta(\mathbf{k}_1 + \mathbf{k}_2 + \mathbf{k}_3)  B_{II_1}(k_1,k_2,k_3)  = i \int_{-\infty}^{0} d\tau \langle [\hat{ \mathcal{R} }_{\mathbf k_1} \hat{ \mathcal{R} }_{\mathbf k_2} \hat{ \mathcal{R} }_{\mathbf k_3} , \hat H_{II_1}(\tau) ] \rangle.
\ee
Upon expanding the commutator and keeping only the fully contracted terms, we arrive at the final integral which is
\be \label{3pt}
 B_{II_1} =  2{\rm Im} \Bigg[ \frac{M_{\rm Pl}^2 \epsilon }{H^2} \frac{\Lambda_{\rm UV}^2}{H^2}\frac{k_1^2 - k_2^2 - k_3^2}{2 k_1^2} \mathcal{R}^{(0)}_{k_1} \mathcal{R}^{(0)}_{k_2} \mathcal{R}^{(0)}_{k_3} \int^{0}_{-\infty} \frac{d\tau}{\tau^3} \mathcal{R}'^*_{k_1} \mathcal{R}^*_{k_2} \mathcal{R}^*_{k_3} + \text{perm} \Bigg],
\ee
where $\mathcal{R}_k$ is given by \eqref{pi-solution}.
 
Let us first focus on the integral
$$I = \int^{0}_{-\infty} \frac{d\tau}{\tau^3} \mathcal{R}'^*_{k_1} \mathcal{R}^*_{k_2} \mathcal{R}^*_{k_3}.$$
Changing the integration variable form $\tau$ to $z = \frac{H}{2 \Lambda_{\rm UV} } k^2 \tau^2$ and rewriting the solution as
\be \label{zeta,y,A}
\mathcal{R}^* (z) = k_1^{-3/2} {\cal C} \left( \frac{\Lambda_{\rm UV}}{H} \right)^{1/4} z^{5/4}H_{5/4}^{(2)}( z ); \quad {\cal C} = - 2^{1/4} \frac{H}{( M_{\rm Pl}^2 \epsilon  )^{1/2}} \sqrt{\frac{\pi}{4}} ,
\ee
%
we obtain
\be \nn
I = k_1^{-3/2} {\cal C}^3  \left( \frac{H}{\Lambda_{\rm UV}} \right)^{3/4} x_2 x_3 \int^{0}_{\infty} dz z^{9/4} H_{1/4}^{(2)}(z)H_{5/4}^{(2)}(x_2^2 z)H_{5/4}^{(2)}(x_3^2 z),
\ee
where $x_2 =k_2/k_1, \; x_3 = k_3/k_1$.
Analytically continuing $z\mapsto -iz$, so that $H_\nu^{(2)}(-iz) = \dfrac{2}{\pi} (-i)^{-\nu-1} K_\nu(z)$, with $K_\nu$ the modified Bessel function of the second kind, yields
\be \label{integral_II1}
I = k_1^{-3/2} {\cal C}^3  \left( \frac{H}{\Lambda_{\rm UV}} \right)^{3/4}  \left( \frac{2}{\pi} \right)^3 e^{i\pi/4}  x_2 x_3 \int_{0}^{\infty} dz z^{9/4} K_{1/4}(z)K_{5/4}(x_2^2 z)K_{5/4}(x_3^2 z).
\ee
%
We may now substitute \eqref{R^0},\eqref{integral_II1} back to \eqref{3pt} and obtain the three-point correlator for the operator $\mathcal{L}_{II_1}$.

\begin{figure}[tbh!!!]
\centering
\includegraphics[scale=0.6]{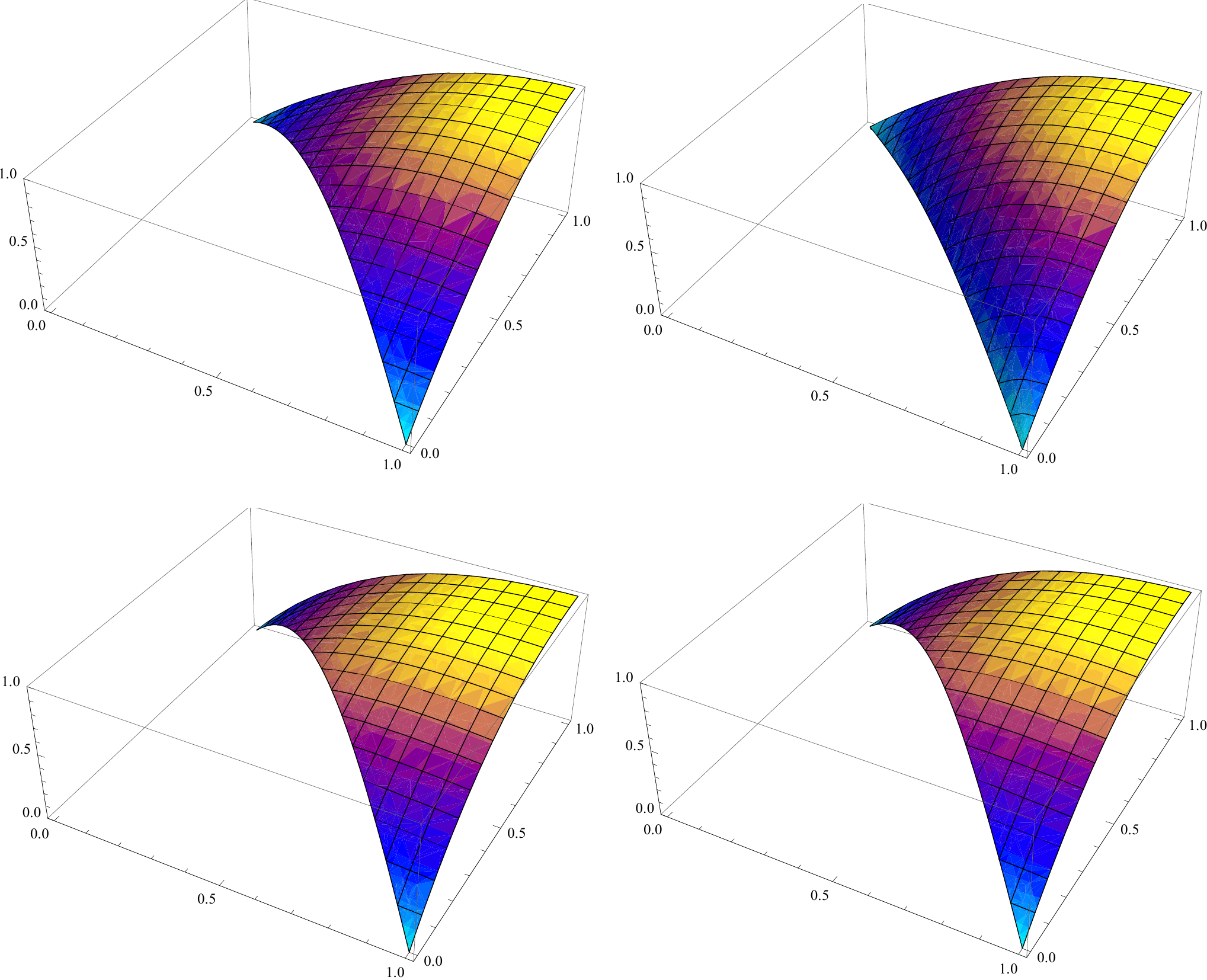}
\caption[{\sf The bispectra of the effective theory}.]{\sf The bispectra $x_2^2 x_3^2 S(1,x_2,x_3)$ of the effective theory \eqref{EFT-new-physics-3}, normalized to one in the equilateral configuration. Clockwise from top left: $S_{I},S_{II_1},S_{II_2},S_{III}$. $S_{II_2}$ and $S_{III}$ are highly degenerate but evaluation at the flattened triangle $x_2=x_3=1/2$ reveals their difference.}
\label{fig:shapes}
\end{figure}

In complete analogy, we may derive the expressions for the rest of the cubic contributions in \eqref{cubics}. Upon defining
\be \nn
f_{\mathrm{NL}}^i = \frac{B_\Phi^i(1,1,1)}{6 k^6 P^2_{\Phi}(k)},
\ee
and using the relation $\Phi= \dfrac{3}{5} \mathcal{R}$,
the three-point functions for the Newtonian potential $\Phi$ read
\be \label{b-shapes}
\begin{split}
  B_{\Phi}^I =  6 P^2_{\Phi}(k) f_{\rm NL}^{I}  S_I^{\rm eq}(1,x_2,x_3),& \quad 
   B_{\Phi}^{II_1} = 6 P^2_{\Phi}(k) f_{\rm NL}^{II_1} S_{II_1}^{\rm eq}(1,x_2,x_3), \\
 B_{\Phi}^{II_2} = 6 P^2_{\Phi}(k) f_{\rm NL}^{II_2}  S_{II_2}^{\rm eq}(1,x_2,x_3),& \quad
 B_{\Phi}^{III} =  6 P^2_{\Phi}(k) f_{\rm NL}^{III} S_{III}^{\rm eq}(1,x_2,x_3),
\end{split}
\ee
%
where $S^{\rm eq}$ is used to denote the shape function normalized at the equilateral limit $x_2=x_3=1$, and the power spectrum of the Newtonian potential satisfies $P_{\Phi}(k)= \frac{18}{25} \frac{\pi^2}{k^3} \mathcal{P}_{\mathcal{R}}(k)$, with $\mathcal{P}_\mathcal{R}$ given in \eqref{power-spec-new-phys-0}. The non linearity parameters read
%
\be \label{fnls}
\begin{split}
f_{\rm NL}^{I} = \frac{5}{18} \frac{2^{1/4} }{\pi \Gamma[5/4]} \times 0.3549 &, \quad f_{\rm NL}^{II_1} = - \frac{5}{72} \frac{2^{1/4} }{\pi \Gamma[5/4]}  \times 7.9071 \dfrac{1}{v_{\rm ph}^{*2}}, \\ f_{\rm NL}^{II_2} = \frac{5}{54} \frac{2^{1/4} }{\pi \Gamma[5/4] } \times 0.5369 \dfrac{c_3}{v_{\rm ph}^{*2}} &, \quad f_{\rm NL}^{III} = \frac{5}{36} \frac{2^{1/4} }{\pi \Gamma[5/4]}  \times 0.4999 \dfrac{\tilde{c}_3}{v_{\rm ph}^{*4}},
\end{split}
\ee
where the reader may recall that $\dfrac{1}{v_{\rm ph}^{*2}} = \dfrac{\Lambda_{\rm UV}}{H}$. 
Finally, the shape functions $S$ are given by
\be \label{Ss}
\begin{split}
S_I(1,x_2,x_3) &= \frac{x_2^2 x_3^2 + x_2^2 +  x_3^2}{\sqrt{x_2 x_3}}  \int_{0}^{\infty} dz z^{5/4+2} K_{1/4}(z)K_{1/4}(x_2^2 z)K_{1/4}(x_3^2 z), \\
S_{II_1}(1,x_2,x_3) &= \frac{1 - x_2^2 - x_3^2}{\sqrt{x_2 x_3}} \int_{0}^{\infty} dz z^{5/4+1} K_{1/4}(z)K_{5/4}(x_2^2 z)K_{5/4}(x_3^2 z) + \text{2 perm}, \\
S_{II_2}(1,x_2,x_3) &= \frac{1 + x_2^2 + x_3^2}{\sqrt{x_2 x_3}} \int_{0}^{\infty} dz z^{5/4+1} K_{1/4}(z)K_{1/4}(x_2^2 z)K_{1/4}(x_3^2 z), \\
S_{III}(1,x_2,x_3) &= \frac{1}{\sqrt{x_2 x_3}}  \int_{0}^{\infty} dz z^{5/4} K_{1/4}(z)K_{1/4}(x_2^2 z)K_{1/4}(x_3^2 z),
\end{split}
\ee
and their graphs are depicted in Fig.~\ref{fig:shapes}.
%

In order to make contact with observations, it is instructive to project our predictions to the templates actually used by experiments. Following \cite{Babich:2004gb}, an inner product between two shapes $S_i(1,x_2,x_3)$ and $S_j(1,x_2,x_3)$ may be defined as
\be
S_i(1,x_2,x_3) * S_j(1,x_2,x_3) = \int dx_2 dx_3 (x_2x_3)^4 S_i(1,x_2,x_3) S_j(1,x_2,x_3),
\ee
which leads to a correlator of the two bispectra:
\be
\cos(S_i,S_j) = \frac{S_i*S_j}{\sqrt{S_i*S_i}\sqrt{S_j*S_j}}.
\ee
The projected non linearity parameters can now be computed as \cite{Senatore:2009gt}
\be
\left( \begin{array}{c} f_{\rm NL}^{\rm equil} \\ f_{\rm NL}^{\rm ortho} \\ f_{\rm NL}^{\rm flat} \end{array} \right) = \left( \begin{array}{cccc} \frac{S_{I} * S_{\rm equil}}{S_{\rm equil}*S_{\rm equil}} & \frac{S_{II_1} * S_{\rm equil}}{S_{\rm equil}*S_{\rm equil}} & \frac{S_{II_2} * S_{\rm equil}}{S_{\rm equil}*S_{\rm equil}} & \frac{S_{III} * S_{\rm equil}}{S_{\rm equil}*S_{\rm equil}} \\ \frac{S_{I} * S_{\rm ortho}}{S_{\rm ortho}*S_{\rm ortho}} & \frac{S_{II_1} * S_{\rm ortho}}{S_{\rm ortho}*S_{\rm ortho}} & \frac{S_{II_2} * S_{\rm ortho}}{S_{\rm ortho}*S_{\rm ortho}} & \frac{S_{III} * S_{\rm ortho}}{S_{\rm ortho}*S_{\rm ortho}} \\ \frac{S_{I} * S_{\rm flat}}{S_{\rm flat}*S_{\rm flat}} & \frac{S_{II_1} * S_{\rm flat}}{S_{\rm flat}*S_{\rm flat}} & \frac{S_{II_2} * S_{\rm flat}}{S_{\rm flat}*S_{\rm flat}} & \frac{S_{III} * S_{\rm flat}}{S_{\rm flat}*S_{\rm flat}} \end{array} \right)  \left( \begin{array}{c} f_{\rm NL}^{I} \\ f_{\rm NL}^{II_1} \\ f_{\rm NL}^{II_2} \\ f_{\rm NL}^{III} \end{array} \right).
\ee
Using the templates \cite{Creminelli:2005hu,Senatore:2009gt,Meerburg:2009ys}
\be
\begin{split}
S_{\rm equil}(x_1,x_2,x_3) & = 6 \left( -\frac{1}{x_1^3x_2^3} -\frac{1}{x_1^3x_3^3} -\frac{1}{x_2^3x_3^3} - \frac{2}{x_1^2x_2^2x_3^2} + 
\left[ \frac{1}{x_1x_2^2x_3^3} + 5 \;\rm{perm} \right] \right), \\
S_{\rm ortho}(x_1,x_2,x_3) & = 6  \left( -\frac{3}{x_1^3x_2^3} -\frac{3}{x_1^3x_3^3} -\frac{3}{x_2^3x_3^3} - \frac{8}{x_1^2x_2^2x_3^2} + 3 \left[ \frac{1}{x_1x_2^2x_3^3} + 5 \;\rm{perm} \right]  \right), \\
S_{\rm flat}(x_1,x_2,x_3) & = 6 \left( \frac{1}{x_1^3x_2^3} +\frac{1}{x_1^3x_3^3} + \frac{1}{x_2^3x_3^3} + \frac{3}{x_1^2x_2^2x_3^2} - \left[ \frac{1}{x_1x_2^2x_3^3} + 5 \;\rm{perm} \right]  \right),
\end{split}
\ee
we obtain
\be
\begin{split}
f_{\rm NL}^{\rm equil}(v_{\rm ph},c_3,\tilde{c}_3) & =  0.0157 + 1.8961 v_{\rm ph}^{*-2} + 0.0128  c_3 v_{\rm ph}^{*-2} + 0.0167  \tilde{c}_3 v_{\rm ph}^{*-4} , \\
f_{\rm NL}^{\rm ortho}(v_{\rm ph},c_3,\tilde{c}_3) & = 0.0005 + 0.1719 v_{\rm ph}^{*-2} - 0.0004 c_3 v_{\rm ph}^{*-2} - 0.0003 \tilde{c}_3 v_{\rm ph}^{*-4}, \\
f_{\rm NL}^{\rm flat}(v_{\rm ph},c_3,\tilde{c}_3) & = 0.0028 + 0.3182 v_{\rm ph}^{*-2} + 0.0024 c_3 v_{\rm ph}^{*-2} + 0.0031 \tilde{c}_3 v_{\rm ph}^{*-4}.
\end{split}
\ee
Inverting these expression is the final step which connects the predictions of the theory to the observational data:
\be \label{constraint-eqs}
\begin{split}
 v_{\rm ph}^{*-2} & = -0.0009 + 38.4502 f_{\rm NL}^{\rm equil} - 29.577 f_{\rm NL}^{\rm ortho} - 209.997 f_{\rm NL}^{\rm flat}, \\
c_3 v_{\rm ph}^{*-2} & = 3.5240 + 46461.8 f_{\rm NL}^{\rm equil} - 41701.4  f_{\rm NL}^{\rm ortho} - 254330 f_{\rm NL}^{\rm flat}, \\
\tilde{c}_3 v_{\rm ph}^{*-4} & =  - 3.54037 - 39917.2 f_{\rm NL}^{\rm equil} + 35320.9   f_{\rm NL}^{\rm ortho} + 218778  f_{\rm NL}^{\rm flat} .
\end{split}
\ee
From this form one may proceed to input the {\sc Planck} values for each $f_{\rm NL}$ and constrain the UV scale relative to the Hubble scale. However, since the {\sc Planck} covariance matrix has not been published and the current error bars still leave a fairly sized parameter space available, it is, currently, hard to draw any conclusion from \eqref{constraint-eqs}, apart from the fact that the value $\Lambda_{\rm UV}=100H$, which is expected from scaling arguments -- see discussion around \eqref{scaling-value}, is perfectly consistent with the {\sc Planck} bounds. We hope that in the near future higher experimental resolution will allow for more precise statements about the UV theory of inflation.

A desirable feature of the EFT \eqref{EFT-new-physics-3}, would be to generate a new distinguishable shape of non Gaussianities but evidently this is not the case. All these shapes are unfortunately identical to the ones obtained in \cite{Senatore:2009gt} for the standard EFT \eqref{Seft-simple}, so effects of massive fields on the inflaton perturbations are difficult to distinguish using the three-point correlator. What is important though is the meaning of the observables when heavy fields are considered. 


One way to lift this shape degeneracy might be the following: recall that we have assumed that the scale of the dispersive regime $\Lambda_{\rm new}$, is much lower than the Hubble scale $H$ so we have neglected any effect of order $\Lambda_{\rm new}/H$. Taking corrections with respect to this ratio might reveal new signatures but the complicated dynamics \eqref{eq-motion-u-lim-1} of the theory are a considerable obstacle towards that direction. Numerical study is always an option but analytical results would be preferable in order to gain insight into the EFT structure from the way this parameter would appear in the observables.

We now close this Chapter with a short summary and finally proceed to our conclusions.
%

\chapter{Summary of Part \ref{part:eft}}

In this Part, we focused on an effective field theory designed to capture effects of heavy scalars on the dynamics of the inflaton perturbations. In the framework of the EFT of inflation developed in \cite{Cheung:2007st}, we considered insertions of a specific class of operators of the form $\frac{ M^2 }{M^2 - \nabla^2} $, with $M$ the heavy mass, which parametrise the low energy couplings of the EFT as arising from the mediation of heavy particles in the UV.

We saw how these operators modify the dispersion relation of the low energy mode $\pi$, the inflaton perturbation, leading to a non linear relation of the form $\omega \propto p^2$. Assuming that the modes crossed the Hubble radius within this dispersive regime, these operators generically lead to $\Lambda_{\rm sb} \sim \Lambda_{\rm sc} \sim \Lambda_{\rm UV}$, where $\Lambda_{\rm sb}$ is the symmetry breaking scale, $\Lambda_{\rm sc}$ is the strong coupling scale and $\Lambda_{\rm UV}$ is the UV scale.

Finally, we saw how the power spectrum, the tensor-to-scalar ratio, the non linearity parameter and the three-point functions are affected by the presence of massive scalars. Even though distinct non Gaussian signatures are not generated, an important feature is that these quantities depend on the parameters $\{\epsilon,H,\Lambda_{\rm UV}\}$, as opposed to $\{\epsilon,H,c_{\rm s}\}$ in the case of the EFT without the $\frac{ M^2 }{M^2 - \nabla^2} $ insertions. Thus, the scale of heavy physics appears directly in the observables and can be constrained from current astrophysical surveys such as {\sc Planck}.  
 \part{Concluding remarks \& future directions} \label{part:conc}
  In this thesis, we focused on two main topics of inflationary physics: the observational signatures of cosmic superstrings formed at the end of string inflationary models, and the properties of primordial scalar perturbations on a background of heavy scalar fields, using a universal effective field theory. 

Cosmic superstrings could, in principle, provide a unique observational window into string theory and the way inflation works within this UV complete framework. The compactifications of string theory to four dimensions yield effective models with unknown couplings, and observational constraints on cosmic superstrings may be used to determine the values of these parameters. Moreover, knowledge of the radiative mechanisms of cosmic superstrings, within each class of string inflationary models, is crucial for the detection of such structures by astrophysical surveys. 

Primordial curvature perturbations provide an insight in both the inflationary dynamics and the process of large scale structure formation in the universe. Constructing an effective field theory for scalar perturbations, allows one to study their dynamics in a model independent way and then use the observational constraints to bound the unknown coefficients of the action. For this to work, it is important to know the precise connection between the experimentally measured quantities and the theoretical predictions. 

Specifically, in Ch.~\ref{ch:paper-1} we studied the $D3/D7$ inflationary model \cite{Dasgupta:2002ew,Dasgupta:2004dw,Haack:2008yb}, which has a $D$-term potential driving inflation. In this class of models, inflation ends when a scalar field develops a tachyonic mass, destabilising the inflationary vacuum and driving the inflaton to a new vanishing vev. This is known as the waterfall stage. In this new stable vacuum, the waterfall fields acquire a non zero vev which spontaneously breaks the $U(1)$ symmetry under which they are charged, hence cosmic superstrings are expected to form at the end of inflation. 

Our aim was to study the nature of these strings. We showed that the $U(1)$ symmetry under which the waterfall fields of $D3/D7$ are charged, is anomalous in the sense that their charges do not sum to zero. In string theory, this anomaly is automatically cancelled by a term in the effective supersymmetric Lagrangian, originating from the reduction of the ten dimensional Chern-Simons part of the D7-brane worldvolume action. Effects of this term on the cosmic superstrings had not been previously considered in this model. 

In the low energy theory, this term has two effects: it results in an axion field that couples to the cosmic superstrings, and in addition, it yields a field dependent Fayet-Iliopoulos term which is a function of this axion. Such axionic strings are known to have long range interactions which lead to their decay. However, we argued that in the $D3/D7$ model, cosmic superstrings do not exhibit these long range forces. As shown in \cite{Davis:2005jf}, when the axion field is allowed to vary in space, it contributes to the string energy in such a way, so that it remains confined in the string core and no long range interactions are induced. In the $D3/D7$ model, the expectation value of this axion field, which we identified as the modulus which controls the volume of the compactification space, naturally has such a spatial dependence. 
This is because of the following reason: the Fayet-Iliopoulos term is field dependent so the volume modulus needs to be stabilised in a vacuum expectation value. The stabilising potential involves the waterfall scalars and hence, the vacuum expectation value of the volume modulus depends on these fields. Since these fields form the cosmic string, the axion inherits a spatial dependence on the string background. 

Furthermore, the fermionic superpartner of the volume modulus, provides a chiral zero mode that may stabilise a string loop and form a vorton. Such configurations are undesirable in cosmological models, since they are inconsistent with observational results. However, in the model under consideration, vortons decay and thus do not have catastrophic cosmological implications which could rule it out.


Apart from our results for the cosmic superstrings produced in the model, we argued that the inflationary process itself seems problematic. As shown in \cite{Komargodski:2009pc,Dienes:2009td,Komargodski:2010rb}, supergravity theories with constant FI terms cannot have all moduli fixed, since this renders the theory gauge variant. As already mentioned, the $D3/D7$ model assumes constant vacuum expectation values for all the moduli fields, and hence a constant FI term. Therefore, it is incompatible with these theoretical constraints and the slow roll inflationary phase needs to be revisited.

In Ch.~\ref{ch:paper-2}, we focused on radiative processes of cosmic F-strings and D1-branes on warped backgrounds, which are geometries where the four dimensional metric has a dependence on the compact directions. In previous studies \cite{Vilenkin:1986ku}, it was shown that since a 2-form field in four dimensions may be equivalently described by a scalar degree of freedom, cosmic superstrings that couple to 2-form fields can radiate scalar particles. In \cite{Firouzjahi:2007dp}, it was claimed that the main radiative channel of type IIB superstrings --which couple to the RR 2-form-- on a warped background, is the scalar radiation, since gravitational radiation is suppressed due to the warping. However, it is well known \cite{Giddings:2001yu}, that in string theory models, the only way to obtain a warped geometry is to include orientifold planes. In presence of orientifold planes, the spectrum of a theory is truncated to the subspace that contains states that are invariant under the orientifold action. We argued that since the RR 2-form is projected out of the four dimensional spectrum, a superstring cannot radiate scalar particles from such a source. The same holds for the NSNS 2-form to which a fundamental string could couple.

Furthermore, we placed our study in the context of a well known type IIB inflationary model on a warped background, since this is a natural context for cosmic superstring formation. This is the $D3/\bar{D}3$ inflationary model \cite{Kachru:2003sx} on a compact version \cite{Giddings:2001yu} of the Klebanov-Strassler throat \cite{Klebanov:2000hb}. Cosmic superstrings on a KS throat were studied in \cite{Firouzjahi:2006vp}, where it was shown that they may be described as a D3-brane wrapping a 2-cycle of the internal manifold, while its two remaining dimensions extend in spacetime. Such a brane couples to all the lower rank form fields, namely the RR 4-form and 2-form, as well as the NSNS 2-form. We argued that, even though the 2-forms are projected out of the four dimensional spectrum, since the D3-brane wraps an internal cycle, the resulting string may couple to fields that arise in four dimensions, from the RR and NSNS forms when they have legs along the compactification manifold. This means that scalar particles resulting from the 2-forms when they have both legs in the internal space, as well as the 2-form that arises form the 4-form when it has two legs along the internal space, could in principle result in scalar radiation from the superstring. However, the equations of motion for these fields are coupled and the analysis of their dynamics is complicated. Therefore, due to the limited progress in the literature, we cannot conclude whether such radiation dominates over gravitational waves or not.

In Ch.~\ref{ch:paper-3}, we changed our perspective from background dynamics to the study of scalar perturbations along a general homogeneous and isotropic, inflating background, within the framework of the effective field theory of inflation. In \cite{Creminelli:2006xe,Cheung:2007st}, it was shown that scalar perturbations transform non trivially under temporal diffeomorphisms, and they can be described as Goldstone bosons that arise from this symmetry breaking. The effective action can be constructed as a polynomial over all the operators that are consistent with the reduced symmetry, with unknown dimensionful coefficients. Observational results can then be used in order to constrain the values of these parameters.

Motivated by string theory models which contain a vast variety of heavy scalar fields, $e.g.$ moduli fields, we identified a class of operators that parametrise the low energy couplings of the effective action of curvature perturbations, as arising from the mediation of massive scalars in the UV. This was accomplished by writing a generic action with multiple heavy fields and then integrating them out in order to obtain an effective action for the light modes, $i.e.$ the inflaton perturbations.  

As in any effective theory, there are three important energy scales that define it. The first scale is the characteristic scale of the system under consideration, which is the energy scale around which the experimental measurement is performed. For the case of curvature perturbations, this is the Hubble scale $H$, where the modes freeze in time, leaving their imprint on the CMB. The second scale is the energy where the effective theory breaks down because all the effects that we neglected become important. This is what we call $\Lambda_{\rm UV}$, and for us, it signifies the scale where the massive fields have to be incorporated in the theory as dynamical degrees of freedom. Lastly, the third scale is the one below which the theory can be treated perturbatively, or equivalently, it is weakly coupled. This ensures that the effective theory at hand can be used as a computational tool; this is what we call the strong coupling scale $\Lambda_{\rm sc}$. For an effective theory to work, there is an obvious requirement: $H \ll \left( \Lambda_{\rm UV},\;\Lambda_{\rm sc} \right)$.

We demonstrated that in the presence of these operators this inequality is satisfied and especially that the weakly coupled regime of the theory is extended towards the scale $\Lambda_{\rm UV}$. Furthermore, we showed how these operators affect the physics of the low energy modes by reducing their propagation speed and consequently modifying their dispersive behaviour. Specifically, the dispersion relation of the inflaton perturbations displays two regimes -- a linear and a non linear, dispersive regime -- separated by another important energy scale of the theory, $\Lambda_{\rm new}$. Assuming that the light modes reach the Hubble scale $H$ within the dispersive regime, $i.e.$ $\Lambda_{\rm new} \ll H$, the observables of the theory have a different meaning compared to the ones of the effective action without these operators. Namely, in the presence of heavy fields, the power spectrum, the tensor-to-scalar ratio and the non linearity parameter are directly related to $H$, the slow roll parameter $\epsilon$, and $\Lambda_{\rm UV}$. We also computed the shapes of the three-point correlators to find that they are indistinguishable with respect to the lower derivative EFT although the momentum dependence of the integrals in the two cases is quite different. 

This thesis has taken a few steps toward the aforementioned directions, but there is still a lot of work to be done. Firstly, it would be interesting to explicitly compute the axionic wave function in the $D3/\bar{D}3$ model and solve it, at least numerically, in order to obtain the power spectrum of radiation. Radiative signals of cosmic superstrings are of considerable importance since if detected, they would support string theory as a UV complete physical theory.

Another interesting topic that combines the UV models of Part~\ref{part:strings} and the effective field theory approach of Part~\ref{part:eft}, is to search for the set of operators in the effective action for inflaton perturbations, that parametrise stringy inflationary models. Such a study could open a new observational window into string theory via the non Gaussian signatures of string inflationary models. Although non Gaussianities of certain models, such as DBI inflation, have been previously studied, a model independent systematic description of stringy effects, to the best of our knowledge, has not been attempted. We took the first step toward this direction by studying operators that capture massive scalars. One could enrich these results by including massive gauge fields and fermions that couple to the inflaton perturbations and study their effects along the lines of Ch.~\ref{ch:paper-3}. 

In view of the \emph{precision era} that cosmology enters, understanding the connection between observable quantities and free parameters of inflationary models is of crucial importance. Equipped with such knowledge, even if observational surveys disfavour non Gaussian signatures or cosmic strings, one may still seek an answer to the persisting question that drives research throughout human history: {\it ``Why?"}.

\baselineskip=18pt plus1pt
\appendix
\chapter{Effective action to second order in slow roll} \label{app:slow-roll}

In this appendix we solve the free, flat gauge effective theory, discussed in Sec.~\ref{subsec:eft-flat}, to second order in the slow roll parameters. Recall that in the unitary gauge, where time diffeomorphism invariance is non linearly realised through $\pi$, the effective action is given by \eqref{S[pi,g]}, 
%
%
which after performing an ADM decomposition of spacetime \eqref{adm},
may be rewritten as
\be \label{S-pi-app-adm}
\begin{split}
S_2[\pi] &\subset M_{\rm Pl}^2\int dx^3 dt a^3 N  \Bigg[ \frac{1}{2} \left( R(\gamma_{ij})+K_{ij}K^{ij}- K^2 \right) - 3H^2(t+\pi) \\ &+ \dot H(t+\pi) \left(-1-\dfrac{1}{N^2}(1+\dot \pi)^2 +2(1+\dot\pi)\partial_i \pi \dfrac{N^i}{N^2} + \frac{(\partial\pi)^2}{a^2} -\dfrac{N^iN^j\partial_i \pi\partial_j \pi}{N^2}\right) \\ &+  \frac{M_2 ^4(t+\pi)}{2!M_{\rm Pl}^2} \left(-\dfrac{1}{N^2}(1+\dot \pi)^2 +2(1+\dot\pi)\partial_i \pi \dfrac{N^i}{N^2} + \left( \gamma^{ij}-\dfrac{N^iN^j}{N^2} \right) \partial_i \pi\partial_j \pi\right)^2 \Bigg],
\end{split}
\ee
where $K_{ij}$ is the extrinsic three-dimensional curvature tensor defined in \eqref{Kij}, only this time we will compute it in the spatially flat gauge using the metric \eqref{flat-gauge}. 
Let us calculate the Einstein-Hilbert term first. Note that the Ricci scalar of the spatial slice vanishes since without loss of generality one can decompose spacetime into flat foliations. Next, from the curvature square term we get
$$
K_{ij}K^{ij}=\frac{1}{N^2}\left(3H^2-2H \partial N+\frac{1}{2}\left( \partial_iN_j\partial^iN^j+\partial_iN_j\partial^jN^i \right) \right),
$$
while the trace term yields
$$
K^2=\frac{1}{N^2}\left(9H^2 + (\partial N)^2 - 6H \partial N\right).
$$
Therefore the Einstein-Hilbert contribution to the action reads
\begin{equation}\label{SEH}
S_{\rm EH}= \frac{M_{\rm Pl}^2}{2}\int dx^3 dt \frac{a^3}{N}  \left( -6H^2 + 4H\partial N - (\partial N)^2  + \frac{1}{2} \left( \partial_iN_j\partial^iN^j+\partial_iN_j\partial^jN^i \right) \right).
\end{equation}
Next, let us concentrate on terms that are quadratic in $\pi$ knowing beforehand that the constraint equations will give solutions that are first order in $\pi$, $i.e.$ $\left( \delta N,\;N^i \right) \sim \mathcal{O}(\pi)$. Moreover, let us assume, as we did in Sec.~\ref{sec:known-models}, that the shift vector is the derivative of a scalar, 
$N^i=\partial^i\psi_1$, so that the term 
$$
-(\partial N)^2+\frac{1}{2} \left( \partial_iN_j\partial^iN^j+\partial_iN_j\partial^jN^i \right)
$$
can be integrated by parts to yield a vanishing contribution. Moreover, we set $\delta N=N-1$, so that $ \dfrac{1}{N}=1-\delta N + (\delta N)^2+\cdots $. At this stage the action \eqref{S-pi-app-adm} reads
\be \label{S-app-adm-exp}
\begin{split} 
S[\pi] &= M_{\rm Pl}^2\int dx^3 dt a^3  \Big[ -3\frac{H^2}{N} - 3H^2(t+\pi)N + 2H\partial N(1-\delta N) -\dot H(t+\pi)\dot\pi^2 \\ &- \dot H(t+\pi)(\delta N)^2 + 2\dot\pi\dot H(t+\pi)\delta N -2\dot H(t+\pi)(1+\dot\pi) \\ &- 2\dot H(t+\pi) \pi \partial N(1-\delta N) + \dot H(t+\pi) \frac{(\partial\pi)^2}{a^2}N + 2\frac{M_2 ^4(t+\pi)}{M_{\rm Pl}^2}(\delta N-\dot\pi)^2 \Big].
\end{split}
\ee
In order to write this action as a series in slow roll, one needs to expand all the time dependent quantities as in \eqref{taylor}.
%
%
Defining the slow roll parameters as 
\begin{eqnarray}\label{sr}
\epsilon  \equiv  -\frac{d \ln H }{Hdt},\;
s \equiv  \frac{d \ln c_{\rm s} }{Hdt},\;
\eta \equiv \frac{d \ln\epsilon }{Hdt},\;
t \equiv  \frac{d \ln s}{H dt}, \;
\xi  \equiv  \frac{d \ln\eta}{Hdt},
\end{eqnarray}
the Taylor expansions to second order in $\pi$ read
%
\begin{equation}
H(t+\pi)=H\left[1 - H \pi \epsilon + 
  H^2 \pi^2 \left( \epsilon^2 - \frac{\epsilon\eta}{2} \right) \right],
\end{equation}
\begin{equation}
\dot H(t+\pi)=\dot H\left[1+ \pi H(\eta-2 \epsilon )+\pi^2H^2 \left( 3 \epsilon^2 -\frac{7}{2} \epsilon\eta + \frac{1}{2} \eta^2 + \frac{1}{2} \eta \xi \right) \right],
\end{equation}
%
%
\begin{equation}
M_2^4(t+\pi)=M_2^4+\dot M_2^4\pi+ \frac{1}{2}\ddot M_2^4\pi^2,
\end{equation}
where all quantities on the RHS are time dependent, $e.g.$ $H \equiv H(t)$. Before performing this expansion, we note that the term $-2\dot H(t+\pi)(1+\dot\pi)$ in the second line of \eqref{S-app-adm-exp}, can be rewritten as $-2\frac{\partial}{\partial t} H(t+\pi)$, in which form it may then be integrated by parts to yield $6a^3 H H(t+\pi)$.
We can now substitute the expanded expressions in the action to get
%
\be
S[\pi] = -M_{\rm Pl}^2\int dx^3 dt a^3  \Big[ 3 (H\delta N + \pi\dot H)^2  + \frac{\dot H}{c_{\rm s}^2} \left( \delta N - \pi \right)^2 - \dot H \frac{(\partial\pi)^2}{a^2} + 2\partial N (H\delta N + \pi\dot H)  \Big] ,
\ee
where we have neglected total derivative terms and made use of the relation \eqref{sof} between the speed of sound and the ratio $\frac{M_2^4}{\dot H M_{\rm Pl}^2}$. 
At this point we need to solve for the lapse and shift constraints. After simple algebra, we get
\begin{equation}
\delta N=\epsilon H\pi \quad \text{and} \quad \partial N=-\frac{\epsilon}{c_{\rm s}^2}  \frac{d}{dt}(H\pi).
\end{equation}
Substituting these back into the action we arrive at the expression
\begin{equation}
S_2[\pi] = -M_{\rm Pl}^2\int dx^3 dt a^3 \frac{\dot H}{c_{\rm s}^2} \Big[\dot\pi^2 - c_{\rm s}^2\frac{(\partial\pi)^2}{a^2} -\epsilon\dot H \pi^2 - 2\epsilon H \pi \dot\pi \Big],
\end{equation}
which can be further integrated by parts, to yield the final action to second order in slow roll
\begin{equation} \label{Seft2-sr}
S_2[\pi] = -M_{\rm Pl}^2\int dx^3 dt a^3 \frac{\dot H}{c_{\rm s}^2} \Big[\dot\pi^2 - c_{\rm s}^2\frac{(\partial\pi)^2}{a^2} +\epsilon H^2 \left(3 - \epsilon + 2 \eta - 2 s  \right) \pi^2 \Big]. 
\end{equation}
This action contains the first order result \eqref{Seft2}. 

Let us now compute the dynamics of the Goldstone mode. The equation of motion in conformal time is
\be \label{pi-eom-sr}
\pi'' + a H (2 - 2s - 2\epsilon + \eta) \pi' + \left( c_{\rm s}^2 k^2 - 2\epsilon a^2H^2 \left(3 - \epsilon + 2 \eta - 2 s  \right)  \right) \pi = 0 .
\ee
Since we are interested in higher order results it is better to define the variable $y=\dfrac{c_{\rm s} k }{a H}$, over which the equation of motion reads
\be \label{pi-eom-sr-y}
\pi_{yy} +  \left( \sigma_1 + \sigma_2 - 2 \right) \frac{\pi_y}{y} + \left( \sigma_3^2 -2\frac{\sigma_4}{y^2}  \right) \pi = 0 ,
\ee
where we have defined 
\be 
\sigma_1 = (s + \eta) \sigma_3,\;\; \sigma_2 = (t s + \eta \epsilon) \sigma_3^2,\;\; \sigma_3 = \frac{1}{s + \epsilon -1},\;\; \sigma_4 = \epsilon (3 + 2 \eta - \epsilon -2 s) \sigma_3.
\ee
The solution is given in terms of a linear combination of Hankel functions of the first and second kind, but requiring positive frequency modes at asymptotic infinity $\tau \to - \infty$ forces us to keep the Hankel function of the first kind
\be \label{pi-sr-app}
\pi = C y^{\frac{1}{2}(3-\sigma_1-\sigma_2)} H^{(1)}_{\nu}(-y\sigma_3), \quad \nu = \dfrac{\sqrt{(3-\sigma_1-\sigma_2)^2 +8 \sigma_4 }}{2}.
\ee
The constant $C$ is fixed by the requirement that the solution satisfies the canonical commutation relation $[\pi(x),P_\pi(y)] = i \delta(x-y) $. The conjugate momentum of $\pi$ is $P_\pi = 2 a^3 \dfrac{\epsilon H^2}{c_{\rm s}^2} \dot \pi$ so that upon converting to conformal time, the commutator reads
\be 
[\pi(x), \pi'(y)] = i \frac{ c_{\rm s}^2 }{ 2 \epsilon ( a H )^2 } \delta(x-y) .
\ee
This commutator can be translated into a Wronskian condition on the solution \eqref{pi-sr-app}, namely
\be 
C^2 (y\sigma_3)' y^{3-\sigma_1-\sigma_2} \mathcal{W} \left[ H^{(1)},H^{(2)} \right] = i \frac{ c_{\rm s}^2 }{ 2 \epsilon ( a H )^2 },
\ee
where $\mathcal{W} \left[ H^{(1)}(x),H^{(2)}(x) \right] = i \dfrac{4}{\pi x}$ denotes the Wronskian of the Hankel function. Using the fact that $(y \sigma_3)' = c_{\rm s} k (1-\sigma_2) $, after some algebra we obtain the final solution for $\mathcal{R} = -H\pi$ as
\be 
\mathcal{R}_k(\tau) = \frac{ \sqrt{\pi} }{2\sqrt{2}} \frac{1}{a(\tau)} \frac{c_{\rm s}}{\sqrt{\epsilon}} \sqrt{\frac{-\sigma_3}{aH(1-\sigma_2)}} H^{(1)}_{\nu}(-y\sigma_3),
\ee
with $\nu$ as in \eqref{pi-sr-app}.
To first order in slow roll, $\sigma_2=0$ and $-\sigma_3 = 1+s+\epsilon$, while $-\frac{1}{aH} = \tau (1-\epsilon)$, so that the above expression reads 
\be 
\mathcal{R}_k(\tau) = \frac{ \sqrt{\pi} }{2\sqrt{2}} \frac{1}{a(\tau)} \sqrt{- \frac{ (1+s) \tau}{\epsilon c_{\rm s}^{-2}}} H^{(1)}_{\nu}(-c_{\rm s}k (1+s) \tau), \quad \nu = \frac{3}{2} + \epsilon + \frac{\eta}{2} + \frac{s}{2},
\ee
in agreement with the one found in \cite{Burrage:2011hd}. 						
\chapter{Integration of several massive fields} \label{app:several-massive-fields}
Here we justify the form of the interaction terms that appear in the generalised effective action \eqref{eft-modified}, by explicitly integrating out several heavy fields.
Let us write the simplest action coupling multiple heavy fields to $\delta g^{00} \equiv g^{00} + 1 $. We will consider an action quadratic in the heavy fields, but to all orders in $\delta g^{00}$. To lowest order in $\delta g^{00}$, we have
\be \label{heavy-1-app}
S = -\frac{1}{2}\int d^3xdt \sum_a  \Bigg\{     \mathcal F_a \left[  -\Box \mathcal   + M_a^2 -  B_a \delta g^{00}  \right]  \mathcal F_a +  2 A_a \delta g^{00} \mathcal F_a   + \sum_{b} C_{ab}  \mathcal F_a \dot{ \mathcal F_b}  \Bigg\} ,
\ee 
where $A_a$, $B_b$, $C_{ab}$ are background quantities and ${\cal F}_a$ are scalar fields of mass $M_a$. The matrix $C_{ab}$ is an anti-symmetric matrix, $\Box$ corresponds to the FLRW version of the D'Alambertian operator 
\be
\Box =  -\partial_t^2 - 3H\partial_t + \tilde\nabla^2 ,
\ee
and the couplings have mass dimensions $[A]=3, \ [B]=2, \ [C]=1$. Notice that we have excluded non diagonal mass terms, which may be eliminated by field redefinitions.

To proceed, we neglect the friction terms coming from the volume factor $a^3$ in $d^3xdt$, and focus on the general structure stemming from integrating out the massive fields $\mathcal F_a$. The more elaborate case in which the friction term is incorporated is completely analogous. The equations of motion are
\be 
 \left(  - \Box + M_a^2 - B_a  \delta g^{00} \right){\mathcal F_a} + \sum_b C_{a b} \dot {\mathcal F_b} = -A_a \delta g^{00} .
\ee
We are interested in the low energy behaviour of this system. Therefore, following the reasoning of Sec.~\ref{sec:integration-1-massive-field}, we disregard the time derivative $\partial_t^2 + 3 H \partial_t$ when compared to the operator $M_a^2 - \nabla^2$. On the contrary, we do not neglect the time derivative in the interaction term, as its role is to couple different massive fields and its contribution depends on the strength of $C_{ab}$. These considerations lead to the equation
\be 
\Omega_a {\mathcal F_a} + \sum_b C_{a b} \dot {\mathcal F_b} = -A_a \delta g^{00} , \label{Lagrange-heavy-fields}
\ee
where
\be \label{omegaexpansion}
\Omega_a \equiv  M_a^2 - \nabla^2  - B_a  \delta g^{00} .
\ee
Since in this limit the heavy fields $\mathcal F_a$ are non dynamical, we may treat them as Lagrange multipliers and insert them back into the action without kinetic terms. This leads to an effective action with the following contribution due to the heavy fields:
\be \label{int-action}
S = -\frac{1}{2}\int d^3xdt \sum_a     \delta g^{00}  A_a \mathcal F_a ,
\ee 
where the $\mathcal F_a  $ are the solutions of \eqref{Lagrange-heavy-fields}. To obtain $\mathcal F_a$, notice first that \eqref{Lagrange-heavy-fields} may be reexpressed as
\be
\left(\begin{array}{cccc} - \Omega_1 & - C_{12} \partial_t & - C_{13} \partial_t & \cdots \\  C_{12} \partial_t & - \Omega_2 & - C_{23} \partial_t & \cdots \\ C_{13} \partial_t &  C_{23} \partial_t & - \Omega_3   & \cdots \\ \vdots &  \vdots & \vdots &  \ddots  \end{array}\right) \left(\begin{array}{c} \mathcal F_1 \\ \mathcal F_2 \\ \mathcal F_3 \\  \vdots \end{array}\right) = \left(\begin{array}{c} A_1 \\ A_2 \\ A_3 \\  \vdots \end{array}\right) \delta g^{00} .
\ee
To deal with this equation, we assume that the off-diagonal terms are subleading when compared to the diagonal terms $\Omega_a$. This allows us to invert the matrix operator perturbatively, leading to the first order result:
\be \label{heavy-eom}
 \left(\begin{array}{c} \mathcal F_1 \\ \mathcal F_2 \\ \mathcal F_3 \\  \vdots \end{array}\right) = 
 \left(\begin{array}{cccc} -\Omega_1^{-1} &  \Omega_1^{-1} C_{12}  \partial_t  \Omega_2^{-1} &  \Omega_1^{-1} C_{13} \partial_t \Omega_3^{-1} & \cdots \\  -\Omega_2^{-1} C_{12} \partial_t  \Omega_1^{-1} &  -\Omega_2^{-1} &  \Omega_2^{-1} C_{23} \partial_t \Omega_3^{-1}  & \cdots \\ - \Omega_3^{-1} C_{13} \partial_t \Omega_1^{-1} & - \Omega_3^{-1} C_{23} \partial_t  \Omega_2^{-1} & - \Omega_3^{-1}   & \cdots \\ \vdots &  \vdots & \vdots &  \ddots  \end{array}\right)
 \left(\begin{array}{c} A_1 \\ A_2 \\ A_3 \\  \vdots \end{array}\right)  \delta g^{00} ,
\ee
which may be rewritten as
\be
\mathcal F_a = - \frac{A_a }{\Omega_a } \delta g^{00} + \sum_{b} C_{a b}  \frac{1}{\Omega_a} \partial_t   \frac{1}{\Omega_b} A_b \delta g^{00} .
\ee
We now plug this solution back into the action \eqref{int-action} to obtain
\be 
S = \frac{1}{2}\int d^3xdt   \left\{   \sum_a A_a  A_a   \delta g^{00} \frac{ 1}{\Omega_a } \delta g^{00} - \sum_{ab}A^a C_{a b}  \delta g^{00}  \frac{1}{\Omega_a} \partial_t   \frac{1}{\Omega_b} A^b \delta g^{00} \right\} .
\ee 
To simplify this expression notice that due to the antisymmetry of $C_{ab}$, the second term vanishes whenever the time derivative $\partial_t$ acts on a quantity that does not carry the label $b$. This means that the only non vanishing contributions coming from the second term are those proportional to $\dot A_b$, $\dot B_b$ and $\dot M^2_b$. For definiteness, and to keep our discussion simple, let us assume that both $B_b$ and $M_b^2$ are constants and consider only a time dependence of the $A_a$ coefficients. In this case we obtain the formal result
\be 
S = \frac{1}{2}\int d^3xdt   \left\{  \sum_a A_a  A_a \delta g^{00}  \frac{ 1 }{\Omega_a } \delta g^{00} - \sum_{a b} ( C_{a b} A_a \dot A_b ) \delta g^{00} \frac{1}{\Omega_a  \Omega_b}   \delta g^{00}  \right\} .
\label{action-for-integrated-heavy-fields}
\ee 
As discussed in Sec.~\ref{sec:integration-1-massive-field}, the inverse of $\Omega_a$ is an operator which has the following expansion
\be
\Omega_a^{-1} = \frac{1 }{M_a^2 -  \nabla^2} \left[1  -  \delta g^{00} \frac{ B_a }{M_a^2 -  \nabla^2} \right]^{-1}  =   \frac{1 }{M_a^2 -  \nabla^2} \sum_n \left[   \delta g^{00} \frac{ B_a }{M_a^2 -  \nabla^2} \right]^{n}.
\ee
Inserting this expansion back into the action (\ref{action-for-integrated-heavy-fields}) and keeping  terms up to cubic order, we finally arrive at the expression
\be
\begin{split}
S &= \frac{1}{2}\int d^3x dt   \Bigg\{  \delta g^{00} \left[  \sum_a  \frac{ A_a^2 }{M_a^2 - \nabla^2 }  - \sum_{a b}  \frac{C_{a b}A_a \dot A_b }{( M_a^2 - \nabla^2 )( M_b^2 - \nabla^2 )} \right] \delta g^{00}   \\  
& 
+ \sum_a  A_a^2 B_a  \delta g^{00}  \frac{ 1 } {M_a^2 - \nabla^2 }   \left[ \delta g^{00}  \frac{ 1 } {M_a^2 - \nabla^2 }  \delta g^{00} \right]    \\ 
& - \sum_{a b } C_{a b}A_a \dot A_b B_b \delta g^{00}  \frac{1}{M_a^2 - \nabla^2 }  \left[ \delta g^{00} \frac{1}{( M_b^2 - \nabla^2 )( M_c^2 - \nabla^2 )}   \delta g^{00} \right]  \\ 
& - \sum_{a b } C_{a b}A_a \dot A_b B_a  \delta g^{00} \frac{1}{( M_b^2 - \nabla^2 )( M_c^2 - \nabla^2 )}   \left[   \delta g^{00}  \frac{1}{M_a^2 - \tilde \nabla^2 }  \delta g^{00} \right]     
 + \cdots  \Bigg\} . 
\label{action-for-integrated-heavy-fields-2}
\end{split}
\ee
This implies that the general quadratic action for the Goldstone boson $\pi$ takes the form
\be
\begin{split}
S^{(2)} & = - M_{\rm Pl}^2 \int d^3 x dt a^3 \dot H \bigg[\dot \pi \bigg (1 +  \sum_a \frac{\beta_a}{ M_a^2 - \tilde \nabla^2}  +  \sum_{ab}\frac{ \beta_{ab} }{( M_a^2 - \tilde \nabla^2 )(  M_b^2  - \tilde \nabla^2) }   + \cdots \bigg )\dot \pi \\ &  - ( \tilde \nabla \pi)^2  \bigg],
\end{split}
\ee
where $\beta_{ab...}$ collectively denote combinations of the $A,B,C$ parameters of \eqref{heavy-1-app}. For instance, $\beta_a$ parametrises the coupling to a heavy field with index $a$, and $\beta_{ab}$ parametrises the interactions between heavy fields carrying labels $a$ and $b$ etc.
In momentum space, the action reads
\be
\begin{split}
S^{(2)} & = - M_{\rm Pl}^2 \int d^3 k dt a^3 \dot H \bigg[\dot \pi \bigg (1 +  \sum_a \frac{\beta_a}{ M_a^2 + p^2}  +  \sum_{ab}\frac{ \beta_{ab} }{( M_a^2 + p^2)(  M_b^2  + p^2) }   + \cdots \bigg )\dot \pi \\ &  + p^2 \pi^2 \bigg].
\end{split}
\ee
The equation of motion for the $\pi$ field is therefore given by
\be
\ddot \pi + 3 H \dot  \pi - c_{\rm s}^2 (p^2) p^2 \pi = 0 ,
\ee
where
\be \label{eee}
c_{\rm s}^2(p) = \dfrac{\prod_a (M_a^2 + p^2)}  {\prod_a (M_a^2 + p^2) \! + \! \sum _a \beta_a \prod_{b \neq a} (M_b^2 \!+\! p^2) \! + \! \sum_{a < b} \beta_{ab} \prod_{c \neq a, b}  (M_c^2 \! + \! p^2) \! + \! \ldots \! + \! \beta_{12\ldots N}}.
\ee
The inverse speed of sound squared is defined as the limit
\be \label{cs_in_multi_int_fields}
c_{\rm s}^{-2}\equiv\lim_{p\rightarrow 0}c_{\rm s}^{-2}(p)=1+\sum _a\frac{ \beta_a}{M_a^2} + \sum_{a < b} \frac{\beta_{ab}}{M_a^2M_b^2} + \ldots + \frac{\beta_{12\ldots N}}{M_1^2 M_2^2 \ldots M_N^2},
\ee
where $N$ is the number of heavy fields and the indices run from $1\ldots N$.
To analyse this, let us consider the short wavelength regime where the friction term can be disregarded.  The dispersion relation is then 
\be
\omega^2 (p) = c_{\rm s}^2(p) p^2.
\ee
For the case of one additional heavy field we get
\be
c_{\rm s}^2(p) = \frac{M^2 + p^2}{M^2 + p^2 + \beta},
\ee
which reduces to the expression \eqref{full-modified-dispersion} when $\beta = \dfrac{2M_2^4 M^2}{M_{\rm Pl}^2|\dot H|}$, in consistency with a speed of sound given by \eqref{sof}. 
For multiple non interacting fields where $\beta_{ab\ldots}=0$, \eqref{eee} becomes
\be
c_{\rm s}^2(p) = \prod_{a} (M_a^2 + p^2)\left[\prod_{a} (M_a^2 + p^2) + \sum _{a} \beta_a\prod_{b \neq a} (M_b^2 + p^2) \right]^{-1},
\ee
with the inverse speed of sound squared given by
\be \label{cs_in_multi_non_int_fields}
c_{\rm s}^{-2}=1+\sum _a\frac{ \beta_a}{M_a^2}.
\ee
%
Recall that we are restricted to the low energy regime $$ \omega^2 \ll M_a^2 + p^2 $$ in order for the expansion (\ref{omegaexpansion}) to be valid. Without loss of generality we can consider two cases: one where the $M_a$ are all comparable, and the other where there exists some hierarchy among these heavy masses. Both can be studied using a representative mass $M_l^2$ which is either equal to or lower than any other. In the former case we require the inequality to hold for all $a$, while in the latter we require  $ \omega^2 \ll M_l^2 + p^2 $. 
The generic UV scale for arbitrary number of fields with different masses will be a complicated function of the speed of sound and the mass scales of the problem. We will thus only study in some detail the case where all the heavy masses $M_a$ are comparable: $M_a^2 \approx M^2 \, \, \, \forall\, \,  a$. The dispersion relation then reads
\be \label{dispersion_multi_same_mass}
\omega^2(p) = \frac{ (M^2 + p^2) p^2 }{ M^2 + p^2 + \sum _a \beta_a + (M^2 + p^2)^{-1}\sum_{a < b} \beta_{ab}  + \ldots + \beta_{12\ldots N}(M^2 + p^2)^{1-N} }.
\ee
From this expression we can read off the low energy regime as an upper bound in the momentum
\be
\label{LER}
p^2 \ll M^2 + \sum _a \beta_a + (M^2 + p^2)^{-1}\sum_{a < b} \beta_{ab}  + \ldots + \beta_{12\ldots N}(M^2 + p^2)^{1-N}.
\ee
We see that in general this is a polynomial inequality of degree $N$ in squared momentum
$$G^N(p^2)\ll 0 \ .$$ Therefore the solution is $p^2\ll p^2_{\rm UV}(M,c_{\rm s},\beta)$ with $p^2_{\rm UV}$ representing the degenerate positive root of the polynomial $G^N$.
The energy scale $\Lambda_{\rm UV}$ is then given by substituting $p^2_{\rm UV}$ into the dispersion relation. Since this is the root of the polynomial $G^N(p^2)$ the denominator of Eq.~\eqref{dispersion_multi_same_mass} is just proportional to $p^2_{\rm UV}$ and the expression simplifies to
\be \label{luv_multi}
\Lambda_{\rm UV}^2 \sim M^2 + p^2_{\rm UV} \ .
\ee
We also see a modification of the dispersion relation in the multiple heavy field case. For small values of $p^2$ compared to the mass squared, the low energy regime condition (\ref{LER}) becomes 
\bea
p^2 & \ll &  M^2 +  \sum_a \beta_a + \sum_{a\neq b} \beta_{ab} M^{-2} + .... + \beta_{12....N} M^{2(1-N)}. 
\eea
This inequality is automatically satisfied when $p^2 \ll M^2$. The dispersion relation in this regime becomes
\be
\label{MFdisprelnlowmom} \omega^2(p) = c_{\rm s}^2 p^2  \left( 1 + \frac{p^2}{M^2} \right)^N\, \,,
\ee
where $c_{\rm s}^2$ is given in \eqref{cs_in_multi_int_fields}. 
The expansion \eqref{MFdisprelnlowmom} includes terms depending on $p^4, p^6...$, but these are suppressed by increasing powers of $p^2/M^2$, so that we recover the usual $\omega^2 \sim p^2$ dispersion relation. 
For large values of $p^2$ compared to $M^2$, the low energy condition becomes
\bea
\label{LEClargep} p^{2N} & \ll & \sum_a \beta _a p^{2(N-1)} + \sum_{a \neq b} \beta_{ab} p^{2(N-2)} + .... + \beta_{123...N}.
\eea
The dispersion relation in this regime is given by 
\be
\omega^2(p) = \frac{1}{ \sum \beta_a p^{-4} + \sum \beta_{ab} p^{-6} + ... + \beta_{1...N} p^{-2(N+1)} }.
\ee
We see that many powers of $p$ can enter. However, for large $p^2 \gg M^2$, the higher terms in the denominator are suppressed, and the dominant behaviour of the dispersion relation is 
\be
\omega^2(p) \approx \frac{p^4}{\sum \beta_a}.
\ee
%

\baselineskip=12pt plus1pt

\bibliographystyle{jhep}
\renewcommand{\bibname}{References} 
\cleardoublepage
\phantomsection
\addcontentsline{toc}{chapter}{References} 
\bibliography{biblio} 				

\end{document}